\documentclass[amsmath,amssymb,amsfonts,twocolumn,english,showpacs,superscriptaddress,prb]{revtex4-1}
\pdfoutput=1

\usepackage[T1]{fontenc}
\usepackage[latin9]{inputenc}
\usepackage{times}
\usepackage{graphicx}
\usepackage{color}
\usepackage{mathrsfs}
\usepackage{float}
\usepackage{indentfirst}
\usepackage{babel}
\usepackage{wasysym}
\usepackage{amsthm}
\usepackage{array}
\usepackage{verbatim}
\usepackage{tabularx}
\usepackage{tabulary}
\usepackage{mwe}

\usepackage{booktabs}
\usepackage{array, multirow}
\usepackage{subfigure}

\usepackage[colorlinks,bookmarks=false,citecolor=blue,linkcolor=blue,urlcolor=blue]{hyperref}
\usepackage[sort&compress]{natbib}
\setcitestyle{numbers,square}

\newcommand{\ket} [1] {\vert #1 \rangle}
\newcommand{\bra} [1] {\langle #1 \vert}
\newcommand{\braket}[1]{\langle #1 \rangle}

\theoremstyle{definition}
\newtheorem{property}{Property}

\makeatother
\graphicspath{{.}}

\begin{document}

\title{Commuting-projector Hamiltonians for chiral topological phases built from parafermions}

\author{Jun Ho Son}
\affiliation{Stanford Institute for Theoretical Physics, Stanford University, Stanford, CA 94305, USA}

\author{Jason Alicea}
\affiliation{Department of Physics and Institute for Quantum Information and Matter, California Institute of Technology, Pasadena, CA 91125, USA}
\affiliation{Walter Burke Institute for Theoretical Physics, California Institute of Technology, Pasadena, CA 91125, USA}

\begin{abstract}

We introduce a family of commuting-projector Hamiltonians whose degrees of freedom involve $\mathbb{Z}_{3}$ parafermion zero modes residing in a parent
fractional-quantum-Hall fluid. The two simplest models in this family emerge from dressing Ising-paramagnet and toric-code spin models with parafermions; we study their edge properties, anyonic excitations, and ground-state degeneracy.  We show that the first model realizes a symmetry-enriched topological phase (SET) for which $\mathbb{Z}_2$ spin-flip symmetry from the Ising paramagnet permutes the anyons.  Interestingly, the interface between this SET and the parent quantum-Hall phase realizes symmetry-enforced $\mathbb{Z}_3$ parafermion criticality with no fine-tuning required.  The second model exhibits a non-Abelian phase that is consistent with $\text{SU}(2)_{4}$ topological order, and can be accessed by gauging the $\mathbb{Z}_{2}$ symmetry in the SET.  Employing Levin-Wen string-net models with $\mathbb{Z}_{2}$-graded structure, we generalize this picture to construct a large class of commuting-projector models for $\mathbb{Z}_{2}$ SETs and non-Abelian topological orders exhibiting the same relation. Our construction provides the first commuting-projector-Hamiltonian realization of \textit{chiral} bosonic non-Abelian topological order. 

\end{abstract}

\maketitle

\section{Introduction} \label{sec:Intro}

Historically, exactly solvable models played an important role in understanding topological phases of matter (see, e.g., Refs.~\onlinecite{Kitaev2003, Kitaev2006, Levin2006}).  Such models typically sacrifice microscopic realism in favor of analytical tractability that facilitates extraction of topological information, including anyon content and entanglement characterizations of ground states.  Furthermore, many exactly solvable models describe renormalization-group fixed points with zero correlation length.  Studying their physical properties can thus reveal useful insights into more realistic systems that flow to the same fixed point.

Commuting-projector Hamiltonians comprise a widely studied class of exactly solvable models. As the name suggests, these Hamiltonians consist of sums of projectors that commute with each other, so that \emph{all} energy eigenstates are simultaneous eigenstates of each projector. Classic examples include Kitaev's quantum-double models \cite{Kitaev2003} and Levin-Wen string-net models \cite{Levin2006}, which capture a wide variety of non-chiral topologically ordered phases (i.e., with chiral central charge $c=0$).  More recent works have introduced commuting-projector Hamiltonians for symmetry-protected topological phases (SPT's) obtained by dressing domain walls with lower-dimensional SPT's \cite{Chen2014}, and for symmetry-enriched topological phases from string-net models \cite{Heinrich2016, Cheng2016}.
   
In all of the above commuting-projector Hamiltonians, bosons form the microscopic constituents.  
Recently, novel commuting-projector Hamiltonians for topological phases of fermions  have been developed \cite{Gu2014, Gu2015, Tarantino2016, Ware2016, Bhardwaj2017, Wang2017,Aasen2017}.  These models realize topologically ordered states and SPT's that are intrinsically fermionic, i.e., they display properties that cannot be emulated in any known bosonic systems.  In this paper we go a step further and construct commuting-projector models built from \emph{fractionalized} degrees of freedom that bind to defects in a topologically ordered host system.  Searching for exactly solvable models for ``topological phases inside topological phases'' using such defects represents largely uncharted territory.  (For some related works see Refs.~\onlinecite{Bais2009, Ludwig2011, Poilblanc2011, Burrello2013, Mong2014, Stoudenmire2015, Barkeshli2015}.)  Notably, this strategy can be expected to circumvent constraints faced by Hamiltonians describing non-fractionalized constituents, paving the way to a richer class of exactly solvable models.  We will indeed show that our models capture \emph{chiral} topological orders that would be impossible to obtain from either bosonic or fermionic commuting-projector Hamiltonians.
 
Our work specifically generalizes the constructions of Refs.~\onlinecite{Tarantino2016} and \onlinecite{Ware2016}.  In Ref.~\onlinecite{Tarantino2016} Tarantino and Fidkowski developed commuting-projector models---obtained by decorating domain walls of an Ising paramagnet with Kitaev chains \cite{Kitaev2001}---for the fermionic SPT's considered by Gu and Levin \cite{Gu2014a}.  We henceforth refer to their Hamiltonians and our generalization as decorated-domain-wall models.  In Ref.~\onlinecite{Ware2016} Ware et al.~introduced a commuting-projector model for a fermionic cousin of Ising topological order with a fully gapped edge.  This result is surprising given that Ising topological order in a bosonic system necessarily carries a gapless edge and nontrivial thermal Hall conductance; conventional wisdom thus dictates that a parent commuting-projector Hamiltonian does not exist.  For fermionic systems, however, it turns out that the $c=1/2$ chiral edge state of bosonic Ising topological order can be gapped out by adding a $c=-1/2$ $p-ip$ superconductor with suitable interactions \cite{Walker_unpub}. 
A commuting-projector description is then possible, which can be understood as arising from the toric code dressed with Kitaev chains.  We thus refer to the latter Hamiltonian and its generalization as decorated-toric-code models.  

Both the decorated-domain-wall and decorated-toric-code models from Refs.~\onlinecite{Tarantino2016,Ware2016} couple spins to a two-dimensional array of Majorana fermions, which famously appear at defects---e.g., domain walls or vortices---in topological superconductors \cite{Kitaev2001,Read2000}.  Our generalizations essentially promote the Majorana fermions in these models to more exotic cousins known as `$\mathbb{Z}_3$ parafermions' \cite{FradkinKadanoff,Fendley2012,AliceaFendleyReview}.  Importantly, the latter can also arise at defects, but only (to the best of our knowledge) in a fractionalized medium.  Possible host platforms include quantum-Hall bilayers \cite{Barkeshli2012, Barkeshli2013, Barkeshli2013a,Barkeshli:2014b}, quantum-Hall/superconductor hybrids \cite{Lindner2012, Clarke2013, Cheng2012,Vaezi:2013,Mong2014}, and cold atoms \cite{Maghrebi}.   For concreteness, we will focus throughout on $\mathbb{Z}_3$ parafermions realized at defects in a bosonic (221) fractional quantum Hall fluid.  The (221) state is similar to a $\nu = \frac{1}{3}$ Laughlin state in the sense that there are three anyon charges, and the anyons obey $\mathbb{Z}_{3}$ fusion rules. We choose (221) over fermionic quantum Hall platforms---which can also host $\mathbb{Z}_3$ parafermions---to sidestep subtleties arising when dealing with fermionic topological orders.
 
\begin{figure}
	\includegraphics[width=0.8\linewidth]{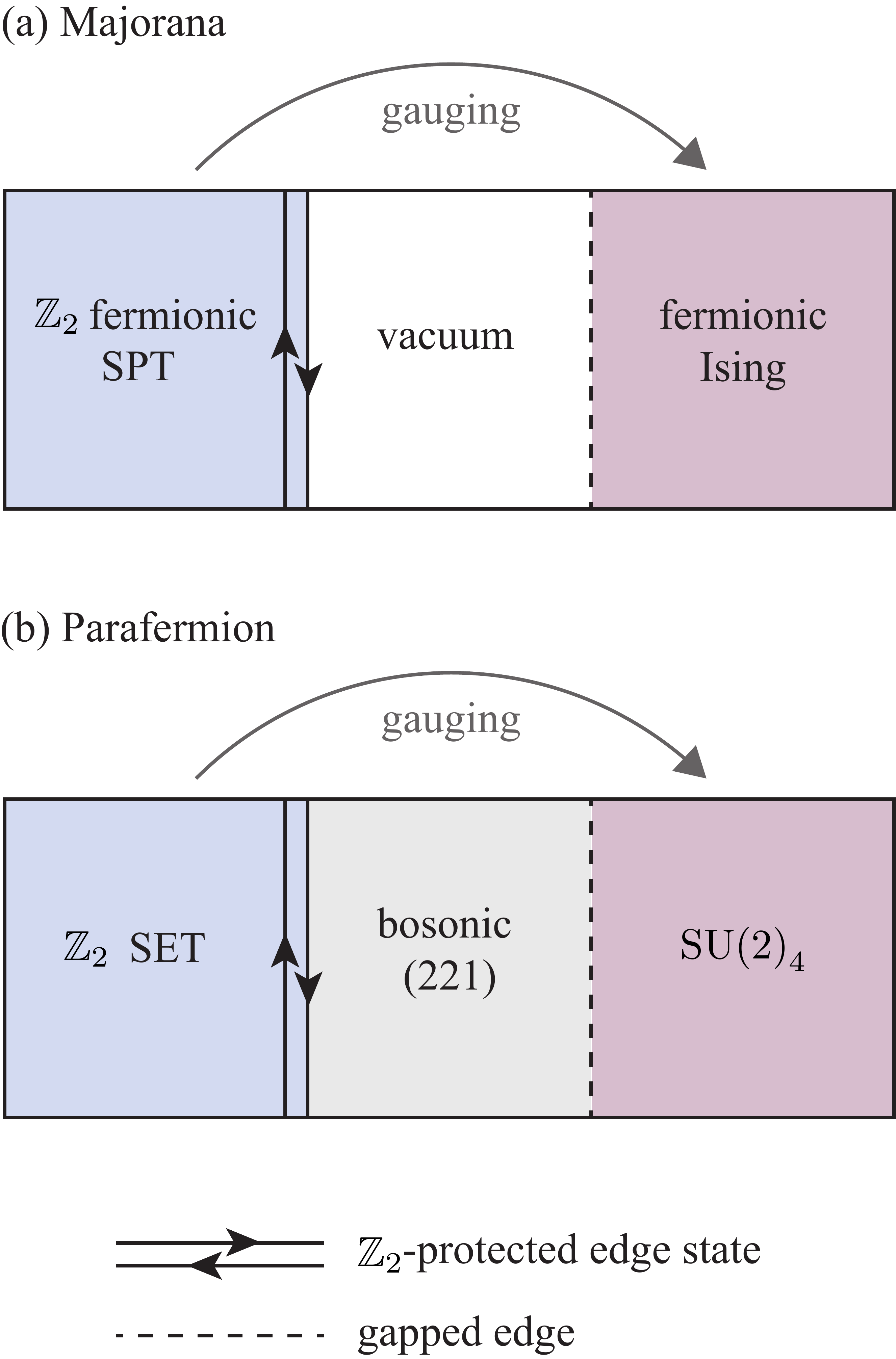}
	\caption{Overview of the phases in the (a) Majorana \cite{Tarantino2016,Ware2016} and (b) $\mathbb{Z}_3$ parafermionic versions of the decorated-domain-wall and decorated-toric-code models.  The decorated-domain-wall construction gives the SPT and SET on the left side, while the decorated-toric-code construction gives topologically ordered states on the right.}
\label{fig:summary}
\end{figure}

Before delving into the detailed construction and analysis, let us outline rough guesses for the topological phases that emerge from our parafermion models. The Majorana constructions from Refs.~\onlinecite{Tarantino2016} and \onlinecite{Ware2016} exhibit the following properties:
\begin{enumerate}
\item The SPT of the decorated-domain-wall model is protected by an on-site $\mathbb{Z}_2$ spin-flip symmetry; the boundary with vacuum hosts a gapless $\mathbb{Z}_2$-protected edge state.  Upon breaking $\mathbb{Z}_2$, this phase becomes adiabatically connected to a trivial state.   
\item In the decorated-toric-code model, fermionic Ising topological order admits a gapped boundary with vacuum.
\item Gauging the on-site $\mathbb{Z}_2$ symmetry in the decorated-domain-wall model yields the same fermionic Ising topological order as the decorated-toric-code model.
\end{enumerate}
Figure~\ref{fig:summary}(a) summarizes these relations.   

What are the $\mathbb{Z}_3$-parafermion analogues of these properties?  Since our parafermions require a fractionalized host system, it is essential that `vacuum' in properties 1 and 2 instead becomes the bosonic (221) state.  Furthermore, the SPT from the decorated-domain-wall model should be elevated to a symmetry enriched topological phase (SET), again reflecting the background quantum Hall fluid.  That SET naturally inherits the topological order from the parent (221) state, with the on-site $\mathbb{Z}_2$ spin-flip symmetry acting nontrivially on the anyons.  In analogy with the Majorana case, the parafermionic decorated-toric-code model ought to yield a \emph{non-Abelian} topological order that, crucially, can exhibit a gapped boundary with the parent (221) state.   The simplest guess for such a non-Abelian state corresponds to SU(2)$_4$ topological order.  Certain anyons in this theory are known to be closely related to $\mathbb{Z}_3$ parafermions \cite{Fern}.  Intriguingly, it is also known that condensing a boson in SU(2)$_4$ produces the same topological order exhibited by the (221) state; thus, there is indeed no need for a gapless interface between these two topological phases \cite{Bais2009}.  As another sanity check, $G$-crossed category formalism \cite{Barkeshli2014, Teo2015, Tarantino2016b} indicates that gauging the $\mathbb{Z}_2$ symmetry in the proposed SET yields SU(2)$_4$ topological order---consistent with a straightforward generalization of property 3 from the Majorana constructions.  

Summarizing, we expect the following characteristics from our parafermion models:
\begin{enumerate}
\item The parafermion decorated-domain-wall model yields an SET protected by an on-site $\mathbb{Z}_2$ spin-flip symmetry; the boundary with the bosonic (221) state hosts a gapless $\mathbb{Z}_2$-protected edge state.  Upon breaking $\mathbb{Z}_2$, the SET becomes adiabatically connected to the (221) state.
\item The parafermion decorated-toric-code model yields SU(2)$_4$ topological order that admits a gapped boundary with the bosonic (221) state.  
\item Gauging the on-site $\mathbb{Z}_2$ symmetry in the parafermion decorated-domain-wall model yields the same SU(2)$_4$ topological order as the decorated-toric-code model.
\end{enumerate}
See Fig.~\ref{fig:summary}(b) for an illustration.  In the following sections we will confirm the properties anticipated above by explicitly constructing and analyzing commuting-projector parafermion Hamiltonians.  For the SET, we explicitly show that $\mathbb{Z}_2$ spin-flip symmetry interchanges the two nontrivial anyons from the parent (221) state, and that the minimal gapless interface with the `undecorated' parent quantum-Hall fluid is described by a non-chiral $\mathbb{Z}_3$ parafermion conformal field theory.  Remarkably, no fine-tuning is required to access this critical theory: $\mathbb{Z}_2$ spin-flip symmetry acts as a duality for the boundary degrees of freedom, enforcing criticality by default.  For the decorated-toric-code model, we uncover an intuitive physical picture for all of the nontrivial particles in SU(2)$_4$ in terms of hybrids of toric-code and (221) anyons, thus providing useful insight into the structure of this exotic non-Abelian topological order.  We further construct parafermion-decorated string-net models with $\mathbb{Z}_2$-graded structure to obtain commuting-projector Hamiltonians for other topological orders and SETs.

The remainder of the paper is organized as follows.  Section~\ref{sec:Review} briefly reviews $\mathbb{Z}_3$ parafermions in bosonic $(221)$ quantum Hall states, establishing formalism necessary for our subsequent analysis. 
Section~\ref{sec:Setup} defines our commuting-projector Hamiltonians, while Sec.~\ref{sec:Prop} analyzes their physical properties. In Sec.~\ref{sec:Torus}, we show that the decorated-domain-wall and decorated-toric-code models yield the ground-state degeneracy on the torus expected from the respective topological phases hypothesized above. 
In Sec.~\ref{sec:LW}, we briefly discuss generalizations to parafermion-decorated string-net models.  Concluding remarks appear in Sec.~\ref{sec:Conc}.

\section{Overview of $\mathbb{Z}_3$ Parafermions} 
\label{sec:Review}

We start by reviewing formalism of $\mathbb{Z}_3$ parafermions that will be employed extensively throughout this paper.  To streamline the presentation we follow a largely phenomenological treatment; for a more detailed bosonization analysis of a very similar setup see Ref.~\onlinecite{Mong2014}.  Along the way we also illustrate how to properly treat parafermions residing in a host system defined on a torus, which will be essential in later sections.  

\subsection{Review}

\begin{figure}
  \includegraphics[width=0.8\linewidth]{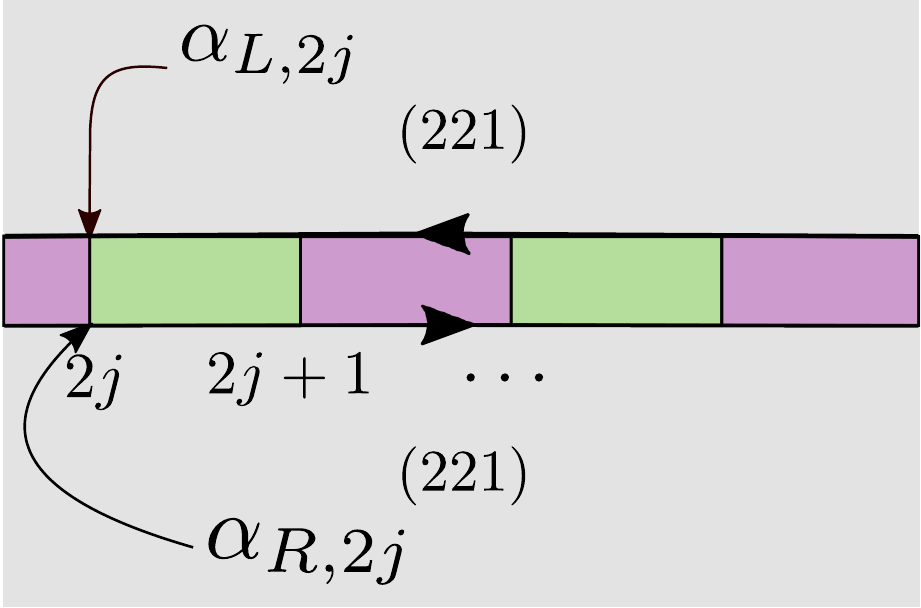} 
  \caption{Counterpropagating edge states residing at opposite ends of a trench in a bosonic (221) quantum Hall state. Sequentially gapping the edge states via boson tunneling (green) and Cooper pairing (purple) yields an array of $\mathbb{Z}_3$ parafermion zero modes localized at each domain wall. }
  \label{fig:fqhtopf}
\end{figure}

Consider a (221) fractional quantum Hall state formed by charge-$e$ bosons.  The (221) state is purely chiral with central charge $c = 2$.  Cutting a `trench' through the quantum-Hall fluid as in Fig.~\ref{fig:fqhtopf} thus generates two left-moving edge states at the upper side of the trench and two right-moving edge states at the lower end.  These counterpropagating edge states can acquire a gap via $(i)$ ordinary boson tunneling across the trench and $(ii)$ condensing charge-$2e$ `Cooper pairs' assembled from bosons residing at opposite ends of the trench.  Domain walls separating regions gapped by these competing mechanisms realize non-Abelian defects with quantum dimension $\sqrt{3}$.  

Next consider a one-dimensional domain-wall array as sketched in Fig.~\ref{fig:fqhtopf}.  The topological degeneracy associated with these non-Abelian defects can be understood as follows.  In a pairing-gapped domain labeled by $j$, the total charge $Q^+_j$ for the edge modes fluctuates wildly due to the Cooper-pair condensation.  The quantity $e^{i \pi Q^+_j}$, however, locks to one of three possible values,
\begin{equation}
  1, \omega = e^{i 2\pi/3},~\text{or}~\bar \omega = e^{-i 2\pi/3},
\end{equation}
each of which yields the same energy.  In other words, the pairing-gapped domain can absorb fractional charge $2e/3$ without energy penalty.  Similarly, in a tunneling-gapped domain the charge difference $Q^-_j$ between the left- and right-moving edge modes fluctuates wildly, though $e^{i \pi Q^-_j}$ pins to $1,\omega$, or $\bar\omega$.  These regions can thus absorb $e/3$ dipoles without energy cost.  The set of $e^{i \pi Q^+_j}$ and $e^{i \pi Q^-_j}$ operators do not commute with each other, which is ultimately a consequence of fractional statistics exhibited by the host (221) system.  To capture their commutation relations it is convenient to write
\begin{equation}
  e^{i \pi Q^+_j} = \tau_j,~~~~e^{i \pi Q^-_j} = \sigma_j^\dagger \sigma_{j+1},
  \label{clock_rep}
\end{equation}
where $\sigma_j, \tau_j$ are unitary $\mathbb{Z}_3$ clock operators that satisfy $\sigma_j^3 = \tau_j^3 = 1$ and $\sigma_j \tau_j = \omega \tau_j \sigma_j$.  One can thus label ground states by either $e^{i \pi Q^+_j}$ or $e^{i \pi Q^-_j}$ eigenvalues, but not both simultaneously.  

$\mathbb{Z}_3$ parafermion operators cycle the system through the degenerate manifold by adding fractional charge to the domain-wall defects, thereby incrementing $e^{i \pi Q^\pm}$ for the adjacent domains.  Since fractional charge can not directly pass across the trench, parafermions come in two `flavors' that we denote by $\alpha_{L}$ and $\alpha_{R}$.  Specifically, $\alpha_{L}$ adds charge $2e/3$ to the \emph{upper} side of a domain wall, cycling the adjacent $e^{i \pi Q^+}$
and $e^{i \pi Q^-}$ operators by $\omega$, while $\alpha_{R}$ adds charge $2e/3$ to the \emph{lower} end, cycling the adjacent $e^{i \pi Q^+}$ by $\omega$ but $e^{i \pi Q^-}$ by $\bar \omega$ (see Fig.~\ref{fig:fqhtopf}).  In terms of clock operators, we explicitly have
\begin{equation}
\label{eq:pftoclock}
\begin{split}
\alpha_{R, 2j-1} = \sigma_{j} \prod_{k < j} \tau_{k}, ~~~~~\alpha_{R, 2j} = \sigma_{j} \prod_{k \leq j} \tau_{k} \\ 
\alpha_{L, 2j-1} = \sigma_{j} \prod_{k < j} \tau_{k}^{\dagger}, ~~~~~\alpha_{L, 2j} = \sigma_{j} \prod_{k \leq j} \tau_{k}^{\dagger},
\end{split}
\end{equation}
which imply the hallmark $\mathbb{Z}_3$-parafermion relations
\begin{equation}
\begin{split}
&\alpha_{L/R,j}^3 = 1, ~~~~~\alpha_{L/R,j}^{\dagger} = \alpha_{L/R,j}^{2} \\
&\alpha_{R,j}\alpha_{R,j'} = e^{i\frac{2\pi}{3}\text{sgn}(j'-j)}\alpha_{R,j'}\alpha_{R,j} \\
&\alpha_{L,j}\alpha_{L,j'} = e^{-i\frac{2\pi}{3}\text{sgn}(j'-j)}\alpha_{L,j'}\alpha_{L,j}.
\end{split}
\label{eq:pfdef}
\end{equation}
One can readily verify using Eqs.~\eqref{clock_rep} and \eqref{eq:pftoclock} that $\alpha_{L/R}$ indeed cycle ground states as outlined above. 

While the zero modes can absorb fractional charges without energy cost, creating fractionally charged quasiparticles in the bulk of the (221) host system costs energy. The ground state manifold must therefore satisfy 
\begin{equation}
  e^{i\pi \sum_{j}Q_{j}^{+}} = \prod_{j}\tau_{j} =1.
  \label{Qplus_constraint}
\end{equation}  
(By contrast, $e^{i\pi \sum_{j}Q_{j}^{-}}$ can vary if it is possible for charges to redistribute between the upper and lower sides of the trench; see Sec.~\ref{sec:torus}.)  Equation~\eqref{Qplus_constraint} is fruitfully viewed as a constraint on the system's global `triality'---the generalization of global fermion parity.  We will strictly enforce such constraints throughout this paper, even when we incorporate parafermion interactions.  This assumption is justified provided the scale for parafermion interactions is small compared to the bulk quasiparticle gap for the host quantum-Hall fluid.  

\begin{figure}
  \includegraphics[width=1\linewidth]{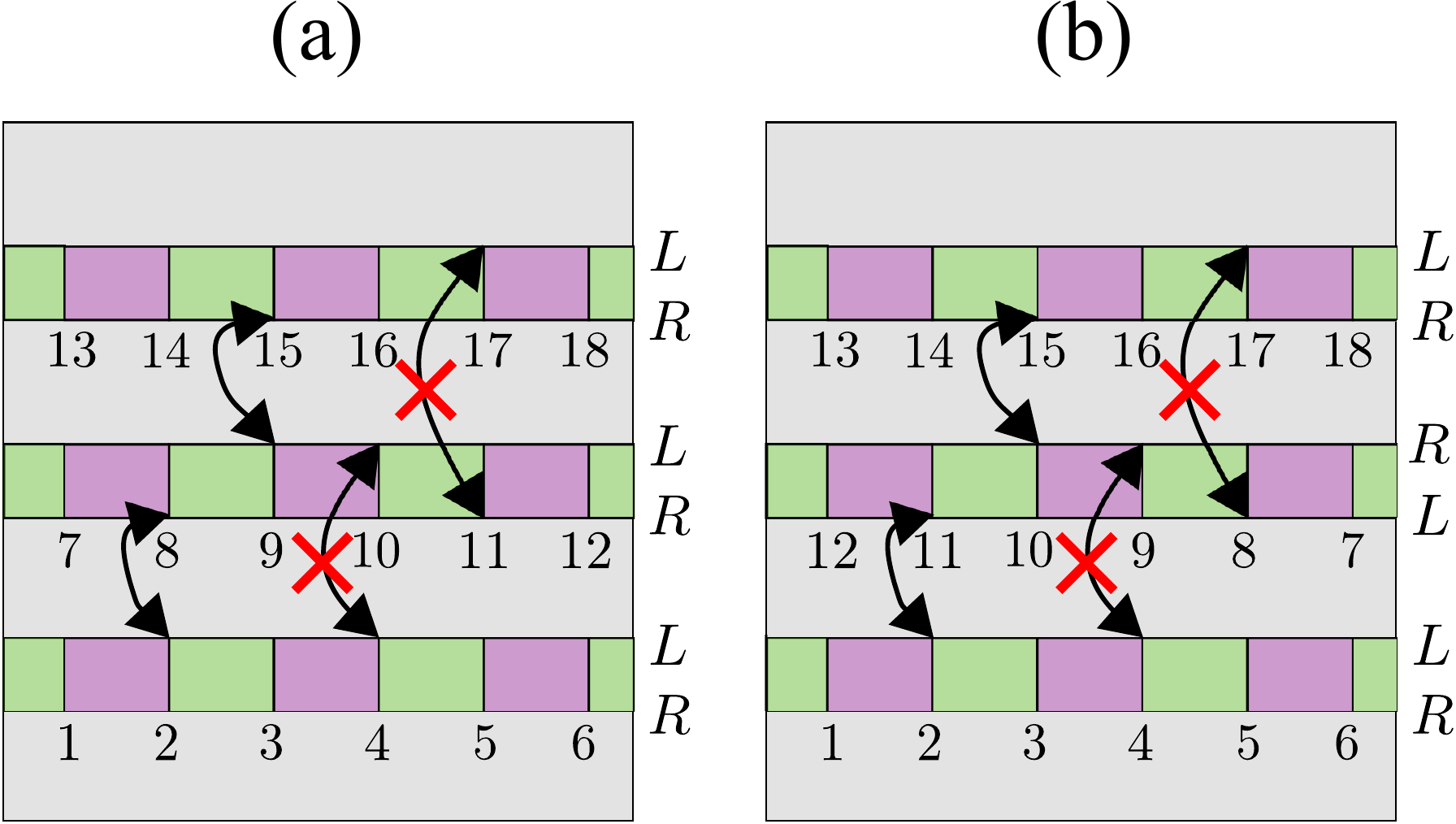} 
  
  \caption{Two-dimensional parafermion arrays as stacked trenches. In (a) we view the right part of trench $n$ as stitched to the left part of trench $n+1$---hence the `typewriter-style' labeling of domain walls.  In (b) we instead stitch the right end of trench $2n-1$ to the right end of trench $2n$, and the left end of trench $2n$ to the left end of trench $2n+1$, yielding a `snake-like' domain-wall labeling.  Either convention allows us to maximally bootstrap off of single-chain formalism to describe the two-dimensional array.  Examples of physical and unphysical inter-trench parafermion couplings are also shown.  Note that allowed inter-trench couplings necessarily hybridize $\alpha_L$ with $\alpha_R$ in (a), but hybridize either two $\alpha_L$'s or two $\alpha_R$'s in (b).    }
  \label{fig:2Dpf}
\end{figure}

We now make the leap to two-dimensional parafermion arrays, which one can of course view simply as stacks of 1D trenches.  For theoretical convenience we will additionally stitch together the ends of neighboring trenches, thereby joining them into a single long chain.  This can be done, for example, by stitching the right end of trench $n$ with the left end of trench $n+1$ as in Fig.~\ref{fig:2Dpf}(a).  Alternatively, one can stitch the right end of trench $2n-1$ with the right end of trench $2n$ and the left end of trench $2n$ with the left end of trench $2n+1$ as in in Fig.~\ref{fig:2Dpf}(b).  Either scheme allows us to directly import the commutation relations specified in Eq.~\eqref{eq:pfdef} to the two-dimensional array; moreover, the global triality constraint from Eq.~\eqref{Qplus_constraint} continues to apply without modification.  Different stitching procedures do give rise to different parafermion orderings, however, as is clear from Fig.~\ref{fig:2Dpf}.  Since there is no canonical choice of which pairs of ends must be sewed, parafermion models should be defined consistently for any choice of stitching scheme, or equivalently, parafermion ordering. 

It is useful to envision interactions among parafermions in the array as arising from dynamical processes that shuttle fractional charges from one domain wall to another.  As an important example, 
\begin{equation}
  e^{i \pi Q^+_j} = \tau_j = \alpha_{R,2j-1}^{\dagger}\alpha_{R,2j} = \alpha_{L,2j}^{\dagger}\alpha_{L,2j-1}
  \label{Qpj}
\end{equation}
describes migration of fractional charge across a single Cooper-paired domain in a given trench.  Parafermion couplings arising from all \emph{other} physical processes will be denoted by
\begin{equation}
  F_{ij} = \alpha_{C_{i},i}\alpha_{C_{j},j}^{\dagger},
  \label{Fij}
\end{equation} 
where $i, j$ are site indices and $C_{i}, C_{j}$ are $R$ and $L$ labels specified below.  Clearly Eq.~\eqref{Qpj} could also be described in terms of $F_{ij}$ operators.  However, separating out $e^{i \pi Q^+_j}$ as we have done clarifies the necessity of introducing a generalized Kasteleyn orientation in our models later on.  

Several comments are in order.   First, for intra-trench couplings, processes whereby fractional charge moves along the upper versus lower end of the trench are not independent.  Equation~\eqref{Qpj} provides one illustration; another follows from $F_{2j, 2j+1} = e^{i \pi Q^-_j} = \alpha_{R,2j}\alpha_{R,2j+1}^{\dagger} = \alpha_{L,2j}\alpha_{L,2j+1}^{\dagger}$.  Second, inter-trench parafermion couplings are highly constrained.  Couplings between parafermions on nearest-neighbor trenches can arise from the transfer of fractional charge between adjacent trenches via the intervening quantum-Hall fluid.  Interactions that couple parafermions on further-neighbor trenches are disallowed since fractional charge can not pass through the trenches.  Third, obtaining physical nearest-neighbor inter-trench couplings requires an appropriate choice for $C_{i,j}$ in Eq.~\eqref{Fij}, again to avoid fractional charge from illegally crossing a trench.  For concreteness let us assume that site $i$ resides on the trench just below that of site $j$.  The conventions in Fig.~\ref{fig:2Dpf}(a) then require $C_i = L$ and $C_j = R$, while the conventions in Fig.~\ref{fig:2Dpf}(b) yield either $C_i = C_j = L$ or $C_i = C_j = R$ depending on which pair of trenches couple.  Figure~\ref{fig:2Dpf} illustrates examples of physical and unphysical processes in both schemes.  Fourth, when the quantum-Hall state is defined on a torus, certain inter- and intra-trench parafermion couplings must be supplemented by additional phase factors and operators related to global properties of the system.  Details appear in the next subsection.  Finally, one can explicitly show that two inter-trench parafermion couplings that describe non-intersecting hopping processes commute with each other.  We refer to Appendix~\ref{app:proofs} for the proof.  This commutation is the key property that enables us to define commuting-projector parafermion Hamiltonians later on.

\subsection{Torus formalism}
\label{sec:torus}

\begin{figure}
	\includegraphics[width=\linewidth]{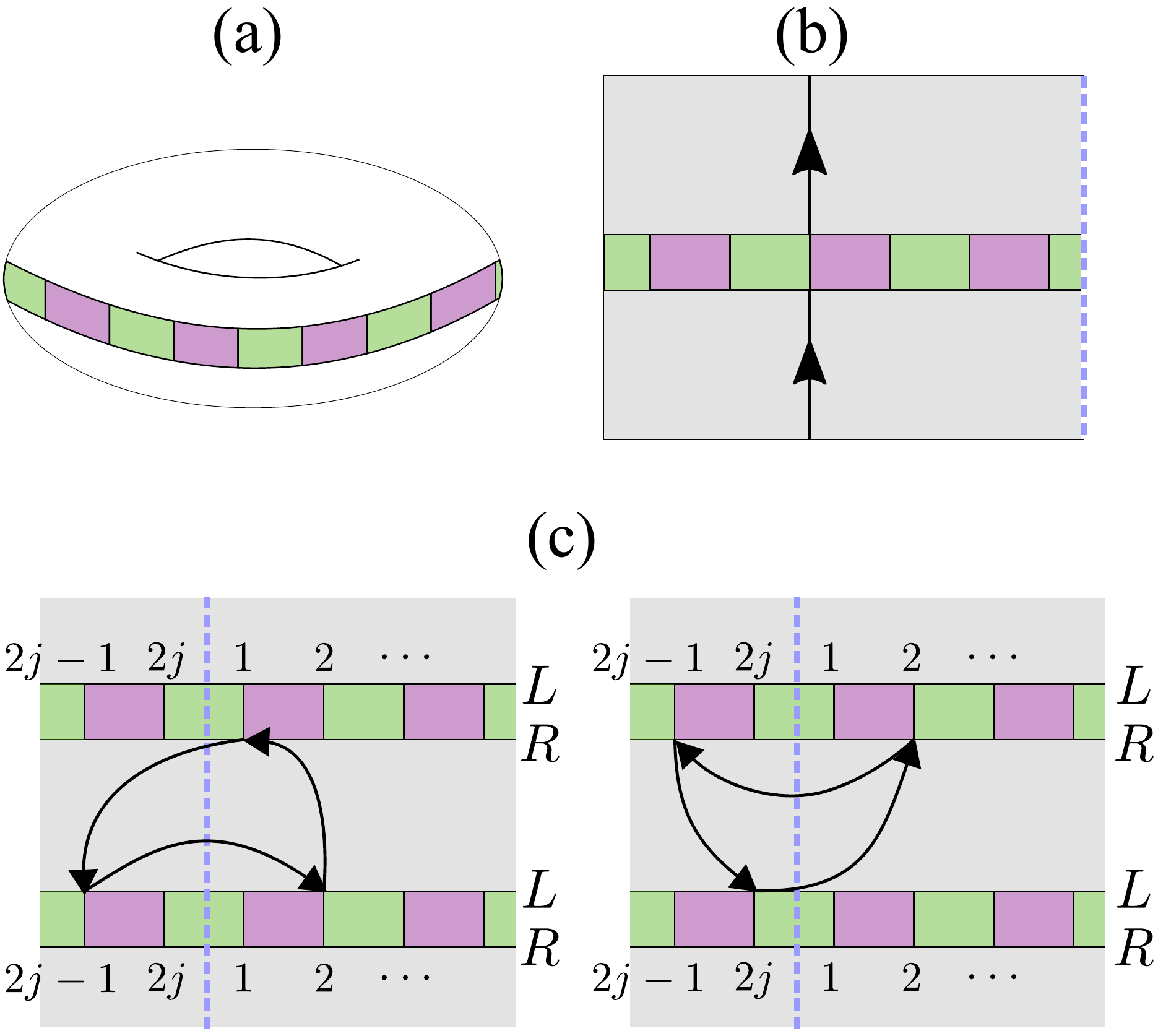}
	\caption{(a) Linear parafermion array wrapped around a cycle of a torus.  (b) Moving charge $2e/3$ along the vertical path shown changes the global charge difference across the trench [Eq.~\eqref{Qminus_constraint}].  The dashed line on the right side illustrates a $T_y$ branch cut that is useful for properly keeping track of parafermion couplings on the torus.  (c) Examples of fractional-charge hopping processes that form a contractible triangle and thus should satisfy Eq.~\eqref{eq:triid}.}   \label{fig:pftorus2_rev}
\end{figure}

So far we have neglected boundary conditions entirely---burying subtleties that we now wish to exhume.  Imagine that the (221) state is defined on a torus, with a linear domain-wall array wrapping along a nontrivial cycle as in Fig.~\ref{fig:pftorus2_rev}(a).  (We return to two-dimensional arrays shortly.)  If the system hosts $N$ tunneling-gapped domains, then according to Eq.~\eqref{clock_rep} we have introduced a chain of $N+1$ $\sigma_j$'s to describe $N$ $e^{i \pi Q^-_j}$ operators.  Clearly we must fix boundary conditions on the $\sigma_j$ chain to maintain a faithful bookkeeping of states.  To this end, define an operator $\mathcal{O} = e^{i \pi \sum_j Q^-_j}$ that counts the charge difference across the \emph{entire} trench.  Assigning naive periodic boundary conditions with $\sigma_{N+1} \equiv \sigma_1$ turns out to be inadequate.  This boundary condition would force $\mathcal{O} = 1$, whereas in our torus setup we can also access configurations with $\mathcal{O} = \omega$ or $\bar \omega$ by shuttling fractional charge between the top and bottom sides of the trench via the vertical path illustrated in Fig.~\ref{fig:pftorus2_rev}(b).   

Accounting for these physical processes requires introducing an additional pair of \emph{global} $\mathbb{Z}_3$ clock operators $T_x$ and $T_y$ that satisfy $T_{x,y}^3 =1$ and $T_x T_y = \omega T_y T_x$.  We impose boundary conditions such that
\begin{equation}
  \sigma_{N+1} = T_y \sigma_1,
\end{equation}
yielding 
\begin{equation}
  e^{i\pi \sum_{j}Q_{j}^{-}} = T_y.
  \label{Qminus_constraint}
\end{equation}  
Thus $T_y$ specifies the global charge difference across the trench, and any operator that changes this quantity should be accompanied by an appropriate power of $T_x$ to correspondingly cycle $T_y$.  In practice it is useful to introduce a branch cut, as in Fig.~\ref{fig:pftorus2_rev}(b), to keep track of $T_y$ \footnote{Notice that in Fig.~\ref{fig:pftorus2_rev}(b) the branch cut slices through a tunneling-gapped region.  We choose this convention throughout so that the cut influences $F_{ij}$ but never $e^{i \pi Q^+_j}$ operators.}  

In addition to tracking powers of $T_{x}$ and $T_{y}$ with branch cuts, we must introduce one additional physical constraint to make the problem well-defined on a torus.   Specifically, three fractional-charge hopping processes $F_{ij}$, $F_{jk}$, $F_{ki}$ that form a contractible triangle without crossing a trench must satisfy
\begin{equation}
F_{ij}F_{jk}F_{ki} = 1.
\label{eq:triid}
\end{equation}
Equation~\eqref{eq:triid} simply asserts that moving an anyon around a loop that does not enclose any nontrivial charge is equivalent to the identity.  
To satisfy this constraint one must fix the ordering of $T_{x}$ and $T_{y}$ (when both are present in a particular fractional-charge hopping process) and also add additional phase factors to the definition of $F_{ij}$.  Only then can one appropriately extend these operators to the torus, and in turn define commuting-projector Hamiltonians consistently.

Fractional charge hoppings can be divided into three cases, depending on which parafermion representations ($R$ or $L$) are involved in the process.  The following rules describe one consistent choice for the assignment of $T_{x,y}$ operators and phase factors that yield $F_{ij}$ operators conforming to the above criteria: 
\begin{enumerate}
  \item Parafermion couplings arising from the transfer of $2e/3$ charge from the lower end of the trench ($R$) to the upper end of the trench ($L$) are accompanied by $T_{x}$. In addition, if the fractional-charge hopping path crosses the $T_y$ branch cut $n$ times from left to right, and $m$ times from right to left, then one attaches $T_{y}^{n-m}$ \textit{behind} $T_{x}$:
\begin{equation}
\label{eq:rule1}
F_{ij} = \alpha_{L,i}\alpha_{R,j}^{\dagger} T_{x} T_{y}^{n-m}.
\end{equation}
Parafermion couplings corresponding to fractional-charge transfer from the upper to the lower end of the trench are obtained by Hermitian conjugation. 
  \item Parafermion couplings between two $R$ parafermions arising from the transfer of $2e/3$ charge along a path that crosses the $T_y$ branch cut $n$ times from left to right, and $m$ times from right to left are accompanied by $T_{y}^{n-m}$: 
\begin{equation}
\label{eq:rule2}
F_{ij} = \alpha_{R,i}\alpha_{R,j}^{\dagger} T_{y}^{n-m}.
\end{equation}

   \item Parafermion couplings between two $L$ parafermions arising from the transfer of $2e/3$ charge along a path that crosses the $T_y$ branch cut $n$ times from left to right, and $m$ times from right to left are accompanied by $( \omega T_{y})^{n-m}$:
\begin{equation}
\label{eq:rule3}
F_{ij} = \alpha_{L,i}\alpha_{L,j}^{\dagger} (\omega T_{y})^{n-m}.
\end{equation}
\end{enumerate}

Let us illustrate the construction of parafermion couplings with two examples sketched in Fig.~\ref{fig:pftorus2_rev}(c).  For ease of visualization the figure depicts the torus in a `repeated-zone scheme', with the branch cut arbitrarily re-positioned relative to Fig.~\ref{fig:pftorus2_rev}(b). 
According to the above rules, the three fractional-charge hoppings on the left side of Fig.~\ref{fig:pftorus2_rev}(c) are written as 
\begin{equation}
\begin{split}
F_{2,2j-1} &= \alpha_{L,2}\alpha_{L,2j-1}^{\dagger}(\omega T_{y}) \\
F_{1,2} &= \alpha_{R,1}\alpha_{L,2}^{\dagger}T_{x}^{\dagger} \\ 
F_{2j-1,1} &= \alpha_{L,2j-1}\alpha_{R,1}^{\dagger} T_{x} T_{y}^{\dagger} .
\end{split}
\end{equation}
One can explicitly check that $F_{2j-1,1}F_{1,2}F_{2,2j-1} = 1$ in agreement with Eq.~\eqref{eq:triid}.  Similarly, fractional-charge hoppings rom the right side of Fig.~\ref{fig:pftorus2_rev}(c) become
\begin{equation}
\begin{split}
F_{2j-1,2} &= \alpha_{R,2j-1}\alpha_{R,2}^{\dagger} T_{y}^{\dagger} \\
F_{2j,2j-1} &= \alpha_{L,2j}\alpha_{R,2j-1}^{\dagger}T_{x} \\ 
F_{2,2j} &= \alpha_{R,2}\alpha_{L,2j}^{\dagger}  T_{y} T_{x}^{\dagger},
\end{split}
\end{equation}
yielding $F_{2,2j}F_{2j,2j-1}F_{2j-1,2} = 1$ as desired.

\begin{figure}
	\includegraphics[width=\linewidth]{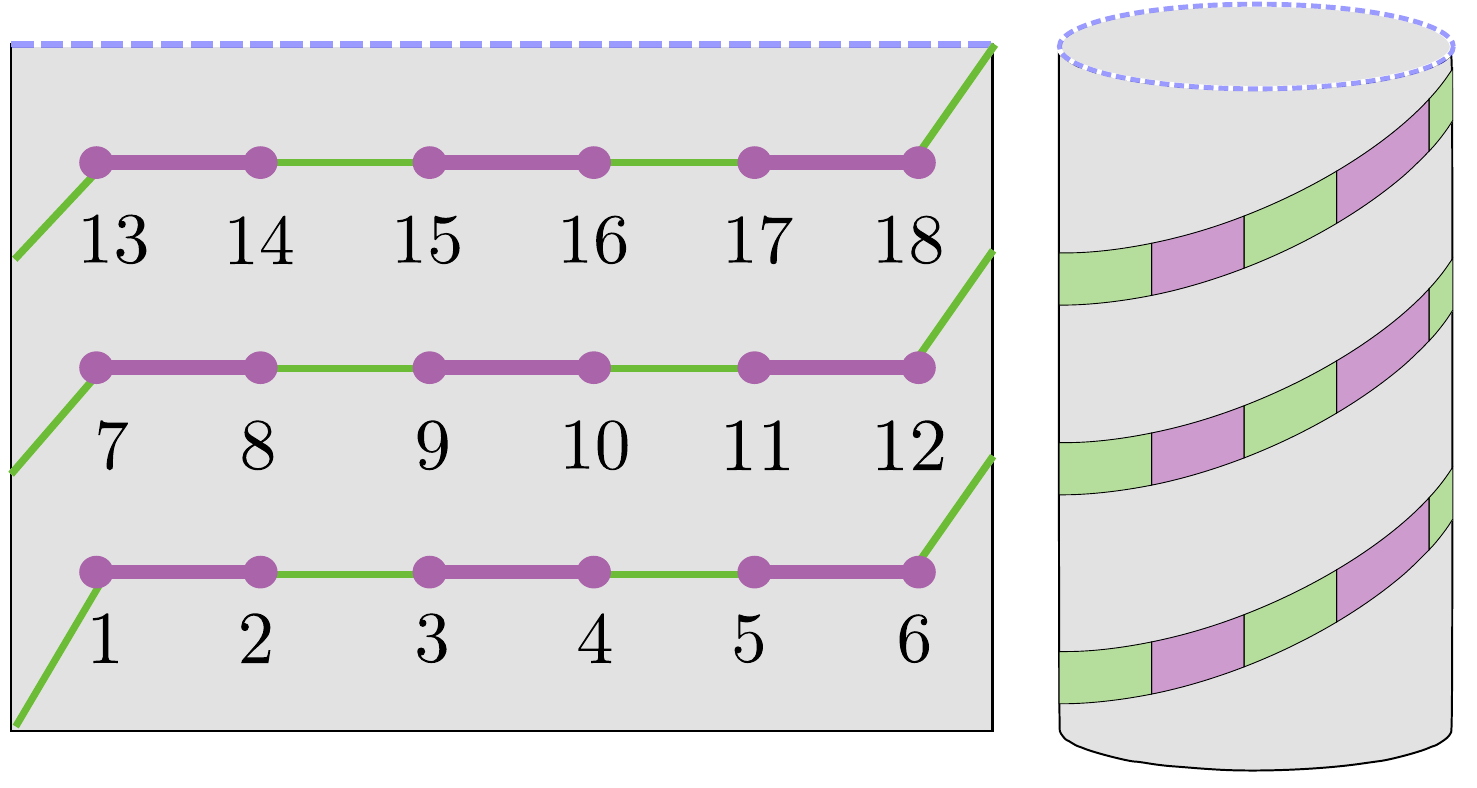}
	\caption{Two-dimensional parafermion array obtained from a single long trench snaked around the torus.  (The torus arises from gluing the upper and lower parts of each figure.)  Dashed line indicates the $T_y$ branch cut.} 
	\label{fig:pftorus_rev}
\end{figure}

Two-dimensional parafermion arrays can be understood as a trench wrapped around the torus in a snake-like manner, as illustrated in Fig.~\ref{fig:pftorus_rev}. Here, too, one can introduce a $T_y$ branch cut (dashed line in the figure) around a non-contractible cycle of the torus and attach proper $T_{x}$ and $T_{y}$ operators to parafermion couplings according to the same rules above.  We stress that these rules, with no modifications, can be applied for different parafermion orderings as well.

\section{Commuting Projector Hamiltonians} \label{sec:Setup}

This section introduces our commuting-projector models. To motivate the constructions, we will first describe the wavefunctions that our models will exhibit as exact ground states.  Writing down the wavefunctions precisely requires specifying two important sets of data: parafermion ordering and a generalized Kasteleyn orientation.  Given these data, we will show that it is indeed possible to define parent commuting-projector Hamiltonians.  In an effort to keep this section intuitive for readers, most technical proofs are relegated to appendices. 
  
\subsection{Ground-state wavefunctions}
\label{sec:wavefunctions}

\begin{figure}
	\includegraphics[width=\linewidth]{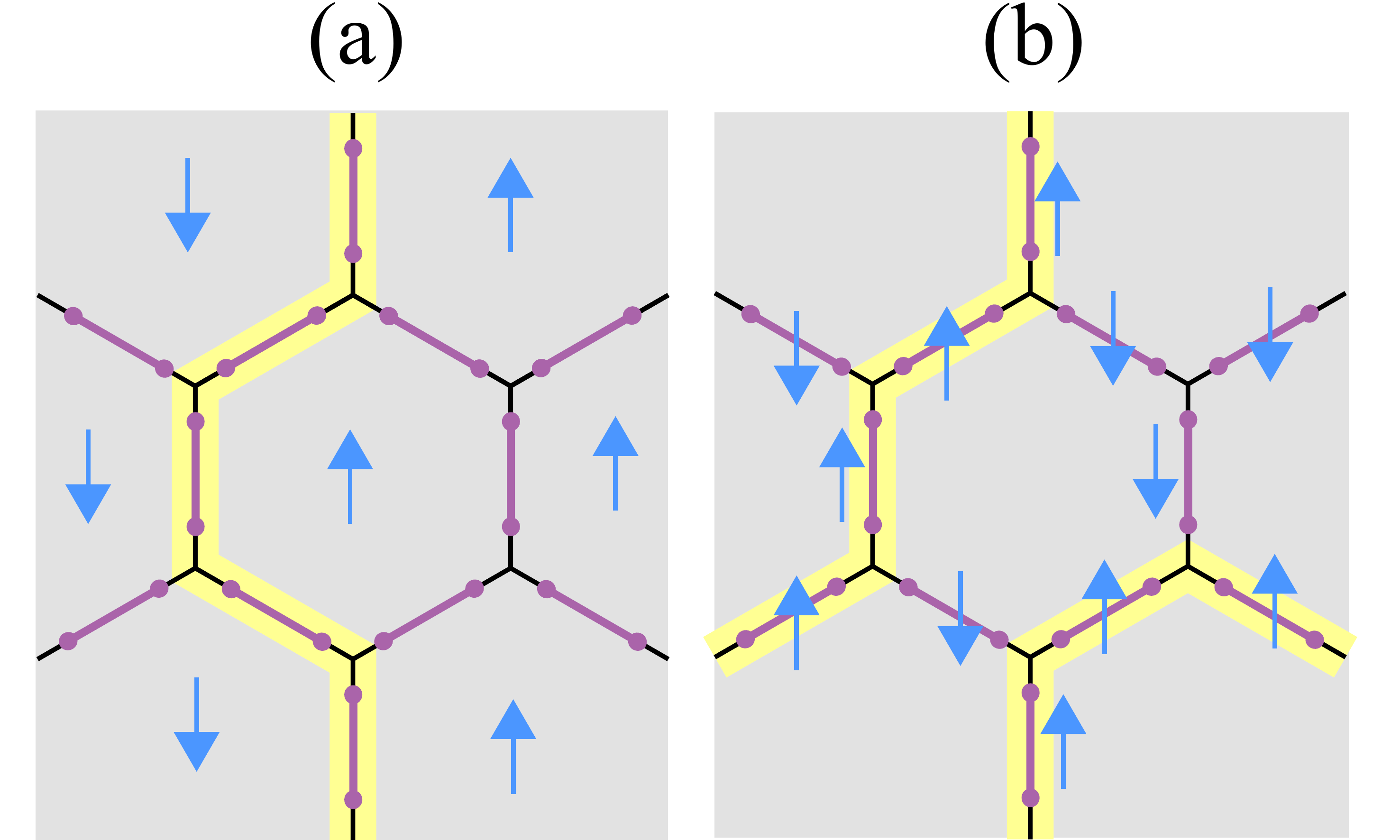}
	\caption{Degrees of freedom in (a) the decorated-domain-wall model and (b) the decorated-toric code model. Ising spins, delineated by blue arrows, live on plaquettes in the former and edges in the latter.  In (a) the yellow line indicates a domain wall separating regions where Ising spins point up and down; in (b) the yellow line indicates a section of a `toric-code loop' corresponding to a line of up spins in a sea of down spins.}
\label{fig:dof}
\end{figure}

We will primarily work with the honeycomb lattice.  We stress, however, that most of statements made in this paper can be straightforwardly extended to any trivalent lattice.  Each edge of the trivalent lattice contains two parafermions connected by a superconducting domain.  For the decorated-domain-wall model, we additionally include an Ising spin on each plaquette; for the decorated-toric-code model we instead incorporate an Ising spin on each edge of the trivalent lattice.  Figure~\ref{fig:dof} illustrates the degrees of freedom for both models.   We focus on planar and torus manifolds, to which the formalism described in Sec.~\ref{sec:Review} applies.  The next subsection briefly comments on potential extensions to arbitrary manifolds.  

To sketch ground-state wavefunctions, it is useful to start from the spin sector and temporarily ignore the parafermions.  In the context of the decorated-domain-wall model we take the spins to form an Ising paramagnet (IP) described by the spin wavefunction $\ket{\Psi_{\rm IP}} = \sum_s \ket{s}$ consisting of a superposition of all possible Ising spin configurations $s$.  (Throughout this paper, spin wavefunctions $\ket{s}$ explicitly refer to product states with definite $\sigma^z$ eigenvalues for each spin; we sometimes use the qualifier `Ising' to emphasize this property.)  In the decorated-toric-code setting we take the spins to form a toric-code (TC) ground state corresponding to $\ket{\Psi_{\rm TC}} = \sum_{s \in \{ s_v \} } \ket{s}$.  Here the sum runs over the restricted set of Ising spin configurations $\{ s_v \}$ that satisfy the rule that an even number of spins adjacent to each vertex point up.  One can profitably view the wavefunction as describing a sea of down spins dressed with fluctuating closed loops of up spins, which we refer to as `toric-code loops' below.  
 
\begin{figure}
   \includegraphics[width=1\linewidth]{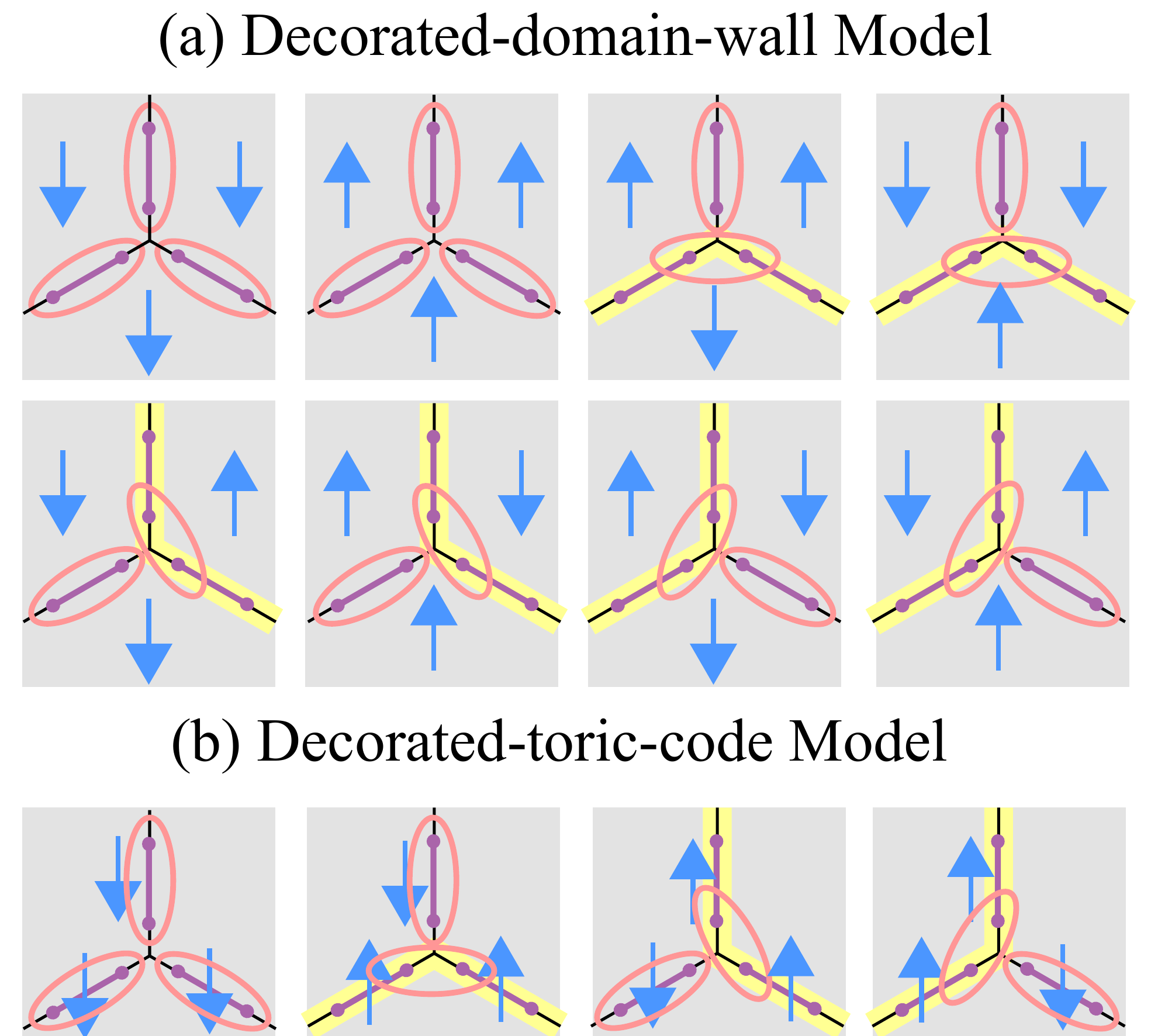}  
  \caption{Local rules that determine parafermion pairing from spins for (a) the decorated-domain-wall construction and (b) the decorated-toric-code construction.  Ovals indicate parafermions that pair up---as defined by Eqs.~\eqref{eq:interedge} and \eqref{eq:intraedge}---in a given spin configuration.  (We only encircle pairs that involve sites completely within the regions displayed.)  Along a domain wall or toric-code loop (yellow lines), two neighboring parafermions from different edges are paired. Otherwise, two parafermions from the same edge are paired.  Parafermion pairings thus have one-to-one correspondence with domain walls in (a) and toric-code loops in (b).}
  \label{fig:localrule}
\end{figure}
 
Next, we restore the full Hilbert space and envision assigning parafermion `pairings' to each spin configuration.  Consider first toric-code loops in $\ket{\Psi_{\rm TC}}$ and domain walls between spin-up and spin-down regions in $|\Psi_{\rm IP}\rangle$; for an illustration of each see yellow lines in Fig.~\ref{fig:dof}.  Along these toric-code loops/domain walls, we pair up parafermions residing on neighboring edges of the lattice.  Elsewhere we pair up parafermions with their partner on the same edge of the lattice.  These parafermion pairings follow from \textit{local vertex rules} illustrated in Fig.~\ref{fig:localrule}.  More quantitatively, inter-edge pairing of parafermions at sites $a$ and $b$ means that states $|\psi\rangle$ in the parafermionic Hilbert space satisfy
\begin{equation}
\omega^{n_{ab}} F_{ab}\ket{\psi} = \ket{\psi}.
\label{eq:interedge}
\end{equation} 
Here $F_{ab}$ is a parafermion bilinear arising from fractional-charge hopping as defined in Sec.~\ref{sec:Review}, while $n_{ab} = 0,\pm1$ is a fixed number assigned to each possible inter-edge pairing with directionality, i.e., $n_{ab} = -n_{ba}$.  (We specify the $n_{ab}$'s below.)  Similarly, intra-edge pairing means that these states satisfy
\begin{equation}
e^{i \pi Q_{\mu}^{+}} \ket{\psi} = \ket{\psi},
\label{eq:intraedge}
\end{equation} 
where $e^{i \pi Q_{\mu}^{+}}$ characterizes the superconducting region linking the parafermions that are paired.  One can view Eqs.~\eqref{eq:interedge} and \eqref{eq:intraedge} as fixing the fusion channel for pairs of non-Abelian defects in a way that depends on the spin configuration.  Schematically, we can then write down the target ground-state wavefunctions for the decorated-domain-wall and decorated-toric-code models as
 \begin{eqnarray}
\ket{\Psi_{\rm DDW}} &=& \sum_{s} \ket{s} \ket{{\rm PF}(s)} \label{psiDIP} \\ 
\ket{\Psi_{\rm DTC}} &=& \sum_{s \in \{ s_v \} } \ket{s} \ket{{\rm PF}(s)}, \label{psiDTC}
\end{eqnarray}
respectively.
Here, $\ket{{\rm PF}(s)}$ denotes a parafermionic state that satisfies Eqs.~\eqref{eq:interedge} and \eqref{eq:intraedge} as appropriate given the corresponding spin configuration $\ket{s}$.

To define these states precisely rather than schematically, we must $(i)$ specify a parafermion ordering to unambiguously define parafermion pairings through Eqs.~\eqref{eq:interedge} and ~\eqref{eq:intraedge} and $(ii)$ choose the integers $n_{ab}$ for each possible inter-edge pairing according to a generalized Kasteleyn orientation.  
In the next subsection we tackle issue $(i)$.  We will observe that the two data above constitute a gauge choice in the sense that there exists a massive number of allowed parafermion orderings and generalized Kasteleyn orientations that lead to the same physics.
 
\subsection{Ordering of parafermions} \label{sec:po}

\begin{figure}
  \includegraphics[width=0.8\linewidth]{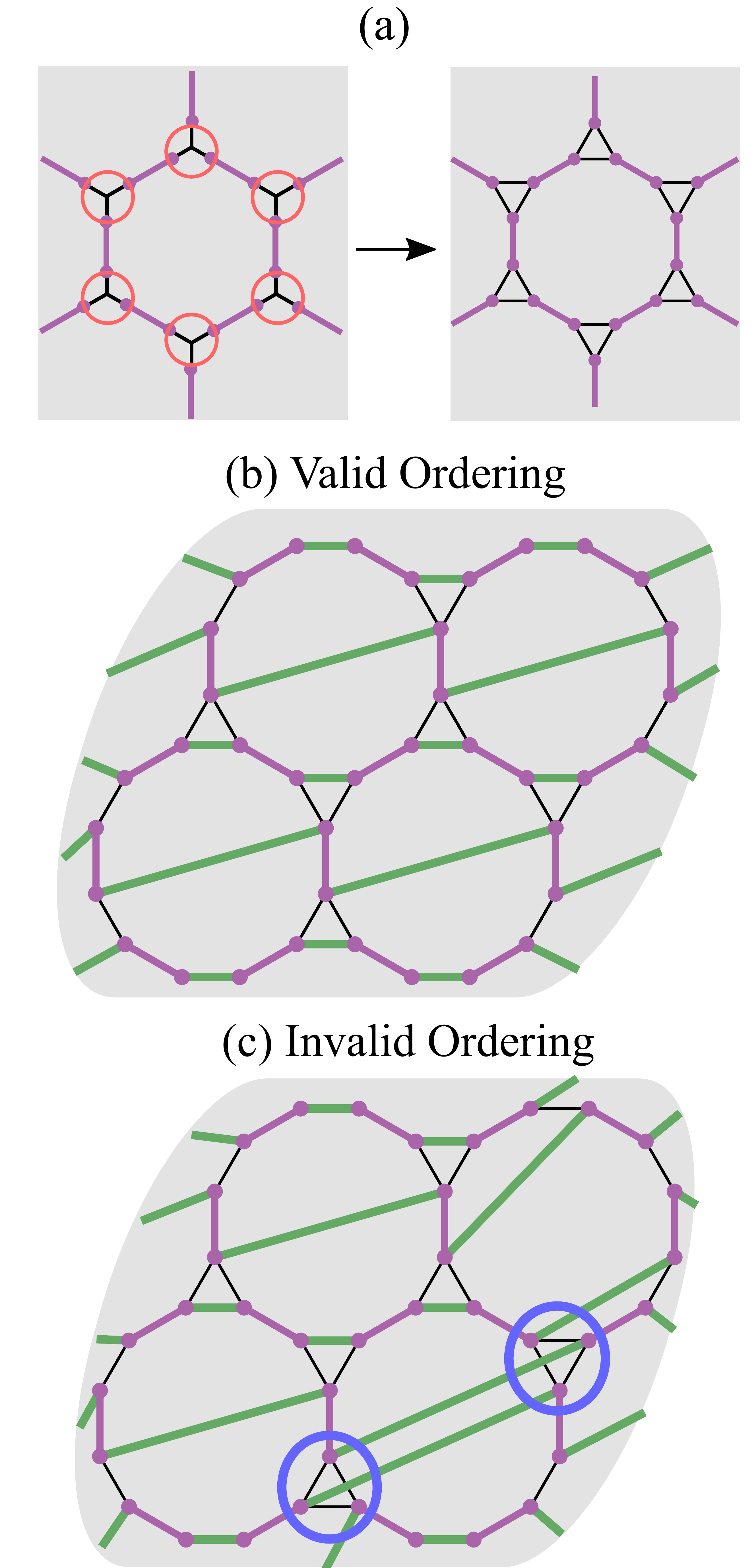} 
  \caption{(a) Substituting the neighborhood of vertices of the original honeycomb lattice (circled in red) with triangles gives the pairing lattice.  (b,c) Examples of valid and invalid ordering of parafermions on the pairing lattice.  Green lines indicate tunneling-gapped regions while purple regions indicate pairing-gapped regions.  The valid ordering shown in (b) allows all possible parafermion pairings arising from the local rules sketched in Fig.~\ref{fig:localrule}; in (c) the trenches that cut through the triangles preclude certain pairings within the blue circles---rendering the ordering invalid.}  
  \label{fig:ordering}
\end{figure}
  
To specify parafermion ordering, we will view the 2D lattice of parafermions as arising from a single trench cut through the parent quantum-Hall state---allowing us to directly import formalism developed in Sec.~\ref{sec:Review}.  We already asserted that each edge of the lattice contains a superconducting domain; thus we need only specify how these domains are connected via tunneling-gapped regions.  Recall that inter-edge pairings employ fractional-charge-hopping operators $F_{ab}$ through Eq.~\eqref{eq:interedge}, and that fractional charge cannot hop across the trench.  It is therefore essential that the ordering path is defined in a way that does not preclude inter-edge parafermion pairings that arise in the wavefunctions $\ket{\Psi_{\rm DDW}}$ and $\ket{\Psi_{\rm DTC}}$.  That is, \emph{the trench should not cross any possible nearest-neighbor inter-edge pairing bonds}.  This is the only criterion that filters out some parafermion orderings; our models should be and actually are consistently defined under any ordering that satisfies this property.  All valid orderings lead to the same physics---thus, parafermion ordering is merely a gauge choice.   
 
 To cast the assignment of parafermion ordering into a more formal language, let us define a new graph dubbed the \textit{pairing lattice}.  Vertices of the pairing lattice correspond to parafermion sites, while edges correspond to all nearest-neighbor inter-edge and intra-edge pairings on the plane or torus.  The pairing lattice can be obtained from the original trivalent lattice by cutting out the neighborhood of each original vertex and inserting a triangle in its place.  For example, this procedure turns the honeycomb lattice into the Fisher lattice, as shown in Fig.~\ref{fig:ordering}(a). Specifying parafermion ordering is tantamount to finding a path that connects all intra-edge pairings in a single line without intersecting triangles.  Figures~\ref{fig:ordering}(b) and (c) illustrate examples of valid and invalid orderings. 
 For the plane, drawing this path alone suffices to specify parafermion orderings that enable all inter-edge pairings; for the torus, one needs to additionally draw a $T_y$ branch cut to mark the start and the end of the ordering (see Sec.~\ref{sec:torus}).  
 
Given such a parafermion ordering, we can now specify five important properties satisfied by the fractional-charge hopping operators $F_{ab}$ associated with inter-edge pairings:
 
\begin{property}
\label{eq:prop1}
$F_{ab}^{3} =1$, $F_{ab}^{\dagger} = F_{ba} = F_{ab}^{2}$
\end{property}
 
\begin{property}
$F_{ab}$ and $e^{i \pi Q_{\mu}^{+}}$ exhibit the commutation relation
\begin{equation}
\label{eq:prop2}
F_{ab}e^{i \pi Q_{\mu}^{+}} = \begin{cases}
e^{i \pi Q_{\mu}^{+}} F_{ab} &\quad \parbox[t]{3cm}{\text{neither $a$ nor $b$ on Cooper-} \\ \text{pairing region $\mu$}} \\
\omega e^{i \pi Q_{\mu}^{+}} F_{ab} &\quad \text{$a$ on the region $\mu$} \\
\bar{\omega} e^{i \pi Q_{\mu}^{+}} F_{ab} &\quad \text{$b$ on the region $\mu$}
\end{cases}
\end{equation}
\end{property}

\begin{property}
$ [ F_{ab}, F_{a'b'} ] =0$ if $a \neq a', b'$ and $b \neq a',b'$
\end{property}

\begin{property}
For an elementary triangle of the pairing lattice with sites $a,b,c$ labeled in a clockwise order, one has
\begin{equation}
\label{eq:prop4}
\begin{split}
F_{ab}F_{bc} &= \omega F_{bc}F_{ab} \\
F_{bc}F_{ca} &= \omega F_{ca}F_{bc} \\
F_{ca}F_{ab} &= \omega F_{ab}F_{ca} \\
F_{ab}F_{bc}& F_{ca} = 1
\end{split}
\end{equation}
\end{property}
\begin{property}
  Let $a_{1}$, $a_{2}$, $\cdots$, $a_{2m}$ label clockwise-oriented parafermion sites on a non-triangular plaquette of the pairing lattice, with $a_{2l-1}$ and $a_{2l}$ connected by a Cooper-paired region for any $l$.  (In the Fisher-lattice context, the non-triangular plaquette corresponds to the $12$-gon in Fig.~\ref{fig:ordering}.)  Denote the edge connecting $a_{2l-1}$ and $a_{2l}$ by $\mu_{l}$.  The following then holds:
\begin{equation}
\label{eq:prop5}
\prod_{l=1}^{m} e^{i\pi Q_{\mu_{l}}^{+}} = \omega^2 F_{a_{2m}a_{1}}\prod_{l=1}^{m-1}F_{a_{2l} a_{2l+1}}
\end{equation}
\end{property}

\begin{figure}
\includegraphics[width=1\linewidth]{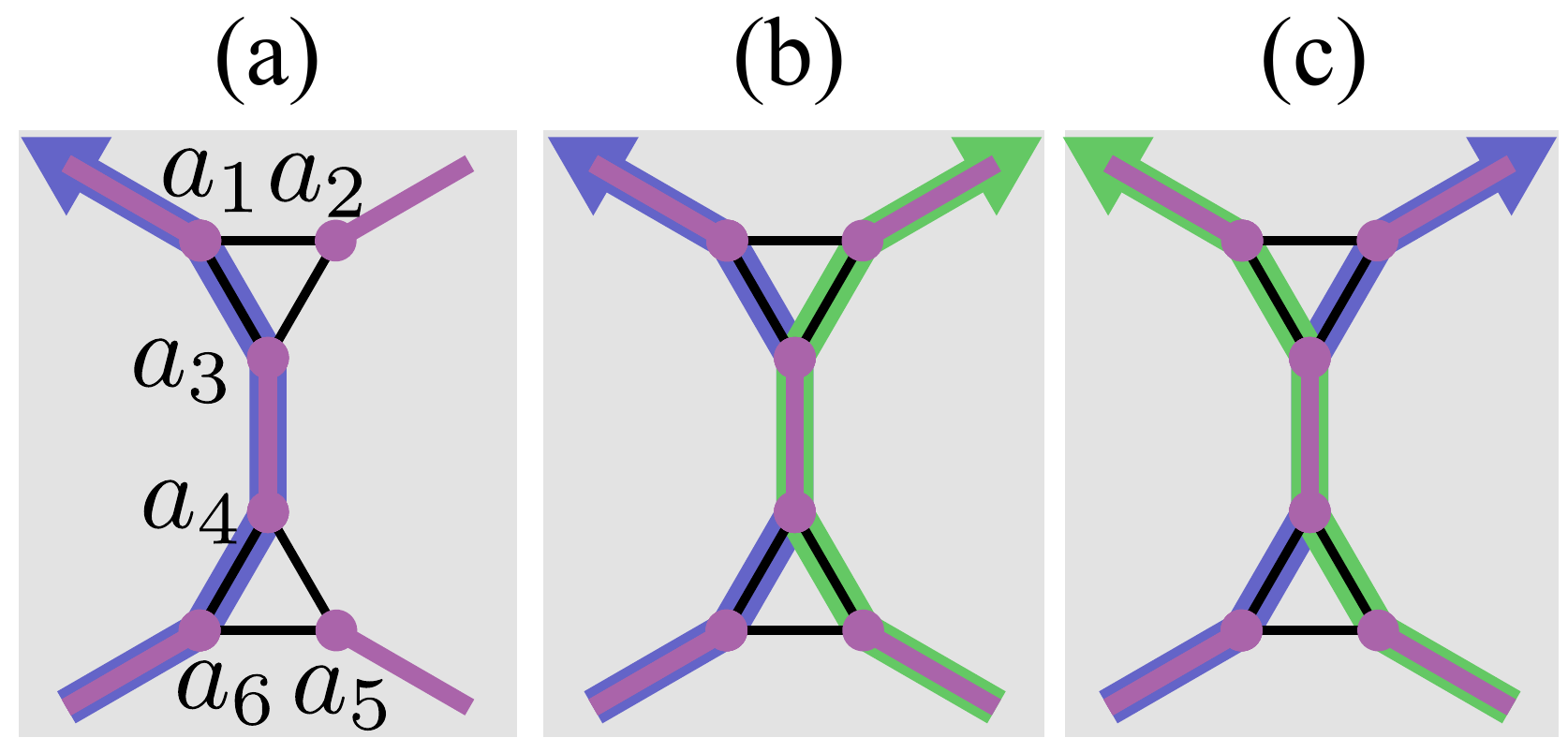}
\caption{(a) Labeling of parafermion sites invoked in Eqs.~\eqref{eq:worldlineg} through \eqref{eq:worldlinenc}.  The worldline of a $2e/3$ charge described by  $F_{a_{1}a_{3}}F_{a_{4}a_{6}}$ is sketched in blue.  (b) Worldlines corresponding to $F_{a_{1}a_{3}}F_{a_{4}a_{6}}$ (blue) and $F_{a_{2}a_{3}}F_{a_{4}a_{5}}$(green) touch the same Cooper-pairing region but can be deformed in a way that they do not cross; these operators thus commute.  (c) Worldlines corresponding to $F_{a_{2}a_{3}}F_{a_{4}a_{6}}$ (blue) and $F_{a_{1}a_{3}}F_{a_{4}a_{5}}$ (green) necessarily cross in a topologically nontrivial way.  Their commutation relation must therefore encode the anyonic braiding statistics of $2e/3$ fractional charges. }
\label{fig:worldline}
\end{figure}

Section~\ref{sec:Review} already provided the physical origins for properties 1, 2, 3, and the last line in Eq.~\eqref{eq:prop4} for property 4.  The remaining identities also follow from our ordering criterion and parafermion algebra specified earlier, though it will be helpful to now provide some more physical motivation.  

The first three lines of Eq.~\eqref{eq:prop4} directly relate to properties of anyon worldlines in the parent quantum-Hall fluid.  To see this, consider parafermion sites $a_{1}, a_{2}, \cdots, a_{6}$ around neighboring triangles as labeled in Fig.~\ref{fig:worldline}(a). We will be concerned with the commutation relations
\begin{equation}
\label{eq:worldlineg}
\begin{split}
F_{a_{1}a_{3}} F_{a_{2}a_{3}} &= \omega^{m_{1}} F_{a_{2}a_{3}} F_{a_{1}a_{3}} \\
F_{a_{4}a_{5}} F_{a_{4}a_{6}} &= \omega^{m_{2}} F_{a_{4}a_{6}} F_{a_{4}a_{5}}  \\
\end{split}.
\end{equation}
We can use one of the first three lines of property 4, together with the relations $F_{a_{1}a_{3}} = F_{a_{3}a_{1}}^{2}$ and $F_{a_{4}a_{6}} = F_{a_{6}a_{4}}^{2}$ from property 1, to deduce that $m_{1} = m_{2} = 1$.  Moreover, enforcing $m_{1} = m_{2} = 1$ for neighboring triangles that are rotated by $2\pi/3$ compared to those in Fig.~\ref{fig:worldline}(a) naturally yields the remainder of the first three lines from property 4, since they follow from cyclic permutations of $a,b,c$. In this sense, property 4 and satisfying $m_{1} = m_{2} = 1$  in Eq.~\eqref{eq:worldlineg} are equivalent.  

Interpreting $F_{ab}$'s as fractional-charge hopping operators in fact requires this choice for $m_{1,2}$ as we now argue.  The combination $F_{a_{1}a_{3}}F_{a_{4}a_{6}}$ [blue line in Fig.~\ref{fig:worldline}(a)] shuttles $2e/3$ charge from $a_{6}$ to $a_{4}$, and then from $a_3$ to $a_1$.  Fractional charge added to the Cooper-pairing region between sites $a_3,a_4$ is readily soaked up by the condensate, so that one can interpret the net process as a $2e/3$ worldline from $a_6$ to $a_1$ along our trivalent lattice.  The combinations $F_{a_{1}a_{3}}F_{a_{4}a_{5}}$, $F_{a_{2}a_{3}}F_{a_{4}a_{5}}$, and $F_{a_{2}a_{3}}F_{a_{4}a_{6}}$ admit similar worldline interpretations.  As seen in Fig.~\ref{fig:worldline}(b), worldlines corresponding to $F_{a_{1}a_{3}}F_{a_{4}a_{6}}$ and $F_{a_{2}a_{3}}F_{a_{4}a_{5}}$ can be deformed such that they do not touch---hence these combinations should commute.  Upon rearrangement using property 3, we obtain
\begin{equation}
\label{eq:worldlinec}
(F_{a_{1}a_{3}}F_{a_{2}a_{3}})(F_{a_{4}a_{6}}F_{a_{4}a_{5}}) = (F_{a_{2}a_{3}}F_{a_{1}a_{3}})(F_{a_{4}a_{5}}F_{a_{4}a_{6}}),
\end{equation}
which yields $m_{1} = m_{2}$. In contrast, Fig.~\ref{fig:worldline}(c) shows that worldlines corresponding to $F_{a_{2}a_{3}}F_{a_{4}a_{6}}$ and $F_{a_{1}a_{3}}F_{a_{4}a_{5}}$ necessarily cross, implying a nontrivial commutator that encodes the anyonic braiding statistics for $2e/3$ fractional charges.  One can similarly rearrange this commutator as
\begin{equation}
\label{eq:worldlinenc}
(F_{a_{2}a_{3}}F_{a_{1}a_{3}})(F_{a_{4}a_{6}}F_{a_{4}a_{5}}) = \omega (F_{a_{1}a_{3}}F_{a_{2}a_{3}})(F_{a_{4}a_{5}}F_{a_{4}a_{6}}),
\end{equation}
which further constrains $m_2 + m_{1} = 2 \pmod{3}$ so that $m_1 = m_2 = 1$ as claimed.  The anyon worldline interpretation of $F_{ab}$'s indeed \textit{requires} property 4.

To motivate property $5$, consider a special parafermion ordering in which sites $a_{1},a_{2},\cdots, a_{2m}$ in Eq.~\eqref{eq:prop5} are ordered consecutively along the trench. For this special case, we can explicitly write the operators in Eq.~\eqref{eq:prop5} as 
\begin{equation}
\begin{split}
e^{i \pi Q_{\mu_{l}}^{+}} &= \alpha_{R,a_{2l-1}}^{\dagger}\alpha_{R,a_{2l}} \\
F_{a_{2l}a_{2l+1}} &= \alpha_{R,a_{2l}}\alpha_{R,a_{2l+1}}^{\dagger}
\end{split}.
\end{equation}
Inserting this decomposition into Eq.~\eqref{eq:prop5} yields 
\begin{equation}
\label{eq:easyprop5}
\prod_{l=1}^{m} \alpha_{R,a_{2l-1}}^{\dagger}\alpha_{R,a_{2l}} = \omega^{2} \alpha_{R,a_{2m}}\alpha_{R,a_{1}}^{\dagger} \prod_{l=1}^{m-1} \alpha_{R,a_{2l}}\alpha_{R,a_{2l+1}}^{\dagger} .
\end{equation}
The above `special-case equation' can be easily proven by starting from the left-hand side and using parafermion commutation relations to push $\alpha_{R,a_{2m}}$ from front to back.  We stress, however, that property 5 holds also for general valid orderings in our two-dimensional parafermion arrays.  

Regarding actual proofs, properties $1$ and $2$ trivially follow from the definition of parafermion operators, though the others involve technical details that we provide in Appendix~\ref{app:proofs}. 
These five properties, combined with the generalized Kasteleyn orientation that we turn to next, are sufficient for establishing the characteristics of our models given later in this section and in Sec.~\ref{sec:Prop}. Thus, an interesting alternative viewpoint is possible: One can use the five properties as a \textit{definition} of $\{ F_{ab} \}$, the set of operators associated with inter-edge pairings.  When our system is defined on the plane or torus, parafermion operators introduced in Sec.~\ref{sec:Review} provide one physically motivated family of solutions that underlie these properties.  We expect that one can find such $\{ F_{ab} \}$ for any trivalent lattice on arbitrary orientable manifolds; however, we do not yet have a clear physical picture for how these operators arise in the general case, contrary to the situation described in Sec.~\ref{sec:Review} for the plane and torus.

\subsection{Generalized Kasteleyn orientation} 
\label{sec:ko}

We now tackle the second issue needed to define our models precisely: specifying $n_{ab} = 0, \pm1$ that determine inter-edge parafermion pairings via Eq.~\eqref{eq:interedge}.  An appropriate choice for $n_{ab}$ is needed to ensure that the target wavefunctions in Eqs.~\eqref{psiDIP} and \eqref{psiDTC} are actually physical.  In particular, the wavefunctions must arise from superpositions of states that all respect the global triality constraint $e^{i\pi \sum_{j}Q_{j}^{+}} = 1$ from Eq.~\eqref{Qplus_constraint}.  We can illustrate the basic issue by starting from a reference spin configuration $|s_{\rm ref}\rangle$ with no domain walls or toric-code loops.  According to the rules in Sec.~\ref{sec:wavefunctions}, the corresponding parafermion part $|{\rm PF}(s_{\rm ref})\rangle$ contains only intra-edge pairings, thereby satisfying Eq.~\eqref{eq:intraedge} for all bonds and thus trivially conforming to Eq.~\eqref{Qplus_constraint}.  Next imagine a second spin configuration $|s'\rangle$ with a single domain wall or toric-code loop that yields inter-edge parafermion pairing around a 12-gon plaquette of the Fisher lattice but preserves the intra-edge pairing elsewhere.  Along the inter-edge pairing bonds, the parafermionic part $|{\rm PF}(s')\rangle$ satisfies Eq.~\eqref{eq:interedge}.  Using Eq.~\eqref{eq:prop5} one readily finds that the global triality for this configuration is then $e^{i\pi \sum_{j}Q_{j}^{+}} = \omega^{2 - \sum n_{ab}}$, where the sum runs over the inter-edge pairing bonds with $n_{ab}$'s directed clockwise.  Preserving global triality therefore constrains $n_{ab}$'s such that the sum `cancels out' the $\omega^2$ factor.  

\begin{figure}
	\includegraphics[width = \linewidth]{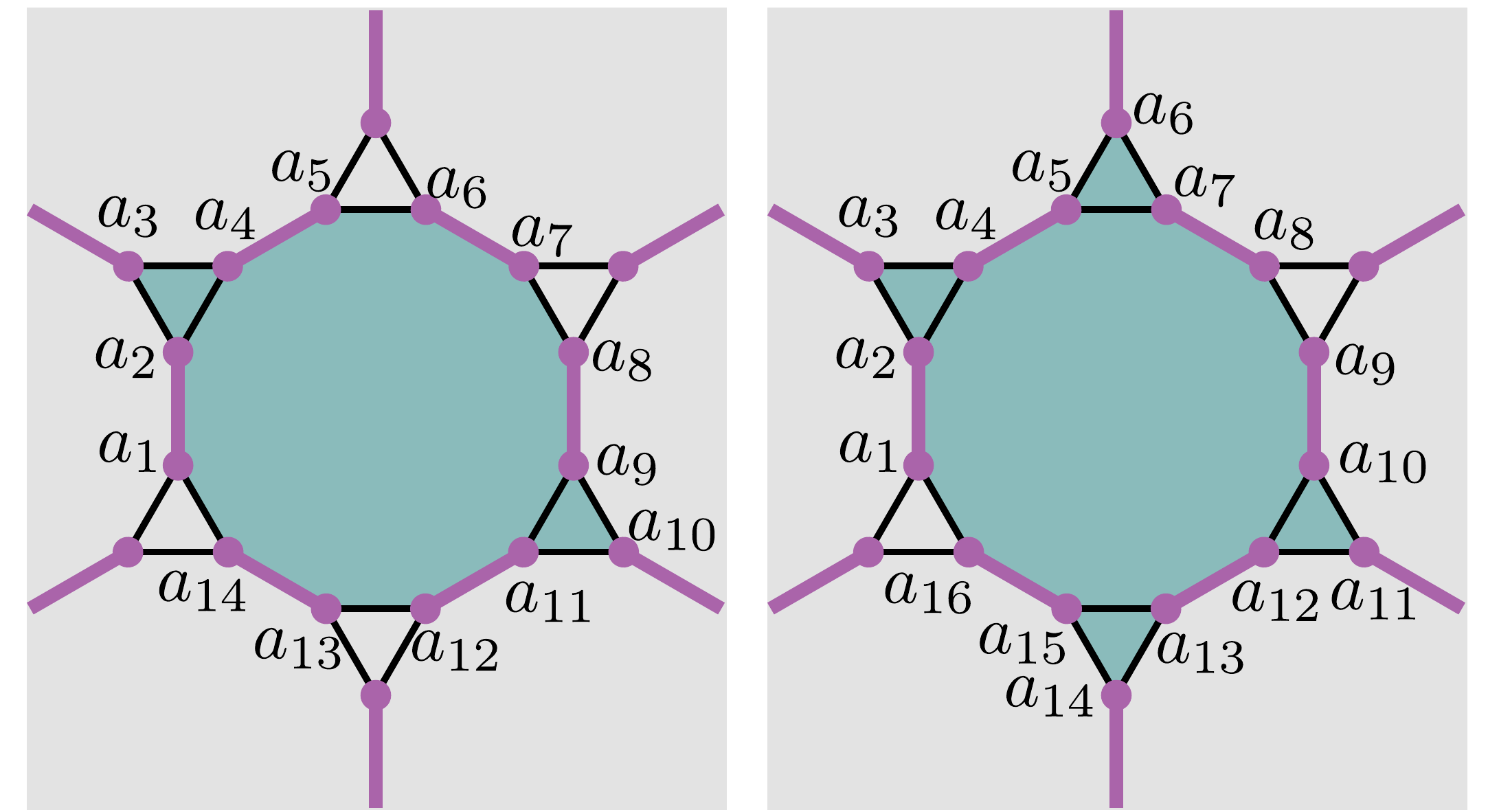}
	\caption{Examples of parafermion labelings used in Eq.~\eqref{eq:trieq}, which applies to plaquettes formed by the central 12-gon combined with an even number of neighboring triangles.  }
	\label{fig:triexample}
\end{figure}

More generally, \emph{all} possible domain-wall/ toric-code-loop configurations that are connected by local moves yield analogous constraints.  For attacking the general case it will be convenient to shift to a Hamiltonian-based viewpoint rather than explicitly tracking the global triality.  We will seek local, triality-preserving `flip operators' whose action cycles the system among all configurations in our target ground-state wavefunctions.  By default the resulting wavefunctions must then satisfy the global triality constraint.  Below we simply deduce general properties of the flip operators that suffice for determining the $n_{ab}$'s; the next subsection constructs the flip operators explicitly.

As a first step we define
 \begin{equation}
\mathcal{F}_{ab} = \begin{cases}
e^{-i \pi Q_{\mu}^{+}} &\quad \text{for $a,b$ on a Cooper-paired region $\mu$} \\
\omega^{n_{ab}} F_{ab} &\quad \text{else}
\end{cases}.
\label{mathcalF}
\end{equation}
Note that $\mathcal{F}_{ab} = \mathcal{F}_{ba}$ if $a$ and $b$ reside on a Cooper-paired region; otherwise $\mathcal{F}_{ab} = \mathcal{F}_{ba}^{\dagger}$. This notation therefore treats $F_{ab}$ and $e^{-i \pi Q_{\mu}^{+}}$ differently---which is the reason for our earlier choice in Sec.~\ref{sec:Review} to not incorporate $e^{-i \pi Q_{\mu}^{+}}$ into the definition of $F_{ab}$.  
Consider clockwise-ordered parafermion sites $a_{1},a_{2},\cdots, a_{2k}$ located around a combination of a polygonal face and an even number of adjacent triangular faces of the pairing lattice; see Fig.~\ref{fig:triexample} for two examples.  (The reason for choosing these particular plaquettes will be given shortly.)  Using properties 4 and 5 from the previous subsection, we prove in Appendix~\ref{app:gkas} that
\begin{equation}
  \prod_{l=1}^{k}\mathcal{F}^\dagger_{a_{2l-1},a_{2l}} = \omega^n \mathcal{F}_{a_{2k},a_{1}}\prod_{l=1}^{k-1}\mathcal{F}_{a_{2l},a_{2l+1}},
\label{eq:trieq}
\end{equation}
where $n=0,\pm1$ follows from the specific choice of $n_{ab}$'s.  
Similar to Eq.~\eqref{eq:prop5}, the above identity relates parafermion pairings that are shifted with respect to each other.  
[In fact, in the special case where the sites enclose no triangular faces, Eqs.~\eqref{eq:prop5} and \eqref{eq:trieq} are completely equivalent.]  

Next, take spin configurations $\ket{s_{e}}$ and $\ket{s_{o}}$ that impose identical parafermion pairings except along the plaquette formed by $a_1,a_2,\cdots,a_{2k}$.  In the case of $\ket{s_e}$, parafermions along this plaquette pair up between sites $a_{2l-1}$ and $a_{2l}$, while for $\ket{s_o}$ the pairings are shifted to $a_{2l}$ and $a_{2l+1}$.   To obtain a local Hamiltonian whose ground state superposes these two configurations, we would like to construct a flip operator that sends $\ket{s_e} \rightarrow \ket{s_o}$ and simultaneously cycles the parafermion pairings along the plaquette yet leaves all other pairings intact.  In general, the flip operator implementing this process can be built from operators acting on the spins together with $\mathcal{F}_{a_{i},a_{i+1}}$'s; 
the latter indeed preserve all other pairings by virtue of properties 2 and 3.   (Other terms such as $\mathcal{F}_{a_{m},a_{n}}$ with $|n-m|>1$ fall into two cases: they are either disallowed in a given parafermion ordering because they cross the trench, or they can be decomposed into products of `neighbor-hopping' $\mathcal{F}_{a_{i},a_{i+1}}$ operators, due to property 4.)

Finally, let us define 
\begin{equation}
  R = \prod_{l=1}^{k}\mathcal{F}_{a_{2l-1},a_{2l}},
  \label{local_triality}
\end{equation}  
which is just the Hermitian conjugate of the left side of Eq.~\eqref{eq:trieq}.  One can view $R$ as the `local triality' for the plaquette formed by sites $a_1,a_2,\cdots,a_{2k}$.  All $\mathcal{F}_{a_i,a_{i+1}}$ operators commute with $R$.  [Commutation is obvious for $\mathcal{F}_{a_{2i-1},a_{2i}}$.  For the shifted operators $\mathcal{F}_{a_{2i},a_{2i+1}}$ commutation can be seen by re-expressing $R$ using the right side of Eq.~\eqref{eq:trieq}.]  It follows that the associated flip operator built from $\mathcal{F}_{a_i,a_{i+1}}$ commutes with $R$ as well.  Given the pairings specified above,  $\ket{{\rm PF}(s_{e})}$ by definition has $R = 1$.  Moreover, because the flip operator takes $\ket{{\rm PF}(s_e)} \rightarrow \ket{{\rm PF}(s_o)}$ and commutes with $R$, $\ket{{\rm PF}(s_0)}$ must also have $R = 1$.  Using the right side of Eq.~\eqref{eq:trieq}, however, one obtains $R\ket{{\rm PF}(s_o)} = \omega^{-n}\ket{{\rm PF}(s_o)}$.  Consistency thus dictates that we choose $n_{ab}$'s so that $n = 0$ for all such plaquettes.  

As we will see explicitly in the next subsection, the spin configurations $\ket{s_o}$ and $\ket{s_e}$ are related by flipping spins locally at a single plaquette.  In the decorated-domain-wall model, such local spin flips in fact generate all possible spin configurations that appear in our target ground-state wavefunction.  Thus, the condition on $n_{ab}$ deduced above is necessary and sufficient for ensuring global triality conservation.  The case for the decorated-toric-code model is more subtle. Local spin flips do not quite generate all permissible spin configurations; instead, starting from a given spin configuration, local spin flips generate all spin configurations within the same topological sector of the toric code.  (Topological sectors of the toric code are characterized by $\mathbb{Z}_{2}$ winding numbers of non-contractible toric code loops---see Sec.~\ref{sec:Torus} or Ref.~\onlinecite{Kitaev2003} for more details).  Thus, the condition on $n_{ab}$ describes a necessary and sufficient condition in a slightly different sense: the condition guarantees global triality conservation within the same topological sector.  At this point, we have not established yet whether \emph{different} topological sectors share the same global triality in general.  We will observe that ensuring global triality conservation across all topological sectors takes an important role in determining the ground-state degeneracy of the decorated-toric-code model in Sec.~\ref{sec:Torus}.

Assignment of $n_{ab}$ can be conveniently represented by drawing arrows on edges of the pairing lattice that represent inter-edge pairing bonds, e.g., triangles in Fig.~\ref{fig:ordering}.  An arrow from $a$ to $b$ corresponds to $n_{ab} =1$, an arrow from $b$ to $a$ corresponds to $n_{ab} = -1$, and the absence of an arrow signifies $n_{ab} = 0$.  Appendix~\ref{app:gkas} shows that the following rule yields $n = 0$ in Eq.~\eqref{eq:trieq} as desired: 
Traverse each elementary plaquette of the pairing lattice clockwise, and add $+1$ when encountering an arrow parallel to the travel direction and add $-1$ when encountering an antiparallel arrow. Then assign arrows so that the total number is $-1 \pmod{3}$ when traversing any elementary plaquette (triangular or non-triangular).

 Arrow assignments satisfying this rule define a generalization of the Kasteleyn orientation that was required for the analogous Majorana models studied in Refs.~\onlinecite{Tarantino2016,Ware2016} (see also Ref.~\onlinecite{Wang2017}).  In the latter context, the Kasteleyn orientation 
similarly preserved the local fermion parity $\prod_j i\gamma_{2j-1}\gamma_{2j}$ around domain-wall/toric-code-loop configurations connected by local moves, which in turn guaranteed that the Majorana analogues of Eqs.~\eqref{psiDIP} and \eqref{psiDTC} involved a superposition of states with common global fermion parity.  Note, however, that the Majorana case is substantially simpler because subtleties with ordering do not exist.

\begin{figure}
  \includegraphics[width=0.8\linewidth]{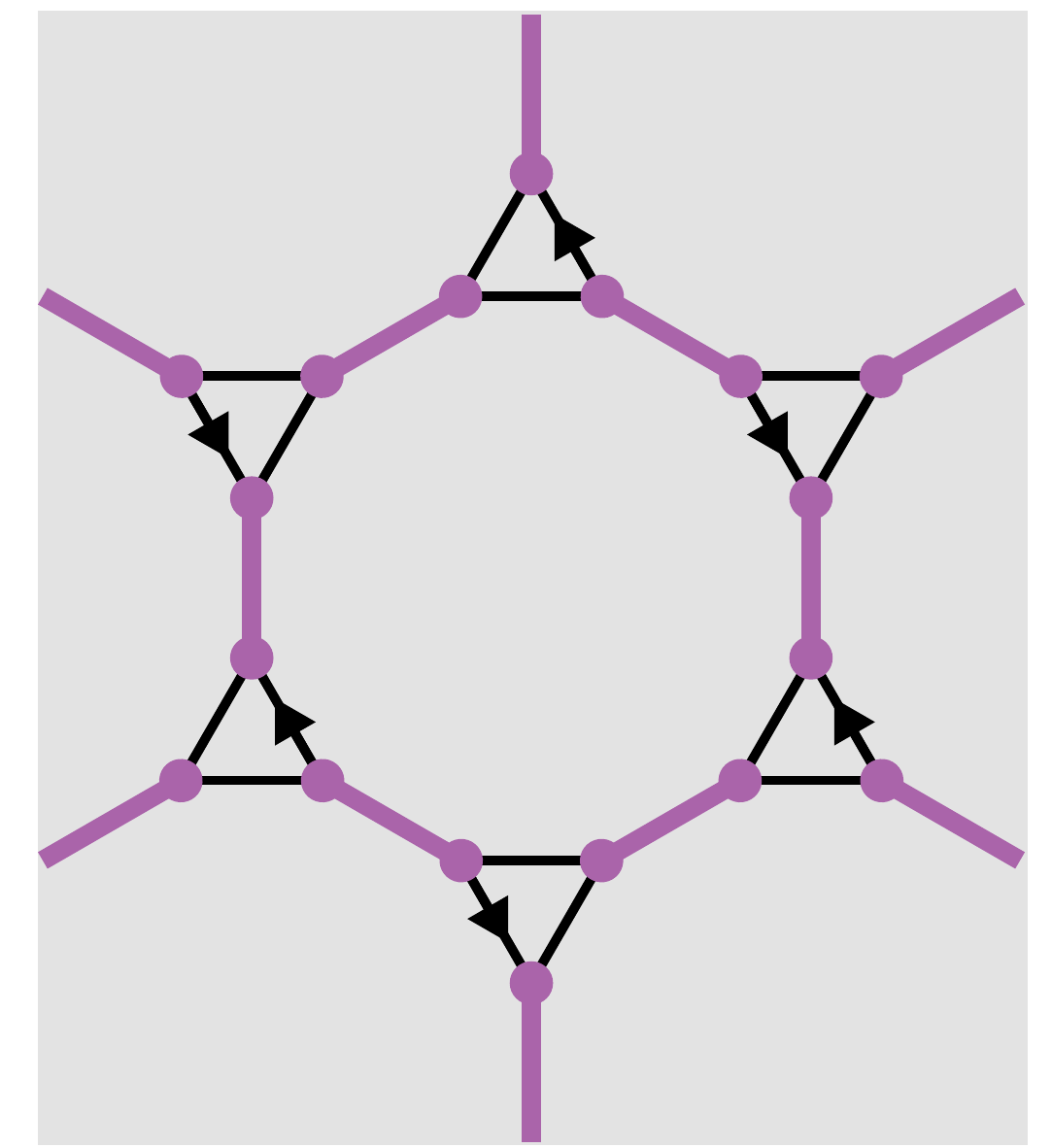} 
  \caption{Example of arrow assignments that satisfy our two conditions for a generalized Kasteleyn orientation.}
  \label{fig:arrowexample}
\end{figure}

Figure~\ref{fig:arrowexample} illustrates one valid generalized Kasteleyn orientation for our parafermion system.  Arrow configurations that satisfy the two conditions above are certainly not unique, though all such configurations yield models with identical physics.  Many---and we conjecture all---valid arrow configurations are in fact connected by gauge transformations.  For the Majorana case, one can generate all Kasteleyn orientations with a series of local modifications associated with the gauge transformation $\gamma_{a} \rightarrow s_{a}\gamma_{a}$ ($s_{a} = \pm 1$) and global boundary-condition changes \citep{Cimasoni2007}. In the parafermion case, one can similarly generate a large class of (and potentially all) allowed orientations using local transformations for the plane and a combination of global and local transformations for the torus. Local modifications are associated with the parafermion gauge transformation
\begin{equation}
\label{eq:gaugetransform}
\begin{split}
&\alpha_{L/R, a_{\mu}} \rightarrow \omega^{p_{\mu}} \alpha_{L/R, a_{\mu}} \\
&\alpha_{L/R, b_{\mu}} \rightarrow \omega^{p_{\mu}} \alpha_{L/R, b_{\mu}}
\end{split},
\end{equation}
where $a_{\mu}$ and $b_{\mu}$ are sites belonging to edge $\mu$ of the original trivalent lattice and $p_{\mu} = 0, \pm 1$.  This transformation alters neither the defining properties of parafermion operators nor the five properties of $F_{ab}$ given in Sec.~\ref{sec:po}.  The $\mathcal{F}_{ab}$ operators do, however, change form; their modified form can equivalently be recovered by leaving the parafermion operators intact and instead transforming
\begin{equation}
\label{eq:gaugetransform1}
\begin{split}
&n_{a_{\mu},c} \rightarrow n_{a_{\mu},c} + p_{\mu} \pmod{3} \\
&n_{b_{\mu},d} \rightarrow n_{b_{\mu},d} + p_{\mu} \pmod{3} \\
\end{split}
\end{equation}
for all $c,d$ adjacent to $a_\mu, b_\mu$.
 One can explicitly show that the new resulting arrow configuration still satisfies the consistency conditions given above.  
Gauge transformations in Eq.~\eqref{eq:gaugetransform} therefore generate local arrow reconfigurations that yield equally valid generalized Kasteleyn orientations.  
 
On the torus, we can also gauge transform $T_{x}$ and $T_{y}$ operators as
\begin{equation}
\begin{split}
\label{eq:gaugetransform2}
&T_{x} \rightarrow \omega^{p_{x}} T_{x} \\ 
&T_{y} \rightarrow \omega^{p_{y}} T_{y}
\end{split}
\end{equation}
with $p_{x},p_{y} = 0, \pm1$. Similarly to the gauge transformation of parafermions, the phase factors can be absorbed into the definition of $n_{ab}$.  Due to the nature of $T_{x}$ and $T_{y}$, however, all arrows that intersect with \textit{global, non-contractible cycles} are transformed.  These transformations are the analogue of global boundary-condition changes in the related Majorana models, with one important difference:  In the Majorana models, such changes cannot be associated with gauge transformations due to the absence of global operators such as $T_{x}$ and $T_{y}$; global changes of arrow configurations may therefore change the physical properties of the Majorana system.  Indeed, Ref.~\onlinecite{Ware2016} observed that tweaking boundary conditions changes the fermion parity of the ground states. On the other hand, global transformations in our parafermion models are associated with gauge transformation and preserve all physical properties of the system.

\subsection{Definition of commuting-projector Hamiltonians} \label{sec:cmh}

Now we are ready to define our commuting-projector Hamiltonians. Below we will frequently employ projectors
\begin{equation}
\label{eq:projdef}
P_{ab} = \frac{1 + \mathcal{F}_{ab} + \mathcal{F}_{ab}^{\dagger} }{3}
\end{equation}
associated with the bond between parafermions at sites $a$ and $b$. From the definition of $\mathcal{F}_{ab}$ in Eq.~\eqref{mathcalF} along with Eqs.~\eqref{eq:interedge} and \eqref{eq:intraedge}, we see that $P_{ab}$ projects onto intra-edge and inter-edge parafermion pairings for appropriate $ab$'s.
These projectors thus naturally comprise basic building blocks of our Hamiltonians, as well as many other operators that will be constructed throughout this paper.

Both the decorated-domain-wall and decorated-toric-code models take the form
\begin{equation}
H = -\sum_{v} A_{v} - \sum_{p} B_{p} ,
\label{H}
\end{equation}
where $v$ and $p$ respectively label vertices and hexagonal plaquettes of the original honeycomb lattice.  
The first piece represents a vertex term that energetically imposes the spin-dependent parafermion pairings sketched in Fig.~\ref{fig:localrule}.  Explicitly, we have
\begin{equation}
  A_{v} = \sum_{{\rm allowed}~s}\left(\prod_{ab~{\rm given}~s,v} P_{ab} \right)\ket{s,v}\bra{s,v}.
  \label{eq:defav}
\end{equation} 
Here $s$ runs over all permissible configurations for the three spins adjacent to a given vertex $v$, with $\ket{s,v}$ the corresponding state those three spins.  In the decorated-domain-wall model, the sum includes all eight possible spin configurations. The decorated-toric-code model, however, includes only half of the configurations since each vertex is constrained to have an even number of adjacent up spins.  Finally, the term in parenthesis contains a product of projectors $P_{ab}$ that enforce the desired parafermion fusion channels given a spin configuration $s$.  For spin configurations with no domain walls or toric-code loops, a product of three intra-edge-pairing projectors is required; otherwise the product involves one intra-edge and one inter-edge projector, as seen from Fig.~\ref{fig:localrule}.  

\begin{figure}
  \includegraphics[width=0.9\linewidth]{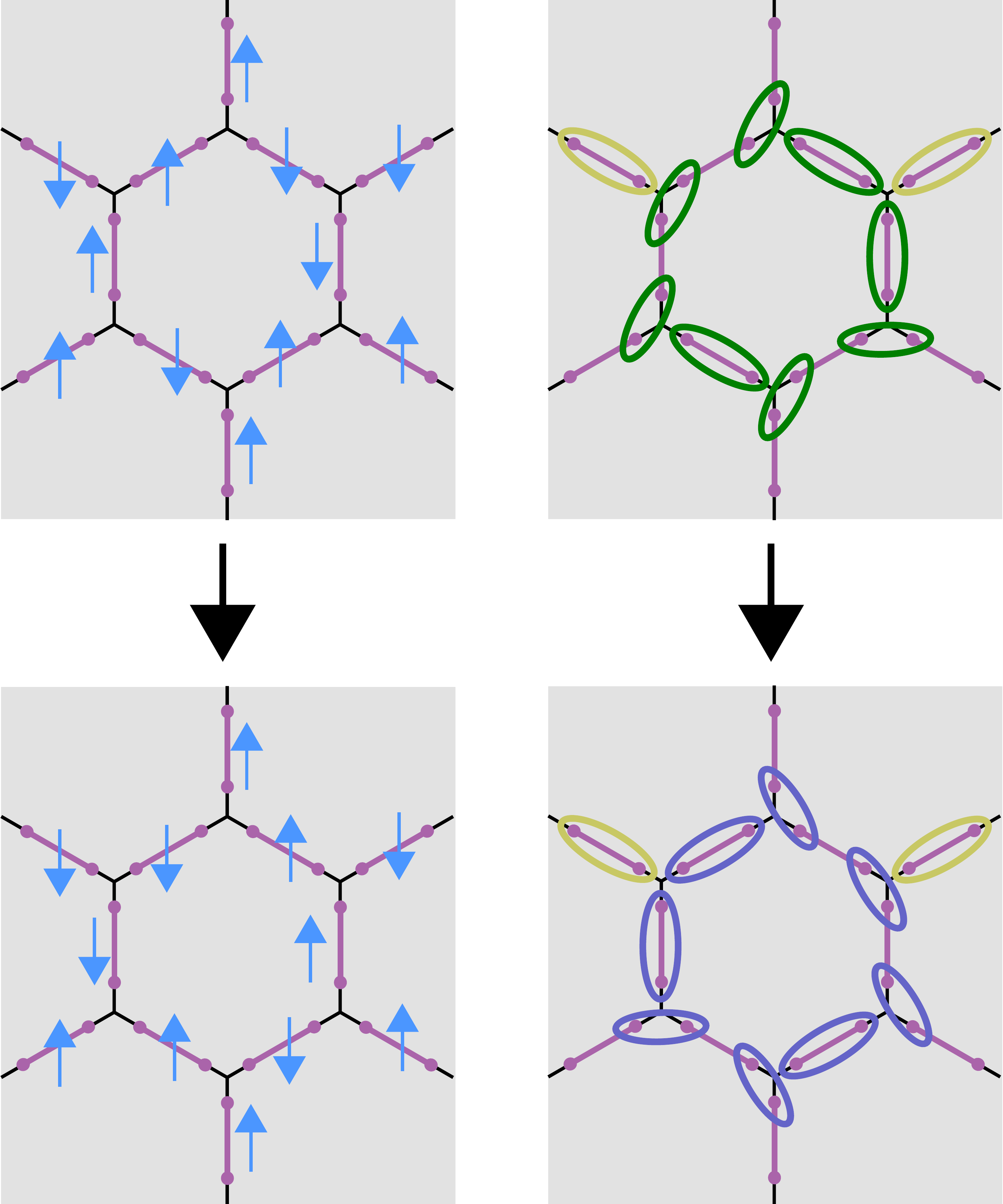} 
  \caption{Illustration of $B_{p}$ for the decorated-toric-code construction. For clarity, we separate out the spin and parafermion degrees of freedom into the left and right panels, respectively.  The starting configuration at the top transforms into the final configuration on the bottom under the action of $B_p$.  In the upper right, green ovals indicate initial parafermion pairings enforced by the first set of projectors in Eq.~\eqref{eq:defbps}.  In the lower right, blue ovals indicate  parafermion pairings enforced by the second set of projectors in Eq.~\eqref{eq:defbps}, as appropriate for the new spin configuration.  Yellow ovals denote special `branch parafermions pairings'; $B_p$ does \emph{not} enforce proper pairing eigenvalues for those bonds, which are instead imposed by the vertex term $A_v$.  
  }
  \label{fig:plaqutteaction1}
\end{figure}

The second piece in Eq.~\eqref{H} is a plaquette-flip term.  Specifically, $B_p$ toggles the spins---thus modifying the structure of domain walls or toric-code loops---and \emph{also} appropriately reconfigures the parafermion pairings.  We write this term as
\begin{equation}
  B_p = S_p \sum_{{\rm allowed}~s} \mathcal{B}_p^s \ket{s,p}\bra{s,p}.
  \label{eq:defbp}
\end{equation}
In the decorated-domain-wall model $S_p$ merely flips the spin in the center of plaquette $p$, while in the decorated-toric-code case $S_p$ instead flips all six spins along the edges of the hexagonal plaquette. The $\ket{s,p}\bra{s,p}$ projector projects onto some allowed configuration $s$ for the spins at plaquette $p$ and adjacent plaquettes/edges; $\mathcal{B}_p^s$ shifts the parafermion pairings to match the new resulting spin configuration.  For the decorated-domain-wall model the $s$ sum runs over all possible spin configurations for the spin at plaquette $p$ and the six surrounding spins on the adjacent plaquettes.  For the toric-code system we instead sum over allowed configurations for the six spins on the boundary of plaquette $p$, as well as the six spins on the edges emanating from that plaquette (see Fig.~\ref{fig:plaqutteaction1}).  Similar to the $A_v$ term, in the latter model the allowed configurations contain an even number of up spins adjacent to each vertex.  

We can explicitly write $\mathcal{B}_{p}^{s}$ in Eq.~\eqref{eq:defbp} as
\begin{equation}
  \mathcal{B}_{p}^{s} =  P_{a_{2k},a_{1}} \prod_{i=1}^{k-1} \sqrt{3}P_{a_{2i},a_{2i+1}} \prod_{j=1}^{k} P_{a_{2j-1},a_{2j}} .
  \label{eq:defbps}
\end{equation}
Here, $a_{1},a_{2},\cdots, a_{2k}$ form a clockwise-ordered loop of parafermions sites around the plaquette.  More precisely, the sites forming the loop implicitly depend on $s$ and must be chosen such that the initial spin state $|s,p\rangle$ pairs up parafermions at sites $a_{2j-1}$ and $a_{2j}$. The first string of projectors $\prod_{i=1}^{k} P_{a_{2i-1},a_{2i}} $ ensures the correct parafermion pairings given the initial spin configuration.  The second string $P_{a_{2k},a_{1}} \prod_{i=1}^{k-1} \sqrt{3}P_{a_{2i},a_{2i+1}}$ projects onto the state with parafermion pairings consistent with the new spin configuration arising from the application of the spin-flip operator $S_p$.  (The factor of $\sqrt{3}$ simply ensures unitarity of $B_{p}$ on the subspace in which it acts nontrivially.)

Figure~\ref{fig:plaqutteaction1} illustrates the action of $B_{p}$ for the decorated-toric-code model.  Notice that parafermions within yellow ovals in the figure must pair up to satisfy the $A_v$ term for the adjacent vertices.  Projectors for these `branch parafermion pairings' are nevertheless absent in $\mathcal{B}_{p}^{s}$, since the corresponding sites do not reside in the loop formed by $a_1,a_2,\cdots,a_{2k}$.  Consequently, $B_{p}$ does not filter out the `wrong' parafermion pairings on the `branches' of the plaquette loop.  This convention differs from string-net-type constructions in which $B_{p}$ annihilates any state with an $A_{v}$ violation.  We deliberately choose this unorthodox convention to simplify construction of wavefunctions corresponding to deconfined fractionalized excitations, which we undertake in Sec.~\ref{sec:Prop}.
 
The Hamiltonians defined above exhibit the following properties: 
\begin{itemize}
\item $A_{v}$ has eigenvalues $0$ or $1$. 
\item $B_{p}$ has eigenvalues $0$ or $\pm 1$.
\item The Hamiltonians are commuting-projector models in the sense that 
\begin{equation}
[ A_{v}, A_{v'} ] = [ B_{p}, B_{p'} ] = [ A_{v}, B_{p} ] =0.
\label{ABcommutators}
\end{equation}
\end{itemize}
Technically, $B_p$ is not quite a projector since it can also have eigenvalue $-1$. While it can be made into a projector with minor modification, we choose $B_p$ in the current form for simplicity.  

 Using these properties we can re-express the ground-state wavefunctions in Eqs.~\eqref{psiDIP} and \eqref{psiDTC} in a more explicit form.  Let $\ket{\psi_\downarrow}$ denote a `root state' containing only down spins, and with parafermions exhibiting intra-edge pairing such that $e^{i \pi Q_\mu^+} = 1$ for all edges $\mu$ of the honeycomb lattice.  In either model, this wavefunction trivially satisfies every $A_v$ vertex term.  Applying $B_p$ operators to $\ket{\psi_\downarrow}$ allows the spins and parafermion pairings to fluctuate, but by construction preserves the norm of the state (i.e., $\ket{\psi_\downarrow}$ is orthogonal to the kernel of $B_p$).  The ground states can then be written
\begin{equation}
  \ket{\Psi_{\rm DDW~or~DTC}} = \prod_p(1+B_p)\ket{\psi_\downarrow}.
  \label{GroundStatesExplicit}
\end{equation}
Because $A_v$ and $B_p$ commute, Eq.~\eqref{GroundStatesExplicit} continues to satisfy all vertex terms; moreover, the $(1+B_p)$ factors project away any elements with $B_p = -1$ eigenvalue, ensuring that the wavefunction maximally satisfies all $B_p$ terms as well.  Ground states for the decorated-domain-wall and decorated-toric-code models differ only in the Hilbert space for the spins in $\ket{\psi_\downarrow}$ and the precise action of $B_p$ in the spin sector.  Note, however, that for the decorated-domain-wall case we can instead use a root state $\ket{\psi_\uparrow}$ with only up spins to equivalently write 
\begin{equation}
  \ket{\Psi_{\rm DDW}} = \prod_p(1+B_p)\ket{\psi_\uparrow},
  \label{GroundStatesExplicit2}
\end{equation}
reflecting the (unbroken) $\mathbb{Z}_2$ spin-flip symmetry present in that model.  The same is \emph{not} true in the decorated-toric-code model---$\ket{\Psi_{\rm DTC}} \neq \prod_p(1+B_p)\ket{\psi_\uparrow}$ since $\ket{\psi_\uparrow}$ does not represent a valid toric-code spin configuration.  
Wavefunctions of a similar form to Eqs.~\eqref{GroundStatesExplicit} and \eqref{GroundStatesExplicit2} will often be employed to construct anyonic excitations in the next section.  

We will now sketch proofs of the above properties except for $[ B_{p}, B_{p'} ] =0$, which involves some technicalities and is postponed to Appendix~\ref{app:pcom}.
First, according to Eq.~\eqref{eq:defav} $A_{v}$ arises from a product of parafermion projectors and spin projectors.  Since each projector has eigenvalues $0$ or $1$, and the products of projectors commute for different $s$ in the sum, it naturally follows that $A_{v}$ admits eigenvalues $0$ or $1$ as well.

To prove the second property, we will show that $B_{p}$ acts as a \textit{unitary operator} on the subspace of $B_{p}$ orthogonal to its kernel. Hermiticity of $B_{p}$ then guarantees that its eigenvalues can only be $\pm 1$ (from the aforementioned subspace) and $0$ (from the kernel).  Expanding the second string of projectors in Eq.~\eqref{eq:defbps} yields
\begin{eqnarray}
  \mathcal{B}_{p}^{s} &=& \frac{\sqrt{3}^{k-1}}{3^{k}} \sum_{l_{k} = 0,1,2} (\mathcal{F}_{a_{2k}a_{1}})^{l_{k}}\prod_{i=1}^{k-1}\sum_{l_{i} = 0,1,2} (\mathcal{F}_{a_{2i}a_{2i+1}})^{l_{i}} 
\nonumber \\
&\times& \prod_{j=1}^{k} P_{a_{2j-1},a_{2j}}.
\label{Bps_expansion}
\end{eqnarray}
We can eliminate $\mathcal{F}_{a_{2k}a_1}$ using Eq.~\eqref{eq:trieq} with $n = 0$, as appropriate given our generalized Kasteleyn orientation; some algebra gives
\begin{eqnarray}
\mathcal{B}_{p}^{s} &=& \frac{\sqrt{3}^{k-1}}{3^{k-1}} \prod_{i=1}^{k-1} \sum_{l_{i} = 0,1,2}(\mathcal{F}_{a_{2i}a_{2i+1}})^{l_{i}} \prod_{j=1}^{k} P_{a_{2j-1},a_{2j}}  \nonumber \\
&=& \prod_{i=1}^{k-1} \sqrt{3}P_{a_{2i},a_{2i+1}} \prod_{j=1}^{k} P_{a_{2j-1},a_{2j}}.
\label{Bps_expansion2}
\end{eqnarray}
Note that if we deviated from our generalized Kasteleyn orientation and used Eq.~\eqref{eq:trieq} with $n = \pm 1$, then passing from Eq.~\eqref{Bps_expansion} to \eqref{Bps_expansion2} would actually yield zero.  
Next, observe that $P_{a_{3}a_{4}} = \mathcal{F}_{a_{3}a_{4}}P_{a_{3}a_{4}} = (\mathcal{F}_{a_{3}a_{4}})^{2}P_{a_{3}a_{4}}$ [see Eq.~\eqref{eq:projdef}]. This identity allows us to rewrite $\mathcal{B}_p^s$ as
\begin{eqnarray}
\mathcal{B}_{p}^{s} &=&  (\sqrt{3}P_{a_{2k-2}a_{2k-1}})(\sqrt{3}P_{a_{2k-4}a_{2k-3}})\cdots (\sqrt{3}P_{a_{4}a_{5}}) \nonumber \\
& \times& U_{a_2a_3a_4}\prod_{j=1}^{k} P_{a_{2j-1}a_{2j}},
\label{Bps_expansion3}
\end{eqnarray}
where
\begin{equation}
  U_{a_{2}a_{3}a_{4}} = \frac{1+\mathcal{F}_{a_{2}a_{3}}\mathcal{F}_{a_{3}a_{4}} + (\mathcal{F}_{a_{2}a_{3}})^{2}(\mathcal{F}_{a_{3}a_{4}})^{2}}{\sqrt{3}}.
\end{equation}
[Because of the projectors on the far right in Eq.~\eqref{Bps_expansion3}, we can simply drop the $\mathcal{F}_{a_{3}a_{4}}$ pieces above, which replaces $U_{a_2a_3a_4} \rightarrow \sqrt{3} P_{a_2a_3}$ and recovers Eq.~\eqref{Bps_expansion2}.]
Using the fact that $\mathcal{F}_{a_{2}a_{3}}\mathcal{F}_{a_{3}a_{4}} = \omega^p \mathcal{F}_{a_{3}a_{4}}\mathcal{F}_{a_{2}a_{3}}$ for $p = \pm 1$, one can prove that $U_{a_2a_3a_4}$ is unitary.  We can similarly replace all projectors from the first line of Eq.~\eqref{Bps_expansion3} with unitary operators.  We thereby obtain \begin{equation}
\mathcal{B}_{p}^{s} = \prod_{i=1}^{k-1} U_{a_{2i}a_{2i+1}a_{2i+2}}\prod_{j=1}^{k} P_{a_{2j-1}a_{2j}} .
\end{equation}
From this form it is clear that the action of the full $B_p$ operator is indeed unitary on the subspace orthogonal to its kernel, as claimed above.

Turning next to Eq.~\eqref{ABcommutators}, $[A_{v}, A_{v'}]=0$ follows readily from the fact that $P_{ab}$ and $P_{cd}$ commute when $a \neq c,d$ and $b \neq c,d$, a direct consequence of properties 2 and 3 from Sec~\ref{sec:po}. Furthermore, $[ A_{v}, B_{p} ] = 0$ can be proven by observing that if $A_{v}$ annihilates some state $\ket{\psi}$, it also annihilates the state $B_{p}\ket{\psi}$, whereas if $A_{v}$ acts as identity on $\ket{\psi}$, it also acts as identity on $B_{p}\ket{\psi}$.  That is, $B_p$ never `corrects' a vertex violation, or produces a vertex violation that wasn't there to begin with.

\section{Physical Properties} \label{sec:Prop}

Our goal now is to validate the properties of our commuting-projector Hamiltonians quoted in the introduction.  For the decorated-domain-wall model, we will show that adding a perturbation that explicitly breaks $\mathbb{Z}_2$ spin-flip symmetry allows one to connect the ground state to a ``trivial parafermionic product state''---implying that the model realizes an SET with the same topological order as the parent quantum-Hall fluid.  To further back up this assertion we will explicitly construct anyon wavefunctions and a symmetry-action operator that explicitly permutes anyons. In the case of the decorated-toric-code model, we will construct wavefunctions corresponding to each anyon in the $\text{SU}(2)_{4}$ topological field theory.  Finally, we will show that gauging $\mathbb{Z}_{2}$ symmetry in the decorated-domain-wall model leads to the decorated-toric-code model.

\subsection{Anyons in the decorated-domain-wall model}
\label{sec:anyonsDIP}

\subsubsection{The decorated-domain-wall model adiabatically connects to the parent quantum-Hall state}
\label{sec:TrivialParafermionInsulator}

Below we follow similar logic to that introduced by several recent works \cite{Chen2014, Heinrich2016, Cheng2016}.  The decorated-domain-wall model admits a $\mathbb{Z}_2$ spin-flip symmetry that is not spontaneously broken in the ground state.  Imagine now modifying the Hamiltonian by adding Zeeman field that explicitly breaks this $\mathbb{Z}_2$, yielding a deformed model
\begin{equation}
H(\tau)  = -\sum_{v} A_{v} - \sum_{p} [ \tau B_{p} + (1-\tau) \sigma_{p}^{z} ].
\end{equation}
Note that $\sigma_{p}^{z}$, which is the spin operator for the spin at plaquette $p$, also commutes with $A_{v}$. Thus, the Hamiltonian remains a commuting-projector model at any $\tau$.  At $\tau=1$ we obtain the original decorated-domain-wall model, while at $\tau = 0$ the Hamiltonian reduces to
\begin{equation}
\label{eq:Hamtrivial}
H(0) = -\sum_{v} A_{v} - \sum_{p} \sigma_{p}^{z}.
\end{equation}
Let us discuss the ground state of Eq.~\eqref{eq:Hamtrivial}.  All spins clearly point up to minimize the energy from the $\sigma^z_p$ plaquette term.  Given this spin configuration, minimizing the $A_{v}$ vertex term requires that all parafermions pair up with their neighbors on the same edge of the honeycomb lattice.  In this sense the ground state at $\tau = 0$ forms a ``trivial parafermionic product state''.  Because the parafermions live in a fractionalized medium, however, the system still realizes a nontrivial topological order given by that of the parent quantum-Hall fluid.  

To prove the statement in the section heading, one only needs to demonstrate that the gap remains finite upon tuning $\tau$ from $0$ to $1$.  We will show that this is indeed the case by $(i)$ explicitly constructing the subspace $\mathcal{A}$ that satisfies the $A_v$ vertex term, $(ii)$ obtaining an effective Hamiltonian projected into that subspace, and $(iii)$ using the effective Hamiltonian to bound the excitation gap as a function of $\tau$.  

First, label the orthonormal ground states of $H(0)$ as $\ket{\psi_{\uparrow}^{(1)}}$, $\ket{\psi_{\uparrow}^{(2)}}$, $\cdots$, $\ket{\psi_{\uparrow}^{(n)}}$.  These states exhibit a spin configuration $|s_\uparrow\rangle$ with all spins pointing up and accordingly contain only intra-edge parafermion pairing; the superscripts account for ground-state degeneracy in multi-genus manifolds arising from the parent quantum-Hall fluid.  (In our framework, the degeneracy can be understood from analogues of the global $T_{x,y}$ operators discussed in Sec.~\ref{sec:torus} for the torus.)  From these root states we can construct wavefunctions satisfying the $A_v$ term for general Ising spin configurations $\ket{s}$ as follows: 
\begin{equation}
\begin{split}
\text{If }\ket{s} = \prod_{p}(\sigma_{p}^{x})^{n_{s,p}}\ket{s_{\uparrow}} \text{, then} \\
\ket{\psi_s^{(k)}} = \prod_{p} (B_{p})^{n_{s,p}} \ket{\psi_{\uparrow}^{(k)}}.
\end{split}
\end{equation}
In the first line $n_{s,p}$ are binary numbers for each plaquette $p$ that determine the final spin state $\ket{s}$; in the second the $B_p$ operators yield the same spin state and also reconfigure the parafermion pairings accordingly.  The set $\{ \ket{\psi_s^{(k)}} \}$ in fact spans the full subspace $\mathcal{A}$ of the Hilbert space orthogonal to the kernel of $A_{v}$.  Since $[B_{p},A_v] = 0$, it is easy to see that $A_{v}$ acts as identity on all $\ket{\psi_s^{(k)}}$ states; hence $\ket{\psi_s^{(k)}} \in \mathcal{A}$.  Next, take an arbitrary state $\ket{\tilde \psi_s} \in \mathcal{A}$ with an associated spin configuration $\ket{s}$ (which is always possible because $A_v$ commutes with $\sigma_p^z$).  By acting with $B_p$ operators we can revert to a state $\ket{\tilde \psi_\uparrow}$ for which all spins again point up and only intra-edge parafermion pairings appear: $\ket{\tilde \psi_\uparrow} = \prod_p(B_p)^{n_{s,p}}|\tilde \psi_s\rangle$.  Crucially, $\ket{\tilde \psi_\uparrow}$ can be expressed as a linear combination of $\ket{\psi_\uparrow^{(k)}}$ states defined above.  If this was not true, then we would obtain a contradiction with the assertion that $\{ \ket{\psi_{\uparrow}^{(k)}} \}$ spans the full ground-state subspace of $H(0)$.  We can therefore write
\begin{eqnarray}
\ket{\tilde \psi_s} &=& \prod_{p} (B_{p})^{n_{s,p}} \ket{\tilde \psi_{\uparrow}} = \prod_{p} (B_{p})^{n_{s,p}} \sum_{k = 1}^n a_{k}\ket{\psi_{\uparrow}^{(k)}} 
\nonumber \\
&=& \sum_{k = 1}^n a_{k}\ket{\psi_s^{(k)}}
\end{eqnarray}
for some complex numbers $a_k$, implying that any element of $\mathcal{A}$ with spin configuration $s$ can be expressed as a linear combination of $\ket{\psi_s^{(k)}}$'s.  The states $\ket{\psi_s^{(k)}}$ are also orthonormal, so that the set $\{ \ket{\psi_s^{(k)}} \}$ forms an orthonormal basis for $\mathcal{A}$.  

All states belonging to $\mathcal{A}$ automatically exhibit the `correct' parafermion pairings dictated by a given spin configuration.  The effective Hamiltonian projected into this subspace thus simplifies dramatically.  The $A_v$ vertex term projects to a constant by definition and will be discarded, while the Zeeman field $\sigma^z_p$ remains unmodified.  More importantly, we can replace the $B_p$ term simply by $\sigma^x_p$ within this subspace, yielding an effective Hamiltonian
\begin{equation}
H_{\rm eff}(\tau) =   - \sum_{p}[ \tau \sigma_{p}^{x} + (1-\tau)\sigma_{p}^{z}].
\label{eq:effH}
\end{equation}
We can maximally satisfy both remaining terms above by aligning all spins along a $\tau$-dependent direction in the $(x,z)$ plane.  Since the $A_v$ term is also maximally satisfied, we have thus established that $H(\tau)$ admits frustration-free ground states for any $\tau$, and that the ground state degeneracy is $\tau$-independent.

Finally, let us put a bound on the spectral gap.  Violation of a single $A_{v}$ term yields an energy penalty of $1$.  A single plaquette-term violation, as seen from \eqref{eq:effH}, costs an energy $2\sqrt{\tau^{2} + (1-\tau)^2}$. The energy gap $\Delta$ therefore 
remains finite for any $\tau$, precluding a phase transition.  We conclude that the decorated-domain wall model can be adiabatically deformed to the trivial parafermion product state on breaking $\mathbb{Z}_{2}$ spin-flip symmetry.  Thus, its topological order should be identical to that of the parent quantum-Hall state.

\subsubsection{Anyonic excitations} \label{sec:ae1}

\begin{figure}
  \includegraphics[width=1\linewidth]{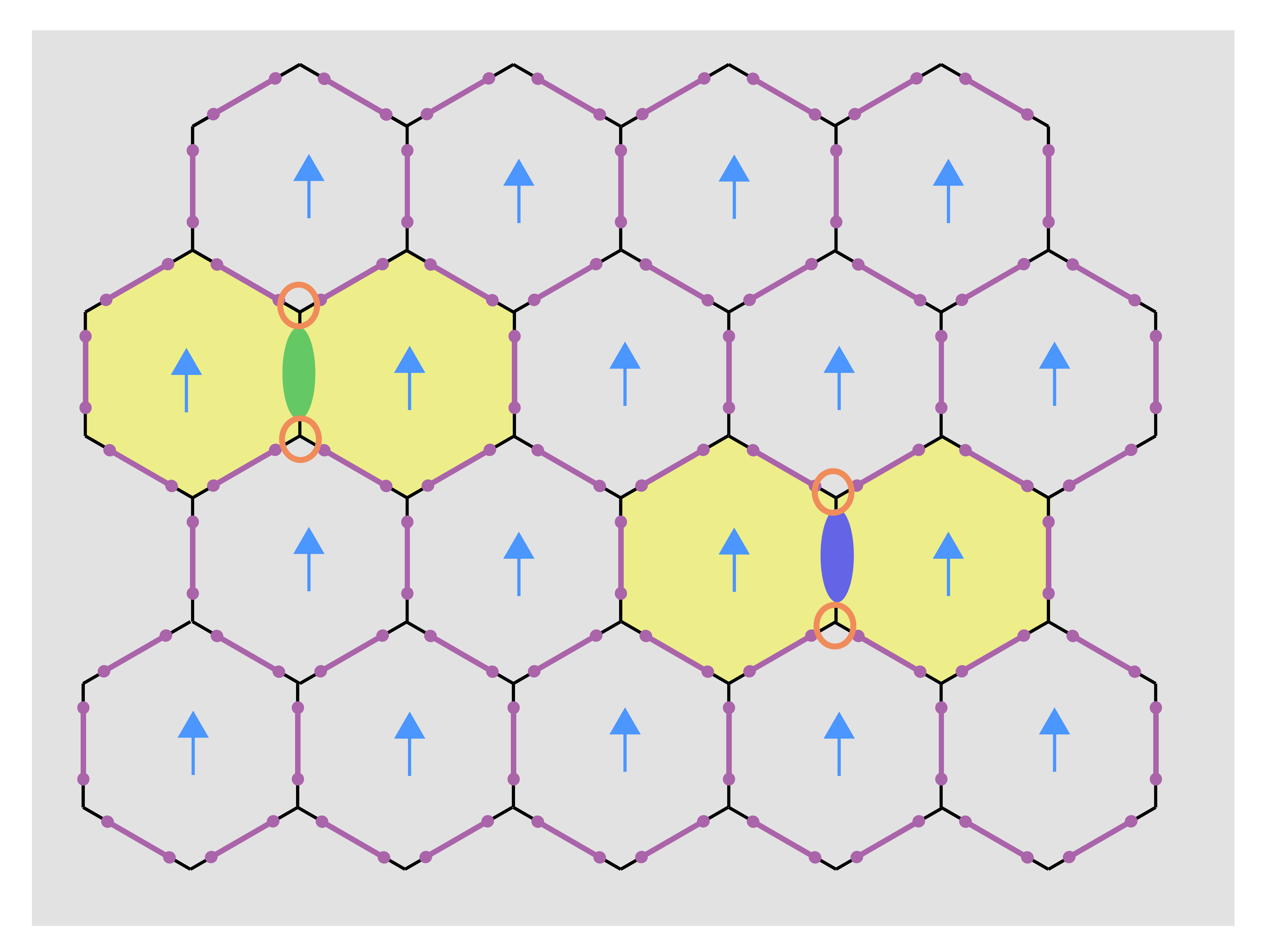} 
  \caption{Illustration of the wavefunction $\ket{\psi_\uparrow; \omega}$ that describes a trivial parafermion insulator with a pair of anyon excitations at bonds $\mu_{1,2}$.  All spins point up, and parafermions correspondingly exhibit only intra-edge pairing.  Anyons are created by assigning the `wrong' eigenvalue $e^{i\pi Q_{\mu_{1}}^{+}} = \omega$ to the green bond and $e^{i\pi Q_{\mu_{2}}^{+}} =\bar{\omega}$ to the blue bond.  This state is annihilated by the four $A_{v}$ vertex terms denoted by orange circles, as well as the four $B_{p}$ plaquette terms marked as yellow.}
  \label{fig:anyon1}
\end{figure} 
 
Next we explicitly construct Hamiltonian eigenstates that correspond to anyonic excitations of the decorated-domain-wall model.  As a primer, we will discuss anyons in the `trivial parafermion insulator' described by Eq.~\eqref{eq:Hamtrivial}---which again adiabatically connects to the decorated-domain-wall Hamiltonian.  Recall that the ground state of the trivial parafermion insulator has all spins up, with parafermions paired in a way that $e^{i \pi Q_{\mu}^{+}} = 1$ for all edges. To create a pair of anyons, simply change the $e^{i \pi Q_{\mu}^{+}}$ eigenvalue for edge $\mu_1$ to $\omega$, and the eigenvalue for a sufficiently far away edge $\mu_2$ to $\bar{\omega}$.  Figure~\ref{fig:anyon1} illustrates the resulting state, which we denote $\ket{\psi_\uparrow; \omega}$.  This wavefunction violates the four $A_{v}$ terms that involve edges $\mu_{1,2}$---see blue circles in Fig.~\ref{fig:anyon1}---yielding an excitation energy of $+4$.  Increasing the separation between $\mu_1$ and $\mu_2$ does not change the energy cost; hence the excitations are deconfined.  
 
Due to global triality conservation, $\ket{\psi_\uparrow;\omega}$ cannot be obtained from the ground state by applying local operators acting in the vicinity of $\mu_{1,2}$; that is, one cannot create charge $\omega$ or $\bar \omega$ locally from the vacuum.    
Instead one must locally create a \emph{pair} of charges $\omega$ and $\bar{\omega}$, and then pull them apart via a string operator.  Different local charges of $e^{i \pi Q_{\mu}^{+}}$ should thus be viewed as different superselection sectors of the trivial parafermion insulator.  Indeed, the three superselection sectors associated with $e^{i \pi Q_\mu^+} = 1,\omega,\bar\omega$ correspond to the three anyon charges of the parent (221) fractional-quantum-Hall state (discussed further below).  
 
To construct an eigenstate that contains a pair of anyons in the full decorated-domain-wall model, we start from $\ket{\psi_\uparrow; \omega}$ and allow spins to fluctuate by applying $B_{p}$ operators.  Denote this decorated-domain-wall excited state by $\ket{\Psi_{\rm DDW}; \omega\uparrow}$; explicitly, we have
\begin{equation}
\label{eq:Anyon1}
\ket{\Psi_{\rm DDW}; \omega\uparrow} =  \prod_{p}(1+B_p) \ket{\psi_\uparrow;\omega}.
\end{equation}
(The reason for appending an $\uparrow$ label next to $\omega$ on the left-hand side will become clear shortly.)
Clearly $\ket{\Psi_{\rm DDW}; \omega\uparrow}$ is an eigenstate of all $B_{p}$ terms. Also, since $B_{p}$ commutes with $A_{v}$ and $\ket{\psi_\uparrow; \omega}$ is an eigenstate of each $A_{v}$, $\ket{\Psi_{\rm DDW}; \omega\uparrow}$ is an eigenstate of every $A_{v}$ term as well.  So $\ket{\Psi_{\rm DDW}; \omega\uparrow}$ is indeed an eigenstate of the decorated-domain-wall Hamiltonian. This state violates the four $A_{v}$ terms involving edges $\mu_{1,2}$, exactly as for $\ket{\psi_\uparrow; \omega}$. Additionally, at the four plaquettes touching edges $\mu_{1,2}$---colored yellow in Fig.~\ref{fig:anyon1}---$B_p$ actually annihilates $\ket{\psi_\uparrow; \omega}$ and thus $\ket{\Psi_{\rm DDW}; \omega\uparrow}$.  
(Note that $B_{p}$ annihilates states with the `wrong' parafermion pairings around plaquette $p$---in $\ket{\psi_\uparrow; \omega}$ such pairings appear at edges $\mu_{1}$ and $\mu_{2}$.)   The total excitation energy for $\ket{\Psi_{\rm DDW}; \omega\uparrow}$ is then $+8$.  This energy cost again does not change upon increasing the separation between the anyons, implying that they remain deconfined here.

 The fact that there are $B_{p}$ terms that annihilate $\ket{\psi_\uparrow; \omega}$ yields an interesting consequence: each anyon has an associated Ising spin.  For the decorated-domain-wall state $\ket{\Psi_{\rm DDW}; \omega\uparrow}$, this property implies that the two plaquette spins neighboring edges $\mu_{1,2}$ are frozen to spin up, hence the $\uparrow$ label added in the ket.  
We can similarly define a state $\ket{\psi_\downarrow; \omega}$ that is identical to $\ket{\psi_\uparrow; \omega}$ except that all spins point down instead of up.  The state $\ket{\Psi_{\rm DDW}; \omega\downarrow} =  \prod_{p}(1+B_p) \ket{\psi_\downarrow;\omega}$ obtained from a trivial generalization of Eq.~\eqref{eq:Anyon1} corresponds to anyons carrying down spins.  Wavefunctions $\ket{\Psi_{\rm DDW}; \omega\uparrow}$ and $\ket{\Psi_{\rm DDW}; \omega \downarrow}$, together with their cousins $\ket{\Psi_{\rm DDW}; \bar\omega\uparrow}$ and $\ket{\Psi_{\rm DDW}; \bar\omega\downarrow}$, describe states with a pair of deconfined anyons in the decorated-domain-wall model that fuse to identity.

\subsubsection{Symmetry action on anyons} \label{sec:sa1}

As noted above the parent (221) state supports three anyon types: a trivial particle $I$ and two nontrivial particles $a$ and $a'$.  The underlying topological field theory is invariant under interchanging $a \leftrightarrow a'$, which will be essential in what follows.  At this point we have established that the decorated-domain-wall model realizes the same topological order, and that excited states $\ket{\Psi_{\rm DDW}; \omega,\sigma^z}$ and $\ket{\Psi_{\rm DDW}; \bar\omega,\sigma^z}$ contain a nontrivial anyon carrying Ising spin $\sigma^z$ at edge $\mu_1$.  We have \emph{not}, however, identified these four wavefunctions with a particular anyon type $a$ versus $a'$.  Our goal now is to determine this correspondence and also to infer how the global $Z_2$ spin-flip symmetry enjoyed by the decorated-domain-wall model acts on the anyons.  

These two objectives in fact closely relate to each other.  The $Z_2$ spin-flip symmetry acts very simply on the wavefunctions,
\begin{eqnarray}
  Z_2: \ket{\Psi_{\rm DDW}; \omega,\uparrow} &\leftrightarrow& \ket{\Psi_{\rm DDW}; \omega,\downarrow},
  \nonumber \\
  \ket{\Psi_{\rm DDW}; \bar\omega,\uparrow} &\leftrightarrow& \ket{\Psi_{\rm DDW}; \bar\omega,\downarrow}.
  \label{Z2action}
\end{eqnarray}
We must distinguish between the following two scenarios.  (A)~The states $\ket{\Psi_{\rm DDW}; \omega,\uparrow}$ and $\ket{\Psi_{\rm DDW}; \omega,\downarrow}$ both correspond to anyon type $a$ while $\ket{\Psi_{\rm DDW}; \bar\omega,\uparrow}$ and $\ket{\Psi_{\rm DDW}; \bar\omega,\downarrow}$ correspond to $a'$.  In this scenario the $Z_2$ symmetry would act trivially on the anyons.  As a corollary, it would then be possible to flip the Ising spin carried by the anyons via a local operator (without changing $\omega$ or $\bar \omega$).  (B) Alternatively, the $a \leftrightarrow a'$ symmetry of the topological field theory allows for the possibility that $\ket{\Psi_{\rm DDW}; \omega,\uparrow}$ and $\ket{\Psi_{\rm DDW}; \bar \omega,\downarrow}$ correspond to anyon type $a$ while $\ket{\Psi_{\rm DDW}; \bar\omega,\uparrow}$ and $\ket{\Psi_{\rm DDW}; \omega,\downarrow}$ correspond to $a'$.  Here the $Z_2$ symmetry action from Eq.~\eqref{Z2action} would permute the anyons, implying that the decorated-domain-wall model realizes a nontrivial SET.  In this case flipping the anyon Ising spins via a local operator would require additionally sending $\omega \leftrightarrow \bar \omega$.  

We will show that scenario B prevails.  We do so by first attempting to transform between $\ket{\Psi_{\rm DDW}; \omega \uparrow}$ and $\ket{\Psi_{\rm DDW}; \omega \downarrow}$ via operators acting solely in the vicinity of edges $\mu_{1,2}$ where the anyons reside.  Doing so would require not only flipping the two spins around each anyon, but also modifying parafermions around the adjacent plaquettes to ensure pairings consistent with the flipped anyon Ising spins.  As we will see, however, reconfiguring the parafermion pairings faces a fundamental obstruction---ruling out scenario A.  We will then show explicitly that it is possible to transform between $\ket{\Psi_{\rm DDW}; \omega \uparrow}$ and $\ket{\Psi_{\rm DDW}; \bar \omega \downarrow}$ via local operators, consistent with scenario B.  

Take two spin configurations $s$ and $\bar s$ whose only difference is that the two spins neighboring edges $\mu_{1,2}$ orient up in $s$ and down in $\bar s$. Parafermion states $\ket{{\rm PF}'(s)}$ and $\ket{{\rm PF}'(\bar s)}$ exhibit pairings consistent with these spin states, except for edges $\mu_1$ and $\mu_2$ which have eigenvalues $e^{i \pi Q_{\mu_1}^+} = \omega$ and $e^{i \pi Q_{\mu_2}^+} = \bar \omega$.  
We will now focus on the anyon at $\mu_1$ for concreteness.  Let $b_{1}$, $b_{2}$, $\cdots$, $b_{2k}$ denote clockwise-ordered parafermion sites around the double plaquette adjacent to $\mu_1$ (yellow regions in Fig.~\ref{fig:anyon1}) and an even number of triangles; in $\ket{{\rm PF}'(s)}$ parafermions pair up between sites $b_{2l-1}$ and $b_{2l}$, while in $\ket{{\rm PF}'(\bar s)}$ pairs occur between $b_{2l}$ and $b_{2l+1}$.  One can prove the following modified version of Eq.~\eqref{eq:trieq} relevant for this parafermion loop:
\begin{equation}
\label{eq:trieq2}
\prod_{l=1}^{k}\mathcal{F}^\dagger_{b_{2l-1}b_{2l}} =  e^{i \pi Q_{\mu_1}^{+}}\mathcal{F}_{b_{2k}b_{1}}\prod_{l=1}^{k-1}\mathcal{F}_{b_{2l}b_{2l+1}}.
\end{equation}
On the right side we have $e^{i \pi Q_{\mu_1}^{+}} = {\omega}$ when acting on either $\ket{{\rm PF}'(s)}$ or $\ket{{\rm PF}'(\bar s)}$ due to the anyon present at $\mu_1$.  Also, $\prod_{l=1}^{k}\mathcal{F}_{b_{2l-1}b_{2l}}$ and $\mathcal{F}_{b_{2k}b_{1}}\prod_{l=1}^{k-1}\mathcal{F}_{b_{2l}b_{2l+1}}$ are local triality operators; following similar logic from Sec.~\ref{sec:ko}, they should act as the identity on both $\ket{{\rm PF}'(s)}$ and $\ket{{\rm PF}'(\bar s)}$, \emph{provided that one can generate $\ket{{\rm PF}'(\bar s)}$  from $\ket{{\rm PF}'(s)}$ by a local transformation}.  According to Eq.~\eqref{eq:trieq2} this scenario is impossible.  We conclude that there is no local operator that transforms $\ket{{\rm PF}'(s)}$ to $\ket{{\rm PF}'(\bar s)}$, and hence no local operator that toggles between $\ket{\Psi_{\rm DDW}; \omega \uparrow}$ and $\ket{\Psi_{\rm DDW}; \omega \downarrow}$.  

\begin{figure}
  \includegraphics[width=1\linewidth]{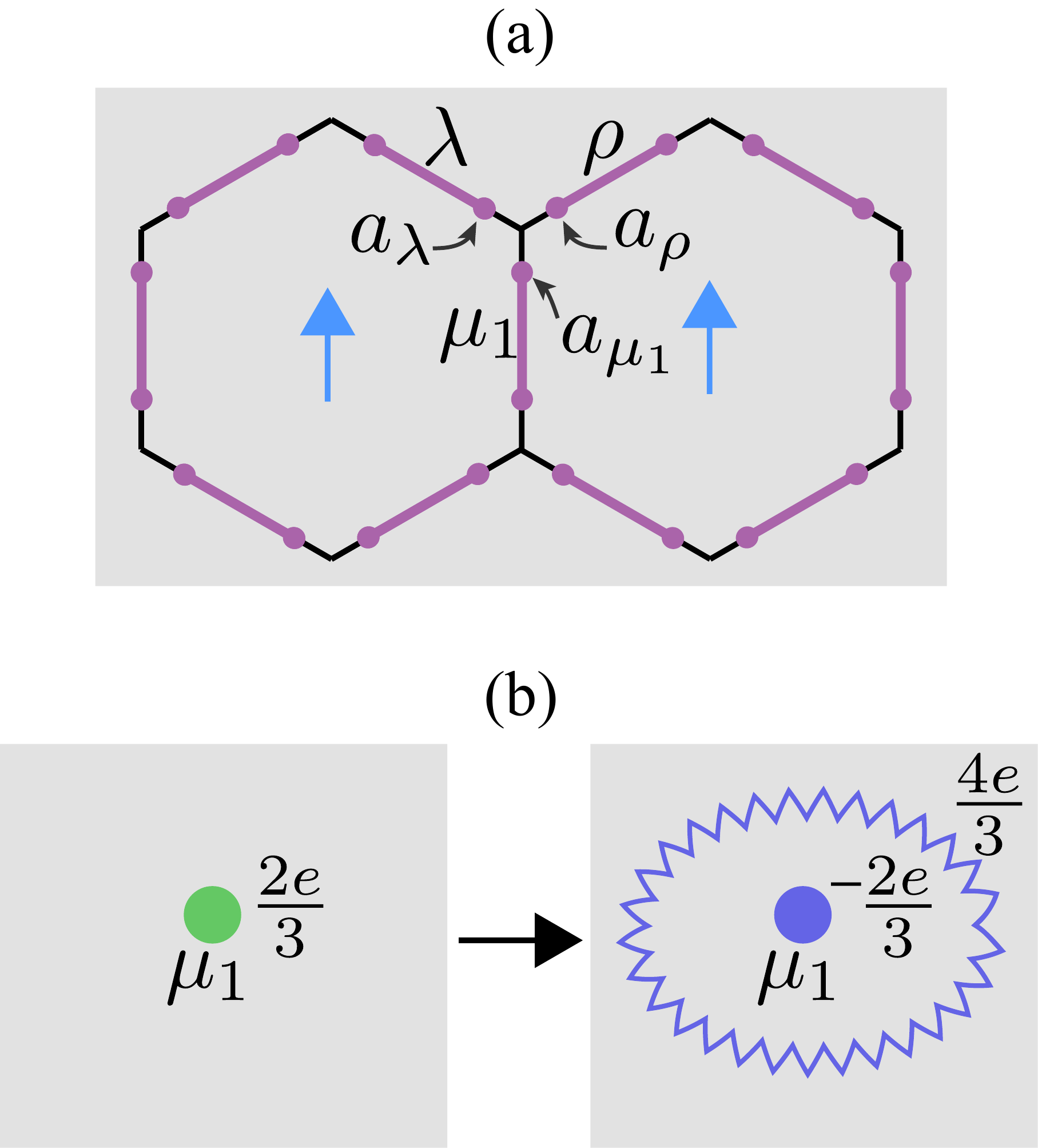} 
  \caption{(a) Definition of the edges $\mu_{1}$, $\rho$, $\lambda$, and associated parafermion sites $a_{\mu_1}, a_{\rho}, a_{\lambda}$ used in the main text. (b) Cartoon picture of the local operation $L_{\mu_{1}, \omega}$ that flips the anyon Ising spins at edge $\mu_1$ and also sends $e^{i\pi Q_{\mu_1}^+}$ from $\omega$ to $\bar\omega$. The initial $2e/3$ charge at $\mu_1$ changes to $-2e/3$; the deficit charge transfers to an adjacent parafermion loop and delocalizes across the system as the domain walls fluctuate. }  
  \label{fig:symact1}
\end{figure} 

It turns out that one can bypass the above obstruction \textit{by allowing the $e^{i \pi Q_{\mu_1}^{+}}$ eigenvalue to transform along with the Ising spins}.  Let us illustrate how this loophole arises. Define parafermion sites $a_{\mu_1}, a_{\rho}, a_{\lambda}$ as in Fig.~\ref{fig:symact1}(a), where in particular $a_{\mu_1}$ belongs to the edge $\mu_1$ hosting the anyon.   Suppose that $\ket{{\rm PF}_{\rm mod}(s)}$ exhibits parafermion pairing consistent with $s$, save for the following amendments: $(i)$ If $s$ enforces intra-edge pairing on edge $\rho$, then the state has $e^{i \pi Q_{\rho}^{+}} = \bar{\omega}$ instead of $1$.  $(ii)$ If $s$ enforces pairing between $a_{\rho}$ and $a_{\lambda}$, then the state has $\mathcal{F}_{a_{\lambda}a_{\rho}} = \omega$ instead of $1$. $(iii)$ Importantly, $e^{i \pi Q_{\mu_1}^{+}} = \bar{\omega}$ instead of $\omega$.  These locally redefined parafermion pairings underlie two key observations:
\begin{itemize}
\item One can generate $\ket{{\rm PF}_{\rm mod}(s)}$ from $\ket{{\rm PF}'(s)}$ by applying $\mathcal{F}_{a_{\rho}a_{\mu_1}}$, the operator moving fractional charge from $a_{\mu_{1}}$ to  $a_{\rho}$. 
\item There is no triality obstruction imposed by Eq.~\eqref{eq:trieq2} on transforming $\ket{{\rm PF}_{\rm mod}(s)}$ to $\ket{{\rm PF}''(\bar{s})}$, a parafermion state with pairings consistent with $\bar{s}$, except $ e^{i\pi Q_{\mu_{1}}^{+}} = e^{i\pi Q_{\mu_{2}}^{+}} = \bar{\omega}$.  
\end{itemize} 
The points above suggest that there exists a local unitary transformation $L_{\mu_{1}, \omega}$ that flips the two Ising spins adjacent to $\mu_{1}$ and alters the eigenvalue of $e^{i\pi Q_{\mu_{1}}^{+}}$ from $\omega$ to $\bar{\omega}$.  The charge at bond $\mu_1$ then changes by $2e/3$ (mod $2e$), with the excess charge transferred to an adjacent parafermion loop as sketched in Fig.~\ref{fig:symact1}(b).  Only the frozen charge at $\mu_1$ is locally conserved, however, so as domain walls fluctuate this excess charge spreads out across the system and becomes `invisible'.  
 
One can in fact define $L_{\mu_{1}, \omega}$ such that it commutes with all Hamiltonian terms, implying that $L_{\mu_{1}, \omega}$ transforms an eigenstate of the Hamiltonian into another eigenstate.  We carry out this exercise in Appendices~\ref{app:symop} and \ref{app:pcom2}.  It follows that acting $L_{\mu_{1}, \omega}$ and $L_{\mu_{2}, \bar{\omega}} = L_{\mu_{2},\omega}^{\dagger}$ implements a \textit{local unitary transformation} from $\ket{\Psi_{\rm DDW}; \omega,\uparrow}$ to $\ket{\Psi_{\rm DDW}; \bar{\omega},\downarrow}$.  These states must then realize the same anyon types as claimed above, proving that the decorated-domain-wall model realizes a nontrivial SET.

\subsection{Anyons in the decorated-toric-code model}

We now wish to similarly analyze the decorated-toric-code model and show that the anyonic excitations can be identified as deconfined quasiparticles of SU(2)$_4$.  Table~\ref{t:su24} summarizes the properties of anyons in SU(2)$_4$ (as well as its cousins, which we briefly discuss in Sec.~\ref{sec:cousins}).  The theory contains a trivial particle $I$, an Abelian self-boson $Z$, a non-Abelian particle $Y$ with quantum dimension $d = 2$, and two other non-Abelian particles $X$ and $X'\sim X \times Z$ with quantum dimension $d = \sqrt{3}$.  In the following our strategy will be to assume SU(2)$_4$ topological order and then identify microscopic incarnations of these particles---beginning with $Z$.  

\begin{table}

\caption{Anyon contents, quantum dimension $d$, and topological spin $\theta$ of $\text{SU}(2)_{4}$ theory and its cousins.}
\label{t:su24}

\begin{center}
\begin{tabularx} {\linewidth} {  >{\centering}X| >{\centering}X | >{\centering}X >{\centering}X >{\centering}X >{\centering}X  X<{\centering} }
\hline \hline

\multicolumn{2}{c|}{Anyon Charge}  & $I$ & $X$ & $Y$ & $X'$ & $Z$ \\
\hline \hline
\multicolumn{2}{c|}{$d$} & $1$ &  $\sqrt{3}$ & $2$ & $\sqrt{3}$ & $1$ \\ 
\hline 
\multirow{4}{*}{$\theta$}& $\text{SU}(2)_{4}$ & $1$ & $e^{i\frac{\pi}{4}}$ & $e^{i\frac{2\pi}{3}}$ & $-e^{i\frac{\pi}{4}}$ & $1$ \\ 
& $\text{JK}_{4}$ & $1$ & $e^{i\frac{\pi}{4}}$ & $e^{i\frac{4\pi}{3}}$ & $-e^{i\frac{\pi}{4}}$ & $1$\\
& $\overline{\text{SU}(2)_{4}}$ & $1$ & $e^{i\frac{3\pi}{4}}$ & $e^{i\frac{4\pi}{3}}$ & $-e^{i\frac{3\pi}{4}}$ & $1$\\
& $\overline{\text{JK}_{4}}$ & $1$ & $e^{i\frac{3\pi}{4}}$ & $e^{i\frac{2\pi}{3}}$ & $-e^{i\frac{3\pi}{4}}$ & $1$\\
\hline
\end{tabularx}
\end{center}

\end{table}

\subsubsection{$Z$ particle = toric-code $m$ particle}
 
The original toric-code model (without parafermion dressing) supports topological $m$-particle excitations characterized by violation of plaquette terms.  A pair of $m$ particles can be created by 
\begin{equation}
\label{eq:mstring}
W_{m} = (-1)^{\rm string~length}\prod_{j} \sigma_{j}^{z},
\end{equation}
where $j$ runs over all spin sites intersected by an open string living on the dual lattice and the prefactor is inserted to simplify signs later on.  The specific path of the string is arbitrary so long as the endpoints remain fixed.  Plaquette-term violations---and hence $m$ particles---reside at the string ends.  In the decorated-toric-code model, precisely the same string operator creates a pair of topological excitations characterized by $B_p$ violation.  
 
In the original toric code, the $m$ particle is a self-boson with quantum dimension $d = 1$.  It is natural to assume that these characteristics are inherited by the analogous topological excitations of the decorated-toric-code model, since Eq.~\eqref{eq:mstring} involves only the spin sector.  The $Z$ particle of SU(2)$_4$ exhibits identical self-statistics and quantum dimension as the toric-code $m$ particle.  Thus, we identify $Z$ with the anyon created by $W_m$ in the decorated-toric-code model.  We can construct an explicit wavefunction with $Z$ particles at plaquettes $p_1$ and $p_2$ as
\begin{eqnarray}
  \ket{\Psi_{\rm DTC};Z} &=& W_m \ket{\Psi_{\rm DTC}}
  \nonumber \\
  &=&\prod_{p = p_1,p_2}(1-B_{p})\prod_{p \neq p_{1,2}}(1+B_p)\ket{\psi_\downarrow}.
\end{eqnarray}
In the second line we expressed the ground state $\ket{\Psi_{\rm DTC}}$ using Eq.~\eqref{GroundStatesExplicit}; recall that $\ket{\psi_\downarrow}$ has only down spins and maximally satisfies all $A_v$ vertex terms.   We also used  the fact that $W_m$ anticommutes with $B_p$ operators residing at the string endpoints but commutes otherwise.  The $(1-B_p)$ factors enforce $B_p = -1$ eigenvalues at the two excited plaquettes, yielding the desired anyonic excitations.

\subsubsection{$Y$ particle = fractional charge of the parent quantum Hall state promoted to non-Abelian anyons}

 In Sec.~\ref{sec:anyonsDIP}, we constructed anyonic excitations of the decorated-domain-wall model, which essentially correspond to intra-edge parafermion bonds with $e^{i\pi Q_{\mu}^{+}} = \omega$ or $\bar{\omega}$. We can similarly construct analogous excitations for the decorated-toric-code model.  Start from the root state $\ket{\psi_\downarrow}$ with all spins pointing downward and parafermion pairings satisfying $e^{i\pi Q_{\mu}^{+}} = 1$ for all edges of the honeycomb lattice. 
 We again stress that, unlike the decorated-domain-wall model, the flipped spin configuration with all spins up is not even a valid toric-code configuration.  Next, create a state $\ket{\psi_\downarrow;\omega}$ by changing $e^{i \pi Q_{\mu_1}^{+}} = \omega$ and $e^{i \pi Q_{\mu_2}^{+}} = \bar\omega$ for some particular bonds $\mu_{1,2}$, and finally define
\begin{equation}
  \ket{\Psi_{\rm DTC}; \omega,0} = \prod_{p} (1+B_{p})\ket{\psi_\downarrow;\omega}.
  \label{Ystate}
\end{equation}
As in the decorated-domain-wall model, the four $A_{v}$ terms that contain edges $\mu_{1,2}$ and the four plaquette terms neighboring $\mu_{1,2}$ annihilate $\ket{\psi_\downarrow;\omega}$ and hence $\ket{\Psi_{\rm DTC}; \omega,0}$; moreover, on the latter state all other $A_{v}$'s and $B_{p}$'s act as identity.  It follows that $\ket{\Psi_{\rm DTC}; \omega,0}$ is an eigenstate of the decorated-toric-code Hamiltonian such that edges $\mu_1$ and $\mu_2$ carry `frozen' down spins and fixed $\mathbb{Z}_3$ charges $\omega$ and $\bar \omega$, respectively.  

The frozen down spin at $\mu_{1}$ allows us to define a topological $\mathbb{Z}_2$ index that counts the total number of toric code loops around this bond mod 2 (and similarly for $\mu_2$).  For the state in Eq.~\eqref{Ystate}, this number is even for both $\mu_1$ and $\mu_2$---hence the `0' label appended to the ket.  By replacing $\ket{\psi_\downarrow;\omega}$ with a different root configuration we can similarly construct an eigenstate $\ket{\Psi_{\rm DTC}; \omega,1}$ where the invariants are both odd.  States $\ket{\Psi_{\rm DTC}; \bar\omega,0}$ and $\ket{\Psi_{\rm DTC}; \bar\omega,1}$ with flipped $\mathbb{Z}_3$ charges can also of course be constructed.  Essentially, the locally distinguishable Ising spins carried by the anyons in the decorated-domain-wall model have been replaced by locally indistinguishable $\mathbb{Z}_2$ numbers.  

\begin{figure}
  \includegraphics[width=0.7\linewidth]{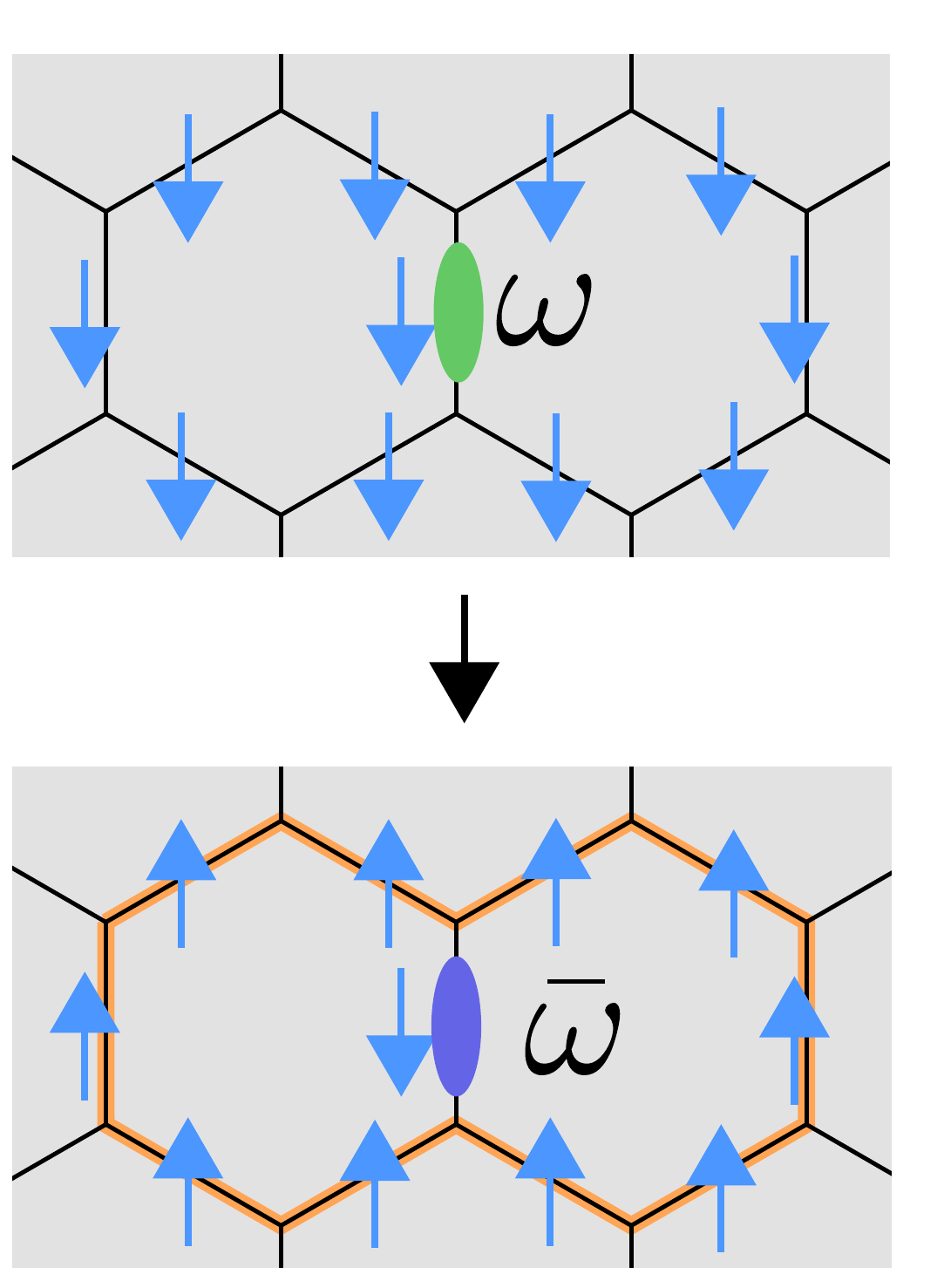} 
  \caption{Illustration of the local transformation around a $Y$ particle that flips the bound $\mathbb{Z}_{3}$ charge from $\omega$ to $\bar \omega$.  The accompanying modification to the spin configuration also necessarily changes the $\mathbb{Z}_2$ index that counts the number of toric code loops surrounding the $Y$ particle mod 2.  In the example above the $\mathbb{Z}_2$ number changes due to the additional toric-code loop highlighted in orange.}
  \label{fig:Yparticle}
\end{figure}

We can again construct a local operator that swaps $\omega \leftrightarrow \bar \omega$, \emph{but only at the expense of changing the $\mathbb{Z}_{2}$ winding number of toric-code loops around the corresponding edge}.  Consider, for example, Eq.~\eqref{Ystate} and focus on edge $\mu_1$.  The local process $(i)$ switches the charge at $\mu_1$ from $2e/3$ to $-2e/3$ while preserving the down spin at that edge, $(ii)$ moves the deficit charge to a surrounding parafermion loop, where it can then delocalize into the bulk, and $(iii)$ flips the spins along the double plaquette enclosing $\mu_1$.  See Fig.~\ref{fig:Yparticle} for an illustration.  Step $(i)$ flips the $\mathbb{Z}_3$ charge while $(iii)$ flips the $\mathbb{Z}_2$ invariant for that edge.  As for the detailed construction of the local operators, one just needs to modify the spin parts of the analogous operators $L_{\mu_{1},\omega}$ constructed for the decorated-domain-wall model; see Appendix~\ref{app:symop}. Attempting to swap $\omega \leftrightarrow \bar \omega$ while preserving the winding numbers, by contrast, faces a fundamental triality obstruction similar to what we encountered in Sec.~\ref{sec:sa1}. 
 
We thus obtain the correspondences 
\begin{eqnarray}
  \ket{\Psi_{\rm DTC}; \omega,0} &\sim& \ket{\Psi_{\rm DTC}; \bar\omega,1},
  \nonumber \\
  \ket{\Psi_{\rm DTC}; \bar \omega,0} &\sim& \ket{\Psi_{\rm DTC}; \omega,1}, 
  \label{eq:Ycorrespondence}
\end{eqnarray}
where the tildes indicate states related by local operators.  The parent (221) quantum-Hall state contains anyons $a$ and $a'$ that carry well-defined fractional charges $2e/3$ and $-2e/3$, respectively, but this charge distinction is evidently obliterated by the  $\mathbb{Z}_{3}$-charge-swapping operators.  Note also that the $Y$ particle exhibits the same topological spin as $a$ and $a'$.  Consequently, we conclude that $a,a'$ lose their identities as two separate topological excitations and merge into $Y$ in the decorated-toric-code model.  The quantum dimension $d = 2$ for the $Y$ particle naturally arises from the topological $\mathbb{Z}_2$ winding number associated with these excitations.  
 
The equivalence classes defined by the $\mathbb{Z}_{2}$ winding number can be more systematically captured using the homology group $H_{1}(M,\mathbb{Z}_{2})$.  Here $M$ denotes the original manifold for our model supplemented by $n$ holes, representing $n$ $Y$ particles.  To see how the quantum dimension of $d=2$ arises from this perspective, let us restrict to the case in which $M$ is the sphere or finite plane with $n$ holes so that there is no extra information coming from non-contractible cycles that appear without the holes. For the sphere, we have $H_{1}(M,\mathbb{Z}_{2}) = \mathbb{Z}_{2}^{n-1}$, whereas for the plane $H_{1}(M,\mathbb{Z}_{2}) = \mathbb{Z}_{2}^{n}$.  Thus, there are $O(2^{n})$ equivalent classes of toric-code loop configurations at asymptotically large $n$, in agreement with the quantum dimension $d=2$ for each $Y$-particle.  Strictly speaking, some of the states counted here might violate global triality conservation and should be excluded.  Such constraints may reduce the actual degeneracy by an $O(1)$ factor, but do not affect the asymptotic Hilbert-space dimension per $Y$ particle.  

Using our construction, we can also gain microscopic insight into fusion involving $Y$ particles.  Consider first the fusion rule $Y \times Z \sim Y$ from the SU(2)$_4$ theory.  In the decorated-toric-code model, bringing $Z$ next to $Y$ clearly preserves the topological winding number for the latter, thus again yielding an anyon with quantum dimension $d = 2$.  This anyon is most naturally associated with another $Y$ particle, consistent with SU(2)$_4$.  
As a second, more nontrivial example, SU(2)$_4$ dictates that a pair of $Y$ particles fuse according to 
\begin{equation}
  Y\times Y \sim I + Z + Y.
  \label{eq:YY}
\end{equation}  
Above we explicitly constructed eigenstates $\ket{\Psi_{\rm DTC}; \omega,0}$, $\ket{\Psi_{\rm DTC}; \omega,1}$, etc.~that contain two $Y$ particles but no other anyons.  In these wavefunctions the pair of $Y$'s exhibit opposite $\mathbb{Z}_3$ charges but the same $\mathbb{Z}_2$ winding numbers \emph{modulo the local equivalence relations summarized in Eq.~\eqref{eq:Ycorrespondence}.} Clearly such excitations must be able to fuse into the vacuum, corresponding to the identity fusion channel $I$ in Eq.~\eqref{eq:YY}.  By fusing one of the $Y$ particles with $Z$ we can also clearly access the $Z$ fusion channel in Eq.~\eqref{eq:YY}.  To recover the $Y$ fusion channel it is useful to return to the root state $\ket{\psi_\downarrow;\omega}$ that contains bonds with $e^{i \pi Q_{\mu_1}^+} = \omega$ and $e^{i \pi Q_{\mu_2}^+} = \bar\omega$.  By shuttling fractional charge from $\mu_2$ to another bond $\mu_3$, we can create a new configuration with $e^{i \pi Q_{\mu_j}^+} = \omega$ for all three excited bonds $\mu_{1,2,3}$.  (Note the preservation of global triality.)  Allowing spins and parafermions to fluctuate using $B_p$ operators then yields a Hamiltonian eigenstate in which one of the $Y$ particles in $\ket{\Psi_{\rm DTC}; \omega,0}$ has splintered into a pair of $Y$'s.  Upon running this process in reverse we see that two $Y$ particles must be able to fuse into another $Y$.  We thereby recover the full SU(2)$_4$ fusion rule in Eq.~\eqref{eq:YY}.

\subsubsection{$X$ and $X'$ particle = deconfined parafermion excitations, or decorated $e,\psi$ particles}

The $X$ and $X'$ particles from SU(2)$_4$ intuitively arise as deconfined decorated-toric-code excitations that carry unpaired $\mathbb{Z}_3$ parafermions, thus encoding the necessary $d = \sqrt{3}$ quantum dimension.  To construct Hamiltonian eigenstates that host a pair of such `deconfined parafermion excitations', we will once again start from the root state $\ket{\psi_\downarrow}$ with all spins down and $e^{i \pi Q_\mu^+} = 1$ everywhere, then deform the spins and parafermions appropriately, and finally superpose states with different spin configurations by applying a series of $B_{p}$ operators.

\begin{figure}
  \includegraphics[width=1\linewidth]{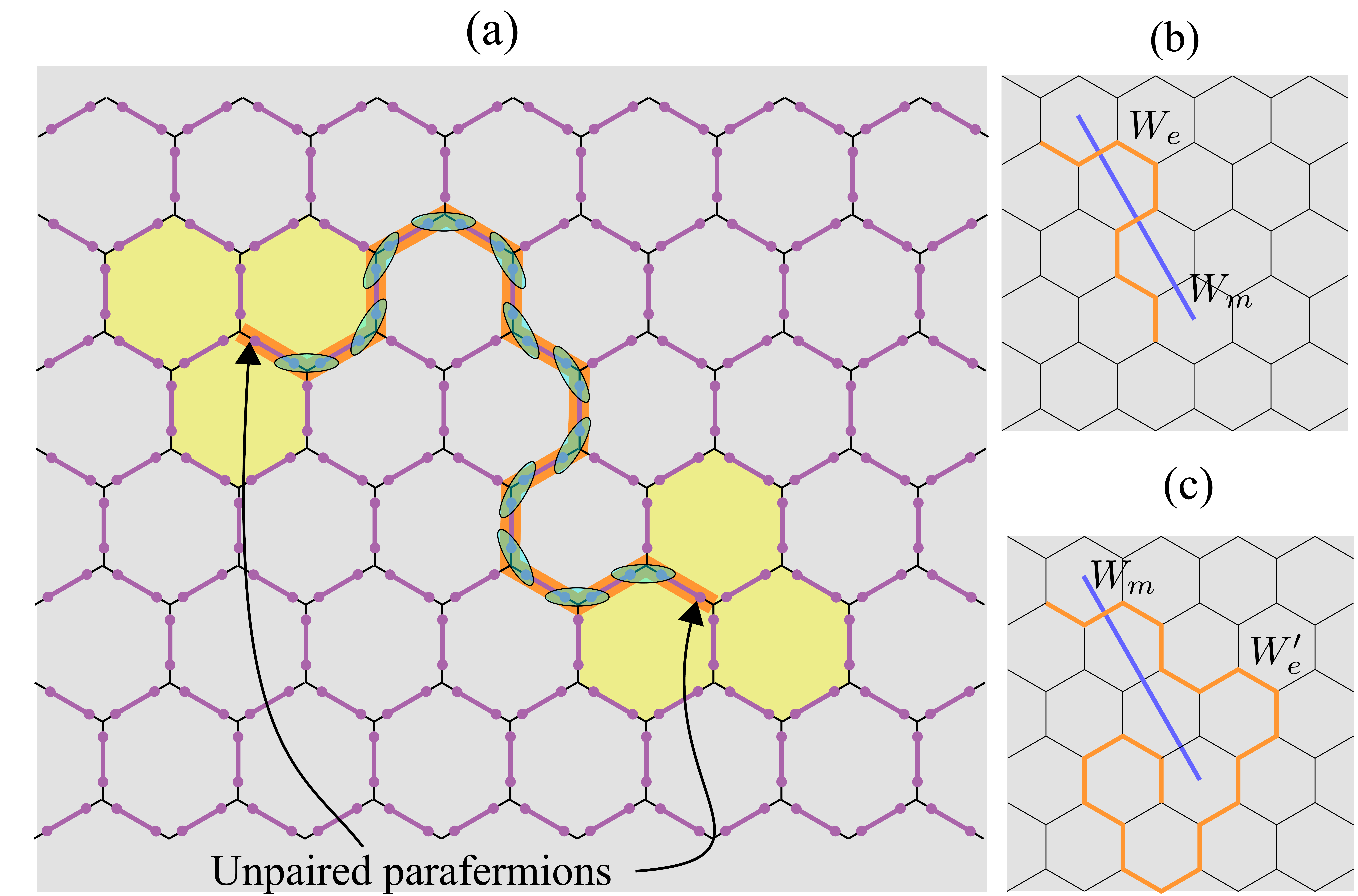} 
  \caption{(a) Action of the string operator $W_{e}$ on the root state $\ket{\psi_\downarrow}$ that contains only down spins and has $e^{i \pi Q_\mu^+} = 1$ for all bonds.  Along the string (orange line), $W_e$ flips spins and applies projectors that pair up parafermions within each oval---leaving an unpaired parafermion at each end.  Importantly, operators $B_p$ acting on any of the six plaquettes marked as yellow annihilate the resulting state.   (b,c) Given an $m$ string created by $W_m$ (blue line), one can define two topologically inequivalent strings $W_e$ and $W_e'$.  In particular, $W_e$ in (b) and $W_e'$ in (c) cross the $W_m$ string an even and odd number of times, respectively.  The parity of this crossing is topologically protected due to the constrained action of $B_p$ at the endpoints of $W_e$ and $W_e'$.}
  \label{fig:Xparticle}
\end{figure}

We will specifically deform $\ket{\psi_\downarrow}$ with the operator
\begin{equation}
W_{e} = \prod_{\mu \in l} \sigma_{\mu}^{x} \prod_{i=1}^{n-1} P_{a_{2i},a_{2i+1}},
\end{equation}
where $l$ is an open string that lives on the original honeycomb lattice and $a_{1}, a_{2}, \cdots, a_{2n}$ denote consecutively ordered parafermion sites along $l$.  The first product above flips all spins on the open string; in the undecorated toric code the exact same process generates $e$ particles at the string endpoints.  The second product reconfigures the parafermion pairings in accordance with the modified spins, \emph{leaving unpaired parafermions at both ends in the sense that neither $a_{1}$ nor $a_{2n}$ appear in any of the projectors}.  
Figure~\ref{fig:Xparticle}(a) illustrates the action of $W_{e}$ on the root state $\ket{\psi_\downarrow}$.  Applying plaquette operators on $W_{e}\ket{\psi_\downarrow}$ via
\begin{equation}
\label{eq:waveX}
\ket{\Psi,1} = \prod_{p}(1+B_{p})W_{e}\ket{\psi_\downarrow}
\end{equation}
yields an eigenstate of the decorated-toric-code model with a pair of particles that we will soon identify as superpositions of $X$ and $X'$.  Equation~\eqref{eq:waveX} violates the $A_v$ vertex term at each edge of the string.  Additionally, the three $B_p$ plaquette terms adjacent to each string end [yellow regions in Fig.~\ref{fig:Xparticle}(a)] annihilate $\ket{\Psi,1}$.  The total excitation energy is then $+8$ and does not change upon increasing the separation between endpoints, indicating deconfinement.

As a consequence of the unpaired parafermions seen above, we can tweak the root state $\ket{\psi_\downarrow}$ to construct two closely related wavefunctions that are exactly degenerate with $\ket{\Psi,1}$.  First observe that $W_e$ trivially commutes with any operator $\mathcal{F}_{a_{2i}a_{2i+1}}$ acting within the string $l$.  Applying $\mathcal{F}_{a_{2i}a_{2i+1}}$'s to $\ket{\psi_\downarrow}$ shuttles fractional charge between adjacent edges in $l$, generating all possible states with the same local string triality $\prod_{j=1}^{n} \mathcal{F}_{a_{2j-1}a_{2j}} = 1$ and $e^{i \pi Q_{\mu}^{+}} =1$ for all other edges.  Using any such state as our root configuration yields a wavefunction identical to Eq.~\eqref{eq:waveX} since the projectors in $W_e$ invariably enforce a fixed parafermion pairing.  Now consider alternate root states  $\ket{\psi_\downarrow^{(\omega)}}$ and $\ket{\psi_\downarrow^{(\bar \omega)}}$ that are the same as $\ket{\psi_\downarrow}$ except with the local string trialities respectively modified to $\prod_{j=1}^{n} \mathcal{F}_{a_{2j-1}a_{2j}} = \omega$ and $\bar \omega$.  
Strictly speaking, these states have the wrong global triality, but this problem can easily be removed by adding an extra compensating $\mathbb{Z}_3$ charge at infinity.  With this fix in mind, for now we will simply relax the global triality constraint and define new Hamiltonian eigenstates
\begin{eqnarray}
\ket{\Psi,\omega} &=& \prod_{p}(1+B_{p})W_{e}\ket{\psi_\downarrow^{(\omega)}}, 
\nonumber \\
\ket{\Psi,\bar\omega} &=& \prod_{p}(1+B_{p})W_{e}\ket{\psi_\downarrow^{(\bar\omega)}}.
\label{eq:waveX2}
\end{eqnarray}
These wavefunctions violate precisely the same terms as $\ket{\Psi,1}$, and hence the trio of states in Eqs.~\eqref{eq:waveX} through \eqref{eq:waveX2} are exactly degenerate.  One can not, however, transform these states into one another by local operators since their respective root states carry different local string trialities.  Thus, they represent three different fusion channels for the pair of deconfined parafermion excitations that we have created, implying that each particle has $\sqrt{3}$ quantum dimension.  The meaning of the fusion channels is clear: When a pair of initially distant deconfined parafermions are brought together, the string triality localizes onto a single edge, which can support three different fractional charges. 

The last piece of the puzzle to be established is the precise relation between our deconfined parafermion excitations and $X,X'$ particles from SU(2)$_4$.  The $X'$ particle actually arises from fusion of $X$ with $Z$, which is reminiscent of the formation of a fermionic $\psi$ particle by binding $e$ and $m$ in the undecorated toric code.  It is therefore illuminating to examine how Eqs.~\eqref{eq:waveX} through \eqref{eq:waveX2} evolve upon adding a $Z$ particle, via $W_m$ from Eq.~\eqref{eq:mstring}, to each end of the $e$ string created by $W_e$.  
To be precise suppose that the original $W_e$ string crosses the $m$ string an even number of times---see Fig.~\ref{fig:Xparticle}(b) for an example---so that $[W_m,W_e] = 0$.  Importantly, this crossing number defines a topological invariant: The open string of up spins created by $W_e$ fluctuates under the action of $B_p$ terms, but the parity of the $e$ and $m$ string crossings can not change since $B_p$ acts as zero on the plaquettes surrounding the $W_e$  endpoints; recall Fig.~\ref{fig:Xparticle}(a).  Furthermore, since $W_m$ also commutes with all $B_p$ operators that act nontrivially in Eqs.~\eqref{eq:waveX} through \eqref{eq:waveX2} we immediately obtain
\begin{equation}
  W_m \ket{\Psi,\omega^q} = \ket{\Psi,\omega^q}
\end{equation}
for $q = 0,1,2$.  To be consistent with SU(2)$_4$ fusion rules, the states we have constructed must therefore involve equal superpositions of $X$ and $X'$ particles so that introducing $Z$ particles returns the same state as found above.  

We can isolate $X$ and $X'$ particles by now introducing a new set of excited states $\ket{\Psi',\omega^q}$ that are identical to Eqs.~\eqref{eq:waveX} through \eqref{eq:waveX2} but with $W_e$ replaced by a string operator $W_e'$ that crosses the $m$ string an odd number of times.  See Fig.~\ref{fig:Xparticle}(c).  In this case $W_m$ and $W_e'$ anticommute, yielding
\begin{equation}
  W_m \ket{\Psi',\omega^q} = -\ket{\Psi',\omega^q}.
\end{equation}
These states must involve a different superposition of $X$ and $X'$ particles such that fusion with $Z$'s produces the original state with an extra overall minus sign.  The specific linear combinations
\begin{eqnarray}
  \ket{\Psi_{\rm DTC}; X,\omega^q} &=& \ket{\Psi,\omega^q} + \ket{\Psi',\omega^q}
  \nonumber \\
  \ket{\Psi_{\rm DTC}; X',\omega^q} &=& \ket{\Psi,\omega^q} - \ket{\Psi',\omega^q}
\end{eqnarray}
transform into one another under $W_m$, and thus are identified with decorated-toric-code excited states hosting a pair of $X$ and $X'$ particles, respectively.

\subsubsection{Cousins of SU(2)$_{4}$}
\label{sec:cousins}

Kitaev's famous 16-fold way tells us that there are $8$ `flavors' of non-Abelian Ising topological order distinguished by the topological spins of Ising anyons and their chiral central charges~\citep{Kitaev2006}. Similarly, SU(2)$_4$ topological order has cousins JK$_4$, $\overline{\text{SU(2)}_{4}}$, and $\overline{\text{JK}_{4}}$ that feature the same anyonic content but with different topological spins for the non-Abelian particles (see Table~\ref{t:su24}). In the next subsection we will show that the interface between the decorated-toric-code phase and the parent (221) state can be fully gapped, implying that the two regions must exhibit identical chiral central charge $c = 2$.  Based on this observation the topological order for the decorated-toric-code model can only be SU(2)$_4$ or $\overline{\text{JK}_{4}}$.  These two possibilities differ in the topological spins for $X$ and $X'$---which we will not attempt to compute in this paper.  Strictly speaking, we thus can not rule out $\overline{\text{JK}_{4}}$ even though we have referred to the topological order as $\text{SU}(2)_{4}$ for convenience.

Assuming that the decorated-toric-code model indeed displays $\text{SU}(2)_{4}$ topological order, one can access $\overline{\text{JK}_{4}}$ by decorating parafermions on the double-semion model \citep{Ware2016} instead of the toric code.  In the double-semion model, the analogue of $m$ particles (i.e., plaquette violations) again carry topological spin 1, while the analogues of $e$ and $\psi$ are semions and anti-semions with topological spin $\pm i$.  Said differently, the topological spins for $e$ and $\psi$ both shift by $+i$ upon passing from the toric code to double-semion model.  The topological spins for $X$ and $X'$---which can be viewed as decorated $e$ and $\psi$ particles---should thus also shift by $i$ in agreement with the $\overline{\text{JK}_{4}}$ theory.  If we instead assume that the decorated-toric-code-model realizes $\overline{\text{JK}_{4}}$ topological order, then decorating the double-semion model yields $\text{SU}(2)_{4}$ by similar reasoning. Thus, dressing two different $\mathbb{Z}_{2}$ gauge theories with parafermions allows us to access both variants of topological order associated with $\text{SU}(2)_{4}$.

\subsection{Edge structure in the decorated-domain-wall and decorated-toric-code models}

 \begin{figure}
  \includegraphics[width=1\linewidth]{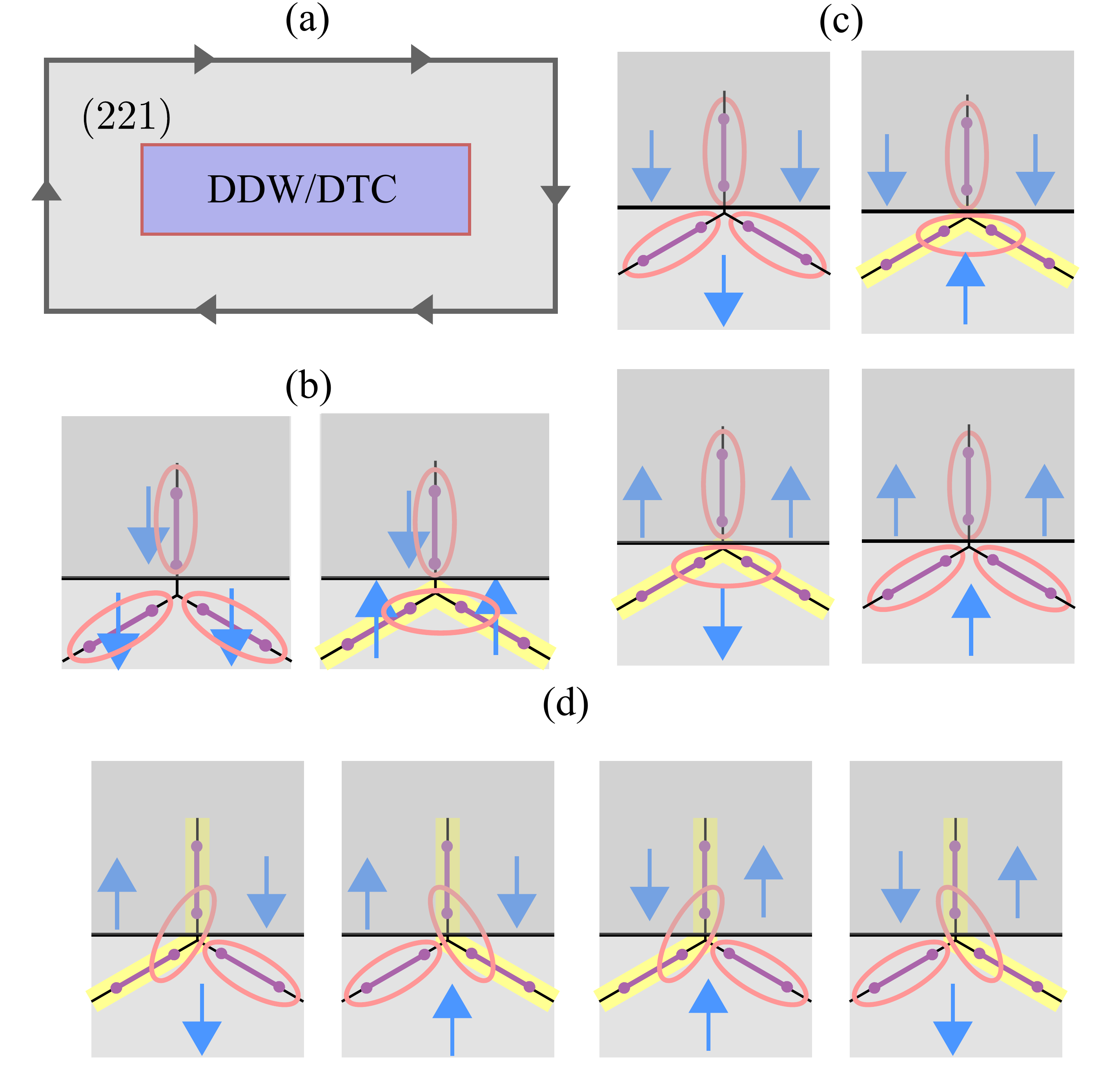} 
  \caption{(a) Setup used to study the interface between a (221) quantum Hall phase and either the decorated-domain-wall (DDW) or decorated-toric-code (DTC) model.  Arrows at the outer boundary indicate chiral edge states that occur between (221) and the vacuum.  (b) Termination for the decorated-toric-code model that yields a fully gapped interface with the (221) state.  The `dangling' edge spins are polarized downward by a Zeeman field.  (c) Possible edge terminations for the decorated-domain-wall model.  As in (b) a Zeeman field polarizes the dangling edge spins down (top panels) or up (bottom panels).  Either case yields a gapped interface with the (221) state, but at the expense of explicitly breaking $\mathbb{Z}_2$ spin-flip symmetry.  When $\mathbb{Z}_2$ symmetry is restored by allowing the boundary spins to fluctuate, a gapless interface is required on general grounds.  (d) Microscopically, one sees that domain walls in the boundary spins harbor unpaired parafermion modes.  The minimal gapless interface then naturally corresponds to a $\mathbb{Z}_3$ parafermion conformal field theory as we show explicitly in the main text.  }  
  \label{fig:boundary}
\end{figure}

Consider the geometry from Fig.~\ref{fig:boundary}(a) in which a region described by either the decorated-domain-wall or decorated-toric-code model is engulfed by a  (221) fractional-quantum-Hall phase.  Our aim here is to understand universal properties of the interface between these two nontrivial regions.  Based on these interfaces, one can easily infer the edge structure between our models and the vacuum by shrinking the outer (221) phase; this procedure merely adds an additional set of chiral (221) edge states.

First we treat the boundary between the decorated-toric-code model and $(221)$ state.  As Fig.~\ref{fig:boundary}(b) illustrates, we model their interface by terminating the decorated toric code with `dangling' edges.  Additionally, we add a perturbation $\delta H = \sum_{j\in {\rm dangling~edges}} \sigma^z_j$ that polarizes the boundary spins downward.  This setup allows vertex terms to enforce parafermion pairings for each spin configuration without generating unpaired parafermions, and the plaquette terms to fluctuate toric-code loops accordingly.  Thus, this edge termination still yields a commuting-projector description without spurious edge zero modes.  Evidently there can be a fully gapped interface between the decorated-toric-code model and $(221)$ state, which is consistent with the fact that condensing the $Z$ particle in SU(2)$_4$ gives (221) topological order.  
The boundary between the decorated toric code and vacuum then hosts exactly the same chiral edge modes as the (221) state.  
 
For the anyon-permuting SET realized by the decorated-domain-wall model, general arguments forbid a $\mathbb{Z}_2$-symmetric gapped boundary with the parent (221) state even though both sides realize the same topological order.  One can study this interface using the usual trick that folds the $(221)$ state underneath the decorated-domain-wall region, yielding an $\text{SET} \times \overline{(221)}$ bilayer state (the overline indicates the conjugate phase with opposite chirality).  A gapped interface between the decorated-domain-wall model and the (221) state translates into a gapped edge between $\text{SET} \times \overline{(221)}$ and vacuum. 
 
Levin observed that a set of quasiparticles that can be condensed to fully gap an edge form a `Lagrangian subgroup' of the full set of quasiparticle types \cite{Levin2013}.  Let $a,a'$ denote nontrivial anyons in the SET, and recall that in Sec.~\ref{sec:sa1} we showed that $\mathbb{Z}_2$ spin-flip symmetry swaps $a \leftrightarrow a'$.  The corresponding particles from $\overline{(221)}$ will be denoted $\bar a,\bar a'$.  For the $\text{SET} \times \overline{(221)}$ bilayer there are two possible Lagrangian subgroups: $\mathcal{L}_{1} =\{ 1, a\bar{a}, a'\bar{a}' \}$ and $\mathcal{L}_{2} = \{ 1, a\bar{a}', a\bar{a} \}$, but neither are closed under $\mathbb{Z}_{2}$ symmetry.  In fact $\mathbb{Z}_2$ sends $\mathcal{L}_{1}\leftrightarrow\mathcal{L}_{2}$.  Consequently, the SET-(221) interface must remain gapless so long as $\mathbb{Z}_{2}$ symmetry is preserved.  

A possible theory for the $\mathbb{Z}_2$ symmetric interface would contain a counterpropagating set of (221) edge states.  These edge states can be described with standard K-matrix formalism, though for brevity we will pursue a more phenomenological treatment that recovers the same physics.  Right-movers encode gapless avatars of the SET anyons $a \sim e^{i \varphi}$ and $a' \sim e^{i \varphi'}$, where $\varphi,\varphi'$ are chiral edge fields; under the anyon-permuting $\mathbb{Z}_2$ symmetry we have $\varphi \leftrightarrow \varphi'$.  Left-movers similarly correspond to anyons $\bar a \sim e^{i \bar\varphi}$ and $\bar a' \sim e^{i \bar\varphi'}$ from the adjacent (221) region.  Although $\mathbb{Z}_2$ symmetry precludes fully gapping the interface, half of the gapless modes can be eliminated without symmetry breaking.  Note that since $a$ and $a'$ fuse to a trivial particle, $a a' \sim e^{i(\varphi + \varphi')} \equiv b_R$ and $\bar a \bar a' \sim e^{i(\bar\varphi + \bar\varphi')}\equiv b_L$ respectively form right- and left-moving local bosons.  The tunneling term $\mathcal{H} = -t \int_x (b_R^\dagger b_L + H.c.)$ is even under $\mathbb{Z}_2$; when relevant, this coupling pins 
\begin{equation}
  (\varphi + \varphi') - (\bar\varphi + \bar\varphi') = 0,
  \label{pinning}
\end{equation}
thus opening a gap for creating isolated local bosons at the interface.  The remaining gapless degrees of freedom are described by a single right-moving chiral field $\varphi_R \equiv \varphi - \varphi'$---which is odd under $\mathbb{Z}_2$---together with a left-mover $\varphi_L \equiv \bar\varphi - \bar\varphi'$.  Interestingly, this theory \emph{still} does not provide the minimal description for the gapless $\mathbb{Z}_2$-symmetric interface.

It is useful to now repackage $\varphi_{R/L}$ into a non-chiral boson field $\Phi = (\varphi_R - \varphi_L)/2$ and its dual $\Theta = (\varphi_R + \varphi_L)/2$ that commute with themselves but obey the nontrivial commutator
\begin{equation}
  [\Phi(x),\Theta(x')] = i\frac{\pi}{3}{\rm sgn}(x-x').
  \label{commutator}
\end{equation}
Equation~\eqref{commutator} reflects the anyonic exchange statistics for (221) quasiparticles, and can be straightforwardly derived from the corresponding K-matrix.  Crucially, since $\varphi_R \rightarrow -\varphi_R$ under $\mathbb{Z}_2$, the non-chiral boson fields transform as
\begin{equation}
  \mathbb{Z}_2: \Phi \rightarrow - \Theta,~~~\Theta \rightarrow - \Phi,
\end{equation}
i.e. \emph{the local spin-flip symmetry implements a duality transformation for the edge modes}.  Let us write the self-dual $\mathbb{Z}_2$-invariant Hamiltonian as $\mathcal{H} = \mathcal{H}_0 + \mathcal{H}_1$.  Here 
\begin{equation}
  \mathcal{H}_0 = \int_x \frac{v}{2\pi}[(\partial_x\Phi)^2 + (\partial_x\Theta)^2]
\end{equation}
describes a fixed point with central charge $c = 1$, while $\mathcal{H}_1$ contains cosine terms that favor instabilities of this fixed point.  For the latter, note that $e^{i(2\varphi-\varphi')}$ creates two $a$-type edge excitations and removes one $a'$-type edge excitation; since two $a$ particles fuse to $a'$, this operator forms another local right-moving boson.  By the same reasoning $e^{i(2\bar\varphi-\bar\varphi')}$ forms a local left-moving boson.  The perturbation 
\begin{eqnarray}
  &-&u\cos(2\varphi-\varphi' - 2\bar \varphi+ \bar \varphi') 
  \nonumber \\
  &=& -u\cos\left[\frac{1}{2}\left(\varphi + \varphi' - \bar\varphi - \bar \varphi'\right) + 3\Phi\right]
\end{eqnarray}
is therefore obviously physical.  Using Eq.~\eqref{pinning} this term reduces simply to $-u \cos(3\Phi)$, which is the lowest allowed harmonic.  Applying $\mathbb{Z}_2$ symmetry yields a second physical term $-u\cos(3\Theta)$.  The leading perturbations to $\mathcal{H}_0$ are therefore given by
\begin{equation}
  \mathcal{H}_1 = -u\int_x[\cos(3\Phi) + \cos(3\Theta)],
\end{equation}
which are known to drive a flow to a new critical point described by a $\mathbb{Z}_3$ parafermion conformal field theory with $c = 4/5$ \cite{LGN}.  This is the minimal possible $\mathbb{Z}_2$-preserving theory for the gapless interface.  Reference~\onlinecite{Mong2014} employed a similar approach to access a $c = 4/5$ parafermion theory from (112) fermionic quantum Hall systems.  There, however, fine-tuning was required to preserve self-duality.  Remarkably, in our case self-duality is symmetry-enforced.  

We can gain microscopic insight into the SET-(221) interface by considering the edge termination shown in Fig.~\ref{fig:boundary}(c).  Suppose first that we uniformly polarize the boundary spins with a Zeeman field.  In this case we retain a commuting-projector description that manifestly realizes a fully gapped edge, similar to what we found for the decorated toric code, but at the expense of explicitly breaking $\mathbb{Z}_2$ spin-flip symmetry.  Figure~\ref{fig:boundary}(c), for example, shows that the parafermion pairings at the edge depend on whether we polarize the boundary spins up versus down.  Next imagine inserting a domain wall at which the boundary spin polarization flips.  From Fig.~\ref{fig:boundary}(d) we see that the domain wall binds an unpaired $\mathbb{Z}_3$ parafermion.  (References~\onlinecite{Chen2014} and \onlinecite{Tarantino2016} pointed out that related decorated-domain-wall models can also host boundary zero modes for certain spin configurations when symmetries are enforced.)  Restoring $\mathbb{Z}_2$ symmetry by allowing the spins to fluctuate then naturally yields gapless parafermion excitations as found above in our continuum treatment.  

Inspired by Grover et al.~\cite{Grover}, we can write down a phenomenological, yet microscopic, boundary Hamiltonian that captures the above properties:
\begin{eqnarray}
  H_{\rm boundary} &=& -t \sum_j[1 + (-1)^j \sigma^z_{j+1/2}](\alpha_j^\dagger \alpha_{j+1} + H.c.)
  \nonumber \\
  &-&\sum_j [J \sigma^z_{j-1/2}\sigma^z_{j+1/2} + h \sigma^x_{j+1/2}].
\end{eqnarray}
Here $t,J,h$ denote non-negative couplings, $\alpha_j$ are $\mathbb{Z}_3$ parafermions that live on integer sites $j$, and $\sigma^{z,x}_{j+1/2}$ denote boundary spin operators that live on half-integer sites.  This Hamiltonian preserves $\mathbb{Z}_2$ spin-flip symmetry implemented as $\sigma^z_{j+1/2}\rightarrow - \sigma^z_{j+1/2}$, $\alpha_j \rightarrow \alpha_{j+1}$.   At $h/J = 0$ the spins order ferromagnetically in the ground state, which in turn causes the parafermions to dimerize and fully gap out via the $t$ term.  Magnetic domain walls produce a kink in the dimerization pattern yielding an unpaired parafermion---precisely as in the decorated-domain-wall model.  And finally, in the opposite limit $h/J \rightarrow \infty$ the transverse field disorders the spins so that $\langle \sigma^z\rangle = 0$, leaving a translation-invariant parafermion chain that indeed realizes a $c = 4/5$ parafermion conformal field theory.  In future work it would be interesting to study properties of the phase transition(s) that arise upon passing between the extreme limits of $h/J$ described above.

\subsection{Gauging equivalence}

Here we follow Ref.~\onlinecite{Levin2013} and \onlinecite{Heinrich2016} to explicitly show that the decorated-toric-code phase emerges upon gauging the $\mathbb{Z}_{2}$ symmetry of the decorated-domain-wall model.  We do so by defining a variant of the decorated-domain-wall model on which gauging can be more conveniently implemented.  This new model's ground state is, up to a local unitary transformation and tensor product of a trivial state, exactly the same as that of the original decorated-domain-wall model. Gauging this cousin model gives a variant of the decorated-toric-code Hamiltonian whose ground state is that same as in the original decorated toric code, again up to a local unitary transformation and stacking of a trivial state. 
  
The Hilbert space for the modified decorated-domain-wall model consists of the usual plaquette spins and parafermions supplemented by an extra spin-1/2 degree of freedom on each edge of the original honeycomb lattice.  We will refer to Ising spin configurations for the supplemental edge spins as $w$, while we continue to denote configurations for  the plaquette spins as $s$.  The ground state of the modified decorated-domain-wall model will take the form
\begin{equation}
\ket{\Psi_{\rm{vDDW}}} = \sum_{s} \ket{s}\ket{w_{s}}\ket{\rm{PF}(s)}.
\end{equation}
Here, the edge-spin configuration $w_{s}$ is determined by $s$ as follows: Supplemental edge spins point up along domain walls in $s$ but point down elsewhere.  

To see how $\ket{\Psi_{\rm{vDDW}}}$ relates to the original decorated-domain-wall ground state $\ket{\Psi_{\rm{DDW}}}$, consider the following unitary transformation defined on each edge $\mu$:
\begin{equation}
U_{\mu} = \frac{1 - \sigma_{p_{\mu}}^{z}\sigma_{p'_{\mu}}^{z}}{2} + \gamma_{\mu}^{x}\frac{1 + \sigma_{p_{\mu}}^{z}\sigma_{p'_{\mu}}^{z}}{2},
\end{equation}
where $p_{\mu}$  and $p'_{\mu}$ label the two plaquettes neighboring edge $\mu$ while $\gamma^a$ denote Pauli matrices that act on supplemental edge spins.  The unitary
$U_{\mu}$ flips the edge spin when there is no domain wall between $p'_{\mu}$ and $p_{\mu}$ but otherwise leaves the edge spin intact.  Consequently, acting $U_{\mu}$ on $\ket{\Psi_{\rm{vDDW}}}$ transforms $\ket{w_{s}}$ to the state $\ket{w = \uparrow}$ with all edge spins up, i.e.,
\begin{equation}
\prod_{\mu} U_{\mu} \ket{\Psi_{\rm{vDDW}}} = \ket{w = \uparrow}\ket{\Psi_{\rm{DDW}}}.
\end{equation}
In this sense $\ket{\Psi_{\rm{vDDW}}}$ describes the same phase as the original decorated-domain-wall model.  

Next we introduce operators $\tilde{A}_{v}^{\rm{DTC}}$ and $\tilde{B}_{p}^{\rm{DTC}}$ that are identical to the vertex and plaquette terms of the decorated-toric-code model, except that the spin parts act now on the $w$ sector.  They can then be regarded as valid operators within our modified decorated-domain-wall model.  Additionally, define an edge term $C_{\mu}$ as  
\begin{eqnarray}
C_{\mu} &=& \frac{1+\gamma_{\mu}^{z}}{2}\frac{1 - \sigma_{p_{\mu}}^{z}\sigma_{p'_{\mu}}^{z}}{2} + \frac{1-\gamma_{\mu}^{z}}{2}\frac{1 + \sigma_{p_{\mu}}^{z}\sigma_{p'_{\mu}}^{z}}{2}
 \nonumber \\
 &=& 1-\gamma_{\mu}^{z}\sigma_{p_{\mu}}^{z}\sigma_{p'_{\mu}}^{z},
 \label{Cmu}
\end{eqnarray}
which acts as identity when $w$ is consistent with the domain-wall configuration dictated by $s$, and zero otherwise. One can explicitly show that
\begin{equation}
H_{{\rm vDDW}} = - \sum_{v} \tilde{A}_{v}^{\rm{DTC}} - \sum_{p} \tilde{B}_{p}^{\rm{DTC}}\sigma^{x}_{p} - \sum_{\mu} C_{\mu}
\label{HvDDW}
\end{equation}
forms a commuting-projector Hamiltonian with $\ket{\Psi_{\rm{vDDW}}}$ as a ground state. 

 Gauging $\mathbb{Z}_{2}$ symmetry can be performed easily and explicitly on the above Hamiltonian. Gauging introduces another spin-1/2 degree of freedom on each edge; let us denote Pauli matrices acting on these new spins by $\eta^a$. To ensure gauge invariance, the gauged model is only defined on the restricted Hilbert space that satisfies the following local constraint for each plaquette $p$: 
\begin{equation}
\label{eq:gaugeconstraint}
\sigma^{x}_{p}\prod_{ \{\mu_{p} \} }\eta^{x}_{\mu_{p}} = 1,
\end{equation}
where $\{\mu_{p} \}$ denotes the set of honeycomb-lattice edges around $p$.  Note that Eq.~\eqref{eq:gaugeconstraint} is invariant under changing the sign of the three $\eta^x$ operators adjacent to any vertex.  The gauged Hamiltonian reads
\begin{equation}
H_{{\rm vDDW}}^{g} =  -\sum_{v} G_{v}- \sum_{v} \tilde{A}_{v}^{\rm{DTC}} - \sum_{p} \tilde{B}_{p}^{\rm{DTC}}\sigma^{x}_{p} - \sum_{\mu} \tilde{C}_{\mu}.
\end{equation}
In the first term we introduced a new vertex operator
\begin{equation}
G_{v} =  \eta^{z}_{\mu_{v}}\eta^{z}_{\rho_{v}}\eta^{z}_{\lambda_{v}}
\end{equation}
with $\mu_{v}$, $\rho_{v}$, $\lambda_{v}$ the three edges meeting at vertex $v$.  This piece---which clearly commutes with the gauge constraint in Eq.~\eqref{eq:gaugeconstraint}---energetically favors vanishing gauge flux around vertices of the honeycomb lattice.  
In the the final term we promoted $C_\mu$ from Eqs.~\eqref{Cmu} and \eqref{HvDDW} to 
\begin{equation}
\tilde{C}_{\mu} = 1 - \gamma_{\mu}^{z}\sigma_{p_{\mu}}^{z}\eta^{z}_{\mu}\sigma_{p'_{\mu}}^{z},
\end{equation}
which also commutes with Eq.~\eqref{eq:gaugeconstraint}.  
One can readily check that the full Hamiltonian $H_{{\rm vDDW}}^{g}$ defines a gauge-invariant commuting-projector model for which the global $\mathbb{Z}_2$ spin-flip symmetry has been promoted to a gauge symmetry associated with \emph{local} signs changes in adjacent $\sigma^z$ and $\eta^z$ operators.

 The ground state of $H_{{\rm vDDW}}^{g}$ can be conveniently investigated by noticing that the plaquette and edge spins bound by the gauge constraint \eqref{eq:gaugeconstraint} can be represented by the combination $\sigma_{p_{\mu}}^{z}\eta^{z}_{\mu}\sigma_{p'_{\mu}}^{z}$, 
which we view as a pseudospin for edge $\mu$. 
Since all Hamiltonian terms can be maximally satisfied, let us consider each piece individually.  The first vertex term, $G_{v}$, favors configurations for which an even number of pseudospins around each vertex point downward; that is, pseudospins form `anti-toric-code' configurations consisting of closed loops of down spins in a background of up spins.  
Our second vertex term, $\tilde {A}_{v}^{\rm{DTC}}$, enforces the supplemental $\gamma^z$ edge spins and parafermions to satisfy the decorated-toric-code local rules summarized in Fig.~\ref{fig:localrule}(b). The plaquette term $\tilde{B}_{p}^{\rm{DTC}}\sigma^{x}_{p}$ flips pseudospins (via the $\sigma^x$) and $\gamma^z$ edge spins and also reconfigures parafermion pairings accordingly. Finally, $\tilde{C}_{\mu}$ forces pseudospins and $\gamma^z$ edge spins to anti-align. Putting everything together, the ground-state wavefunction of $H_{{\rm vDDW}}^{g}$ can be written as
\begin{equation}
\ket{\Psi_{\rm{vDTC}}} = \sum_{w \in \{ s_{v} \} } \ket{w}\ket{\bar{w}} \ket{\rm{PF}(s)},
\end{equation}
where $\bar{w}$ denotes the `anti-toric-code' pseudospin configuration opposite that of the $\gamma^z$ toric-code spin configuration $w$. 

We can expose a simple relation between $\ket{\Psi_{\rm{vDTC}}}$ and the original decorated-toric-code ground state $\ket{\Psi_{\rm{DTC}}}$ using a unitary transformation $V_\mu$ that acts according to 
\begin{equation}
\begin{split}
V_{\mu}: \ket{\uparrow}\ket{\downarrow} \rightarrow \ket{\uparrow}\ket{\uparrow} \\
 \ket{\downarrow}\ket{\uparrow} \rightarrow \ket{\downarrow}\ket{\uparrow}\\
 \ket{\uparrow}\ket{\uparrow} \rightarrow \ket{\uparrow}\ket{\downarrow}\\
 \ket{\downarrow}\ket{\downarrow} \rightarrow \ket{\downarrow}\ket{\downarrow},
\end{split}
\end{equation}
where the first and second kets respectively indicate $\gamma^z$ spins and $\sigma_{p_{\mu}}^{z}\eta^{z}_{\mu}\sigma_{p'_{\mu}}^{z}$ pseudospins for edge $\mu$.  
When acting $V_{\mu}$ on $\ket{\Psi_{\rm{vDTC}}}$, the first two lines above remove all `anti-toric-code' down-spin loops for the pseudospins, so that 
\begin{equation}
\prod_{\mu} V_{\mu} \ket{\Psi_{\rm{vDTC}}} = \ket{{\rm pseudospin} \uparrow}\ket{\Psi_{\rm{DTC}}}.
\end{equation}
(Only the first two lines in the definition for $V_\mu$ are used here; the last two lines are arbitrary provided $V_\mu$ remains unitary.)  
Thus, the ground state of $H_{{\rm vDDW}}^{g}$ is connected to the ground state of the decorated-toric-code model by a local unitary transformation and stacking of a trivial product state. This shows that the topological phase of the decorated toric code can be accessed by gauging the $\mathbb{Z}_{2}$ symmetry of the decorated-domain-wall model.

\section{Ground-State Degeneracy on a Torus} \label{sec:Torus}

 In this section, we will investigate the ground-state degeneracy of the decorated-domain-wall and decorated-toric-code models defined on a torus.  We will observe that our formalism developed in Sec.~\ref{sec:Review} yields the expected degeneracy for both models given the topological orders identified in the previous section.  Additionally, we will see that the decorated-domain-wall ground states exhibit a unique feature related to the nontrivial symmetry action that signals an SET phase.

\subsection{Degenerate ground states in the decorated-domain-wall model}
 
Section~\ref{sec:TrivialParafermionInsulator} briefly discussed ground states for the decorated-domain-wall model on a torus.  There we established that, by adiabatic continuity, the topological degeneracy is identical to that in the trivial parafermion insulator Hamiltonian from Eq.~\eqref{eq:Hamtrivial}.  In the latter Hamiltonian, each term is maximally satisfied by fixing $e^{i\pi Q_{\mu}^{+}} =1$ for each bond and taking all spins up.  The Hamiltonian does not, however, impose any constraint on the global degrees of freedom controlled by $T_x$ or $T_y$.  Thus, we conclude that there should be three ground states, $\ket{\psi^{(1)}_{\uparrow}}$, $\ket{\psi^{(2)}_{\uparrow}}$, and $\ket{\psi^{(3)}_{\uparrow}}$, associated with the three inequivalent eigenvalues of either $T_x$ or $T_y$.  Ground states $\ket{\Psi^{(1,2,3)}_{\rm DDW}}$ of the original decorated-domain-wall model are recovered by letting the spins and parafermion pairings fluctuate using $B_p$ operators, i.e., 
\begin{equation}
\label{eq:gssup}
\ket{\Psi^{(i)}_{\rm DDW}} = \prod_{p} (1+B_{p})\ket{\psi^{(i)}_{\uparrow}}.
\end{equation}

There is an interesting $\mathbb{Z}_{2}$ symmetry property on ground states of the decorated-domain-wall model. In the ground-state subspace, the action of $\mathbb{Z}_2$ is captured by the $3 \times 3$ matrix $M$ defined by
\begin{equation}
\label{eq:gsymact}
M_{ij} = \bra{\Psi^{(i)}_{\rm DDW}} \prod_{p} \sigma_{p}^{x} \ket{\Psi^{(j)}_{\rm DDW}}.
\end{equation}
Equation~\eqref{eq:gssup} allows us to reduce the above expression to
\begin{equation}
\label{eq:prodbp}
M_{ij} = \bra{\psi^{(i)}_{\uparrow}} \prod_{p} B_{p}\sigma_{p}^{x} \ket{\psi^{(j)}_{\uparrow}}.
\end{equation}
We do not know how to calculate $M$ with a general parafermion ordering.  However, given a particular ordering in a minimal $2 \times 2$ system, $M$ can be obtained numerically.  We checked that in three different orderings $M$ is given by
\begin{equation}
\label{eq:defprodbp}
M = \begin{pmatrix} 1 & 0 & 0 \\ 0 & 0 & \omega \\ 0 & \omega^{2} & 0 \end{pmatrix}.
\end{equation}
Since $M$ is related to topological properties of the system (see below), it is natural to expect that this result applies for any valid parafermion ordering and any system size. 

The form of $M$ above implies that \textit{global $\mathbb{Z}_2$ symmetry action permutes two ground states $\ket{\Psi^{(2)}_{\rm DDW}}$ and $\ket{\Psi^{(3)}_{\rm DDW}}$}. This result is consistent with the observation in Ref.~\onlinecite{Lu2016} that an SET characterized by an anyon-permuting symmetry action should have minimally entangled ground states that are also permuted by the symmetry action.  There is one caveat: We do not have access to entanglement entropy in this work and thus, strictly speaking, do not know whether the basis for the ground state subspace that gives off-diagonal symmetry action corresponds to minimally entangled states. Indeed, upon diagonalizing Eq.~\eqref{eq:defprodbp}, one obtains a different linear combination of ground states in which the symmetry action is simply $\pm 1$. Still, we argue that exhibiting anyon-permuting symmetry action in any basis is a highly nontrivial consistency check on both our torus formalism in Sec.~\ref{sec:torus} and our identification of the decorated-domain-wall model as an anyon-permuting SET.

\subsection{Degenerate ground states in the decorated-toric-code model}

 Recall that on a torus, toric-code spin configurations can be classified into four topological sectors labeled by two $\mathbb{Z}_2$ numbers $(a,b)$.  Here $a$ and $b$ are the crossing numbers of toric-code loops across the two non-contractible cycles of the torus, which can not change under any local operator.  Starting from one spin configuration in a given topological sector, all other spin configurations in the same sector can be obtained by applying toric-code plaquette terms appropriately.  
 The four degenerate toric-code ground states arise from superpositions of spin configurations within each of the four topological sectors. 
 
In the decorated-toric-code model, we expect five-fold degeneracy on a torus, reflecting the five anyon types in SU(2)$_4$ topological order.  Recovering this counting poses an interesting puzzle.  Since we have only four topological $(a,b)$ sectors coming from the spin degrees of freedom, consistency requires that at least one of these sectors must support multiple ground states.  Previously we observed that in the decorated-domain-wall model, there are actually three ground states with identical superpositions of spin configurations, but different global properties arising from $T_{x,y}$ operators.  For the decorated toric code, one might similarly expect a triplet of states within each topological $(a,b)$ sector---but this extrapolation predicts an excessive twelve-fold ground-state degeneracy.  We will resolve this conundrum by showing that some consistency conditions intimately related to topological properties of the $(221)$ state and the decorated-domain-wall model eliminate many of these twelve putative ground states, leaving only five as required for SU(2)$_4$.

\subsubsection{$(0,0)$ sector}
\label{00sector}

The $(0,0)$ sector is simplest to examine.  Recall that there are three states with all spins down and all $A_{v}$ terms acting as identity (the trio reflects global $T_{x,y}$ operators, exactly as in the decorated-domain-wall model).  Call these states $\ket{\phi_{\downarrow}^{(1)}}$, $\ket{\phi_{\downarrow}^{(2)}}$, and $\ket{\phi_{\downarrow}^{(3)}}$.  One can attempt to construct three ground states from each of these root configurations.  As we will see, however, there is a fundamental obstruction that allows only two ground states in the $(0,0)$ sector.   

Let $\ket{\psi}$ be any frustration-free ground state, i.e., $B_{p}\ket{\psi} = \ket{\psi}$ and $A_{v}\ket{\psi} = \ket{\psi}$ for all $B_{p}$ and $A_{v}$.  Then for any state $\ket{\chi}$ that satisfies $\prod_{p} B_{p}\ket{\chi} = -\ket{\chi}$, one necessary has $\braket{\psi | \chi}=0$.  This statement follows from the matrix element $\bra{\psi}\prod_{p}B_{p}\ket{\chi}$; $\prod_{p}B_{p}$ acts an identity on $\bra{\psi}$ but yields $-1$ on $\ket{\chi}$, hence $\ket{\psi}$ and $\ket{\chi}$ must be orthogonal.  Crucially, such a $\ket{\chi}$ can then not serve as a root configuration for a decorated-toric-code ground state.

Now let us deduce the action of $\prod_{p}B_{p}$ on $\ket{\phi_{\downarrow}^{(n)}}$. Note that $\prod_{p}B_{p}$ keeps the Ising spin configurations intact (because every spin-flip operator appears twice in the product). Thus, $\prod_{p}B_{p}$ maps the three-dimensional subspace spanned by $\ket{\phi_{\downarrow}^{(1)}}$, $\ket{\phi_{\downarrow}^{(2)}}$ and $\ket{\phi_{\downarrow}^{(3)}}$ into itself.  One can conveniently represent its action as a $3 \times 3$ matrix
\begin{equation}
N_{ij} = \bra{\phi_{\downarrow}^{(i)}} \prod_{p}B_{p} \ket{\phi_{\downarrow}^{(j)}}.
\end{equation}
It turns out that \textit{the $N$ matrix is identical to the $M$ matrix presented in Eq.~\eqref{eq:defprodbp}}.  [The equivalence can be seen from Eq.~\eqref{eq:prodbp}.  There $\prod_p B_p \sigma^x_p$ also leaves the spins intact because the $\sigma^x_p$ cancels the spin flip operators from each $B_p$.  The product then acts only on the parafermions in exactly the same way as in the $N_{ij}$ matrix above.]  Let $\ket{\tilde{\phi}_{\downarrow}^{(1)}}, \ket{\tilde{\phi}_{\downarrow}^{(2)}}, \ket{\tilde{\phi}_{\downarrow}^{(3)}}$ form a basis that diagonalizes $N$, such that $\prod_{p}B_{p}$ acts as identity on $\ket{\tilde{\phi}_{\downarrow}^{(1)}}$ and $\ket{\tilde{\phi}_{\downarrow}^{(2)}}$ and $-1$ on $\ket{\tilde{\phi}_{\downarrow}^{(3)}}$.  We can now readily construct two ground states for the $(0,0)$ sector,
\begin{equation}
\label{eq:gsDTC}
\ket{{\Psi}^{(1,2)}_{\rm DTC};(0,0)} = \prod_{p} (1+B_{p}) \ket{\tilde{\phi}_{\downarrow}^{(1,2)}},
\end{equation}
whereas $\ket{\tilde{\phi}_{\downarrow}^{(3)}}$ can not serve as a root state following the discussion above.  

For further insight into the obstruction encountered with $\ket{\tilde{\phi}_{\downarrow}^{(3)}}$, consider $\prod_p(1+B_p)\ket{\tilde{\phi}_{\downarrow}^{(3)}}$.  Upon expanding the product, one finds two terms with only down spins: $\ket{\tilde{\phi}_{\downarrow}^{(3)}}$ (with prefactor $+1$) and $\prod_p B_p \ket{\tilde{\phi}_{\downarrow}^{(3)}} = -\ket{\tilde{\phi}_{\downarrow}^{(3)}}$, which exactly cancel.  Similarly, every other Ising configuration appears twice with opposite sign and also cancels.  The state $\prod_p(1+B_p)\ket{\tilde{\phi}_{\downarrow}^{(3)}}$ therefore vanishes identically.  This is an intuitive way to see why one cannot obtain a ground state from root configurations with $\prod_p B_{p}$ acting as $-1$.

One can take a more formal approach to show that \textit{the two ground states we constructed above exhaust the ground states in the $(0,0)$ sector}.  In particular, it is possible to explicitly construct the subspace on which all $A_{v}$ and $\prod_{p}B_{p}$ terms act as identity and then map the decorated-toric-code Hamiltonian to the ordinary toric-code Hamiltonian on this subspace, similar to the approach adopted in Sec.~\ref{sec:TrivialParafermionInsulator}.  We simply remark that this more rigorous, subspace-based `proof' can be readily constructed from our formalism, but will be eschewed in favor of the more intuitive picture developed above.

\subsubsection{Other sectors}

In the $(0,1)$, $(1,0)$, and $(1,1)$ topological sectors, we proceed similarly. We start by constructing wavefunctions $\ket{\phi_{s_{r}};(a,b)} = \ket{s_{r}}\ket{{\rm PF}(s_{r})}$ for some root spin configuration $s_r$ belonging to topological sector $(a,b)$ [different from $(0,0)$].  Letting the spins and parafermions fluctuate by applying $\prod_p (1+B_{p})$ will naturally generate ground states, \emph{provided} that $\prod_p B_{p}$ acts as identity on $\ket{\phi_{s_{r}}}$.  In the $(0,0)$ sector, this exercise was relatively painless because we were able to exploit a simple spin configuration, i.e., all down spins, for which the corresponding parafermion state is trivial to write down.  For non-$(0,0)$ sectors, constructing $\ket{\phi_{s_{r}}}$ is more challenging, as the choice of a simple reference state is no longer obvious.  We will nonetheless show that the construction can still be achieved in an analytically tractable way.
 
 As a first step, for each non-$(0,0)$ sector let us pick a root spin state $s_{r}$ whose toric-code loop configuration contains exactly one non-contractible loop around the torus.  In this case, $e^{i \pi Q_{\mu}^{+}} = 1$ for any Cooper-paired region not residing on the toric-code loop.  However, $e^{i \pi Q_{\mu}^{+}}$ fluctuates wildly for bonds on the loop.  To construct $\ket{\phi_{s_{r}};(a,b)}$, we bootstrap off of $\ket{\phi_{\downarrow}^{(1)}}$, $\ket{\phi_{\downarrow}^{(2)}}$, and $\ket{\phi_{\downarrow}^{(3)}}$---the same trio considered in Sec.~\ref{00sector}, which have $e^{i \pi Q_{\mu}^{+}}=1$ for all $\mu$ but different global degrees of freedom controlled by $T_{x}$ and $T_{y}$.  Starting from these states, we aim to generate $\ket{\phi_{s_{r}};(a,b)}$ by flipping spins on the toric-code loop in $s_{r}$ and applying projectors that enforce parafermion pairings consistent with $s_{r}$.  
 
To be more specific, label parafermions on the toric-code loop as $a_{1}$, $a_{2}$, $\cdots$, $a_{2k}$, with $a_{2i-1}$ and $a_{2i}$ belonging to the same Cooper-paired region $\mu_{i}$.  Then we define 
\begin{equation}
\ket{\phi_{s_{r}};(a,b)} = P_{a_{2k}a_{1}}\prod_{i=1}^{k-1} P_{a_{2i}a_{2i+1}} \left( \prod_{i=1}^{k}\sigma_{\mu_{i}}^{x}\right)\ket{\chi_{\downarrow}},
  \label{phisr}
\end{equation}
where $\ket{\chi_{\downarrow}}$ is some linear combination of $\ket{\phi_\downarrow^{(1,2,3)}}$.  
There are three linearly independent choices for $\ket{\chi_{\downarrow}}$, which naively should give three possible candidates for $\ket{\phi_{s_{r}};(a,b)}$.  However, we will see that in some choices of $\ket{\chi_{\downarrow}}$, Eq.~\eqref{phisr} \emph{vanishes}.  There is actually only one possible $\ket{\phi_{s_{r}};(a,b)}$ in each non-$(0,0)$ sector, in sharp contrast to the three root states with the same spin configuration that one obtains in the $(0,0)$ sector.  

We would like to now deduce how the operator
\begin{equation}
  W_{l} = \mathcal{F}_{a_{2k}a_{1}}\prod_{i=1}^{k-1} \mathcal{F}_{a_{2i}a_{2i+1}}
  \label{Wl}
\end{equation}
acts on $\ket{\chi_{\downarrow}}$.  
The commutation relation between $e^{i\pi Q_{\mu}^{+}}$ and $\mathcal{F}_{ab}$ given in Sec.~\ref{sec:po} (property 2) indicates that $W_{l}$ commutes with all $e^{i\pi Q_{\mu}^{+}}$.  It immediately follows that $(i)$ $W_{l}$ maps the three-dimensional subspace $\mathcal{A_{\downarrow}}$ spanned by $\ket{\phi_{\downarrow}^{(1,2,3)}}$ to itself because $W_{l}$ cannot change the eigenvalues of $e^{i\pi Q_{\mu}^{+}}$, and $(ii)$ $W_{l}$ must take the form 
\begin{equation}
\label{eq:wilsonloop}
W_{l} = \omega^{n_{l}} T_{x}^{n_{l,x}}T_{y}^{n_{l,y}}\prod_{\mu} e^{i n_{l, \mu}\pi Q_{\mu}^{+}}
\end{equation}
with $n$'s denoting integers.  
Hereafter we will discard any $e^{i\pi Q_{\mu}^{+}}$ pieces since they act as identity within the subspace $\mathcal{A_{\downarrow}}$.  The branch cuts that determine $n_{l,x}$ and $n_{l,y}$ are non-contractible loops \textit{without self-intersection}---which, crucially, enforces $n_{l,x}$ and $n_{l,y}$ to be mutually prime as detailed in Appendix~\ref{app:wilsonconst}.  In particular, $n_{l,x}$ and $n_{l,y}$ cannot both be multiples of $3$, so that $W_{l}$ always acts nontrivially on the global degrees of freedom $T_{x,y}$.  The above properties 
are actually quite natural: When following the anyon-worldline interpretation we employed to motivate property 4 in Sec.~\ref{sec:po}, $W_{l}$ \textit{can be interpreted as a Wilson-loop operator around a non-contractible cycle of the trivial parafermion insulator, which cycles the system among the degenerate ground states for that phase}.  

Let us choose a basis for $\mathcal{A_{\downarrow}}$ given by states $\ket{\chi_{\downarrow}^{(n)}}$ that are eigenvectors of $W_l$ with eigenvalues $\omega^n$, where $n = 0, 1,$ or 2.  (Because of the constrained form deduced above, $W_l$ necessarily has eigenvalues $1,\omega, \omega^2$.)  The key observation is that the product of projectors $P_{a_{2k}a_{1}}\prod_{i=1}^{k-1} P_{a_{2i}a_{2i+1}}$ appearing in Eq.~\eqref{phisr} annihilates $\ket{\chi_{\downarrow}^{(2)}}$ and $\ket{\chi_{\downarrow}^{(3)}}$ but acts as a unitary operator (modulo factors of $\sqrt{3}$) on $\ket{\chi_{\downarrow}^{(1)}}$.  To see why, notice the similarity between the string of projectors above and the projectors in the plaquette term $B_p$ [Eqs.~\eqref{eq:defbp} and \eqref{eq:defbps}] that reconfigure parafermion pairings.  Additionally, $W_l$ resembles to the local triality operator $R$ defined in Eq.~\eqref{local_triality}; i.e., compare Eq.~\eqref{Wl} to $R$ rewritten using Eq.~\eqref{eq:trieq}.  In Sec.~\ref{sec:cmh} we showed that triality conservation guarantees that $B_p$ acts as a unitary operator on the subspace orthogonal to its kernel, and that, conversely, $B_p$ acts as zero upon violating triality conservation.  The condition $W_l = 1$ is analogous to the triality-conservation condition.  Repeating the logic from Sec.~\ref{sec:cmh}, one thus finds that in this sector $P_{a_{2k}a_{1}}\prod_{i=1}^{k-1} P_{a_{2i}a_{2i+1}}$ acts as a unitary operator but acts as zero in the $W_l = \omega, \omega^2$ sectors---leading to the result quoted above.  
 
Within each non-(0,0) sector, one can therefore construct a single root state from the reference configuration with all spins down, corresponding to Eq.~\eqref{phisr} with $\ket{\chi_{\downarrow}} = \ket{ \chi_{\downarrow}^{(1)}}$.  The remaining question is how $\prod_{p}B_{p}$ acts on this root state.  We checked through small-system numerics on three different orderings that $\prod_{p}B_{p}$ acts as identity on $\ket{\phi_{s_{r}};(a,b)}$ for all three sectors.  Thus, 
\begin{equation}
 \ket{\Psi_{\rm DTC};(a,b)} = \prod_p(1+B_p)\ket{\phi_{s_{r}};(a,b)}
\end{equation}
are ground states of the decorated-toric-code model on the torus in sectors $(a,b) = (0,1), (1,0),$ and $(1,1)$.  Combined with the two previously obtained ground states in the $(0,0)$ sector, Eq.~\eqref{eq:gsDTC}, we find total of five ground states, consistent with number of anyon types in SU$(2)_{4}$.
 
 More formally oriented readers may ask (one may skip this paragraph if you are not so formally oriented): We constructed $\ket{\phi_{s_{r}};(a,b)}$ by applying projectors to states in the rather special subspace $\mathcal{A}_{\downarrow}$.  Is it possible to obtain different root states with the same spin configuration $s_r$, but which are nevertheless orthogonal to what we constructed in Eq.~\eqref{phisr}?  If such exotic root states exist, one must be able to access them by applying projectors on states whose parafermionic part is orthogonal to those in $\mathcal{A}_{\downarrow}$.  It turns out that no new root states can be constructed by this method. Let us consider some basic requirements for a generic parafermionic state on which projectors $ P_{a_{2k}a_{1}}\prod_{i=1}^{k-1} P_{a_{2i}a_{2i+1}}$ act nontrivially: $(i)$ $e^{i \pi Q_{\mu}^{+}} =1$ for edges $\mu$ that are not on the loop in $s_{r}$, and $(ii)$ triality should be strictly conserved. It turns out that states that satisfy $(i)$ and $(ii)$ can be generated from some state in $\mathcal{A}_{\downarrow}$ by applying some sequence of operators $\mathcal{F}_{a_{2i}a_{2i+1}}$ along the loop.  That is, any such state $\ket\tau$ can be generically written as 
\begin{equation}
  \ket{\tau} = (\mathcal{F}_{a_{2k}a_{1}})^{n_{k,\tau}} \prod_{i=1}^{k-1} (\mathcal{F}_{a_{2i}a_{2i+1}})^{n_{i,\tau}} \ket{\chi_{\downarrow}}
\end{equation}
for some integers $n_{i,\tau}$.  
Then we have
\begin{equation}
\begin{split}
P_{a_{2k}a_{1}}  \prod_{i=1}^{k-1} P_{a_{2i}a_{2i+1}}\ket{\tau} 
 = P_{a_{2k}a_{1}}\prod_{i=1}^{k-1} P_{a_{2i}a_{2i+1}} \ket{\chi_{\downarrow}}.
\end{split}
\end{equation}
To obtain the right-hand side, we used $P_{ab}\mathcal{F}_{ab} = P_{ab}$.  Thus, projecting from an arbitrary state $\ket{\tau}$ is equivalent to projecting from some state in the subspace $\mathcal{A}_{\downarrow}$.  It follows that the root states $\ket{\phi_{s_r};(a,b)}$ that we constructed are indeed unique within a given topological sector, so that no additional degeneracies appear.

\section{Extension to $\mathbb{Z}_2$-graded String-net models} 
\label{sec:LW}

\subsection{Review of string-net models and their symmetry-enriched versions}
\label{sec:reviewstringnet}
 
Levin-Wen string-net models~\citep{Levin2006} define a wide class of commuting-projector Hamiltonians. This approach systematically constructs models out of an algebraic input called a unitary fusion category. Roughly speaking, a unitary fusion category contains the following data:  
\begin{enumerate}
\item String types.  The number of different possible strings (oriented or unoriented) determine the degrees of freedom for each edge in the model.  The full Hilbert space is spanned by all possible string types for the edges. 
\item Fusion rules.  These rules specify which three string types are allowed to meet at a trivalent vertex, and enter as a vertex term in the Hamiltonian.  
\item $d_{t}$ and $F$-symbols. Here $d_{t}$ is a number assigned to all string types $t$, while $F$-symbols encode information about associativity relations.  Roughly speaking, $d_t$ corresponds to the quantum dimension of anyons associated with a given string.  Both $d_t$ and $F$-symbols are necessary to define fixed-point wavefunctions and plaquette-exchange terms.   They satisfy a number of consistency conditions, which we will not review here. 
\end{enumerate}

 Schematically, the Hamiltonian for string-net models takes the form
\begin{equation}
\label{eq:stringnetH}
H_{\text{string-net}} = -\sum_{v} A_{v} -\sum _{p ,t} \frac{d_{t}}{\mathcal{D}^2}B_{p}^{t},
\end{equation}
where $\mathcal{D} = \sqrt{\sum_{t}d_{t}^{2}}$ and for concreteness we assume that the strings live on a honeycomb lattice.
In Eq.~\eqref{eq:stringnetH} $A_{v}$ assigns eigenvalue $1$ to string configurations consistent with fusion rules at a given vertex, while $B_{p}^{t}$ is a plaquette term that allows strings to fluctuate. Note that there are plaquette terms for all string types $t$, weighted by $d_t$. Their precise definition involves $F$-symbols, but they can be roughly interpreted as adding a loop of $t$-string on the plaquette. 
Superpositions of all string configurations consistent with fusion rules yield ground states of string-net models. 

A string-net model has a `$G$-graded structure' if one can assign a group element $g$ to each string type such that fusion rules are consistent with group multiplication. Recent works \cite{Heinrich2016, Cheng2016} revealed that one can `ungauge' any string-net model $\mathcal{C}$ with $G$-graded structure. These works further established that the model $\mathcal{C}_G$ obtained by ungauging $\mathcal{C}$ gives a commuting-projector Hamiltonian for topological order enriched by on-site unitary symmetry $G$.  Conversely, gauging the symmetry $G$ of $\mathcal{C}_{G}$ recovers the original string-net model $\mathcal{C}$. 
 
For simplicity and relevance to this paper, we will describe how to construct $\mathcal{C}_{G}$ when $G= \mathbb{Z}_{2}$. 
First, assign labels $\pm 1$ to each string type (in a manner consistent with fusion rules) and introduce two degrees of freedom for each plaquette, corresponding to the two group elements of $\mathbb{Z}_{2}$.  For the latter we will specifically introduce spin-$1/2$'s with spin-up and spin-down associated with group elements $-1$ and $+1$, respectively.  
Next, define a modified Hamiltonian
\begin{equation}
H_{\text{SET-net}} = -\sum_{v} A_{v} -\sum _{p ,t} \frac{d_{t}}{\mathcal{D}^2}\tilde{B}_{p}^{t} -\sum_{l} P_{l} .
\label{HSETnet}
\end{equation}
The vertex term $A_{v}$ is the same as in Eq.~\eqref{eq:stringnetH}, while $\tilde{B}_{p}^{t}$ is given by
\begin{equation}
\label{eq:defbptilde}
\tilde{B}_{p}^{t} = 
\begin{cases}
B_{p} &\quad \text{if $t$ is a string type assigned $+1$} \\
B_{p}\sigma_{p}^{x} &\quad \text{if $t$ is a string type assigned $-1$}
\end{cases}.
\end{equation}
The last term in $H_{\text{SET-net}}$ acts on the pair of spins residing at plaquettes $p_{1,2}$ separated by edge $l$ via
\begin{equation}
\label{eq:defpl}
  P_{l} = \sum_{t}\frac{1 + (-1)^{n_{t}}\sigma_{p_{1}}^{z}\sigma_{p_{2}}^{z}}{2} \ket{t_{l}}\bra{t_{l}} .
\end{equation}
Here, the $t$ sum runs over all string types, $\ket{t_{l}}\bra{t_{l}}$ projects onto string-type $t$ at edge $l$, and $n_{t}$ is an integer defined as
\begin{equation}
\label{eq:nt}
n_{t} = 
\begin{cases}
0 &\quad \text{if $t$ is a string type assigned $+1$} \\
1 &\quad \text{if $t$ is a string type assigned $-1$}
\end{cases}.
\end{equation}
In the ground state, $P_{l}$ enforces that strings assigned $-1$ bind to domain walls between up and down spin configurations, whereas strings assigned $+1$ do not carry domain walls.  
Equation~\eqref{HSETnet} defines a commuting-projector Hamiltonian whose ground state superposes all configurations with valid string and spin assignments. 
Moreover, $H_{\text{SET-net}}$ commutes with $\prod_{p} \sigma_{p}^{x}$ and thus describes a topologically ordered phase with a global $\mathbb{Z}_2$ spin-flip symmetry. 
 
 To understand the topological order in the $\mathcal{C}_{\mathbb{Z}_{2}}$ model described above, consider an alternative plain-vanilla string-net model $\mathcal{C}_{\text{restricted}}$.  In particular, the data used to define $\mathcal{C}_{\text{restricted}}$ are directly inherited from the gauged string-net model $\mathcal{C}$ with $\mathbb{Z}_2$-grading, \textit{except that only strings labeled $+1$ in $\mathcal{C}$ enter as valid string types in $\mathcal{C}_{\text{restricted}}$}. We emphasize that $\mathcal{C}_{\text{restricted}}$ contains neither plaquette spins nor $\mathbb{Z}_{2}$ symmetry---hence the qualifier `plain-vanilla string-net model'.  By turning on a Zeeman field in $H_{\text{SET-net}}$ to polarize the plaquette spins either up or down, one can adiabatically connect the ground states of $\mathcal{C}_{\text{restricted}}$  and $\mathcal{C}_{\mathbb{Z}_{2}}$ \cite{Heinrich2016, Cheng2016}. Thus, $\mathcal{C}_{\text{restricted}}$ and $\mathcal{C}_{\mathbb{Z}_{2}}$ possess exactly the same topological order.  Note the similarity to the scenario described in Sec.~\ref{sec:TrivialParafermionInsulator}; there we showed that breaking $\mathbb{Z}_{2}$ symmetry via a Zeeman field adiabatically connects the decorated-domain-wall model to the trivial parafermion insulator. 
  
The topological order of $\mathcal{C}_{\mathbb{Z}_{2}}$ can be more intuitively viewed by recalling that strings assigned $-1$ in $\mathcal{C}_{\mathbb{Z}_2}$ carry domain walls separating up and down plaquette spins.  Thus the ungauging process in a sense `demotes' strings with $-1$ label, which are associated with deconfined anyons in $\mathcal{C}$, to symmetry defects in $\mathcal{C}_{\mathbb{Z}_2}$.  Only strings with $+1$ label correspond to deconfined anyons in $\mathcal{C}_{\mathbb{Z}_2}$, and the data for those anyons can be equivalently bootstrapped from $\mathcal{C}_{\text{restricted}}$.  In addition, explicitly breaking $\mathbb{Z}_{2}$ symmetry washes away data associated with spin-flip symmetries---including, crucially, symmetry-defect data given by $-1$ strings. 
 
 The nontrivial interplay between $\mathbb{Z}_{2}$ symmetry and topological order can be observed by gauging $H_{\text{SET-net}}$ to recover the gauged string-net model $\mathcal{C}$. If symmetry acts trivially on the anyons, then $H_{\text{SET-net}}$ must be equivalent to the string-net model $\mathcal{C}_{\text{restricted}}$ tensored with a $\mathbb{Z}_{2}$-symmetric product state for spins. In this case gauging $\mathbb{Z}_{2}$-symmetry simply yields topological order for $\mathcal{C}_{\text{restricted}}$ combined with toric code (or double semion) topological order. However, in many cases the topological order of $\mathcal{C}$ takes a richer form, signaling that $\mathbb{Z}_2$ symmetry in $H_{\text{SET-net}}$ nontrivially enriches the topological order for the ungauged model $\mathcal{C}_{\mathbb{Z}_2}$.

\subsection{Binding $\mathbb{Z}_{3}$ parafermion chains to $\mathbb{Z}_2$-graded string-net models}

As noted in the previous subsection, string-net models with $G$-graded structure must exhibit fusion rules that are consistent with group multiplication. When $G = \mathbb{Z}_2$, this condition can be rephrased as follows: Among three strings that meet at a trivalent vertex, an even number of them should be graded as $-1$.  This vertex rule dictates that strings with $-1$ grade necessarily form closed loops.  Such loops are natural objects for integrating with parafermion chains, as done for toric-code loops in the decorated toric-code model.  
We will specifically add two parafermions to each honeycomb-lattice edge and pair up the parafermions according to the vertex rule illustrated in Fig~\ref{fig:localrule}(b), with loops of $-1$ strings now playing the role of spin-up loops.  In other words, $+1$ and $-1$ strings respectively impose intra-edge and inter-edge parafermion pairings.   

The following Hamiltonian describes a string-net model so decorated with parafermions:
\begin{equation}
\label{eq:pfstringnetH}
\begin{split}
H_{\text{PF-string-net}} &= -\sum_{v} A_{v}^{\rm{PF}} \\
&-\sum _{p ,t} \frac{d_{t}}{\mathcal{D}^2}B_{p}^{t} \left( \mathcal{B}_{p}^{s_{t}}\ket{s_{t},p}\bra{s_{t},p} \right)^{n_{t}}.
\end{split}
\end{equation}
The vertex term $A_{v}^{\rm{PF}}$ gives eigenvalue $1$ when three strings that meet at a given vertex are consistent with fusion rules \emph{and} the associated parafermion pairings are consistent with $\mathbb{Z}_{2}$ string labels; otherwise $A_{v}^{\rm{PF}}$ acts as zero.  The second line allows strings to fluctuate and reconfigures parafermion pairings accordingly.  There $B_{p}^{t}$ is identical to the flip operator that appears in Eq.~\eqref{eq:stringnetH}, while $\left( \mathcal{B}_{p}^{s_{t}}\ket{s_{t},p}\bra{s_{t},p} \right)^{n_{t}}$ acts on the parafermions.  The latter operator contains several pieces.  First, the exponent $n_t$ is again given by Eq.~\eqref{eq:nt}; this choice ensures that the parafermions are modified only when $t$ carries a $-1$ string label.  Second, $\mathcal{B}_{p}^{s_{t}}$ is identical to the product of parafermion projectors defined in Eq.~\eqref{eq:defbps}.  The superscript $s_{t}$ is a spin label assigned to each string configuration purely for bookkeeping purposes, and does not represent additional degrees of freedom in the model's Hilbert space.  For a given string configuration, we take $s_t$ to be spin up if string type $t$ carries a $-1$ label and spin up if the string carries a $+1$ label.  Finally, $\ket{s_{t},p}\bra{s_{t},p}$ projects onto string configurations around plaquette $p$ consistent with spin label $s_{t}$.  The rather elaborate form of $\left( \mathcal{B}_{p}^{s_{t}}\ket{s_{t},p}\bra{s_{t},p} \right)^{n_{t}}$ belies the simple interpretation of this operator: again, it merely adjusts parafermion pairings so that they are always consistent with the string configurations.

Commutation between the parafermion parts, which we prove in Sec.~\ref{sec:cmh} and Appendix~\ref{app:pcom}, together with commutation between parts that act on strings inherited from the string-net model, naturally establishes $H_{\text{PF-string-net}}$ as a commuting-projector Hamiltonian.  Additionally, the graded structure built into the model allows one to ungauge $\mathbb{Z}_{2}$ symmetry.  As outlined previously the ungauged model contains supplemental physical spins at the plaquettes, and is described by the modified Hamiltonian
 \begin{equation}
\label{eq:pfSETnetH}
\begin{split}
H_{\text{PF-SET-net}} &=   - \sum_{v} A_{v}^{\rm{PF}} \\
&-\sum _{p ,t} \frac{d_{t}}{\mathcal{D}^2}\tilde{B}_{p}^{t} \left( \mathcal{B}_{p}^{s_{t}}\ket{s_{t},p}\bra{s_{t},p} \right)^{n_{t}} - \sum_{l} P_{l}.
\end{split}
\end{equation}
The definitions of $\tilde{B}_{p}^{t}$ and $P_{l}$ are identical to those given in Eq.~\eqref{eq:defbptilde} and \eqref{eq:defpl}, while $A_{v}^{\rm{PF}}$ and $\left( \mathcal{B}_{p}^{s_{t}}\ket{s_{t},p}\bra{s_{t},p} \right)^{n_{t}}$ are directly inherited from Eq.~\eqref{eq:pfstringnetH}.  Polarizing the plaquette spins with a Zeeman field adiabatically connects $H_{\text{PF-SET-net}}$ to a trivial parafermion insulator tensored with a string-net model built out of strings labeled $+1$---thus we effectively lose the parafermion decoration.  (Note again the similarity to the decorated-domain-wall scenario described in Sec.~\ref{sec:TrivialParafermionInsulator}.)  Equation~\eqref{eq:pfSETnetH} thus clearly realizes the same topological order as the latter string-net model, possibly with nontrivial enrichment from the global $\mathbb{Z}_2$ spin-flip symmetry.

\subsection{Topological properties}

 It is a difficult task to identify the precise topological phases exhibited by the parafermion-decorated string-net models in Eqs.~\eqref{eq:pfstringnetH} and \eqref{eq:pfSETnetH}.  We will proceed below by positing reasonable ansatzes.  

Let us establish some notation as a preliminary step. As in Sec.~\ref{sec:reviewstringnet}, $\mathcal{C}$ denotes some string-net model with $\mathbb{Z}_{2}$-graded structure, $\mathcal{C}_{\mathbb{Z}_{2}}$ denotes the model realized by ungauging $\mathbb{Z}_2$ symmetry, and $\mathcal{C}_{\text{restricted}}$ is the plain-vanilla string-net model that retains only $+1$-graded strings from $\mathcal{C}$.  Similarly, $\mathcal{C}^{\rm{PF}}$ is the string-net model described by the Hamiltonian in Eq.~\eqref{eq:pfstringnetH} while $\mathcal{C}_{\mathbb{Z}_{2}}^{\rm{PF}}$ and $\mathcal{C}_{\text{restricted}}^{\rm{PF}}$ are the associated ungauged and restricted plain-vanilla string-net models, respectively.  In a slight abuse of notation, we will use the same symbols to indicate the topological order realized by the respective models.  Our goal is to understand the nature of $\mathcal{C}^{\rm{PF}}$ and $\mathcal{C}_{\mathbb{Z}_{2}}^{\rm{PF}}$.  
   
 Reference~\onlinecite{Barkeshli2014} proposed that the `inverse' process of gauging a $\mathbb{Z}_{2}$-symmetry-enriched topological phase corresponds to condensing a single boson in the gauged topological phase.  In the context of our string-net models, gauging symmetry in $\mathcal{C}_{\mathbb{Z}_{2}}$---which has the same topological order as $\mathcal{C}_{\text{restricted}}$---gives $\mathcal{C}$; hence there should exist a boson $b$ whose condensation takes $\mathcal{C}$ to $\mathcal{C}_{\text{restricted}}$ \footnote{We thank Meng Cheng for pointing this out.}. Analogously, there should exist a boson in $\mathcal{C}^{\rm{PF}}$ that drives a transition into $\mathcal{C}^{\rm{PF}}_{\text{restricted}}$. The previous subsection established that topological order for the latter is equivalent to that of the trivial parafermion insulator stacked with $\mathcal{C}_{\text{restricted}}$, i.e., $\mathcal{C}^{\rm PF}_{\text{restricted}} = \mathcal{C}_{\text{restricted}} \times (221)$. 
 
We will illustrate an intuitive way to understand $\mathcal{C}^{\rm{PF}}$ and $\mathcal{C}_{\mathbb{Z}_{2}}^{\rm{PF}}$ by considering a `parent topological phase' $\mathcal{C} \times \text{SU}(2)_{4}$.  Note that $\text{SU}(2)_{4}$ contains a self-boson $Z$ that, when condensed, drives a transition to the $(221)$ state.  Condensing the \emph{product} $bZ$ from the parent topological phase yields topological order that we denote as $\mathcal{C}_{\text{intermediate}}$.  In $\mathcal{C}_{\text{intermediate}}$, the boson $b$ becomes equivalent to $Z$ and remains deconfined since $b$ and $Z$ have trivial mutual statistics.  Further condensing $b \simeq Z$ drives a transition into $\mathcal{C}_{\text{restricted}} \times (221)$---which again is equivalent to $\mathcal{C}^{\rm PF}_{\text{restricted}}$.  Following the logic in the previous paragraph, we thus propose that $\mathcal{C}_{\text{intermediate}}$ and $\mathcal{C}^{\rm PF}$ share the same topological order; $\mathcal{C}_{\mathbb{Z}_{2}}^{\rm{PF}}$ is naturally understood as the SET obtained by ungauging $\mathcal{C}^{\rm{PF}}$.  
  
 As a consistency check, by assuming $\mathcal{C}$ to be $\mathbb{Z}_{2}$ topological order or double-semion topological order, $\mathcal{C}_{\text{intermediate}}$ is given either by SU$(2)_{4}$ and $\overline{\text{JK}}_{4}$.  This result agrees with what we observed in Sec.~\ref{sec:Prop}.  We leave verifying our conjecture more explicitly and completely to future work.  

\section{Conclusions} \label{sec:Conc}
 
We introduced commuting-projector Hamiltonians built from bosonic spins decorated with $\mathbb{Z}_3$ parafermions.  Our work generalizes recent studies that obtained exactly solvable models for fermionic topological phases by dressing spins with Kitaev chains \cite{Tarantino2016,Ware2016}.  The extension to parafermions entails two essential modifications: First, the models must be viewed as living within a parent topological liquid since (contrary to Kitaev chains) parafermions can only appear in a fractionalized medium.  
Second, we needed to establish basic algebraic properties of 2D parafermion arrays and introduce a generalized Kasteleyn orientation to define our exactly solvable models.  
We provided concrete evidence that the decorated-toric-code model exhibits $\text{SU}(2)_{4}$ topological order (or a close cousin thereof) and that the decorated-domain-wall model realizes an SET with the same topological order as the parent topological liquid, but enriched by an anyon-permuting global spin-flip symmetry. We also demonstrated that an interface involving the SET harbors an exotic edge state in which spin-flip symmetry acts as a duality transformation and protects $\mathbb{Z}_{3}$ parafermion criticality---which in strictly 1D systems requires fine-tuning. We extended our parafermion-decoration scheme to string-net models with $\mathbb{Z}_2$-graded structure to produce a broader class of commuting-projector Hamiltonians for chiral SETs and their gauged topological phases.

It may appear surprising that we have constructed commuting-projector Hamiltonians for \emph{chiral} topological orders like $\text{SU}(2)_{4}$.   We conclude by providing some additional viewpoints to stress that our work actually fits well into the conventional wisdom and to motivate some possible future directions. 
 
There are two very crude categories to which most explicit models for topological phases belong---continuum field theories like Chern-Simons and explicit lattice models like Levin-Wen string-net models. Our approach actually reflects a `hybrid' of these two approaches in the following sense: We assumed the existence of a fractionalized medium [e.g., the (221) state] together with line defects that host parafermion zero modes to build our lattice models, but the existence of such topological liquids and line defects are justified in continuum field theory.  It is known that almost all symmetry-protected topological phases and topological orders without protected edge state are realizable as exactly-solvable lattice models. Thus, it is natural to similarly expect that given some chiral topological liquid and its defects, 
one should be able to construct all associated SET and topologically ordered phases that transform to the original `background topological liquid' upon condensing a set of bosons.  Our work establishes this line of thought as a promising route to explicit, exactly solvable models for chiral topological phases.
 
There are a number of tools available in explicit lattice models to extract topological information, most notably entanglement entropies \cite{Kitaev2006b, Levin2006b} and modular $S$ and $T$ matrices \cite{Zhang2012, Cincio2013, Zalatel2013}. However, it is not clear how such conventional tools can be applied to our models. The subtleties are especially apparent on the torus, where we introduced non-local degrees of freedom to account for the parent topological liquid.  
It is not entirely clear how these global degrees of freedoms should be `partitioned' for entanglement entropy calculations.  It would be useful to demonstrate how such tools can be applied in models with parafermions or other exotic line defects introduced in the background of topological phases.
 
Tensor-network constructions provide another powerful method for describing ground-state wavefunction.  It is known that Levin-Wen string-net models and various non-chiral topological phases admit tensor-network descriptions \cite{Buerschaper2009, Williamson2017}.  Fermionic tensor-network constructions for Kitaev-chain-decorated models appeared recently as well 
\cite{Aasen2017,Bultinck2018}.  
Might our models also be amenable to tensor networks? Recently, 1D matrix-product states for $\mathbb{Z}_{3}$ parafermionic topological phases appeared \cite{pfMPS}; extending this construction to 2D tensor-network states poses an interesting and nontrivial challenge. 
 
The models that we have constructed are clearly quite far from experiment at this stage.  It is worth noting, however, that assuming the existence of parafermion zero modes, exotic $\text{SU}(2)_{4}$ non-Abelian topological order is a rather natural phase that `only' requires a $\mathbb{Z}_{2}$ gauge theory associated with parafermions.  Alternatively, viewing our models as an effective description, one can try to search for alternative setups that trade exact solvability for experimental realism.  Perhaps a system of strongly interacting anyons can form a $\mathbb{Z}_{2}$ gauge theory of their own.
  
Finally, it is a challenging but interesting task to investigate explicit constructions of more exotic SETs and chiral topological orders using defects in a parent topological liquid.  Parafermion-decorated string-net models pose one concrete direction that we explored only very briefly.  Recently, Aasen et al.~\cite{Aasen2017} formalized a fermion-condensation method that systematically constructs fermionic topological orders; developing an `anyon-condensation' picture of a similar spirit may prove to be a fruitful complementary approach to constructing chiral topological phases.

\acknowledgements

We would like to thank D.~Aasen, B.~Bauer, M.~Cheng, P.~Fendley, L.~Fidkowski, J.~Haegeman, C.-M. Jian, N.~Tarantino, Zitao Wang, and B.~Ware for illuminating discussions.  We gratefully acknowledge partial support from the National Science Foundation through grant DMR-1723367 and the Army Research Office under Grant Award W911NF-17-1-0323.  This research was also supported by the Caltech Institute for Quantum Information and Matter, an NSF Physics Frontiers Center with support of the Gordon and Betty Moore Foundation through Grant GBMF1250, and the Walter Burke Institute for Theoretical Physics at Caltech.

\appendix

\section{Proof of parafermion properties}
\label{app:proofs}

In this appendix, we will provide proofs for properties 3, 4, and 5 of parafermion operators listed in Sec.~\ref{sec:ko}, assuming that the system is defined on a torus. Proofs of these properties for the planar setup can be understood as special cases in which fractional-charge hopping never crosses the $T_{y}$ branch cut; $T_{x}$ and $T_{y}$ operators can then be safely discarded without affecting properties 1 through 5.

\begin{figure}
\includegraphics[width=\linewidth]{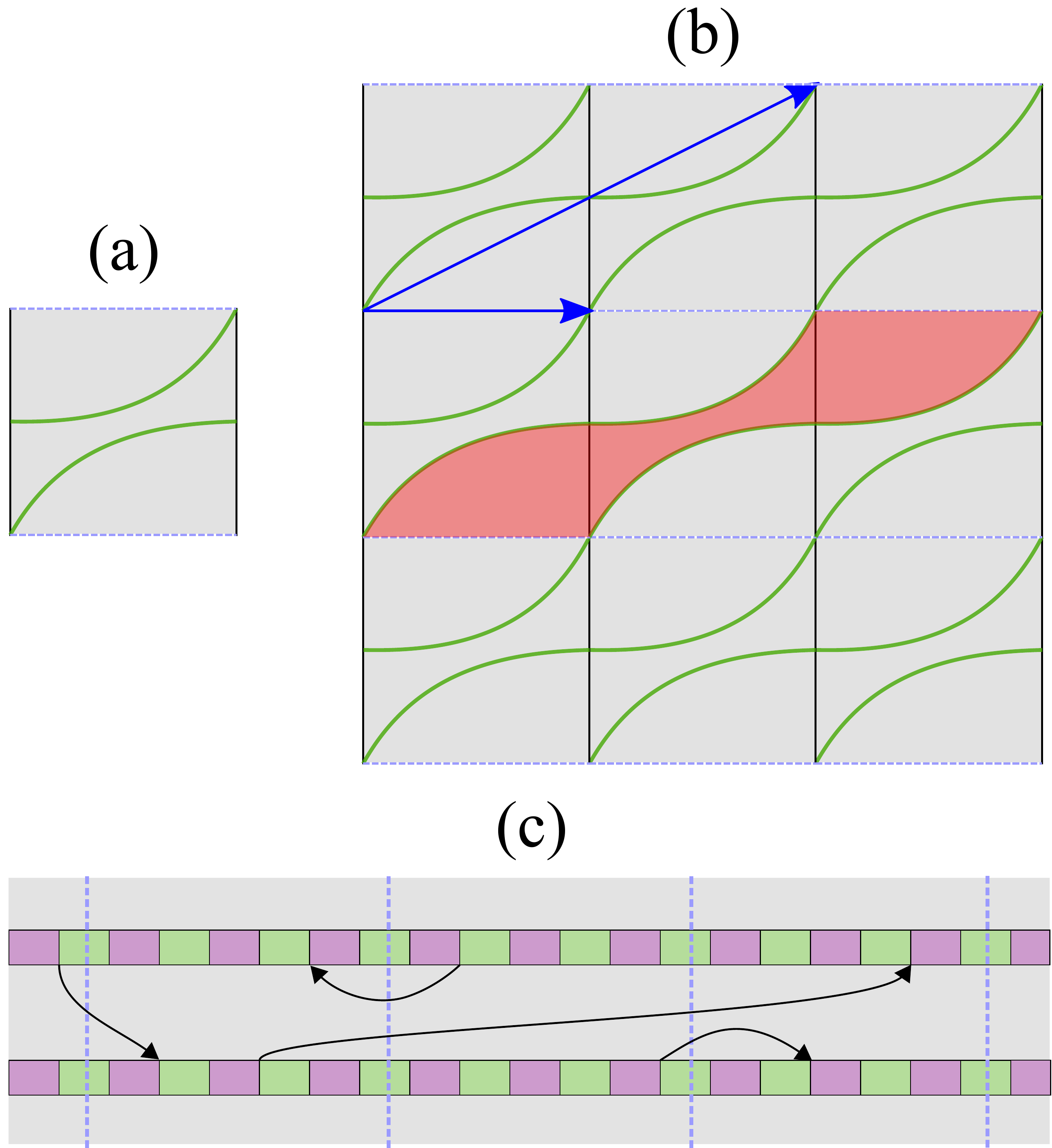}
\caption{(a) Illustration of the trench (thick green line) and $T_{y}$ branch cut (blue dashed line) on which parafermions reside. Gluing the upper and lower edges, and the left and right edges of the square yields the torus. (b) Gluing copies of the square from (a) give an infinite array of green lines and branch cuts on $\mathbb{R}^{2}$. Generators for Deck transformations are illustrated as blue arrows; an example of a zone is shaded in red. (c) Taking a strip in the plane bounded by two neighboring lines gives the repeated-zone geometry.  Black arrows represent possible fractional-charge hopping processes.}
\label{fig:repeatedzone}
\end{figure}

\subsection{Justification for the repeated-zone scheme}

 Before detailing the proofs, let us justify the repeated-zone scheme employed in Fig.~\ref{fig:pftorus2_rev}(c) as a \textit{general method} of representing fractional-charge hopping processes on a torus with any parafermion ordering we consider. As a starting step, we represent the torus as a unit square, as in Fig.~\ref{fig:repeatedzone}(a); upon gluing the top-bottom edges and the left-right edges, one recovers the torus. We respectively indicate the $T_{y}$ branch cut and trench by the blue dashed line and green line. By construction, the green line and the blue dashed line are non-self-intersecting, non-contractible cycles on the torus, and they only cross at a single point that we denote in this section as $a$.
 
Suppose that we now `tile up'  unit squares on the infinite plane, as in Fig.~\ref{fig:repeatedzone}(b). One thereby obtains an infinite stack of green lines and blue dashed lines. In the language of algebraic topology, this `tiling up' process amounts to constructing a universal cover of the torus. Infinite stacks of green lines and blue dashed lines can be understood as $\textit{all possible lifts}$ of the original trench line and $T_{y}$ branch cut on the universal cover $\mathbb{R}^{2}$.
Let us pick a point $a_{(0,0)}$ on the infinite plane where a green and blue line meet. This point represents the unique position $a$ on the torus at which the $T_{y}$ branch cut and the trench line meets. On the infinite plane, however, there are infinite numbers of points representing $a$ that are generated by translation by primitive vectors $e_{x}$ and $e_{y}$.  We will denote the point obtained by translation by the vector $n e_{x} + m e_{y}$ from $a_{(0,0)}$ as $a_{(n,m)}$. 
 
Consider the following two translation vectors: The first shifts $a_{(0,0)}$ to the closest point $a_{(n_{1},m_{1})}$ on the same green line. The second shifts $a_{(0,0)}$ to the closest point $a_{( n_{2},m_{2})}$ on the same blue line; alternatively, $a_{( n_{2},m_{2})}$ can be understood as being placed on the green line `one pitch below' the original green line where $a_{(0,0)}$ is placed.  Blue arrows in Fig.~\ref{fig:repeatedzone}(b) illustrate an example of these two translation vectors, which in algebraic topology language are generators of Deck transformations.  Importantly, the set of points generated by these translations is precisely $\{ a_{(n,m)} \}$.  One can straightforwardly extend the statement about these two translation vectors to any point $b$ on the torus and the set of points $\{ b_{(n,m)} \}$ on the infinite plane that represents $b$.  

Next, think of a region in the infinite plane bounded by two neighboring green lines and two neighboring blue lines---e.g., the red area in Fig.~\ref{fig:repeatedzone}---which we will refer as a `zone'.  As seen in the figure, starting from a point in one zone, each Deck transformation brings you to another point in a different zone.  This means a set of points $\{ b_{(n,m)} \}$ that represent the same point $b$ on the torus belong to different zones---there is no pair of points in the same zone that represents the same point on the torus.  Also, think of any point $c$ on the torus and any point $c_{(0,0)}$ in the infinite plane that represents $c$. One can move $c_{(0,0)}$ to any zone with an appropriate Deck transformation. Thus, every zone necessarily contains a point $c_{(n,m)}$ that represents $c$. In this sense, there is one-to-one correspondence between points in a single zone and points on the torus.

The repeated-zone scheme can be understood as taking a single strip bounded by two green lines from Fig.~\ref{fig:repeatedzone}(b). Alternatively, it can be understood as `tiling up zones' only in one dimension, effectively forming an infinite strip instead of the infinite plane. Upon some deformation, this strip geometry can be illustrated as in Fig.~\ref{fig:repeatedzone}(c). Since fractional-charge hopping processes---indicated by arrows in Fig.~\ref{fig:repeatedzone}(c)---cannot cross the trench, any physical process can be represented on this strip.  The strip in the repeated-zone scheme contains many points that represent the same point on the torus, giving an `overcomplete' picture.  Nevertheless, we will see throughout the proofs that this scheme actually strongly constraints the forms of parafermion operators in a way that is not obvious in a single-zone representation.
 
 We note that if one allows the trench to be self-intersecting, when represented in the infinite plane as in Fig.~\ref{fig:repeatedzone}(b), the green lines on the infinite plane will generically cross each other.  In such settings, it is impossible to define a strip, and the repeated-zone scheme breaks down. Fortunately, our physical setup rules out such cases.

\subsection{More preliminary remarks}
 
 We set up some notations and properties of parafermion operators as the final preliminary step for the proofs. First, it is convenient to classify fractional-charge hopping processes according to which sides of the trench ($L$ or $R$) fractional charge hops from/to.  We will denote $AB$-type hopping ($A,B = L, R$) as a fractional-charge hopping process from side $B$ to side $A$ of the trench.  Below we will frequently need to consider commutation relations between parafermions in different representations (that is, $\alpha_L$ versus $\alpha_R$).  Fortunately they obey the simple commutation relation
\begin{equation}
\label{eq:comLR}
\alpha_{L,i}\alpha_{R,j} = \omega \alpha_{R,j}\alpha_{L,i}  ~~~~(i \neq j) ,
\end{equation}
which can be straightforwardly proved by expressing parafermion operators in terms of clock variables as in Eq.~\eqref{eq:pftoclock}.  
Finally, we emphasize that fractional-charge hopping processes in our consideration do not self-intersect. 

\subsection{Proof of property $3$}

Property 3---$[ F_{ab}, F_{cd} ] =0$ if $a \neq c,d$ and $b \neq c,d$---comes from a more general property: Any two operators that describe fractional-charge hopping without crossing commute.  We prove this statement by separately considering different classes of hopping operators. 
\begin{figure}
\includegraphics[width=\linewidth]{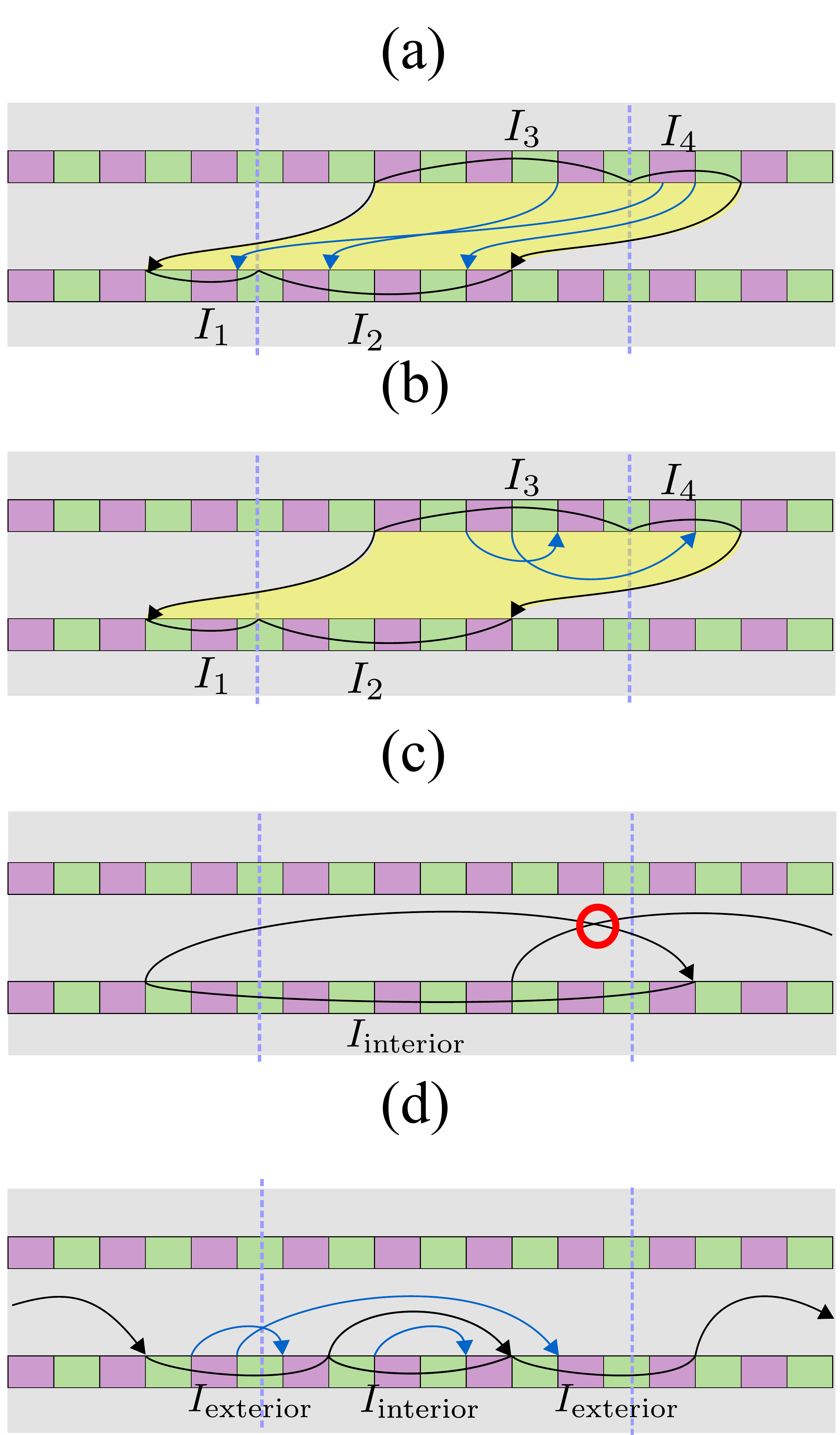}
\caption{(a) Illustration of $F_{ab}$ (black lines) and $F_{cd}$ (blue lines) that correspond to case $(i)$ in our proof of Property 3. The two black lines represent identical fractional-charge hopping processes, and any $F_{cd}$ that does not cross $F_{ab}$ must reside within the yellow region bounded by the black lines, as shown. (b) Illustration of $F_{ab}$ and $F_{cd}$ in case $(ii)$.  Again, $F_{cd}$ must reside within the yellow region bounded by $F_{ab}$ lines.  (c) When the region $I_{\text{interior}}$ encapsulated by a hopping process exceeds the size of a single zone, the non-self-intersection assumption is violated.  One such `illegal' process is illustrated by the black lines, which are two equivalent representations that cross in the red circled region.
(d) Examples of $F_{ab}$ and $F_{cd}$ in case $(iv)$. Blue lines representing $F_{cd}$ start and end within either $I_{\text{interior}}$ or its complement $I_{\text{exterior}}$.}
\label{fig:constraint1}
\end{figure}
 
 $(i)$ Both hopping processes are $LR$-type: We take the first fractional-charge hopping process to be
\begin{equation}
\label{eq:canRL}
F_{ab} = \alpha_{L,a}\alpha_{R,b}^{\dagger}T_{x}T_{y}^{m}.
\end{equation}
(For the assignments of $T_x,T_y$, recall Sec.~\ref{sec:torus}.)
To constrain the $LR$-type operator $F_{cd}$ that does not cross $F_{ab}$, it is useful to draw two lines that represent the same fractional charge hopping process $F_{ab}$ in the repeated zone---see, e.g., black arrows in Fig.~\ref{fig:constraint1}(a). Crucially, \textit{$F_{cd}$ should describe a hopping process within the area bounded by those two lines so that $F_{cd}$ and $F_{ab}$ do not intersect}. In Fig.~\ref{fig:constraint1}(a), this region is colored yellow.  Let us define regions $I_{1}$, $I_{2}$, $I_{3}$, and $I_{4}$ as in Fig.~\ref{fig:constraint1}(a).  The operators $F_{cd}$ of interest should describe hopping from either $(1)$ $I_{3}$ to $I_{2}$, $(2)$ $I_{3}$ to $I_{1}$ or $I_{4}$ to $I_{2}$, or $(3)$ $I_{4}$ to $I_{1}$. These possibilities can be summarized as
\begin{equation}
F_{cd} =  \begin{cases}
\alpha_{L,c}\alpha_{R,d}^{\dagger}T_{x}T_{y}^{m+1} &\quad \text{$(1)$ $c<a$ and $b<d$  } \\
\alpha_{L,c}\alpha_{R,d}^{\dagger}T_{x}T_{y}^{m} &\quad \parbox[t]{3cm}{$(2)$ \text{$a<c$ and $b<d$} \\ \text{or $c<a$ and $d<b$ }} \\
\alpha_{L,c}\alpha_{R,d}^{\dagger}T_{x}T_{y}^{m-1} &\quad \text{$(3)$ $a<c$ and $d<b$  }
\end{cases}
\end{equation}
and are illustrated by blue arrows in Fig.~\ref{fig:constraint1}(a). One can explicitly check that $F_{cd}$ commutes with $F_{ab}$ in all three cases. 
 
 $(ii)$ One hopping process is $LR$-type and the other is $RR$-type: A similar constraint applies here as well.  When the $LR$-type operator $F_{ab}$ and the $RR$-type operator $F_{cd}$ describe non-crossing fractional-charge hopping, $F_{cd}$ should be represented within a region between two lines corresponding to $F_{ab}$; see Fig~\ref{fig:constraint1}(b).  Using Eq.~\eqref{eq:canRL} for $F_{ab}$, one can deduce that $F_{cd}$ should describe hopping $(1)$ from $I_{3}$ to $I_{4}$, $(2)$ within $I_{3}$ or $I_{4}$, or $(3)$ from $I_{4}$ to $I_{3}$.  These possibilities are summarized by
\begin{equation}
F_{cd} =  \begin{cases}
\alpha_{R,c}\alpha_{R,d}^{\dagger}T_{y} &\quad \text{$(1)$ $c<b$ and $d>b$ } \\
\alpha_{R,c}\alpha_{R,d}^{\dagger} &\quad \text{$(2)$ $c,d<b$ or $c,d>b$ } \\
\alpha_{R,c}\alpha_{R,d}^{\dagger}T_{y}^{\dagger} &\quad \text{$(3)$ $c>b$ and $d<b$ }
\end{cases}
\end{equation}
One can again explicitly check that such $F_{cd}$ commutes with $F_{ab}$ in all cases.
 
 $(iii)$ One hopping process is $LR$-type and the other is $LL$-type: This case is a trivial generalization of $(ii)$ above.
 
 $(iv)$ Both hopping processes are $LL$-type: We establish some general constraints that apply to any $LL$-type operators without self-intersection (the same constraints apply for $RR$-type operators). Previously we observed that the repeated-zone scheme gives a highly degenerate representation---there are infinite number of lines on the strip that represent the same fractional-charge hopping process. The condition of non-self-intersection restricts the shape of this infinite array of lines.  In addition to the obvious condition that each line should not self-intersect, two different lines \textit{should not cross each other as well}. 
 
An $LL$-type hopping, when represented in the repeated-zone scheme as a line, encloses some part of the trench on the repeated-zone scheme that we denote as $I_{\text{interior}}$.  If $I_{\text{interior}}$ encapsulates more parafermions than there are on a single zone, then one can draw a second line that represents the same $LL$-type process but necessarily crosses the first line---clearly contradicting the non-self-intersecting condition as sketched in Fig.~\ref{fig:constraint1}(c). 
This constraint on the size of $I_{\text{interior}}$ allows us to categorize parafermions in a single zone into two species---$I_{\text{interior}}$ for parafermions enclosed inside some line that describes an $LL$-type hopping, and $I_{\text{exterior}}$ for those outside. Note that a single zone faithfully represents the torus, hence parafermions in a single zone represent all parafermions on the torus. In this sense, given an $LL$-type process, we can generically classify parafermions into those in $I_{\text{interior}}$ and those in $I_{\text{exterior}}$.
 
 Now consider $I_{\text{interior}}$ and $I_{\text{exterior}}$ defined by $F_{ab}$; additionally we assume $a<b$ without loss of generality.  (If $F_{ab}$ commutes with $F_{cd}$, then $F_{ba} = F_{ab}^{2}$ commutes with $F_{cd}$ as well.)  Parafermions in $I_{\text{exterior}}$ and $I_{\text{interior}}$ carry site indices in the intervals $(a,b)$ and $(0,a) \cup (b,2N+1)$, where $2N$ denotes the last parafermion index in the system. The precise identification of $I_{\text{interior}}$ as $(a,b)$ or $(0,a) \cup (b,2N+1)$ depends on details of $F_{ab}$ and is unimportant for the proof.
  
  As seen in Fig.~\ref{fig:constraint1}(d), if $F_{cd}$ describes hopping process that does not intersect with $F_{ab}$, both $c,d \in I_{\text{interior}}$  or both $c,d \in I_{\text{exterior}}$. This information is sufficient to establish that $F_{cd}$ and $F_{ab}$ commutes when they do not intersect with each other through case-by-case analysis.
 
 $(v)$ Both hopping processes are $RR$-type: This is also a trivial generalization of case $(iv)$.
 
 $(vi)$ One process is $LL$-type and the other is $RR$-type: In this case neither $F_{ab}$ nor $F_{cd}$ contains $T_{x}$.  Equation~\eqref{eq:comLR} alone then suffices to establish that any $LL$-type fractional charge hopping operator commutes with any $RR$-type operator.

\subsection{Proof of property $4$}

\begin{figure}
\includegraphics[width=\linewidth]{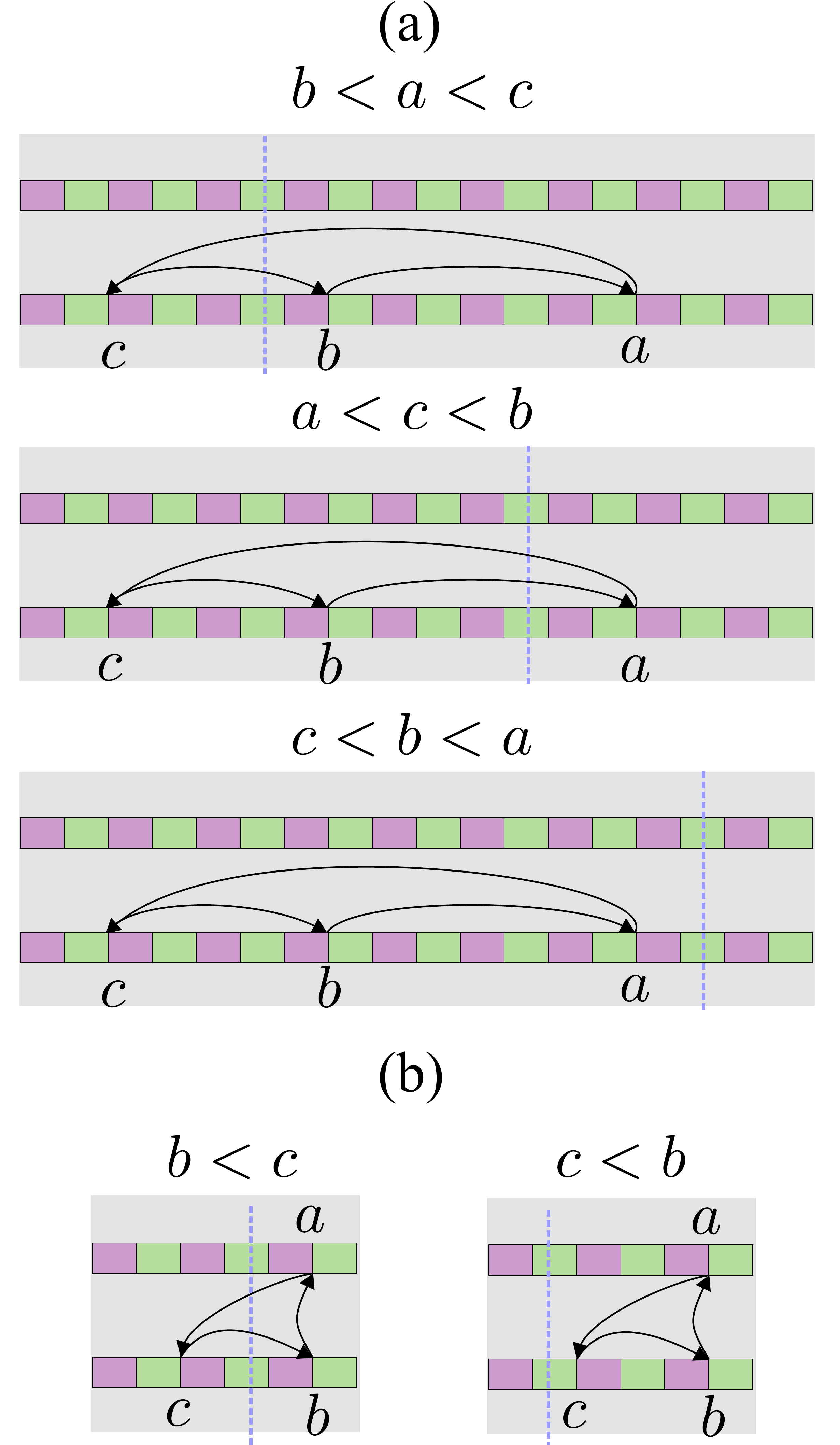}
\caption{(a) Three subtypes of triangles used in case $(i)$ of our Property 4 proof.  (b) Two triangle subtypes used in case $(ii)$.}
\label{fig:constraint2}
\end{figure}

 Next we will prove that any three inter-edge fractional-charge hopping processes that form a contractible triangle satisfy Property 4 defined by Eqs.~\eqref{eq:prop4}. Crucially, contractible triangles on the torus are represented as contractible loops in the repeated-zone scheme as well.  Below we divide triangles into four categories, depending on the types of hopping processes involved.
 
$(i)$ All hopping processes are $LL$-type: 
Denote $a$, $b$, $c$ as three parafermions that form the triangle vertices; they are arranged as $c \rightarrow b \rightarrow a$ in the repeated-zone-scheme representation of the triangle to reflect their clockwise order on the torus.  In other words, hopping is directed \emph{rightward} from $c$ to $b$ to $a$ in the repeated-zone scheme.  [One can instead use $ b \rightarrow a \rightarrow c$ or $a \rightarrow c \rightarrow b$ ordering in the repeated-zone scheme, but Eq.~\eqref{eq:prop4} is insensitive to relabeling them to our chosen $c \rightarrow b \rightarrow a$ ordering. Thus, this choice can be made without loss of generality.]  Figure~\ref{fig:constraint2}(a) shows examples of triangles represented in the repeated-zone scheme according to our conventions.  Notice that the site indices $a,b,c$ within a given zone do not need to satisfy the relation $c<b<a$ in order to conform to our rightward-hopping assumption.  

Let us assume canonical forms of the fractional-charge-hopping operators $F_{ab}$, $F_{bc}$, and $F_{ca}$ as:
\begin{equation}
\label{eq:proof4canon}
\begin{split}
F_{ab} &= \alpha_{L,a}\alpha_{L,b}^{\dagger} (\omega T_{y})^{n_{1}} \\
F_{bc} &= \alpha_{L,b}\alpha_{L,c}^{\dagger} (\omega T_{y})^{n_{2}} \\
F_{ca}  &= \alpha_{L,c}\alpha_{L,a}^{\dagger} (\omega T_{y})^{n_{3}}
\end{split}
\end{equation} 
We will first investigate constraints on $F_{ab}$ that arise from the non-self-intersecting criterion. In case $(iv)$ of the previous subsection, we observed that for any single-line representation of an $LL$-type operator in the repeated-zone scheme, the associated interval $I_{\text{interior}}$ should enclose less than total number of parafermions on the torus. It follows that $a$ and $b$ should appear either in the same zone or in two neighboring zones.  In the former case the exponent in the definition of $F_{ab}$ is $n_{1} = 0$; in the latter $n_1 = 1$ when $b<a$ while $n_1 = -1$ when $a<b$.  Our rightward-hopping assumption further constrains $F_{ab}$: This criterion precludes $n_1 = -1$.  Moreover, when $n_1 = 0$ we must have $b<a$.  These constraints are summarized as
\begin{equation}
\label{eq:constn1}
n_{1} = \begin{cases}
0 &\quad \text{$b<a$} \\
1 &\quad \text{$a<b$} \\
\end{cases}
\end{equation}
One can analogously restrict $F_{bc}$ and $F_{ca}$, yielding 
\begin{equation}
\label{eq:constn2n3}
n_{2} = \begin{cases}
0 &\quad \text{$c<b$} \\
1 &\quad \text{$b<c$} \\
\end{cases}, \qquad
n_{3} = \begin{cases}
0 &\quad \text{$c<a$} \\
-1 &\quad \text{$a<c$} \\
\end{cases}
\end{equation}
Thus the values of $n_{1,2,3}$ fix the ordering of parafermion sites involved in the hopping processes. 

 Finally, contractibility of the triangle requires $n_{1}+n_{2}+n_{3} = 0$.  Equations~\eqref{eq:constn1} and Eq.~\eqref{eq:constn2n3} together with $n_{1}+n_{2}+n_{3} = 0$ leave only three choices of $(n_{1},n_{2},n_{3})$ or, equivalently, relative ordering of parafermion sites $a,b,c$.  The three possible options are given by 
\begin{equation}
(n_{1},n_{2},n_{3}) = \begin{cases}
(0,1,-1) &\quad \text{$b<a<c$} \\
(1,0,-1) &\quad \text{$a<c<b$} \\
(0,0,0) &\quad \text{$c<b<a$}
\end{cases}.
\end{equation}
Figure~\ref{fig:constraint2}(a) illustrates all three triangle subtypes in the repeated-zone scheme.  The information about possible ordering between parafermion indices and values of $(n_{1},n_{2},n_{3})$ in each subtype suffices to establish Eq.~\eqref{eq:prop4} through a case-by-case analysis.

$(ii)$ All hopping processes are $RR$-type:  This case trivially generalizes $(i)$ above.

$(iii)$ Two hopping processes are $RL/LR$-type and one is $LL$-type: Such triangles fall into two subtypes based on how many times the $LL$-type hopping crosses the $T_{y}$ branch cut. As seen in Fig.~\ref{fig:constraint2}(b), $LL$-type hopping (chosen arbitrarily to connect $b$ and $c$) corresponds to 
\begin{equation}
F_{bc} = \alpha_{L,b}\alpha_{L,c}^{\dagger} (\omega T_{y})^{n_{1}}
\end{equation}
 with $b<c$ when $n_{1}=1$ and $c<b$ when $n_{1}=0$. No other cases are possible.  The other hoppings take the form
\begin{equation}
\begin{split}
F_{ab} &= \alpha_{R,a}\alpha_{L,b}^{\dagger} T_{y}^{n_{2}}T_{x}^{\dagger} \\
F_{ca}  &= \alpha_{L,c}\alpha_{R,a}^{\dagger} T_{x}T_{y}^{n_{3}}
\end{split}.
\end{equation}
Since $F_{ab}$, $F_{bc}$, and $F_{ca}$ form a closed loop, we must have $n_{3} = - n_{1} - n_{2}$. This information, along with relative ordering between $b$ and $c$ and the value of $n_{1}$, suffices to establish property $4$ for this type of triangle.
 
$(iv)$ Two hopping processes are $RL/LR$-type and one is $RR$-type: This case trivially generalizes $(iii)$ above.

\subsection{Proof of property $5$}
\label{Property5proof}

\begin{figure}
  \includegraphics[width=0.7\linewidth]{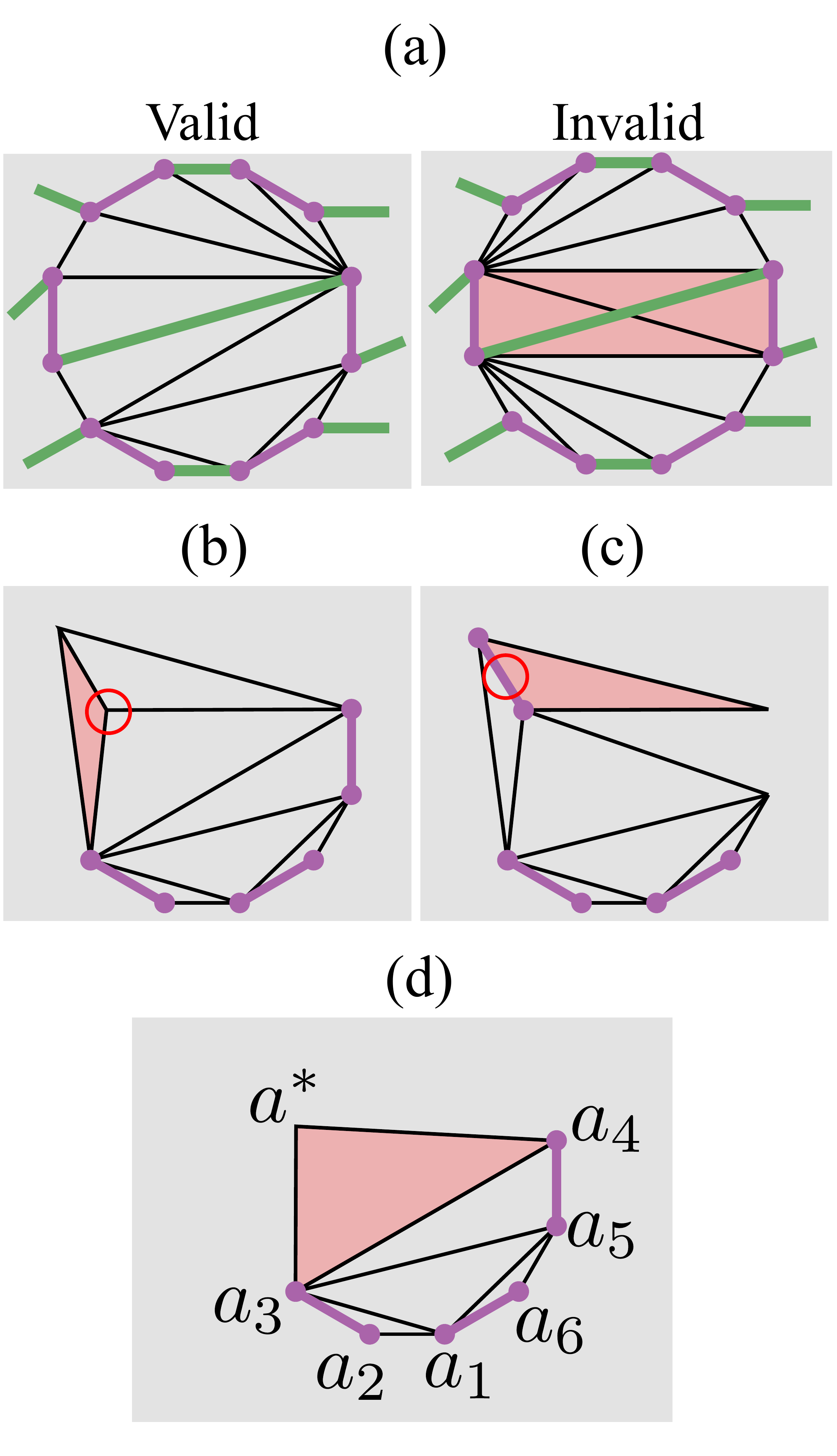} 
  \caption{(a) Examples of valid (left panel) and invalid (right panel) plaquette triangulations.  In the invalid triangulation, two triangle colored red cross the green line that denotes tunneling-gapped regions of a trench. (b) Gluing two edges of the red triangle to the rest of the polygon poses a contradiction in that a parafermion site (circled red) resides within the enlarged polygon. (c) Similarly, gluing a Cooper-pairing region of an $R_{2}$ triangle to a polygon encounters a contradiction that a Cooper-pairing region is encompassed by the new polygon.  (d) Labeling of parafermions used in our induction steps. }
  \label{fig:prop5proof}
\end{figure}

Finally, we prove Property 5, summarized in Eq.~\eqref{eq:prop5}.  Consider a non-triangular elementary plaquette of the pairing lattice such as the 12-gon shown in Fig.~\ref{fig:prop5proof}(a).  Let us triangulate the plaquette such that the vertices of each triangle are parafermion sites and triangles never cross the parafermion ordering path. This triangulation is always possible: Imagine slicing the plaquette along the parafermion ordering path, which decomposes the plaquette into polygons whose sides coincide with the parafermion ordering paths or edges of the original non-triangular plaquette.  By construction, triangulation within these polygons never crosses the parafermion ordering path.  Figure~\ref{fig:prop5proof}(a) illustrates an example of a valid and invalid triangulation.  

Triangles produced from the triangulation procedure fall into two major categories: an `$R_{1}$' triangle contains no Cooper pairing region on its edges, while an `$R_{2}$' triangle features a Cooper-pairing region on one of its edges.  (Triangles can not contain more than one Cooper-pairing segment.)  Denote $a$, $b$, and $c$ as clockwise-directed parafermion sites that form a triangle. $R_{1}$ triangles satisfy Eq.~\eqref{eq:prop4}, as established in the previous subsection through a case-by-case analysis.  One can run a similar case-by-case analysis for $R_2$ triangles to show that when $c$ and $a$ are linked by a Cooper-paired bond $\mu_{ca}$, $F_{ab}$ and $F_{bc}$ satisfy
\begin{equation}
\label{eq:type2treq}
\omega^{2} F_{ab}F_{bc} = e^{i \pi Q_{\mu_{ca}}^{+}}.
\end{equation}
 
 Now we turn our attention to polygons formed by multiple triangles.  In particular, consider a polygon that $(i)$ consists of a subset of triangles obtained from triangulation of a plaquette, $(ii)$ contains at least one $R_{2}$ triangle, and $(iii)$ is simply connected. Label vertices (alternatively, parafermion sites) of this polygon as $a_{1}, a_{2}, \cdots , a_{n}$ in clockwise order, taking $a_{1}$ and $a_{n}$ to form a Cooper-paired bond $\mu_{a_{n}a_{1}}$.  We will prove by an induction-like technique that any polygon conforming to $(i)-(iii)$ obeys
\begin{equation}
\label{eq:canonicallemma4}
e^{i \pi Q_{\mu_{a_{n}a_{1}}}^{+}}\prod_{j} e^{i \pi Q_{\mu_{a_{j}a_{j+1}}}^{+}} = \omega^2 \prod_{k} F_{a_{k}a_{k+1}}.
\end{equation}
On the left side $j$ satisfies $1 \leq j <n$ and runs over Cooper-paired bonds $\mu_{a_{j}a_{j+1}}$.  
On the right, $k$ likewise satisfies $1 \leq k <n$ but runs over bonds that are not Cooper paired. 
Note that possible values of $j$ and $k$ exhaust all integer from $1$ to $n-1$, and that the full non-triangular plaquette of the pairing lattice is a special case of a polygon that satisfies $(i)-(iii)$.  Moreover, Eq.~\eqref{eq:canonicallemma4} applied to the full plaquette corresponds to Eq.~\eqref{eq:prop5}. Showing Eq.~\eqref{eq:canonicallemma4} for any polygon thus automatically proves Eq.~\eqref{eq:prop5}.
  
 The starting case of the induction procedure is a polygon with a single $R_{2}$ triangle. Here, Eqs.~\eqref{eq:canonicallemma4} and \eqref{eq:type2treq} are equivalent.

 Now assume a polygon with parafermion sites $a_{1},a_{2},\cdots,a_{n}$ satisfying Eq.~\eqref{eq:canonicallemma4}. We will investigate whether Eq.~\eqref{eq:canonicallemma4} holds upon attaching more triangles.  Note that any polygon we consider cannot contain any parafermion inside; all parafermions should reside on the polygon edges by assumption.  Thus, all polygons we consider are obtained by attaching \textit{only one side of a triangle}. Figure~\ref{fig:prop5proof}(b) exemplifies the contradiction encountered when one tries attaching two sides of a triangle to the polygon. Thus, without loss of generality, one may label vertices of an attached triangle as $a^{*}$, $a_{k^* + 1}$, and $a_{k^*}$ in clockwise order; $k^{*}$ satisfies $1 \leq k^* < n$, and $a^{*}$ corresponds to the new polygon vertex introduced by attaching the triangle. See Fig.~\ref{fig:prop5proof}(d) for an illustration.
 
Suppose first that one attaches an $R_{1}$ triangle to the polygon. From Eq.~\eqref{eq:prop4}, the following is satisfied:
\begin{equation}
F_{a_{k^*}a^{*}}F_{a^{*}a_{k^* + 1}}F_{a_{k^* + 1}a_{k^*}} = 1.
\end{equation}
Insert the above identity to the right side of Eq.~\eqref{eq:canonicallemma4} as follows:
\begin{equation}
\begin{split}
\omega^2 \prod_{k} & F_{a_{k}a_{k+1}} = \omega^2 (\prod_{k < k^*} F_{a_{k}a_{k+1}})\\
& (F_{a_{k^*}a^{*}}F_{a^{*} a_{k^* + 1}}F_{a_{k^* + 1}a_{k^*}}) F_{a_{k^{*}}a_{k^{*}+1}} (\prod_{k^* < k} F_{a_{k}a_{k+1}}) \\
 & = \omega^2 (\prod_{k < k^*} F_{a_{k}a_{k+1}})F_{a_{k^*}a^{*}}F_{a^{*} a_{k^* + 1}} (\prod_{k^* < k} F_{a_{k}a_{k+1}}).
\end{split}
\end{equation}
When going from the second to the third line, we used $F_{a_{k^* + 1}a_{k^*}} F_{a_{k^{*}}a_{k^{*}+1}}= 1$. Thus, we have a modified identity
\begin{equation}
\begin{split}
e^{i \pi Q_{\mu_{a_{n}a_{1}}}^{+}}& \prod_{j}  e^{i \pi Q_{\mu_{a_{j}a_{j+1}}}^{+}}  \\
& = \omega^2 (\prod_{k < k^*} F_{a_{k}a_{k+1}})F_{a_{k^*}a^{*}}F_{a^{*} a_{k^* + 1}} (\prod_{k^* < k} F_{a_{k}a_{k+1}}).
\end{split}
\end{equation}
Upon relabeling $a_{1},\cdots,a_{k^{*}},a^{*},a_{k^{*}+1},\cdots,a_{n}$ to  $a_{1},\cdots,a_{k^{*}},a_{k^{*}+1},a_{k^{*}+2},\cdots,a_{n+1}$, the above equation simply reduces to Eq.~\eqref{eq:canonicallemma4} for the new polygon obtained by attaching an $R_{1}$ triangle to the old polygon. 

Imagine next attaching an $R_{2}$ triangle. This scenario can be further divided into cases in which $(a_{k^{*}},a^{*})$ or $(a^{*},a_{k^{*}+1})$ forms the Cooper-paired edge. Note that the plaquette and polygons we are considering \textit{do not contain Cooper-pairing regions inside them}.  If $(a_{k^{*}},a_{k^{*}+1})$ forms a Cooper-pairing region, then this Cooper-pairing region will be contained inside the new polygon after gluing the triangle, posing a contradiction [see Fig.~\ref{fig:prop5proof}(c)]. Thus, one can safely rule out the case in which $(a_{k^{*}},a_{k^{*}+1})$ is Cooper-paired. 
 
Assume that $(a^{*},a_{k^{*}+1})$ forms the Cooper-paired edge.  Equation~\eqref{eq:type2treq} then yields the relation
\begin{equation}
\label{eq:type2tmod}
\omega^{2} F_{a_{k^* + 1}a_{k^*}}F_{a_{k^*}a^{*}} = e^{i \pi Q_{\mu_{a^{*}a_{k^{*}+1}}}^{+}}.
\end{equation}
We will insert this identity into Eq.~\eqref{eq:canonicallemma4} for the old polygon to obtain Eq.~\eqref{eq:canonicallemma4} for the new polygon. To proceed, multiply $e^{i \pi Q_{\mu_{a^{*}a_{k^{*}+1}}}^{+}}$ to both sides of the equation. The left side becomes
\begin{equation}
\label{eq:canonlem4midstep1}
\begin{split}
& e^{i \pi Q_{\mu_{a^{*}a_{k^{*}+1}}}^{+}} e^{i \pi Q_{\mu_{a_{n}a_{1}}}^{+}}\prod_{j} e^{i \pi Q_{\mu_{a_{j}a_{j+1}}}^{+}} \\
&= e^{i \pi Q_{\mu_{a_{n}a_{1}}}^{+}}(\prod_{j<k^{*}} e^{i \pi Q_{\mu_{a_{j}a_{j+1}}}^{+}}) e^{i \pi Q_{\mu_{a^{*}a_{k^{*}+1}}}^{+}} (\prod_{j>k^{*}} e^{i \pi Q_{\mu_{a_{j}a_{j+1}}}^{+}}),
\end{split}
\end{equation}
where we reorganized the exponents using the fact that $e^{i \pi Q_{\mu}^{+}}$'s commute with each other.  One can similarly move $e^{i \pi Q_{\mu_{a^{*}a_{k^{*}+1}}}^{+}}$ to the middle of the product on the right side of Eq.~\eqref{eq:canonicallemma4}:
\begin{equation}
\label{eq:canonlem4midstep2}
\begin{split}
&\omega^2 e^{i \pi Q_{\mu_{a^{*}a_{k^{*}+1}}}^{+}}  \prod_{k} F_{a_{k}a_{k+1}} \\
&= (\prod_{k < k^*} F_{a_{k}a_{k+1}}) F_{a_{k^*}a_{k^{*}+1}}e^{i \pi Q_{\mu_{a^{*}a_{k^{*}+1}}}^{+}} (\prod_{k^* < k} F_{a_{k}a_{k+1}}) \\
&= (\prod_{k < k^*} F_{a_{k}a_{k+1}}) F_{a_{k^*}a_{k^{*}+1}}\omega^{2} F_{a_{k^* + 1}a_{k^*}}F_{a_{k^*}a^{*}} (\prod_{k^* < k} F_{a_{k}a_{k+1}}) \\
&= \omega^{2} (\prod_{k < k^*} F_{a_{k}a_{k+1}})F_{a_{k^*}a^{*}} (\prod_{k^* < k} F_{a_{k}a_{k+1}}).
\end{split}
\end{equation} 
 In going from the first to the second line, one should be careful about the commutation relation between $F_{a_{k}a_{k+1}}$ (for $k \leq k^{*}$) and $e^{i \pi Q_{\mu_{a^{*}a_{k^{*}+1}}}^{+}}$. This commutation relation follows from Eq.~\eqref{eq:prop2}, which tells us that $F_{a_{k}a_{k+1}}$ and $e^{i \pi Q_{\mu_{a^{*}a_{k^{*}+1}}}^{+}}$ commute for all $k < k^{*}$. The one nontrivial commutation relation,
\begin{equation}
e^{i \pi Q_{\mu_{a^{*}a_{k^{*}+1}}}^{+}} F_{a_{k^* }a_{k^* + 1}} = \omega F_{a_{k^* }a_{k^* + 1}} e^{i \pi Q_{\mu_{a^{*}a_{k^{*}+1}}}^{+}},
\end{equation}
cancels the factor $\omega^2$ that was present in the first line.  In passing from the second to the third line, we used Eq.~\eqref{eq:type2tmod}. Note that the correct factor of $\omega^{2}$ is restored upon using this relation. 
 Upon taking the last lines of Eq.~\eqref{eq:canonlem4midstep1} and \eqref{eq:canonlem4midstep2} and relabeling $a_{1},\cdots,a_{k^{*}},a^{*},a_{k^{*}+1},\cdots,a_{n}$ to  $a_{1},\cdots,a_{k^{*}},a_{k^{*}+1},a_{k^{*}+2},\cdots,a_{n+1}$, one obtains respectively the left and right side of Eq.~\eqref{eq:canonicallemma4} for the new polygon supplemented by the $R_{2}$ triangle. 
 
 One may proceed very similarly for the case in which the attached $R_{2}$ triangle has $(a_{k^{*}},a^{*})$ as the Cooper-paired edge to prove that Eq.~\eqref{eq:canonicallemma4} still holds upon gluing triangles.  As advertised, starting from a single $R_{2}$ triangle on which Eq.~\eqref{eq:canonicallemma4} holds, one can glue triangles one by one to form the original plaquette of interest.  Equation~\eqref{eq:canonicallemma4} continues to hold throughout the procedure---thus proving Property 5.  
 
\subsection{Comments on global triality}

 In Sec.~\ref{sec:Review}, we used an explicit physical picture to argue that the correct Hilbert space spanned by the parafermion operators should have fixed global triality $\prod_{\mu} e^{i\pi Q_{\mu}^{+}} =1$. Meanwhile, we proposed in Sec.~\ref{sec:Setup} that Properties $1$ through $5$ furnish \textit{the more fundamental definition} of $F_{ij}$ operators. One can then ask: Are the two sectors with $\prod_{\mu} e^{i\pi Q_{\mu}^{+}} = \omega$ and $\omega^{2}$ fundamentally inconsistent with Properties $1$ through 5? The answer is `yes'.  The goal of this section is to explain why sectors with $\prod_{\mu} e^{i\pi Q_{\mu}^{+}} \neq 1$ necessarily violate Property 4 or 5.
 
Imagine triangulating all non-triangular elementary plaquettes of the pairing lattice on a torus such that no triangle crosses parafermion ordering path.  Following the same logic from Appendix~\ref{Property5proof}, this triangulation of the torus is always possible. One can generically classify triangles in such triangulation into three categories:  $R_{1}/R_{2}$ triangles already defined in Appendix~\ref{Property5proof} and `$R_{0}$' triangles that coincide with triangular plaquettes of the pairing lattice. Recall that for clockwise-directed triangular vertices $a$, $b$, $c$,
\begin{equation}
\label{eq:commentid1}
\begin{split}
F_{ab}F_{bc}F_{ca} &= 1 \quad \text{(for $R_{0}/R_{1}$ triangles)} \\
\omega^{2} F_{ab}F_{bc} &= e^{i \pi Q_{\mu_{ca}}^{+}} \quad \text{(for $R_{2}$ triangles)}
\end{split}
\end{equation}
are the key ingredients for proving Property 4 [more specifically, only the last line of Eq.~\eqref{eq:prop4}] and Property 5. We will see that there is a subtlety we glanced over in proving these identities previously. This subtlety is benign for the sector with the correct triality $\prod_{\mu} e^{i\pi Q_{\mu}^{+}} = 1$. However, for the sectors with $\prod_{\mu} e^{i\pi Q_{\mu}^{+}} \neq 1$, there is a unique triangle in a given triangulation in which the `phase-twisted version' of the above identities holds: 
\begin{equation}
\label{eq:commentid2}
\begin{split}
F_{ab}F_{bc}F_{ca} &= \omega^{k} \quad \text{(for $R_{0}/R_{1}$ triangles)} \\
\omega^{2}   F_{ab}F_{bc} &= \omega^{k} e^{i \pi Q_{\mu_{ca}}^{+}} \quad \text{(for $R_{2}$ triangles)},
\end{split}
\end{equation}
where $k=1,2$. This `phase-twisted triangle' also twists the right-hand side of the fourth line of Eq.~\eqref{eq:prop4} and Eq.~\eqref{eq:prop5} for the plaquette that includes the twisted triangle. This means that Property 4 or 5 is always violated for a single plaquette in the sectors with the wrong triality. 

\begin{figure}
	\includegraphics[width=1.0\linewidth]{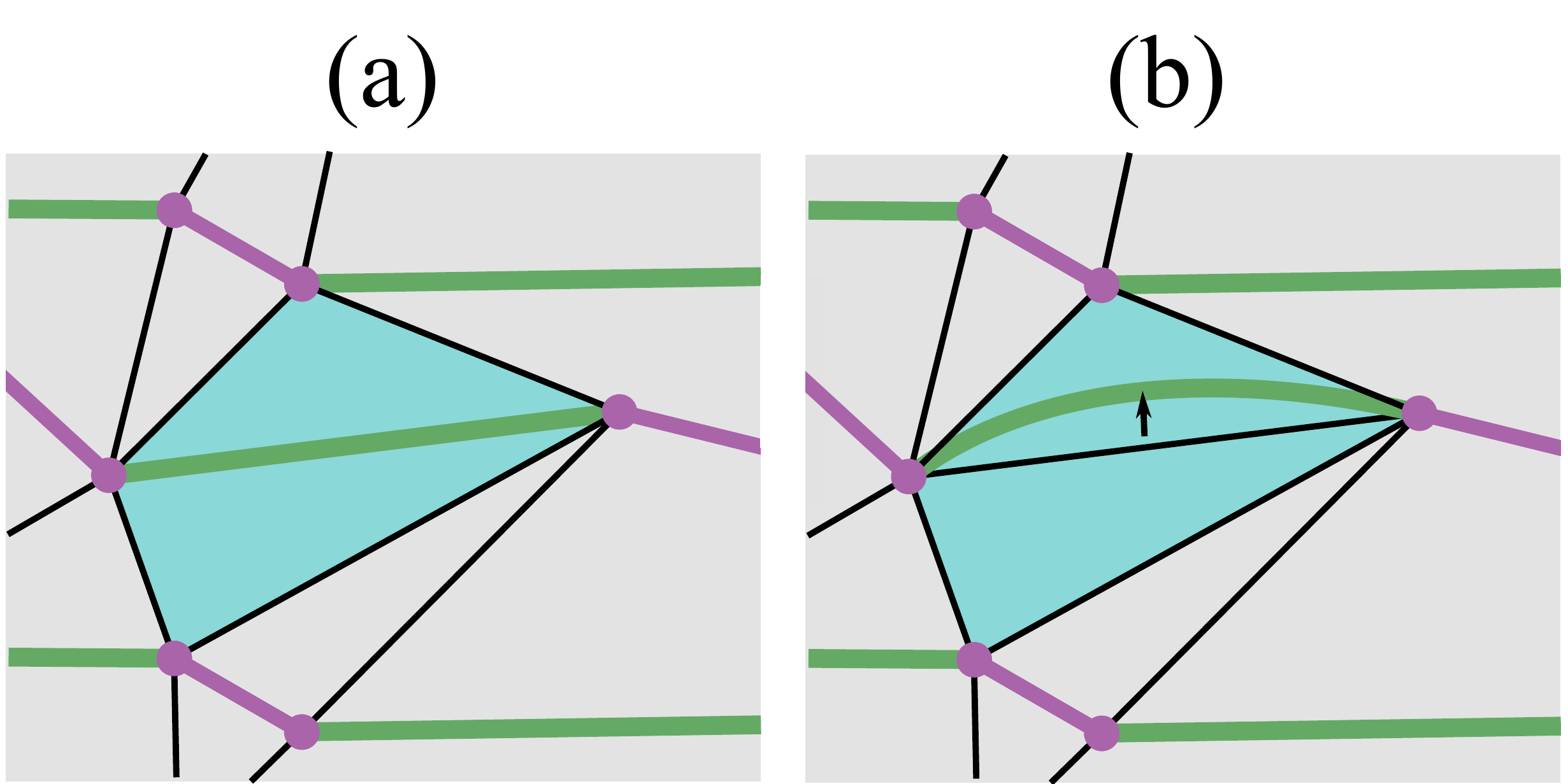} 
  \caption{(a) Two light-blue triangles share a tunneling-gapped region (green line) as their common edge. (b) As a possible `regularization', one may bend the tunneling-gapped region slightly upward, in which case the parafermion ordering path crosses the upper triangle.  Whether Eq.~\eqref{eq:commentid1} generally holds for the upper triangle in this regularization depends on the global triality sector.  } 
  \label{fig:commentgt}
\end{figure}
 
We will now see how this loophole arises. Consider a segment of the parafermion ordering path between two neighboring parafermions that do not form Cooper-pairing regions. This segment corresponds to a tunneling-gapped region in bosonized language. The key observation is that since the parafermion ordering path cannot cross triangles, \textit{for each tunneling gapped region, there is a pair of neighboring triangles that share the tunneling gapped region as their edges}; see Fig.~\ref{fig:commentgt}(a).

 In writing down Eq.~\eqref{eq:commentid1} and Eq.~\eqref{eq:commentid2}, we implicitly assumed that there is a unique fractional-charge hopping operator $F_{ab}$ associated with each edge of the triangulated torus.  For the tunneling-gapped region, however, it is ambiguous whether $F_{ab}$ should be an $LL$-type operator $F_{ab}^{LL}$ or an $RR$-type operator $F_{ab}^{RR}$.  One may `regularize' this ambiguity by sliding up/down the parafermion ordering path slightly so that the problematic tunneling-gapped region does not overlap with the triangle edges.  Doing so fixes $F_{ab} = F_{ab}^{LL}$ or $F_{ab}^{RR}$ depending on the sliding direction. However, choosing either regularization makes the parafermion ordering path cross one of the two neighboring triangles that share the tunneling-gapped region, as illustrated in Fig~\ref{fig:commentgt}(b).  This situation is inherently dangerous since Eq.~\eqref{eq:commentid1} is only proved for triangles that do not cross the parafermion ordering path.
 
 Fortunately, when $(a,b) = (2j, 2j+1)$, we already observed in Sec.~\ref{sec:Review} that $F_{ab}^{LL} = F_{ab}^{RR}$. Thus, Eq.~\eqref{eq:commentid1} is valid for triangles that share the tunneling-gapped region with $(a,b) = (2j, 2j+1)$.  For $F_{1,2N}$ we instead have
\begin{equation}
\label{eq:twisttunnel}
F_{1,2N}^{RR} = \left( \prod_{\mu} e^{i\pi Q_{\mu}^{+}} \right)  F_{1,2N}^{LL},
\end{equation}
i.e., the $LL$- and $RR$-type hoppings across this bond may be nontrivially related depending on the global triality sector.  
For the physical sector with $\prod_{\mu} e^{i\pi Q_{\mu}^{+}} =1 $, there is no issue involving the regularization, guaranteeing that Eq.~\eqref{eq:commentid1} is valid for all triangles. However, this is no longer true for the sectors with $\prod_{\mu} e^{i\pi Q_{\mu}^{+}} = \omega^k$ where $k=1,2$. For example, consider a triangle whose vertices include parafermions with index $1$ and $2N$.  Assume that choosing the regularization corresponding to $F_{1,2N} = F_{1,2N}^{RR}$ makes the parafermion ordering path go through that triangle. This also means that if we choose $F_{1,2N} = F_{1,2N}^{LL}$, the parafermion ordering path will not cross that triangle. Since we proved Eq.~\eqref{eq:commentid1} for triangles that do not cross the parafermion ordering path, choosing $F_{1,2N} = F_{1,2N}^{LL}$ will allow Eq.~\eqref{eq:commentid1} to be satisfied on the triangle we consider.  If we choose $F_{1,2N} = F_{1,2N}^{RR}$, the right-hand side of Eq.~\eqref{eq:commentid1} has to be `twisted' by $\omega^k$---i.e., modifying the relation to be Eq.~\eqref{eq:commentid2}---to account for the identity in Eq.~\eqref{eq:twisttunnel}.

\section{Generalized Kasteleyn orientation and triality conservation equation}
\label{app:gkas}

\subsection{Triality conservation equation for a single plaquette}

 The goal of this section is to prove Eq.~\eqref{eq:trieq}. Before jumping into the proof, we note that for clockwise-oriented sites $a,b,c$ defining a triangular plaquette, 
\begin{equation}
\label{eq:gobrel}
\begin{split}
\mathcal{F}_{ab} \mathcal{F}_{bc}  &= \omega^2 \mathcal{F}_{ac} \\
\mathcal{F}_{bc} \mathcal{F}_{ab}   &= \omega \mathcal{F}_{ac}
\end{split}.
\end{equation}
These relations straightforwardly follow from Eq.~\eqref{eq:prop4} after considering phase factors from our generalized Kasteleyn orientation. Also, we observed in Sec.~\ref{sec:ko} that for parafermions around a non-triangular plaquette $a_{1},a_{2},\cdots,a_{2m}$ in clockwise order, Eq.~\eqref{eq:prop5} holds. Considering the generalized Kasteleyn orientation, Eq.~\eqref{eq:prop5} can be rewritten as
\begin{equation}
\label{eq:eq22rewr}
\prod_{l=1}^{m}  \mathcal{F}_{a_{2l-1}a_{2l}}^{\dagger} =\mathcal{F}_{a_{2lm}a_{1}}\prod_{l=1}^{m-1} \mathcal{F}_{a_{2l}a_{2l+1}}.
\end{equation}  
Note that the above equation is identical to the special case of Eq.~\eqref{eq:trieq} applied to an elementary non-triangular plaquette of the pairing lattice.  We still need to prove that Eq.~\eqref{eq:trieq} holds for the polygon obtained by attaching an even number of triangles to such a non-triangular plaquette. 

Consider attaching $2r$ triangles whose vertices correspond to $(a_{2k_{1}},a_{1}^{*}, a_{2k_{1}+1})$, $(a_{2k_{2}},a_{2}^{*},a_{2k_{2}+1}), \cdots, (a_{2k_{2r}},a_{2r}^{*},a_{2k_{2r}+1})$. Without loss of generality, we assume $k_{1}<k_{2}<\cdots<k_{2r} \leq 2m$.  Note that when $k_{2r} = 2m$, $a_{2k_{2r}+1}$ refers to $a_{1}$. 

\begin{figure}
  \includegraphics[width=0.9\linewidth]{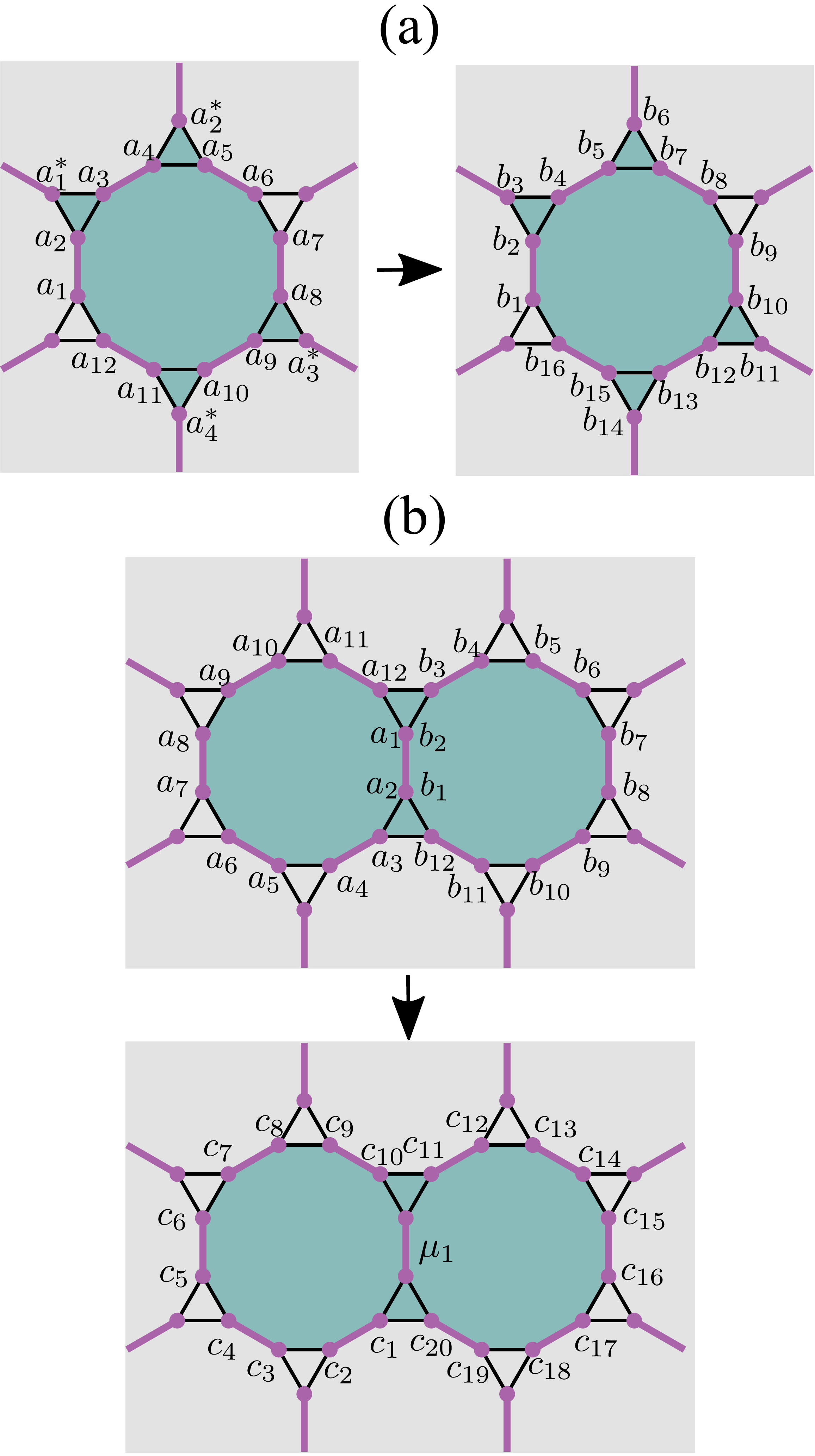} 
  \caption{(a) An example of the relabeling used in our proof of Eq.~\eqref{eq:trieq}.  (b) Relabeling used in our proof of the `base case' of Eq.~\eqref{eq:trieq2}. }
  \label{fig:gkoproof}
\end{figure}

Using Eq.~\eqref{eq:gobrel}, we are free to write
\begin{equation}
\label{eq:triadd}
\mathcal{F}_{a_{2k_{t}}a_{2k_{t}+1}} = \begin{cases}\omega \mathcal{F}_{a_{2k_{t}}a_{t}^{*}} \mathcal{F}_{a_{t}^{*}a_{2k_{t}+1}} & \text{$t$ is odd} \\
\omega^2 \mathcal{F}_{a_{t}^{*}a_{2k_{t}+1}}\mathcal{F}_{a_{2k_{t}}a_{t}^{*}} & \text{$t$ is even}
\end{cases} ;
\end{equation}
treating $t$ even and odd separately in this fashion is done for later convenience.  
Now let us use the relation above to rewrite Eq.~\eqref{eq:eq22rewr} as
\begin{equation}
\label{eq:gkokeystep}
\begin{split}
& \prod_{l=1}^{m}  \mathcal{F}_{a_{2l-1}a_{2l}}^{\dagger} \\
& =\prod_{t=1}^{r} (\mathcal{F}_{a_{2k_{2t-1}}a_{2t-1}^{*}} \mathcal{F}_{a_{2t-1}^{*}a_{2k_{2t-1}+1}})(\mathcal{F}_{a_{2t}^{*}a_{2k_{2t}+1}}\mathcal{F}_{a_{2k_{2t}}a_{2t}^{*}}) \\
&\prod_{l \not\in \{ k_{t} \}} (\mathcal{F}_{a_{2l}a_{2l+1}}).
\end{split}
\end{equation}
Notice that phase factors of $\omega$ and $\omega^{2}$ appear alternatingly in applying the substitution in Eq.~\eqref{eq:triadd} and thus cancel. Utility of the above expression becomes apparent when we relabel parafermion indices $a_{1}, a_{2}, \cdots, a_{2k_{1}}, a_{1}^{*}, a_{2k_{1}+1}, \cdots , a_{2k_{t}}, a_{t}^{*}, a_{2k_{t}+1}, \cdots, a_{2m}$ to $b_{1}, b_{2}, \cdots, b_{2m+2r}$. Here, $b_{1}, \cdots, b_{2m+2r}$ correspond to parafermions arranged clockwise around the new polygon built from the original non-triangular plaquette and $2r$ triangles. 
This relabeling can be expressed more compactly as follows:
\begin{equation}
\label{eq:relabel1}
\begin{split}
a_{k} & \rightarrow \begin{cases}
b_{k} & k \leq 2k_{1}\\
b_{k+t} & 2k_{t} < k \leq 2k_{t+1} \\
b_{k + 2r} & k > 2k_{2r}
\end{cases} \\
a_{t}^{*} & \rightarrow b_{2k_{t} + t}
\end{split}
\end{equation}
Figure~\ref{fig:gkoproof}(a) illustrates such relabeling. 

Equation~\eqref{eq:gkokeystep} can in fact be rearranged to recover Eq.~\eqref{eq:trieq}.  Two important observations underlie the proof. The first is that all terms on the left side of Eq.~\eqref{eq:gkokeystep} commute with each other---thus we are free to rearrange terms in the product to our convenience.  For the right side of Eq.~\eqref{eq:gkokeystep}, any two terms in different parentheses commute with each other as well. However, terms within the same parenthesis do not commute, so their relative order should be retained upon rearrangement to avoid factors of $\omega$.  

  To make the second observation, let us denote $\mathcal{F}_{b_{2l}b_{2l+1}}$ as an `odd $\mathcal{F}$' operator, and $\mathcal{F}_{b_{2l-1}b_{2l}}$ as an `even $\mathcal{F}$' operator. We count $\mathcal{F}_{b_{2m+2r}b_{1}}$ as odd by treating $b_{1} = b_{2m+2r+1}$, as usual. The first product on the right side of Eq.~\eqref{eq:gkokeystep} contains two $\mathcal{F}$'s in each parenthesis. It turns out that, after relabeling, each parenthesis is organized as $(\text{odd $\mathcal{F}$}\times \text{even $\mathcal{F}$})$.
 
These two observations allow one to conclude that, schematically, both sides of Eq.~\eqref{eq:gkokeystep} can be rearranged as
\begin{equation}
\prod (\text{odd $\mathcal{F}^{\dagger}$}) \prod (\text{even $\mathcal{F}^{\dagger}$}) = \prod (\text{odd $\mathcal{F}$}) \prod (\text{even $\mathcal{F}$}).
\end{equation}
By sending $\prod (\text{odd $\mathcal{F}^{\dagger}$})$ to the right side and $\prod (\text{even $\mathcal{F}$})$ to the left, one arrives at
\begin{equation}
\prod_{l=1}^{m+r} \mathcal{F}_{b_{2l-1}b_{2l}} = \mathcal{F}_{b_{2m+2r}b_{1}}\prod_{l=1}^{m+r-1} \mathcal{F}_{b_{2l}b_{2l+1}},
\end{equation}
which is precisely Eq.~\eqref{eq:trieq}. Thus, Eq.~\eqref{eq:trieq} holds generally for any polygon consisting of one non-triangular plaquette and an even number of adjoining triangular plaquettes.

\subsection{Triality conservation equation for a double plaquette}

Next we prove Eq.~\eqref{eq:trieq2}, which is essentially a counterpart of Eq.~\eqref{eq:trieq} for a double plaquette dressed with an even number of triangles. In proving Eq.~\eqref{eq:trieq}, we knew that the `base case' consisting of a single plaquette without any additional triangles naturally follows from Eq.~\eqref{eq:prop5}. 
For the double-plaquette problem of interest here, however, it is not even clear that Eq.~\eqref{eq:trieq2} holds in the analogous `base case'---corresponding to the shaded polygon in Fig.~\ref{fig:gkoproof}(b). 
Once we establish Eq.~\eqref{eq:trieq2} for this case, one can show through almost identical steps presented in the previous subsection that Eq.~\eqref{eq:trieq2} remains satisfied after attaching an even number of triangles to the base-case polygon.
 
 To prove Eq.~\eqref{eq:trieq2} for the base-case polygon, we first label parafermions around the first non-triangular plaquette as $a_{1}, a_{2}, \cdots, a_{2m}$ and the second non-triangular plaquette as $b_{1}, b_{2}, \cdots, b_{2m'}$, both labelings oriented clockwise as shown in the upper panel of Fig.~\ref{fig:gkoproof}(b).  These plaquettes share a Cooper-paired edge, and parafermions on that common edge do not have unique labels; in our convention we take $a_{1} = b_{2}$ and $b_{2} = a_{1}$.  Equation~\eqref{eq:prop5} holds separately for each non-triangular plaquette, i.e.,
\begin{equation}
\begin{split}
&\prod_{l=1}^{m}  \mathcal{F}_{a_{2l-1}a_{2l}}^{\dagger} =\mathcal{F}_{a_{2m}a_{1}}\prod_{l=1}^{m-1} \mathcal{F}_{a_{2l}a_{2l+1}} \\
&\prod_{l=1}^{m'}  \mathcal{F}_{b_{2l-1}a_{bl}}^{\dagger} =\mathcal{F}_{b_{2m'}b_{1}}\prod_{l=1}^{m'-1} \mathcal{F}_{b_{2l}b_{2l+1}}
\end{split}.
\end{equation}

Suppose that we multiply the above equations.  After multiplication, all terms on the left side commute with each other. Meanwhile, on the right side, $\mathcal{F}_{a_{2m}a_{1}}$ and $\mathcal{F}_{b_{2}b_{3}}$ do not commute because they have an overlapping index, and similarly for $\mathcal{F}_{b_{2m'}b_{1}}$ and $\mathcal{F}_{a_{2}a_{3}}$. All other pairs of $\mathcal{F}$'s commute with each other.  Bearing this in mind, we can rearrange the multiplied equations into the form
\begin{equation}
\begin{split}
&\mathcal{F}_{a_{1}a_{2}} (\prod_{l=2}^{m} \mathcal{F}_{a_{2l-1}a_{2l}}^{\dagger})(\prod_{l=2}^{m'} \mathcal{F}_{b_{2l-1}b_{2l}}^{\dagger}) \\
&= (\mathcal{F}_{a_{2m}a_{1}}\mathcal{F}_{b_{2}b_{3}}) (\mathcal{F}_{a_{2}a_{3}}\mathcal{F}_{b_{2m'}b_{1}}) (\prod_{l=2}^{m-1} \mathcal{F}_{a_{2l}a_{2l+1}})(\prod_{l=2}^{m'-1} \mathcal{F}_{b_{2l}b_{2l+1}}) \\
&= \mathcal{F}_{a_{2m}b_{3}} \mathcal{F}_{b_{2m'}a_{3}} (\prod_{l=2}^{m-1} \mathcal{F}_{a_{2l}a_{2l+1}})(\prod_{l=2}^{m'-1} \mathcal{F}_{b_{2l}b_{2l+1}}).
\end{split}
\label{dpeq}
\end{equation}
In passing from the second to the third line, we used the Hermitian conjugate of Eq.~\eqref{eq:gobrel}, which is applicable because $(a_{2m},a_{1}=b_{2},b_{3})$ and $(b_{2m'},b_{1}=a_{2},a_{3})$ form triangles. 

Consider the following relabeling illustrated in the lower panel of Fig.~\ref{fig:gkoproof}(b):
\begin{equation}
\begin{split}
\mathcal{F}_{a_{1}a_{2}}^{\dagger} & \rightarrow e^{i \pi Q_{\mu_{1}}^{+}} \\
a_{3},a_{4},\cdots a_{2m} & \rightarrow c_{1},c_{2},\cdots c_{2m-2} \\
b_{3},b_{4},\cdots b_{2m'} & \rightarrow c_{2m-1},c_{2m},\cdots c_{2m+2m'-4}
\end{split}.
\end{equation} 
After relabeling, Eq.~\eqref{dpeq} becomes
\begin{equation}
e^{- i \pi Q_{\mu_{1}}^{+}} \prod_{l=1}^{m+m'-2} \mathcal{F}_{c_{2l-1}c_{2l}}^{\dagger} = \mathcal{F}_{c_{2m+2m'-4}c_{1}} \prod_{l=1}^{m+m'-3} \mathcal{F}_{c_{2l}c_{2l+1}}
\end{equation}
Upon further minor rearrangement, we obtain precisely Eq.~\eqref{eq:trieq2} for the base-case polygon.
It remains to prove that Eq.~\eqref{eq:trieq2} generically holds after attaching an even number of triangles.  As we noted at the beginning, one can proceed identically as in the previous subsection to establish this result, so we omit the remaining part of the proof here.

\section{Proof for $[B_{p},B_{p'}] = 0$}
\label{app:pcom}

\subsection{Three projector identities}

 Before stating the actual proof, we will introduce three useful projector identities. First, we observed in Sec.~\ref{sec:cmh} that the triality-conservation Eq.~\eqref{eq:trieq} guarantees that removing $P_{a_{2k}a_{1}}$ from Eq.~\eqref{eq:defbps} does not change the form of $\mathcal{B}_{p}^{s}$. Since there is no special meaning to $P_{a_{2k}a_{1}}$ compared to any projector $P_{a_{2l}a_{2l+1}}$ in Eq.~\eqref{eq:defbps}, one may equivalently remove any $P_{a_{2l}a_{2l+1}}$ to obtain the same expression. This logic gives our first identity:
\begin{equation}
\label{eq:projid1}
\begin{split}
\mathcal{B}_{p}^{s} &= \left( P_{a_{2k},a_{1}} \prod_{i=1}^{k-1} \sqrt{3}P_{a_{2i},a_{2i+1}} \right) \left( \prod_{j=1}^{k} P_{a_{2j-1},a_{2j}} \right) \\
&= \left( \prod_{i=1}^{k-1} \sqrt{3}P_{a_{2i},a_{2i+1}} \right) \left( \prod_{j=1}^{k} P_{a_{2j-1},a_{2j}} \right) \\
&= \left( \sqrt{3} P_{a_{2k},a_{1}} \prod_{i=1}^{l-1} \sqrt{3}P_{a_{2i},a_{2i+1}} \right) \left( \prod_{i=l+1}^{k-1} \sqrt{3}P_{a_{2i},a_{2i+1}} \right) \\
& \quad \left( \prod_{j=1}^{k} P_{a_{2j-1},a_{2j}} \right),
\end{split} 
\end{equation} 
where on the right side $l$ indicates the projector that has been removed.  

 Second, for clockwise-oriented parafermion sites $a$, $b$, $c$, we have
\begin{equation}
\label{eq:projid2}
\begin{split}
P_{ab}P_{bc}P_{ca} &= -\frac{i}{\sqrt{3}}P_{ab}P_{ca}\\
P_{ca}P_{bc}P_{ab} &= \frac{i}{\sqrt{3}}P_{ca}P_{ab}. 
\end{split}
\end{equation}
The first line can be proved using the following manipulations:
\begin{equation}
\begin{split}
P_{ab}P_{bc}P_{ca} &= P_{ab}\frac{1+\mathcal{F}_{bc}+\mathcal{F}_{cb}}{3}P_{ca} \\
& =  P_{ab}\frac{1+\mathcal{F}_{ab}\mathcal{F}_{bc}\mathcal{F}_{ca} + \mathcal{F}_{ba}\mathcal{F}_{cb}\mathcal{F}_{ac}}{3}P_{ca} \\
& = P_{ab}\frac{1+ \omega^{2} +\omega^2}{3}P_{ca} = -\frac{i}{\sqrt{3}}P_{ab}P_{ca}.
\end{split}
\end{equation}
In going from the first to the second line, we used $P_{ab} = P_{ab}\mathcal{F}_{ab} = P_{ab}\mathcal{F}_{ba}$ and $P_{ca} = \mathcal{F}_{ca} P_{ca} = \mathcal{F}_{ac} P_{ca}$; in passing from the second to the third, we used Eq.~\eqref{eq:gobrel}. The second line of Eq.~\eqref{eq:projid2} follows easily by Hermitian conjugating the first line.

 Finally, for any three sites $a$, $b$, $c$ in which there can be parafermion pairing between $a$ and $b$ and between $b$ and $c$, one has
\begin{equation}
\label{eq:projid3}
P_{ab}P_{bc}P_{ab} = \frac{1}{3}P_{ab}.
\end{equation}
This identity stems from the relation
\begin{equation}
\label{eq:projid3int}
P_{ab}\mathcal{F}_{bc}P_{ab} = P_{ab}\mathcal{F}_{bc}^{2}P_{ab} =0.
\end{equation}
We can write the left side as
\begin{equation}
\label{eq:projid3int2}
\begin{split}
P_{ab}\mathcal{F}_{bc}P_{ab} &= \frac{1 + \mathcal{F}_{ab} + \mathcal{F}_{ab}^{\dagger}}{3} \mathcal{F}_{bc}P_{ab} \\
&=  \mathcal{F}_{bc} \frac{1 + \omega^{k}\mathcal{F}_{ab} + \bar{\omega}^{k}\mathcal{F}_{ab}^{\dagger}}{3} P_{ab},
\end{split}
\end{equation}
where in the second line we used the commutation relation $\mathcal{F}_{ab}\mathcal{F}_{bc} = \omega^{k}\mathcal{F}_{bc}\mathcal{F}_{ab}$ ($k=1,2$) that arises from Property 2 or 4 in Sec.~\ref{sec:po}.  
Note that $P_{ab}$ projects onto states with $\mathcal{F}_{ab}$ eigenvalue $1$, while $(1 + \omega^{k}\mathcal{F}_{ab} + \bar{\omega}^{k}\mathcal{F}_{ab}^{\dagger})/3$ projects onto states with $\mathcal{F}_{ab}$ eigenvalue  $\bar{\omega}^{k} \neq 1$. Equation~\eqref{eq:projid3int2} thus vanishes.  Almost identical logic shows that $P_{ab}\mathcal{F}_{bc}^{2}P_{ab} = 0$ as well.
Expanding $P_{bc}$ on the left side of Eq.~\eqref{eq:projid3} and using Eq.~\eqref{eq:projid3int} then gives our third identity as claimed,
\begin{equation}
P_{ab}P_{bc}P_{ab} = P_{ab}\frac{1 + \mathcal{F}_{bc} + \mathcal{F}_{bc}^{2}}{3}P_{ab} = \frac{1}{3}P_{ab}.
\end{equation}


\subsection{Strategy to prove $[B_{p},B_{p'}] = 0$}
 
Equation~\eqref{eq:defbp} shows that plaquette terms can be divided into two parts: $S_{p}$ flips the spins, and $\mathcal{B}_{p}^{s}$ reconfigures the parafermion pairings. It is trivial to prove that the spin-flip parts $S_{p}$ and $S_{p'}$ commute in both the decorated-domain-wall and decorated-toric-code models. Also, the parafermion projectors in $\mathcal{B}_{p}^{s}$ and those in $\mathcal{B}_{p'}^{s'}$ trivially commute when $p$ and $p'$ are non-neighboring plaquettes. The nontrivial part is to prove that, for neighboring $p$ and $p'$, $B_{p}B_{p'}$ and $B_{p'}B_{p}$ act identically on the parafermionic part of the Hilbert space.  
 
 To sharpen the goal, think of the following: Fix a starting state with Ising spin configuration $s_{b}$ around two neighboring plaquettes $p$ and $p'$; in the decorated-toric-code model, we enforce $s_{b}$ to satisfy the toric-code vertex rule around $p$ and $p'$ so that $B_{p}B_{p'}$ and $B_{p'}B_{p}$ do not act as zero (in which case the commutation relation again arises trivially). Denote $s_{i_{1}}$ and $s_{i_{2}}$ as spin configurations obtained from $s_{b}$ by applying $S_{p'}$ and $S_{p}$, respectively. When acting on the prescribed starting state, the parafermion parts of $B_{p}B_{p'}$ and $B_{p'}B_{p}$ correspond to $\mathcal{B}_{p}^{s_{i_{1}}}\mathcal{B}_{p'}^{s_{b}}$ and $\mathcal{B}_{p'}^{s_{i_{2}}}\mathcal{B}_{p}^{s_{b}}$.  One can prove $[ B_{p} , B_{p'} ] =0$ by showing that
\begin{equation}
\label{eq:comactualform}
\mathcal{B}_{p}^{s_{i_{1}}}\mathcal{B}_{p'}^{s_{b}} = \mathcal{B}_{p'}^{s_{i_{2}}}\mathcal{B}_{p}^{s_{b}} = e^{i\phi_{s_{b}}}\mathcal{C}_{p,p'}^{s_{b}}
\end{equation} 
for all permissible spin configurations $s_{b}$ around $p$ and $p'$.
Here $e^{i\phi_{s_{b}}}$ is an $s_{b}$-dependent phase factor, while $\mathcal{C}_{p,p'}^{s_{b}}$ denotes a product of parafermion projectors that directly implement parafermion-pairing flips around \textit{the double plaquette consisting of $p$ and $p'$} (rather than the step-by-step, plaquette-by-plaquette parafermion-pairing flips implemented by $\mathcal{B}_{p}^{s_{i_{1}}}\mathcal{B}_{p'}^{s_{b}}$ and $\mathcal{B}_{p'}^{s_{i_{2}}}\mathcal{B}_{p}^{s_{b}}$).

\begin{figure}
  \includegraphics[width=0.5\linewidth]{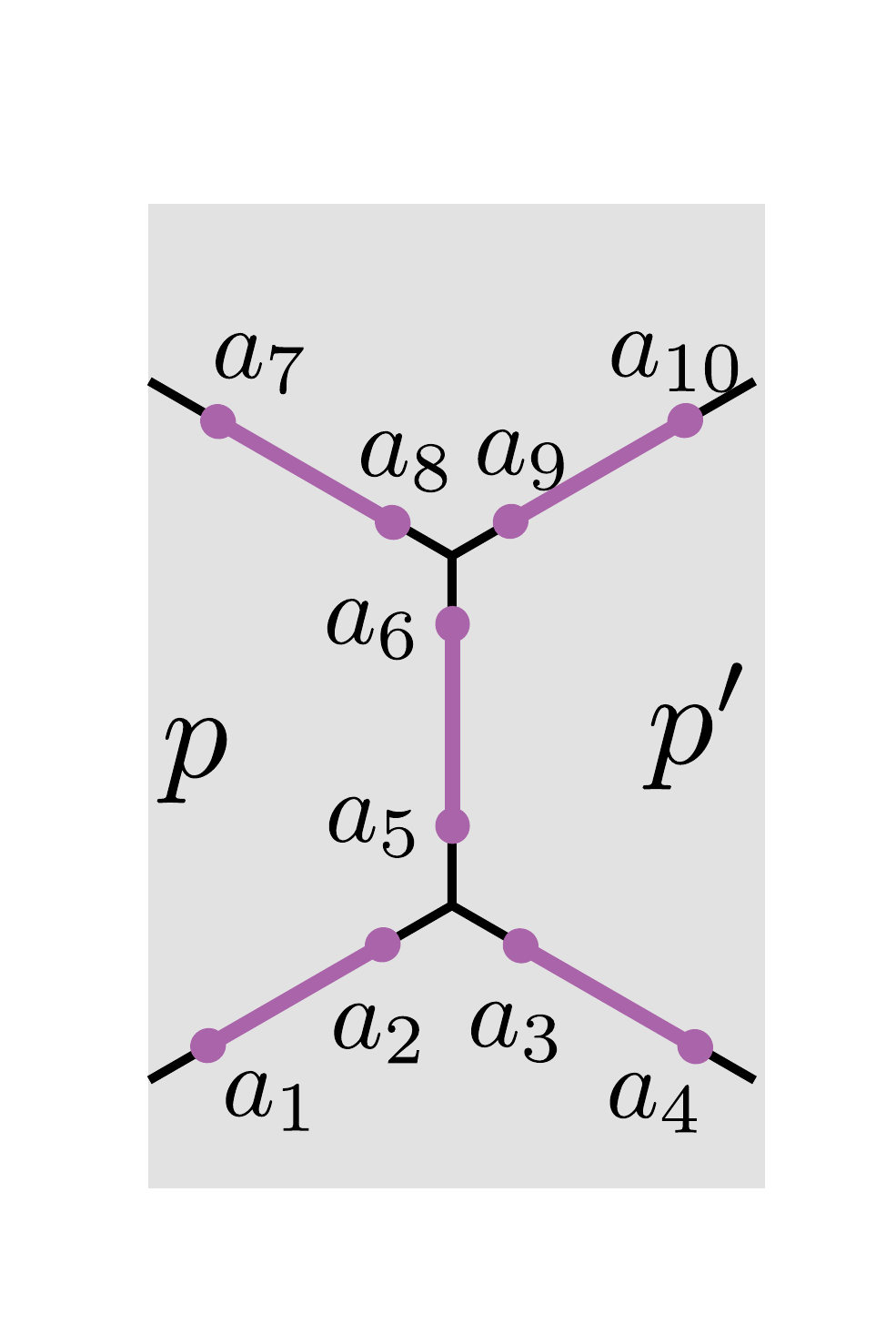} 
  \caption{Illustration of parafermions in the neighborhood of the edge where plaquettes $p$ and $p'$ meet. Only pairings on these parafermions are relevant for proving Eq.~\eqref{eq:comactualform} case-by-case. We also show parafermion labelings used in the proof.}
  \label{fig:appendixcommutation}
\end{figure}

  The nontrivial part of the proof comes from the fact that $\mathcal{B}_{p}^{s_{i_{1}}}\mathcal{B}_{p'}^{s_{b}}$ and $\mathcal{B}_{p'}^{s_{i_{2}}}\mathcal{B}_{p}^{s_{b}}$ contain some projectors that do not appear in $\mathcal{C}_{p,p'}^{s_{b}}$.  We will observe that they can nevertheless be removed with the help of the projector identities listed in the previous subsection. Moreover, such `problematic projectors' only involve parafermions around the edge where $p$ and $p'$ meet. Thus, different Ising spin configurations $s_{b}$ that enforce the same parafermion pairings within the restricted area shown in Fig.~\ref{fig:appendixcommutation} can be treated on equal footings.  A similar observation was made in establishing commutation of plaquette terms in previous Majorana-decorated models~\citep{Ware2016}. 
 
 There are eight possible parafermion pairing configurations within the geometry of Fig.~\ref{fig:appendixcommutation}, which are illustrated in Table~\ref{t:summary}.  We denote the set of spin configurations consistent with pairing patterns in the left and right columns as $\mathcal{S}_{b}$ and $\mathcal{S}_{f}$, respectively (hence the labeling in the table; here and below $b$ stands for base and $f$ stands for final).  
Note that spin configurations in $\mathcal{S}_{b}$ are obtained by acting $S_{p}S_{p'}$ on spin configurations in $\mathcal{S}_{f}$, and vice versa. Consequently, $\mathcal{B}_{p}^{s_{i_{1}}}\mathcal{B}_{p'}^{s_{b}}$ and $\mathcal{B}_{p'}^{s_{i_{2}}}\mathcal{B}_{p}^{s_{b}}$ with $s_{b} \in \mathcal{S}_{f}$ are \textit{exactly the Hermitian conjugate} of the same operators for some $s_{b} \in \mathcal{S}_{b}$.  It then suffices to prove that Eq.~\eqref{eq:comactualform} holds for four species of Ising spin configurations in $\mathcal{S}_{b}$; Eq.~\eqref{eq:comactualform} for the remaining spin configurations in $\mathcal{S}_{f}$ naturally follows by Hermitian conjugation.
 
 The last column of Table~\ref{t:summary} summarizes the phases $e^{i\phi_{s_{b}}}$ obtained below for starting configurations $s_b \in \mathcal{S}_b$; note especially the nontrivial value in the third row.


\begin{table}

\caption{Parafermion pairings within the area shown in Fig.~\ref{fig:appendixcommutation} that are consistent with spin configurations $s_b$ in $\mathcal{S}_{b}$ (first column) and $\mathcal{S}_{f}$ (second column).  The third column shows the phase $e^{i\phi_{s_{b}}}$ [defined in Eq.~\eqref{eq:comactualform}] obtained for each starting configuration in $\mathcal{S}_{b}$.}
\label{t:summary}

\begin{center}
\begin{tabularx}{\linewidth}{|>{\centering\hsize=1.2\hsize}X|>{\centering\hsize=1.2\hsize}X|X<{\centering\hsize=0.6\hsize}|}
\hline 
$\mathcal{S}_{b}$ & $\mathcal{S}_{f}$ & $e^{i\phi_{s_{b}}}$ \\
\hline \hline
\raisebox{-.5\height}{\includegraphics[width = \linewidth]{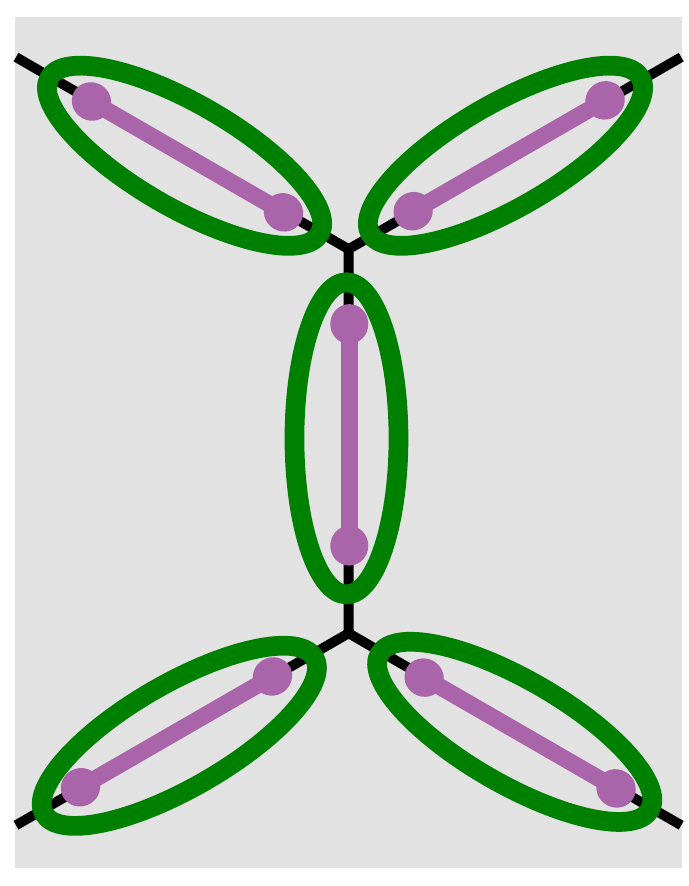}} & \raisebox{-.5\height}{\includegraphics[width = \linewidth]{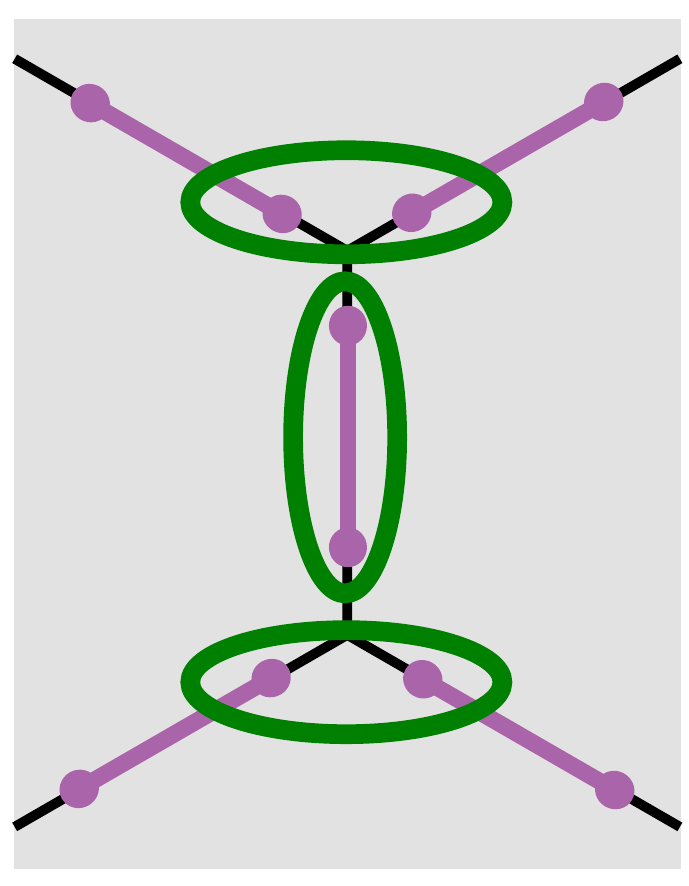}} & 1 \\ \hline
\raisebox{-.5\height}{\includegraphics[width = \linewidth]{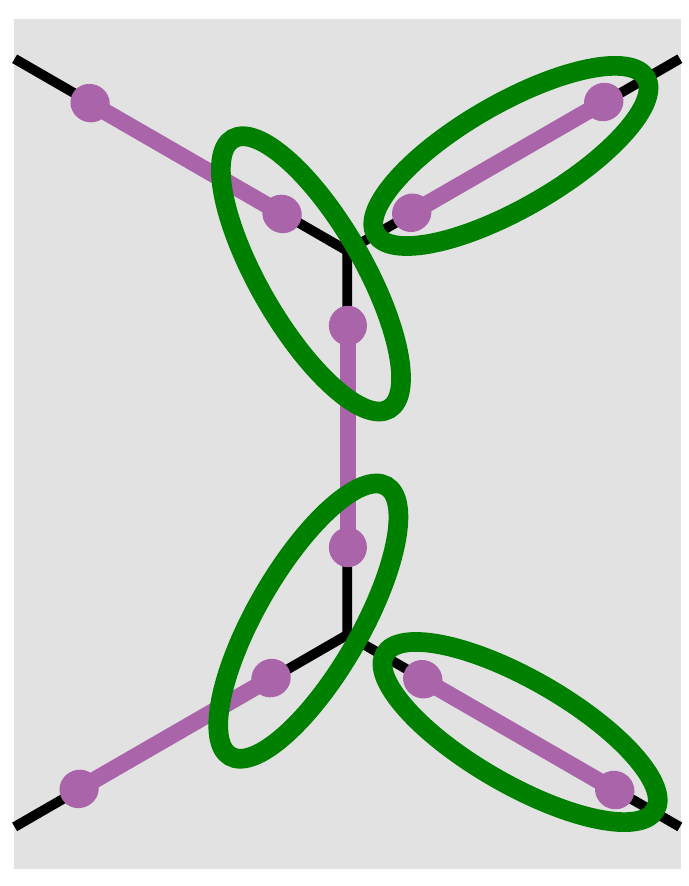}} & \raisebox{-.5\height}{\includegraphics[width =\linewidth]{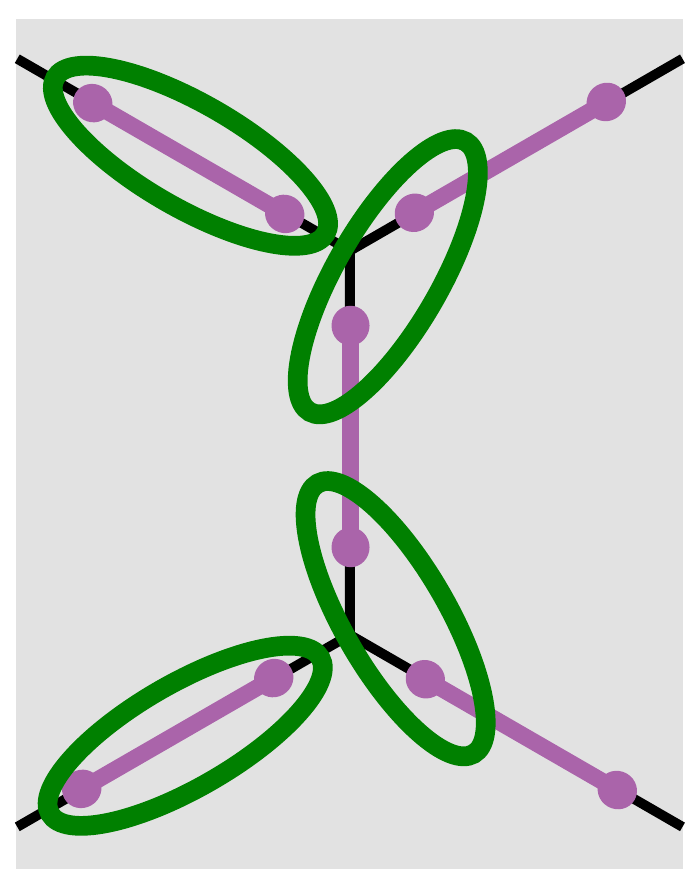}} & 1 \\ \hline
\raisebox{-.5\height}{\includegraphics[width = \linewidth]{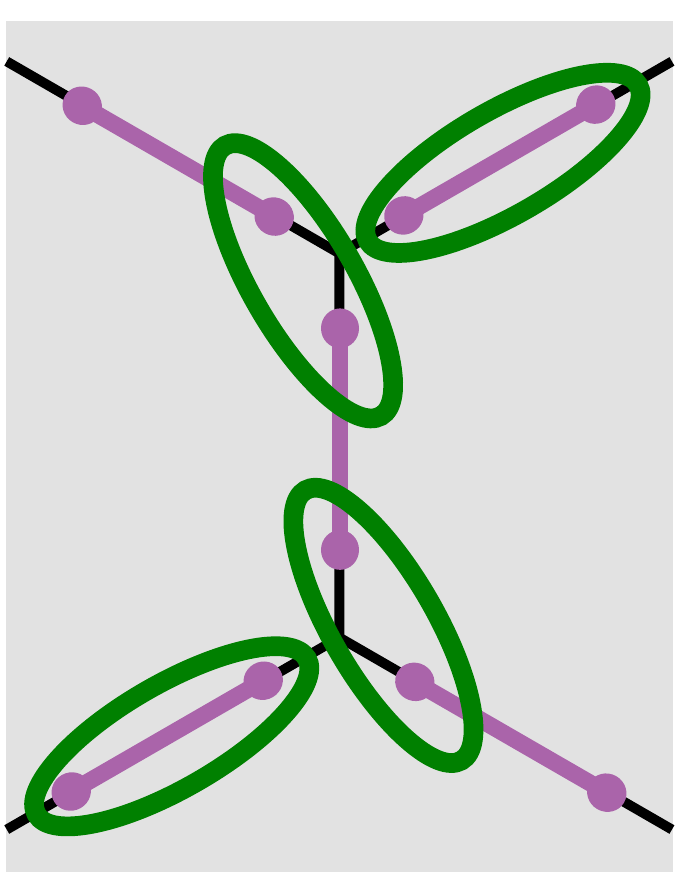}} & \raisebox{-.5\height}{\includegraphics[width = \linewidth]{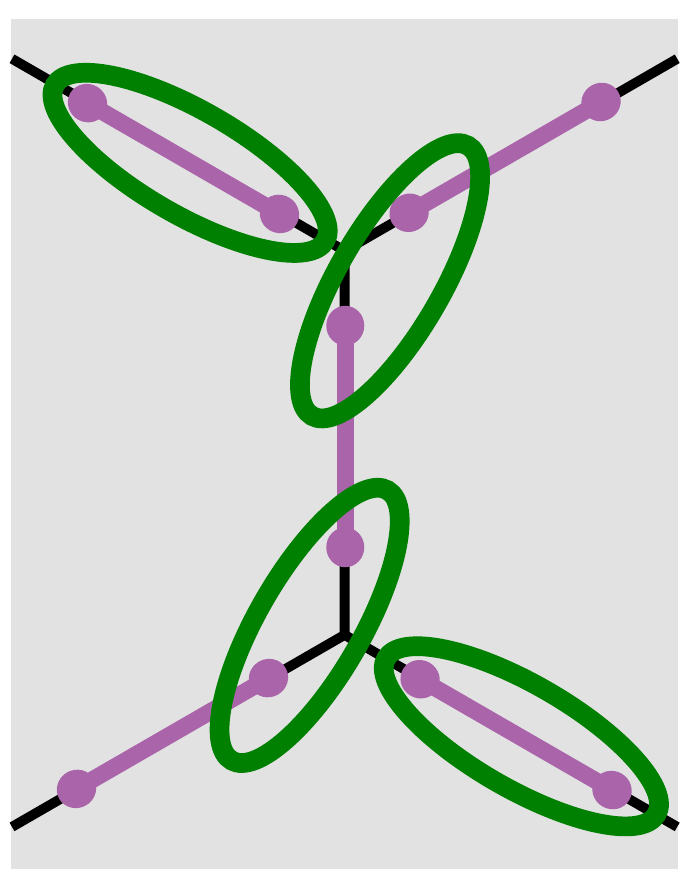}} & $i$ \\ \hline
\raisebox{-.5\height}{\includegraphics[width = \linewidth]{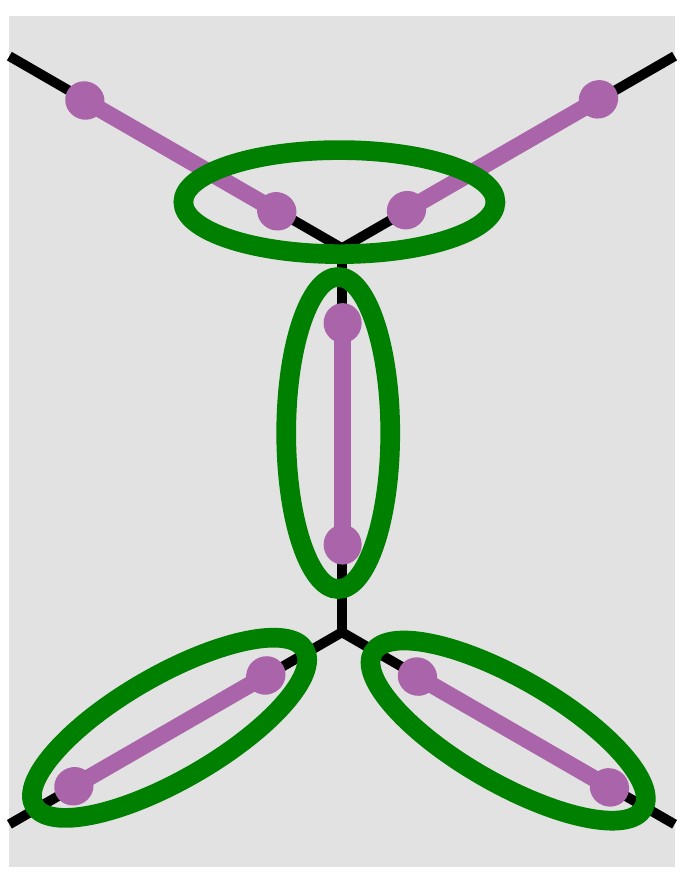}} & \raisebox{-.5\height}{\includegraphics[width = \linewidth]{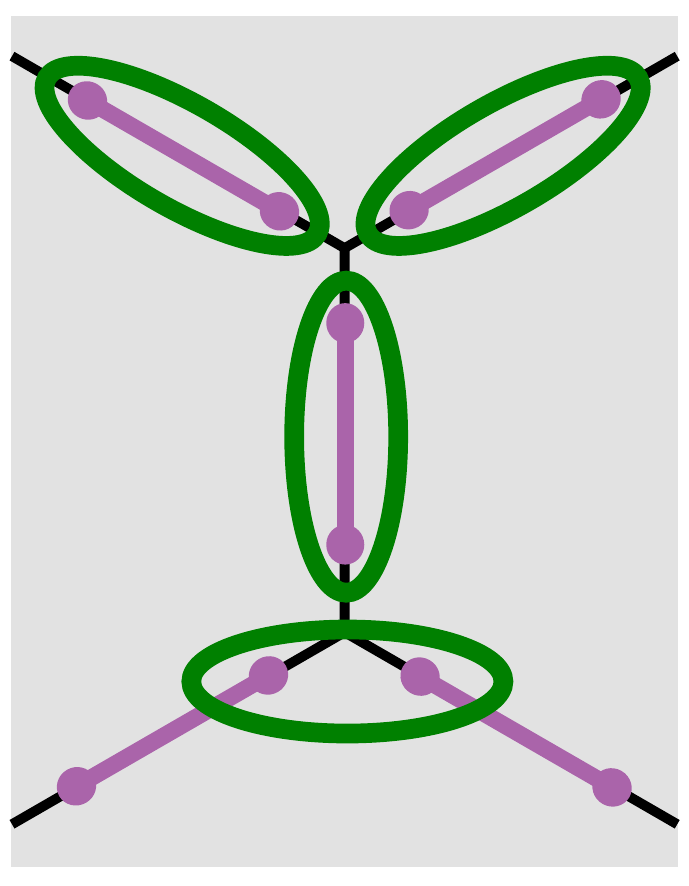}} & 1 \\ \hline
\end{tabularx}
\end{center}

\end{table}

\subsection{The projector algebra}

 Here we work out the projector algebra explicitly for a number of cases to give readers a sense of the techniques employed to show Eq.~\eqref{eq:comactualform}. The general strategy proceeds as follows:
\begin{enumerate}
\item Write down the explicit form of $\mathcal{B}_{p}^{s_{i_{1}}}\mathcal{B}_{p'}^{s_{b}}$ given the definition in Eq.~\eqref{eq:defbps}. Then, $\mathcal{B}_{p}^{s_{i_{1}}}\mathcal{B}_{p'}^{s_{b}}$ can be written as products of projectors that we separate into four groups.

\item Remove projectors in the second group that also appear in the third group using $P^2 = P$ for any projector $P$.

\item Use Eq.~\eqref{eq:projid1} to remove one projector from the third group. 

\item Thereafter, one may remove more `problematic projectors' using Eqs.~\eqref{eq:projid2} and \eqref{eq:projid3}  to obtain $e^{i\phi_{s_{b}}}\mathcal{C}_{p,p'}^{s_{b}}$.

\end{enumerate}

One can repeat the same exercise for $\mathcal{B}_{p'}^{s_{i_{2}}}\mathcal{B}_{p}^{s_{b}}$. The particularly nontrivial part is to check that $\phi_{s_{b}}$ obtained from both expressions match, since using Eq.~\eqref{eq:projid2} introduces a factor of $i$.

\subsubsection{Case 1}

First, let us show the relation for starting spin configurations consistent with the first row, first column of Table~\ref{t:summary}. In this case, $\mathcal{B}_{p}^{s_{i_{1}}}\mathcal{B}_{p'}^{s_{b}}$ can be generically written as
\begin{equation}
\label{eq:pprimecase1}
\begin{split}
& \mathcal{B}_{p}^{s_{i_{1}}}\mathcal{B}_{p'}^{s_{b}} = N_{1} (P_{a_{2}a_{3}}P_{a_{5}a_{6}}P_{a_{8}a_{9}}R^{p}_{a_{1}a_{7}})\\
&(P_{a_{3}a_{5}}P_{a_{6}a_{9}}P_{a_{1}a_{2}}P_{a_{7}a_{8}}R^{p})(P_{a_{3}a_{5}}P_{a_{6}a_{9}}R^{p'}_{a_{4},a_{10}})\\
&(P_{a_{3}a_{4}}P_{a_{5}a_{6}}P_{a_{9}a_{10}}R^{p'}).
\end{split}
\end{equation}
Here, $N_{1}$ is a normalization factor given by some integer power of $\sqrt{3}$.  Unitarity of $\mathcal{B}_{p}^{s_{i_{1}}}\mathcal{B}_{p'}^{s_{b}}$ on the subspace orthogonal to its kernel completely fixes the normalization factor throughout the steps of our projector manipulation---thus, we need not carefully track normalization.
We abbreviated a string of projectors that involve parafermions away from the ten sites shown in Fig.~\ref{fig:appendixcommutation} as $R^{p}$ and $R^{p'}$; subscripts indicate parafermions from the latter ten sites that are additionally involved in these strings.   
The fact that this abbreviation is possible is why $s_{b}$'s consistent with the same parafermion pairings within the area of Fig.~\ref{fig:appendixcommutation} can be treated on equal footing.  

We will explicitly show that $\mathcal{B}_{p}^{s_{i_{1}}}\mathcal{B}_{p'}^{s_{b}} = \mathcal{C}_{p,p'}^{s_{b}}$ with
\begin{equation}
\begin{split}
&\mathcal{C}_{p,p'}^{s_{b}} = N_{3}(P_{a_{2}a_{3}}P_{a_{5}a_{6}}P_{a_{8}a_{9}}R^{p'}_{a_{4}a_{10}}R^{p}_{a_{1},a_{7}})\\
&(P_{a_{1}a_{2}}P_{a_{3}a_{4}}P_{a_{5}a_{6}}P_{a_{7}a_{8}}P_{a_{9}a_{10}}R^{p'}R^{p}).
\end{split}
\end{equation}
Note that $\mathcal{B}_{p}^{s_{i_{1}}}\mathcal{B}_{p'}^{s_{b}}$ contains projectors $P_{a_{3}a_{5}}$ and $P_{a_{6}a_{9}}$ that are not in $\mathcal{C}_{p,p'}^{s_{b}}$. We will see how one can remove these two projectors and reorganize terms to establish the relation above. 

 To proceed, we first note that any two terms with non-overlapping subscripts commute.  We can thus move $P_{a_{3}a_{5}}$ in the second parenthesis of Eq.~\eqref{eq:pprimecase1} to the front of the third parenthesis. By using $P_{a_{3}a_{5}}^{2} = P_{a_{3}a_{5}}$, one can absorb the projector $P_{a_{3}a_{5}}$ that was originally in the second parenthesis to the third parenthesis. Additionally, we observed in Eq.~\eqref{eq:projid1} that erasing one parafermion projector from the plaquette exchange term keeps the expression identical. Thus, one can \textit{also remove $P_{a_{3}a_{5}}$ in the third parenthesis}. After removing all $P_{a_{3}a_{5}}$'s from Eq.~\eqref{eq:pprimecase1} and rearranging we get 
\begin{equation}
\begin{split}
&\mathcal{B}_{p}^{s_{i_{1}}}\mathcal{B}_{p'}^{s_{b}} = N_{1}(P_{a_{2}a_{3}}P_{a_{5}a_{6}}P_{a_{8}a_{9}}R^{p'}_{a_{4}a_{10}}R^{p}_{a_{1},a_{7}})\\
&(P_{a_{6}a_{9}})(P_{a_{1}a_{2}}P_{a_{3}a_{4}}P_{a_{5}a_{6}}P_{a_{7}a_{8}}P_{a_{9}a_{10}}R^{p'}R^{p}).
\end{split}
\end{equation}
Finally, Eq.~\eqref{eq:projid2} implies $P_{a_{5}a_{6}}P_{a_{6}a_{9}}P_{a_{5}a_{6}} = \frac{1}{3}P_{a_{5}a_{6}}$, which allows us to write
\begin{equation}
\label{eq:finalcase1}
\begin{split}
&\mathcal{B}_{p}^{s_{i_{1}}}\mathcal{B}_{p'}^{s_{b}} = \frac{N_{1}}{3}(P_{a_{2}a_{3}}P_{a_{5}a_{6}}P_{a_{8}a_{9}}R^{p'}_{a_{4}a_{10}}R^{p}_{a_{1},a_{7}})\\
&(P_{a_{1}a_{2}}P_{a_{3}a_{4}}P_{a_{5}a_{6}}P_{a_{7}a_{8}}P_{a_{9}a_{10}}R^{p'}R^{p}).
\end{split}
\end{equation}
The above expression is identical to the double-plaquette exchange term $\mathcal{C}_{p,p'}^{s_{b}}$ defined above.

One may similarly write
\begin{equation}
\begin{split}
&\mathcal{B}_{p'}^{s_{i_{2}}}\mathcal{B}_{p}^{s_{b}} = N_{2}(P_{a_{2}a_{3}}P_{a_{5}a_{6}}P_{a_{8}a_{9}}R^{p'}_{a_{4}a_{10}})\\
&(P_{a_{2}a_{5}}P_{a_{6}a_{8}}P_{a_{3}a_{4}}P_{a_{9}a_{10}}R^{p'})(P_{a_{2}a_{5}}P_{a_{6}a_{8}}R^{p}_{a_{1},a_{7}})\\
&(P_{a_{1}a_{2}}P_{a_{5}a_{6}}P_{a_{7}a_{8}}R^{p}).
\end{split}
\end{equation}
Minor modification of the preceding projector manipulations enables removal of $P_{a_{3}a_{5}}$ and $P_{a_{6}a_{9}}$ from $\mathcal{B}_{p'}^{s_{i_{2}}}\mathcal{B}_{p}^{s_{b}}$ to show $\mathcal{B}_{p'}^{s_{i_{2}}}\mathcal{B}_{p}^{s_{b}} = \mathcal{C}_{p,p'}^{s_{b}}$.  

\subsubsection{Case 2}

For $s_{b}$'s consistent with parafermion pairings in the second row, first column of Table~\ref{t:summary}, we have
\begin{equation}
\label{eq:AppInitialexp21}
\begin{split}
&\mathcal{B}_{p}^{s_{i_{1}}}\mathcal{B}_{p'}^{s_{b}} = N'_{2} (P_{a_{1}a_{2}} P_{a_{3}a_{5}} P_{a_{6}a_{9}} P_{a_{7}a_{8}} R^{p})\\
& (P_{a_{2}a_{3}} P_{a_{5}a_{6}} P_{a_{8}a_{9}} R^{p}_{a_{1},a_{7}})(P_{a_{2}a_{3}} P_{a_{5}a_{6}} P_{a_{8}a_{9}} R^{p'}_{a_{4}a_{10}})\\
&(P_{a_{3}a_{4}} P_{a_{2}a_{5}} P_{a_{6}a_{8}} P_{a_{9}a_{10}} R^{p'})
\end{split}
\end{equation}
and
\begin{equation}
\label{eq:AppInitialexp22}
\begin{split}
&\mathcal{B}_{p'}^{s_{i_{2}}}\mathcal{B}_{p}^{s_{b}} = N_{2}(P_{a_{3}a_{5}} P_{a_{6}a_{9}} R^{p'}_{a_{4}a_{10}})\\
&(P_{a_{3}a_{4}} P_{a_{5}a_{6}} P_{a_{9}a_{10}}R^{p'})(P_{a_{1}a_{2}} P_{a_{5}a_{6}} P_{a_{7}a_{8}} R^{p})\\
&(P_{a_{2}a_{5}} P_{a_{6}a_{8}} R^{p}_{a_{1},a_{7}}).
\end{split}
\end{equation}
We aim to show that $\mathcal{B}_{p}^{s_{i_{1}}}\mathcal{B}_{p'}^{s_{b}} = \mathcal{B}_{p'}^{s_{i_{2}}}\mathcal{B}_{p}^{s_{b}} = \mathcal{C}_{p,p'}^{s_{b}}$ with
\begin{equation}
\label{eq:AppFinalexp21}
\begin{split}
&\mathcal{C}_{p,p'}^{s_{b}} = N_{3}(P_{a_{1}a_{2}} P_{a_{7}a_{8}} P_{a_{3}a_{5}} P_{a_{6}a_{9}} R^{p'}_{a_{4}a_{10}}  R^{p}) \\
&(P_{a_{2}a_{5}} P_{a_{6}a_{8}}  P_{a_{3}a_{4}} P_{a_{9}a_{10}}R^{p'} R^{p}_{a_{1},a_{7}}).
\end{split}
\end{equation}

For $\mathcal{B}_{p}^{s_{i_{1}}}\mathcal{B}_{p'}^{s_{b}}$, note that $P_{a_{2}a_{3}}$, $P_{a_{5}a_{6}}$, and $P_{a_{8}a_{9}}$ appear in both the second and third parenthesis. Thus, one may drop these projectors from the second parenthesis.  We further remove $P_{a_{5}a_{6}}$ from the third parenthesis using the procedure justified by Eq.~\eqref{eq:projid1}. After rearranging, one obtains
\begin{equation}
\begin{split}
&\mathcal{B}_{p}^{s_{i_{1}}}\mathcal{B}_{p'}^{s_{b}} = N'_{2} (P_{a_{1}a_{2}} P_{a_{3}a_{5}} P_{a_{6}a_{9}} P_{a_{7}a_{8}} R^{p} R^{p'}_{a_{4}a_{10}} )\\
& (P_{a_{2}a_{3}} P_{a_{8}a_{9}})(P_{a_{3}a_{4}} P_{a_{2}a_{5}} P_{a_{6}a_{8}} P_{a_{9}a_{10}} R^{p'} R^{p}_{a_{1},a_{7}}).
\end{split}
\end{equation}
Equation~\eqref{eq:projid2} implies
\begin{equation}
\begin{split}
&P_{a_{3}a_{5}} P_{a_{2}a_{3}} P_{a_{2}a_{5}} = \frac{-i}{\sqrt{3}} P_{a_{3}a_{5}} P_{a_{2}a_{5}}  \\
& P_{a_{6}a_{9}}P_{a_{8}a_{9}} P_{a_{6}a_{8}} = \frac{+i}{\sqrt{3}} P_{a_{6}a_{9}} P_{a_{6}a_{8}},
\end{split}
\end{equation}
which allows us to remove $P_{a_{2}a_{3}}$ and $P_{a_{8}a_{9}}$ as well, without any additional phase factors introduced. We then obtain an expression identical to the right side of Eq.~\eqref{eq:AppFinalexp21}.
 
Showing $\mathcal{B}_{p'}^{s_{i_{2}}}\mathcal{B}_{p}^{s_{b}} = \mathcal{C}_{p,p'}^{s_{b}}$ is much simpler. Both the second and the third parenthesis of Eq.~\eqref{eq:AppInitialexp22} contain $P_{a_{5}a_{6}}$, so $P^2 = P$ and Eq.~\eqref{eq:projid1} enable us to simply remove $P_{a_{5}a_{6}}$ from both parentheses. By rearranging projectors that commute with each other, one can show that the final expression is identical to $\mathcal{C}_{p,p'}^{s_{b}}$. 

\subsubsection{Case 3}

Moving on to the third row, first column of Table~\ref{t:summary}, here we get
\begin{equation}
\begin{split}
&\mathcal{B}_{p}^{s_{i_{1}}}\mathcal{B}_{p'}^{s_b} = N_{1}(P_{a_{2}a_{5}} P_{a_{6}a_{9}} P_{a_{7}a_{8}} R^{p}_{a_{1}})\\
&(P_{a_{1}a_{2}} P_{a_{5}a_{6}} P_{a_{8}a_{9}} R^{p}_{a_{7}})(P_{a_{3}a_{4}} P_{a_{5}a_{6}} P_{a_{8}a_{9}} R^{p'}_{a_{10}})\\
&(P_{a_{9}a_{10}}P_{a_{3}a_{5}}P_{a_{6}a_{8}}R^{p'}_{a_{4}})
\end{split}
\end{equation}
and
\begin{equation}
\begin{split}
&\mathcal{B}_{p'}^{s_{i_{2}}}\mathcal{B}_{p}^{s_{b}} = N_{2}(P_{a_{2}a_{5}} P_{a_{3}a_{4}} P_{a_{6}a_{9}} R^{p'}_{a_{10}})\\
&(P_{a_{2}a_{3}} P_{a_{5}a_{6}} P_{a_{9}a_{10}} R^{p'}_{a_{4}})(P_{a_{2}a_{3}} P_{a_{5}a_{6}} P_{a_{7}a_{8}} R^{p}_{a_{1}})\\
&(P_{a_{1}a_{2}}P_{a_{3}a_{5}}P_{a_{6}a_{8}}R^{p}_{a_{7}}).
\end{split}
\end{equation}
These expressions are related by $\mathcal{B}_{p}^{s_{i_{1}}}\mathcal{B}_{p'}^{s_b} = \mathcal{B}_{p'}^{s_{i_{2}}}\mathcal{B}_{p}^{s_{b}} = i \mathcal{C}_{p,p'}^{s_{b}}$ with
\begin{equation}
\begin{split}
&\mathcal{C}_{p,p'}^{s_{b}} = N_{3}(P_{a_{2}a_{5}} P_{a_{3}a_{4}} P_{a_{6}a_{9}} P_{a_{7}a_{8}} R^{p'}_{a_{10}} R^{p}_{a_{1}}) \\
& (P_{a_{1}a_{2}} P_{a_{3}a_{5}} P_{a_{6}a_{8}} P_{a_{9}a_{10}} R^{p'}_{a_{4}} R^{p}_{a_{7}}).
\end{split}
\end{equation}

Let us explicitly work out the projector manipulations for $\mathcal{B}_{p}^{s_{i_{1}}}\mathcal{B}_{p'}^{s_b}$.  Both the second and third parenthesis contain $P_{a_{8}a_{9}}$ and $P_{a_{5}a_{6}}$. We will drop these projectors from the second parenthesis, and additionally remove $P_{a_{5}a_{6}}$ from the third parenthesis following the usual justification.  After reorganizing terms, one obtains 
\begin{equation}
\begin{split}
&\mathcal{B}_{p}^{s_{i_{1}}}\mathcal{B}_{p'}^{s_b} =  N_{1}(P_{a_{2}a_{5}} P_{a_{3}a_{4}} P_{a_{6}a_{9}} P_{a_{7}a_{8}} R^{p}_{a_{1}}R^{p'}_{a_{10}})\\
&(P_{a_{8}a_{9}} ) (P_{a_{1}a_{2}}P_{a_{9}a_{10}}P_{a_{3}a_{5}}P_{a_{6}a_{8}} R^{p}_{a_{7}}R^{p'}_{a_{4}}).
\end{split}
\end{equation}
One can remove $P_{a_{8}a_{9}}$ as well with the help of Eq.~\eqref{eq:projid2},
\begin{equation}
P_{a_{6}a_{9}} P_{a_{8}a_{9}} P_{a_{6}a_{8}} =  \frac{i}{\sqrt{3}}P_{a_{6}a_{9}} P_{a_{6}a_{8}},
\end{equation}
yielding the final expression
\begin{equation}
\begin{split}
&\mathcal{B}_{p}^{s_{i_{1}}}\mathcal{B}_{p'}^{s_b} = i \frac{N_{1}}{\sqrt{3}}(P_{a_{2}a_{5}} P_{a_{3}a_{4}} P_{a_{6}a_{9}} P_{a_{7}a_{8}} R^{p}_{a_{1}}R^{p'}_{a_{10}})\\
&(P_{a_{1}a_{2}}P_{a_{9}a_{10}}P_{a_{3}a_{5}}P_{a_{6}a_{8}} R^{p}_{a_{7}}R^{p'}_{a_{4}}) = i \mathcal{C}_{p,p'}^{s_{b}}.
\end{split}
\end{equation}
Almost identical steps establish that $\mathcal{B}_{p'}^{s_{i_{2}}}\mathcal{B}_{p}^{s_{b}}=i \mathcal{C}_{p,p'}^{s_{b}}$.  (One only needs to effectively $\pi$-rotate the parafermion indices used for the projector algebra for $\mathcal{B}_{p}^{s_{i_{1}}}\mathcal{B}_{p'}^{s_b}$.)

\subsubsection{Case 4}

We will leave how the projector algebra works out for the fourth row Table~\ref{t:summary} as an exercise for intellectually stimulated readers.

\section{Explicit construction of $L_{\mu_1,\omega}$ for the decorated-domain-wall model}
\label{app:symop}

In Sec.~\ref{sec:sa1} we introduced a local operator $L_{\mu_1,\omega}$ that flips the anyon Ising spins at bond $\mu_1$ and also changes the eigenvalue for $e^{i \pi Q_{\mu_1}^+}$ from $\omega$ to $\bar \omega$.  Here we will explicitly construct $L_{\mu_1,\omega}$.  

We first define a unitary operator $\mathcal{L}^{(s)}_{\mu_1,\omega}$ that transforms $\ket{{\rm PF}'(s)}$ into $\ket{{\rm PF}''(\bar{s})}$ via the intermediate state $\ket{{\rm PF}_{\rm mod}(s)}$.  (For the precise definition of these states see Sec.~\ref{sec:sa1}.)  It will be convenient to employ the following set of projectors involving sites from Fig.~\ref{fig:symact1}(a): 
\begin{equation}
P_{\mu_{1}}^{t} = \frac{1 + t^{*} e^{i\pi Q_{\mu_{1}}^{+}} + t e^{-i\pi Q_{\mu_{1}}^{+}}}{3} 
\end{equation}
along with 
\begin{equation}
P^{t}_{ab} = \begin{cases}
\frac{1 + t^{*} \mathcal{F}_{ab} + t \mathcal{F}_{ab}^{\dagger}}{3} &\quad \text{if $a$ and $b$ are on $\rho$} \\
\frac{1 + t \mathcal{F}_{a_{\rho}a_{\lambda}} + t^{*} \mathcal{F}_{a_{\rho}a_{\lambda}}^{\dagger}}{3} &\quad \text{if $(a,b) = (a_{\lambda},a_{\rho})$ or $(a_{\rho}, a_{\lambda})$ } \\
P_{ab}  &\quad \text{else}
\end{cases} ,
\label{Ptab}
\end{equation}
where $t = 1$, $\omega$ or $\bar{\omega}$. Note that $P_{\mu_{1}}^{\bar{\omega}}$ enforces eigenvalues of $e^{i\pi Q_{\mu_{1}}^{+}}$ appropriate for the states $\ket{{\rm PF}_{\rm mod}(s)}$ and $\ket{{\rm PF}''(\bar{s})}$, while $P^{\omega}_{ab}$ enforces eigenvalues of bonds other than $\mu_{1}$ to be consistent with $\ket{{\rm PF}_{\rm mod}(s)}$. Bearing this in mind, we then have
\begin{eqnarray}
\label{eq:ldef}
 \mathcal{L}_{\mu_{1},\omega}^{(s)} &=& \left(P_{a_{2k}a_{1}}\prod_{i=1}^{k-1} \sqrt{3} P_{a_{2i} a_{2i+1}}\right)S_{p_{1}}S_{p_{2}}
 \nonumber \\
 &\times& \left(P_{\mu_{1}}^{\bar\omega} \prod_{i=1}^{k} P^{\omega}_{a_{2i-1} a_{2i}}\right)\mathcal{F}_{a_{\rho}a_{\mu_{1}}} \nonumber \\
 &=& \mathcal{D}_{\mu_{1},\omega}^{(s)}\mathcal{F}_{a_{\rho}a_{\mu_{1}}} S_{p_{1}}S_{p_{2}}.
\end{eqnarray}
The second line sends $\ket{{\rm PF}'(s)}$ to $\ket{{\rm PF}_{\rm mod}(s)}$, and then the first line flips the spins at plaquettes $p_{1,2}$ adjacent to edge $\mu_1$ and reconfigures the parafermion pairings to yield the desired final state $\ket{{\rm PF}''(\bar{s})}$.
In the last line, we grouped the string of projectors into $\mathcal{D}_{\mu_{1},\omega}^{(s)}$ for later convenience.

A similar technique to that presented in Sec.~\ref{sec:Setup} establishes that $\mathcal{L}_{\mu_{1},\omega}^{(s)} $ is unitary on the subspace orthogonal to its kernel, and that its unitarity crucially relies on the fact that Eq.~\eqref{eq:trieq2} does not impose a triality obstruction.  

 To construct the full operator $L_{\mu_{1},\omega}$, let us divide spin configurations around the the double plaquette $p_{1,2}$ into two categories, $\{ s_{1} \}$ and $\{s_{2} \}$.  Spin configurations in $\{ s_{1} \}$ impose intra-edge pairing on bond $\rho$, while all others belong to $\{s_2\}$.  
We can then write
\begin{equation}
\label{eq:symactt}
L_{\mu_{1},\omega} =  \sum_{k = 1,2} \sum_{ \{s_{k} \} } \bar{\omega}^{k} \mathcal{L}_{\mu_{1},\omega}^{(s_{k})} \ket{s_{k},\mu_1}\bra{s_{k},\mu_1},
\end{equation} 
where $\ket{s_{k},\mu_1}\bra{s_{k},\mu_1}$ projects onto a particular configuration for spins at the plaquettes near edge $\mu_1$.  
 Proving that $L_{\mu_{1},\omega}$ commutes with $A_{v}$ and $B_{p}$ is generally either trivial or proceeds via a straightforward generalization of techniques we presented in Sec.~\ref{sec:Setup} and Appendix~\ref{app:pcom}.  The exception is commutation between $L_{\mu_{1},\omega}$ and $B_{p_{\rho}}$, $p_{\rho}$ being the plaquette that contains the edge $\rho$ but not $\mu_{1}$.  Appendix~\ref{app:pcom2} presents the projector algebra needed to establish this commutation relation---which in fact requires us to add the phase factor $\bar{\omega}^{k}$ in Eq.~\eqref{eq:symactt}.  
  
 As a final remark, let us define $L_{\mu_{1},\bar\omega}$, the `inverse' operator of $L_{\mu_{1},\omega}$, as 
\begin{equation}
L_{\mu_{1},\bar\omega} =  \sum_{k = 1,2} \sum_{ \{s_{k} \} } \bar{\omega}^{k} \mathcal{L}_{\mu_{1},\bar\omega}^{(s_{k})} \ket{s_{k},\mu_1}\bra{s_{k},\mu_1}
\end{equation}
with
\begin{eqnarray}
 \mathcal{L}_{\mu_{1},\bar\omega}^{(s)} &=& \mathcal{D}_{\mu_{1},\bar\omega}^{(s)} \mathcal{F}_{a_{\rho}a_{\mu_{1}}}^{\dagger} S_{p_{1}}S_{p_{2}} \\
 \mathcal{D}_{\mu_{1},\bar\omega}^{(s)} &=& \left(P_{a_{2k}a_{1}}\prod_{i=1}^{k-1} \sqrt{3} P_{a_{2i} a_{2i+1}}\right) \left(P_{\mu_{1}}^{\omega} \prod_{i=1}^{k} P^{\bar\omega}_{a_{2i-1} a_{2i}}\right).
 \nonumber 
\end{eqnarray}
While it is not clear in the current form that $L_{\mu_{1},\bar\omega} = L_{\mu_{1},\omega}^{\dagger}$, one can show this is indeed true by using 
\begin{equation}
\label{eq:hopprojcom}
\begin{split}
\mathcal{F}_{a_{\rho}a_{\mu_{1}}}P_{\mu_{1}}^{t} &= P_{\mu_{1}}^{ \omega t}\mathcal{F}_{a_{\rho}a_{\mu_{1}}} \\
\mathcal{F}_{a_{\rho}a_{\mu_{1}}}P_{ab}^{t} &= P_{ab}^{\omega t}\mathcal{F}_{a_{\rho}a_{\mu_{1}}},
\end{split}
\end{equation}
which follows from the commutation relation between $\mathcal{F}_{a_{\rho}a_{\mu_{1}}}$ and other $\mathcal{F}_{ab}$'s.

\section{Proof of $[ L_{\mu_1,\omega}, B_{p_{\rho}}] = 0 $}
\label{app:pcom2}

\subsection{Additional projector identities}

We start by providing more projector identities that are relevant to the proof of the commutation relation $[ L_{\mu_1,\omega}, B_{p_{\rho}}] = 0 $ that we seek to establish in this Appendix.  When proving $[B_{p}, B_{p'}] = 0$ we observed that one of the projectors on the right side of Eq.~\eqref{eq:defbps} can be removed to obtain an equivalent expression.  The same modification can be applied to the string of projectors $\mathcal{D}_{\mu_{1},\omega}^{(s)}$ that appears in the definition of $L_{\mu_1,\omega}$ [see Eqs.~\eqref{eq:ldef} and \eqref{eq:symactt}], and also to the operator $\mathcal{B}_{p_{\rho},\omega}^{s}$ obtained by swapping a different set of projectors into $\mathcal{B}_{p_\rho}^s$ [see below].  That is, we can straightforwardly generalize Eq.~\eqref{eq:projid1} for
these operators.  The generalized triality conservation Eq.~\eqref{eq:trieq2} makes such a manipulation possible in $\mathcal{D}_{\mu_{1},\omega}^{(s)}$.  

Second, we will require the following variant of Eq.~\eqref{eq:projid2},
\begin{equation}
\label{eq:projid4}
\begin{split}
P_{a_{8}a_{10}} \mathcal{F}_{a_{7}a_{8}}^{\dagger} P_{a_{10}a_{11}} P_{a_{8}a_{11}} = \frac{-i\omega}{\sqrt{3}} P_{a_{8}a_{10}} \mathcal{F}_{a_{7}a_{8}}^{\dagger} P_{a_{8}a_{11}} \\
P_{a_{8}a_{11}} \mathcal{F}_{a_{7}a_{8}}^{\dagger} P_{a_{10}a_{11}} P_{a_{8}a_{10}} = \frac{i\bar{\omega}}{\sqrt{3}} P_{a_{8}a_{11}} \mathcal{F}_{a_{7}a_{8}}^{\dagger} P_{a_{8}a_{10}},
\end{split}
\end{equation}
where the site labels are shown in Fig.~\ref{fig:appcomL1}.  
Let us show how to recover the first line.  Using $P_{a_{8}a_{10}} = P_{a_{8}a_{10}}\mathcal{F}_{a_{8}a_{10}}$ and $P_{a_{8}a_{11}} = \mathcal{F}_{a_{11}a_{8}}P_{a_{8}a_{11}}$ we can express
\begin{equation}
\begin{split}
&P_{a_{8}a_{10}} \mathcal{F}_{a_{7}a_{8}}^{\dagger}P_{a_{10}a_{11}} P_{a_{8}a_{11}} \\
&= \frac{1}{3}P_{a_{8}a_{10}}(\mathcal{F}_{a_{7}a_{8}}^{\dagger} + \mathcal{F}_{a_{8}a_{10}}\mathcal{F}_{a_{7}a_{8}}^{\dagger}
\mathcal{F}_{a_{10}a_{11}}\mathcal{F}_{a_{11}a_{8}} \\
&+ \mathcal{F}_{a_{10}a_{8}}\mathcal{F}_{a_{7}a_{8}}^{\dagger}
\mathcal{F}_{a_{11}a_{10}}\mathcal{F}_{a_{11}a_{8}}) P_{a_{8}a_{11}}. \\
\end{split}
\end{equation}
Next, Property 2 yields $\mathcal{F}_{a_{8}a_{10}}\mathcal{F}_{a_{7}a_{8}}^{\dagger} = \omega \mathcal{F}_{a_{7}a_{8}}^{\dagger} \mathcal{F}_{a_{8}a_{10}}$, which allows us to write
\begin{equation}
\begin{split}
&P_{a_{8}a_{10}} \mathcal{F}_{a_{7}a_{8}}^{\dagger}P_{a_{10}a_{11}} P_{a_{8}a_{11}} \\
&= \frac{1}{3}P_{a_{8}a_{10}}\mathcal{F}_{a_{7}a_{8}}^{\dagger} ( 1 + \omega \mathcal{F}_{a_{8}a_{10}}
\mathcal{F}_{a_{10}a_{11}}\mathcal{F}_{a_{11}a_{8}} \\
&+ \bar{\omega}\mathcal{F}_{a_{10}a_{8}}
\mathcal{F}_{a_{11}a_{10}}\mathcal{F}_{a_{11}a_{8}}) P_{a_{8}a_{11}}. \\
\end{split}
\end{equation}
And finally, using Eq.~\eqref{eq:gobrel} one can deduce 
\begin{equation}
\mathcal{F}_{a_{8}a_{10}}\mathcal{F}_{a_{10}a_{11}}\mathcal{F}_{a_{11}a_{8}} = \mathcal{F}_{a_{10}a_{8}}
\mathcal{F}_{a_{11}a_{10}}\mathcal{F}_{a_{11}a_{8}} = \bar{\omega},
\end{equation}
so that
\begin{equation}
\begin{split}
P_{a_{8}a_{10}} \mathcal{F}_{a_{7}a_{8}}^{\dagger} & P_{a_{10}a_{11}} P_{a_{8}a_{11}} \\
&= \frac{1}{3}P_{a_{8}a_{10}}\mathcal{F}_{a_{7}a_{8}}^{\dagger}(1 + 1 + \omega) P_{a_{8}a_{11}} \\ 
&= \frac{-i \omega }{\sqrt{3}} P_{a_{8}a_{10}} \mathcal{F}_{a_{7}a_{8}}^{\dagger} P_{a_{8}a_{11}}.
\end{split}
\end{equation}
One can proceed similarly to prove the second line of Eq.~\eqref{eq:projid4}.

Finally, the following generalized versions of Eq.~\eqref{eq:projid3} hold for projectors involving parafermion sites $a_{6}$, $a_{7}$, $a_{8}$, $a_{10}$, and $a_{11}$ in Fig.~\ref{fig:appcomL1}: 
\begin{equation}
\label{eq:projid5}
\begin{split}
P_{a_{7}a_{8}}P_{a_{8}a_{10}}P_{a_{7}a_{8}}^{\omega} &= \frac{1}{3} P_{a_{7}a_{8}}\mathcal{F}_{a_{10}a_{8}}P_{a_{7}a_{8}}^{\omega} \\
P_{a_{7}a_{8}}P_{a_{8}a_{11}}P_{a_{7}a_{8}}^{\omega} &= \frac{1}{3} P_{a_{7}a_{8}}\mathcal{F}_{a_{11}a_{8}}P_{a_{7}a_{8}}^{\omega} \\
P_{a_{6}a_{7}}P_{a_{7}a_{8}}^{t}P_{a_{6}a_{7}}^{\omega} &= \frac{t}{3}P_{a_{6}a_{7}} \mathcal{F}_{a_{7}a_{8}}^{\dagger}P_{a_{6}a_{7}}^{\omega},
\end{split}
\end{equation}
where $P^t_{ab}$ is the modified projector defined in Eq.~\eqref{Ptab}.  
Note that these identities leave $\mathcal{F}_{a_{10}a_{8}}$, $\mathcal{F}_{a_{11}a_{8}}$, or $\mathcal{F}_{a_{7}a_{8}}^{\dagger}$ sandwiched between operators while Eq.~\eqref{eq:projid3} completely removes the projector at the middle.

 We will present the proof for the third line of Eq.~\eqref{eq:projid5} [the first two lines of Eq.~\eqref{eq:projid4} can be shown with almost identical techniques]. Start from
\begin{equation}
P_{a_{6}a_{7}}P_{a_{7}a_{8}}^{t}P_{a_{6}a_{7}}^{\omega} = P_{a_{6}a_{7}}\frac{1+te^{i \pi Q_{\rho}^{+}} + t^{*}e^{-i \pi Q_{\rho}^{+}}}{3}P_{a_{6}a_{7}}^{\omega}.
\end{equation}
Note that $P_{a_{6}a_{7}}P_{a_{6}a_{7}}^{\omega} =0$ since the first projector projects onto a state with $\mathcal{F}_{a_{6}a_{7}} =1$ while $P_{a_{6}a_{7}}^{\omega}$ projects onto a state with  $\mathcal{F}_{a_{6}a_{7}} = \omega$. Similarly, we have
\begin{equation}
\begin{split}
 P_{a_{6}a_{7}} & t^{*}e^{-i \pi Q_{\rho}^{+}}P_{a_{6}a_{7}}^{\omega} \\
 &= t^{*}e^{-i \pi Q_{\rho}^{+}} \frac{1 + \omega\mathcal{F}_{a_{6}a_{7}} + \bar{\omega} \mathcal{F}_{a_{6}a_{7}}^{\dagger}}{3} P_{a_{6}a_{7}}^{\omega} \\
&= t^{*}e^{-i \pi Q_{\rho}^{+}} P_{a_{6}a_{7}}^{\bar{\omega}}P_{a_{6}a_{7}}^{\omega} = 0.
\end{split}
\end{equation}
In going from the first to the second line, we used Property 2.  The last line of Eq.~\eqref{eq:projid5} naturally follows after noting that $e^{i \pi Q_{\rho}^{+}} = \mathcal{F}_{a_{7}a_{8}}^{\dagger}$. 

\subsection{Strategy to prove $[ L_{\mu_1,\omega}, B_{p_{\rho}}] = 0 $}

Our strategy to establish $[ L_{\mu_1,\omega}, B_{p_{\rho}}] = 0 $ is almost identical to the one we used to prove $[B_{p}, B_{p'}] = 0$ in Appendix ~\ref{app:pcom}. We already know that the spin parts of the two operators trivially commute; the nontrivial part of the proof involves showing that their parafermion parts commute as well. Analogous to Eq.~\eqref{eq:comactualform}, proving commutativity reduces to showing that 
\begin{equation}
\label{eq:comL1raw}
\bar{\omega}^{k_{s_{b}}} \mathcal{B}_{p_{\rho}}^{s_{i_{1}}} \mathcal{D}_{\mu_{1},\omega}^{(s_{b})} \mathcal{F}_{a_{\rho}a_{\mu_{1}}} = \bar{\omega}^{k_{s_{i_{2}}}} \mathcal{D}_{\mu_{1},\omega}^{(s_{i_{2}})} \mathcal{F}_{a_{\rho}a_{\mu_{1}}} \mathcal{B}_{p_{\rho}}^{s_{b}} 
\end{equation}
for all starting spin configurations $s_{b}$.
Here, $s_{i_{1}}$ is a spin configuration obtained from $s_{b}$ by applying $S_{p_{1}}S_{p_{2}}$, while $s_{i_{2}}$ is a spin configuration obtained from $s_b$ by applying $S_{p_{\rho}}$.  We also have $k_{s_{b}}, k_{s_{i_2}} =1,2$ depending on whether $s_{b}$ and $s_{i_2}$ are consistent with intra-edge pairing on edge $\rho$; these factors reflect the $\bar{\omega}^{k}$ introduced in Eq.~\eqref{eq:symactt}. 

\begin{figure}
  \includegraphics[width=0.8\linewidth]{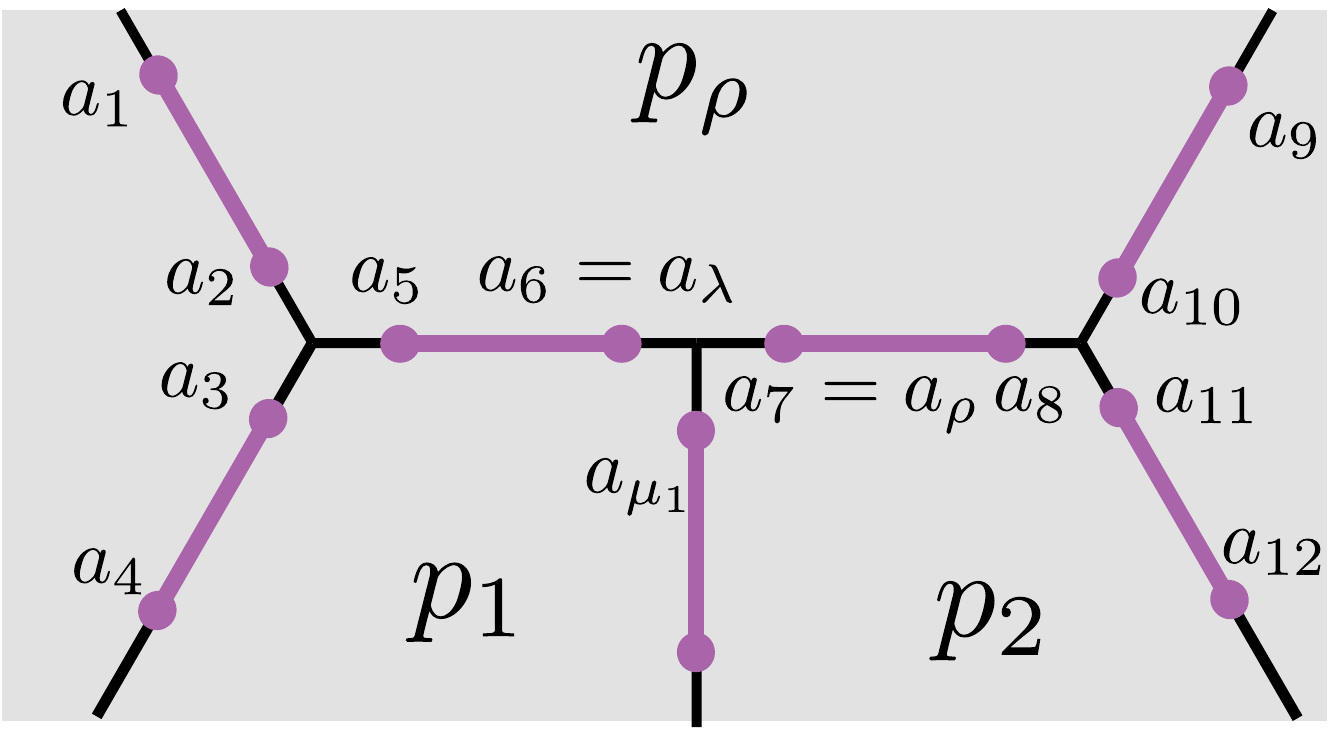} 
  \caption{Illustration of parafermions in the neighborhood of the edges where plaquettes $p_{\rho}$, $p_{1}$ and $p_{2}$ meet.  Note that we have distorted the usual honeycomb geometry for convenience.  Only parafermions labeled with arabic numbers are involved in `problematic projectors' that should be handled with projector identities when proving $[ L_{\mu_1,\omega}, B_{p_{\rho}}] = 0 $.}
  \label{fig:appcomL1}
\end{figure}

Further simplification is possible by defining an operator $\mathcal{B}_{p_{\rho},\omega}^{s}$ that is identical to $\mathcal{B}_{p_{\rho}}^{s}$ but with all projectors $P_{ab}$ replaced by $P_{ab}^{\omega}$.  
Using Eq.~\eqref{eq:hopprojcom}, one can show that $\mathcal{F}_{a_{\rho}a_{\mu_{1}}}\mathcal{B}_{p_{\rho}}^{s} = \mathcal{B}_{p_{\rho},\omega}^{s}\mathcal{F}_{a_{\rho}a_{\mu_{1}}}$. This relation implies that Eq.~\eqref{eq:comL1raw} is equivalent to the following without the annoying $\mathcal{F}_{a_{\rho}a_{\mu_{1}}}$ flying around: 
\begin{equation}
\label{eq:comL1refined}
\bar{\omega}^{k_{s_{b}}} \mathcal{B}_{p_{\rho}}^{s_{i_{1}}} \mathcal{D}_{\mu_{1},\omega}^{(s_{b})} = \bar{\omega}^{k_{s_{i_{2}}}} \mathcal{D}_{\mu_{1},\omega}^{(s_{i_{2}})} \mathcal{B}_{p_{\rho},\omega}^{s_{b}} = e^{i\phi_{s_{b}}} \mathcal{T}^{s_{b}}.
\end{equation}
On the right side we introduced a phase $e^{i\phi_{s_b}}$ and an operator $\mathcal{T}^{s_{b}}$ that directly implements parafermion-pairing flips around the triple-plaquette consisting of $p_{1}$, $p_{2}$, and $p_{\rho}$. 
Once again, nontriviality of the above relation comes from the projectors involving parafermions around the region where $p_{1}$, $p_{2}$, and $p_{\rho}$ meet; thus, the projector algebra for different $s_{b}$'s with the same parafermion pairings around the geometry in Fig.~\ref{fig:appcomL1} can be treated on equal footing. 
 
The eight starting configurations are illustrated in the $\mathcal{S}_{b}$ columns of Table~\ref{t:summary2}.  Yellow bonds indicate pairing consistent with $P^{\omega}_{ab}=1$ rather than $P_{ab}=1$.  Also, the $\mu_{1}$ bond is colored blue, indicating that the starting configuration is always consistent with $P_{\mu_{1}}^{\bar{\omega}}=1$. The operator $\mathcal{T}^{s_{b}}$, as well as the equivalent operators in Eq.~\eqref{eq:comL1refined}, reconfigures parafermion pairings to those illustrated in the $\mathcal{S}_{f}$ columns. Note that configurations in the $\mathcal{S}_{f}$ columns do not have yellow bonds.

We stress that Hermitian conjugating Eq.~\eqref{eq:comL1refined} gives a different equation that cannot be obtained by simply choosing a different $s_{b}$ in Eq.~\eqref{eq:comL1refined}---in sharp contrast to the situation for Eq.~\eqref{eq:comactualform}. 
Thus, Eq.~\eqref{eq:comL1refined} for all eight species of Ising configurations should be considered individually, whereas in Appendix~\ref{app:pcom} only working out half of the cases was sufficient.  


\begin{table*}

\caption{Summary of the results of the projector algebra used for proving Eq.~\eqref{eq:comL1refined}. Columns $\mathcal{S}_{b}$ and $\mathcal{S}_{f}$ respectively illustrate the eight starting parafermion pairing configurations and the resulting configurations after acting $\bar{\omega}^{k_{s_{b}}} \mathcal{B}_{p_{\rho}}^{s_{i_{1}}} \mathcal{D}_{\mu_{1},\omega}^{(s_{b})} = \bar{\omega}^{k_{s_{i_{2}}}} \mathcal{D}_{\mu_{1},\omega}^{(s_{i_{2}})} \mathcal{B}_{p_{\rho},\omega}^{s_{b}}$.  Blue bonds indicate pairing consistent with $P^{\bar{\omega}}_{\mu_{1}}=1$; yellow bonds indicate pairing consistent with $P_{ab}^{\omega}=1$. The operator $\mathcal{O}^{s_{b}}$ [Eq.~\eqref{eq:tripleplaquetteex}] and the phase factor $e^{i\phi_{s_{b}}}$ [Eq.~\eqref{eq:comL1refined}] obtained from the projector algebra for each of eight starting configurations are also shown.}
\label{t:summary2}

\begin{center}
\begin{tabularx}{\linewidth}{|>{\centering\hsize=1.4\hsize}X|>{\centering\hsize=1.4\hsize}X|>{\centering\hsize=0.6\hsize}X|>{\centering\hsize=0.6\hsize}X|>{\centering\hsize=1.4\hsize}X|>{\centering\hsize=1.4\hsize}X|>{\centering\hsize=0.6\hsize}X|X<{\centering\hsize=0.6\hsize}|}
\hline 
$\mathcal{S}_{b}$ & $\mathcal{S}_{f}$ & $\mathcal{O}^{s_{b}}$ & $e^{i\phi_{s_{b}}}$ & $\mathcal{S}_{b}$ & $\mathcal{S}_{f}$ & $\mathcal{O}^{s_{b}}$ & $e^{i\phi_{s_{b}}}$ \\
\hline \hline
\raisebox{-.5\height}{\includegraphics[width = 1.4\linewidth]{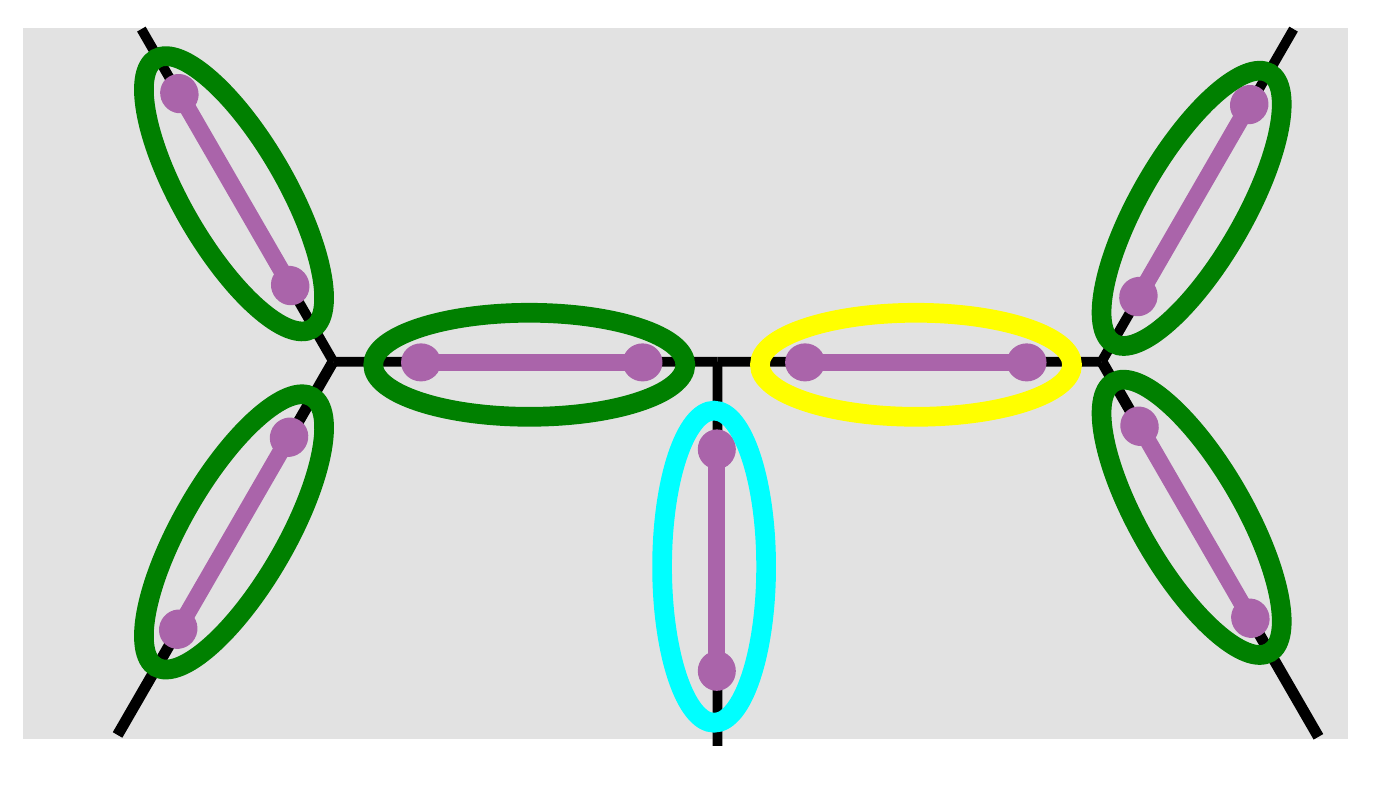}} & \raisebox{-.5\height}{\includegraphics[width = 1.4\linewidth]{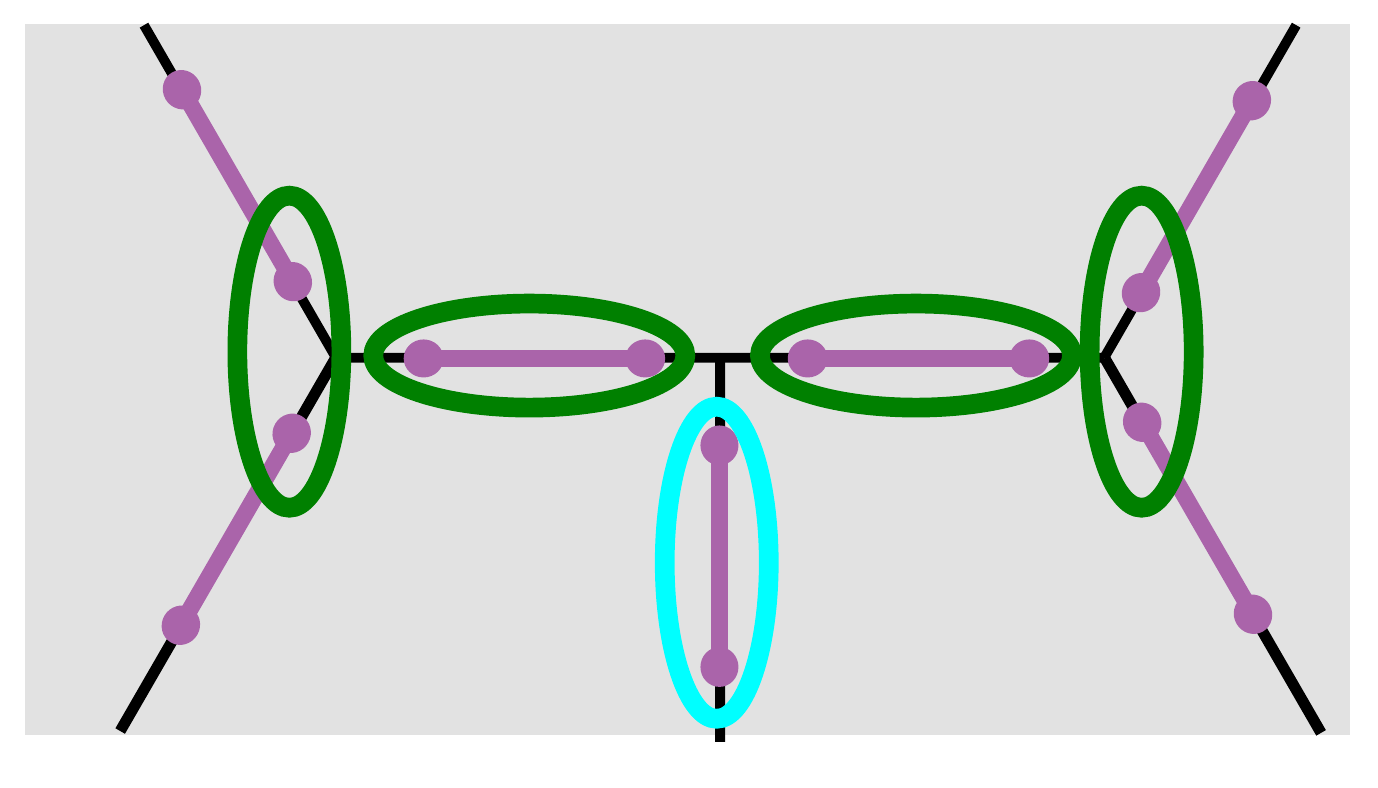}} & $\mathcal{F}_{a_{10}a_{8}}$ & $\omega$ & \raisebox{-.5\height}{\includegraphics[width = 1.4\linewidth]{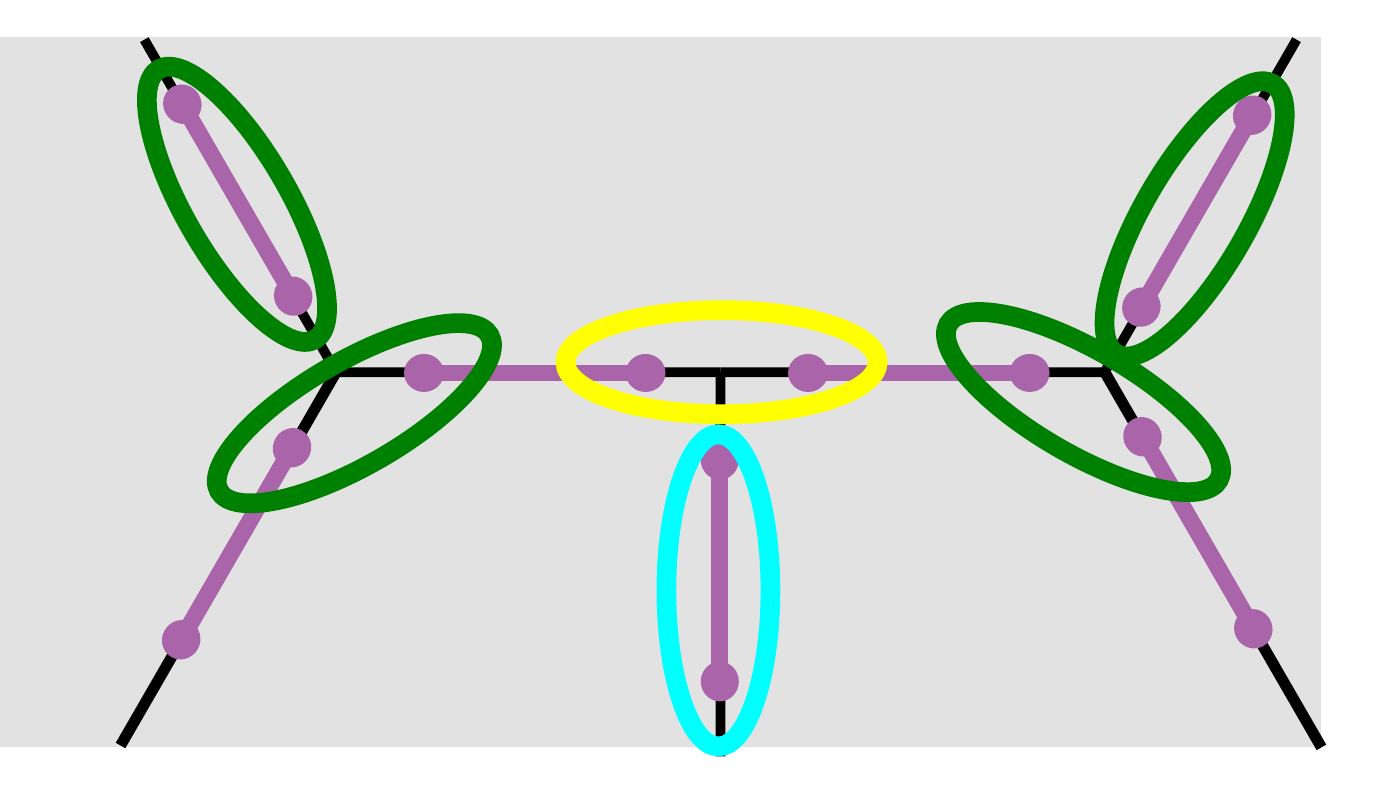}} & \raisebox{-.5\height}{\includegraphics[width = 1.4\linewidth]{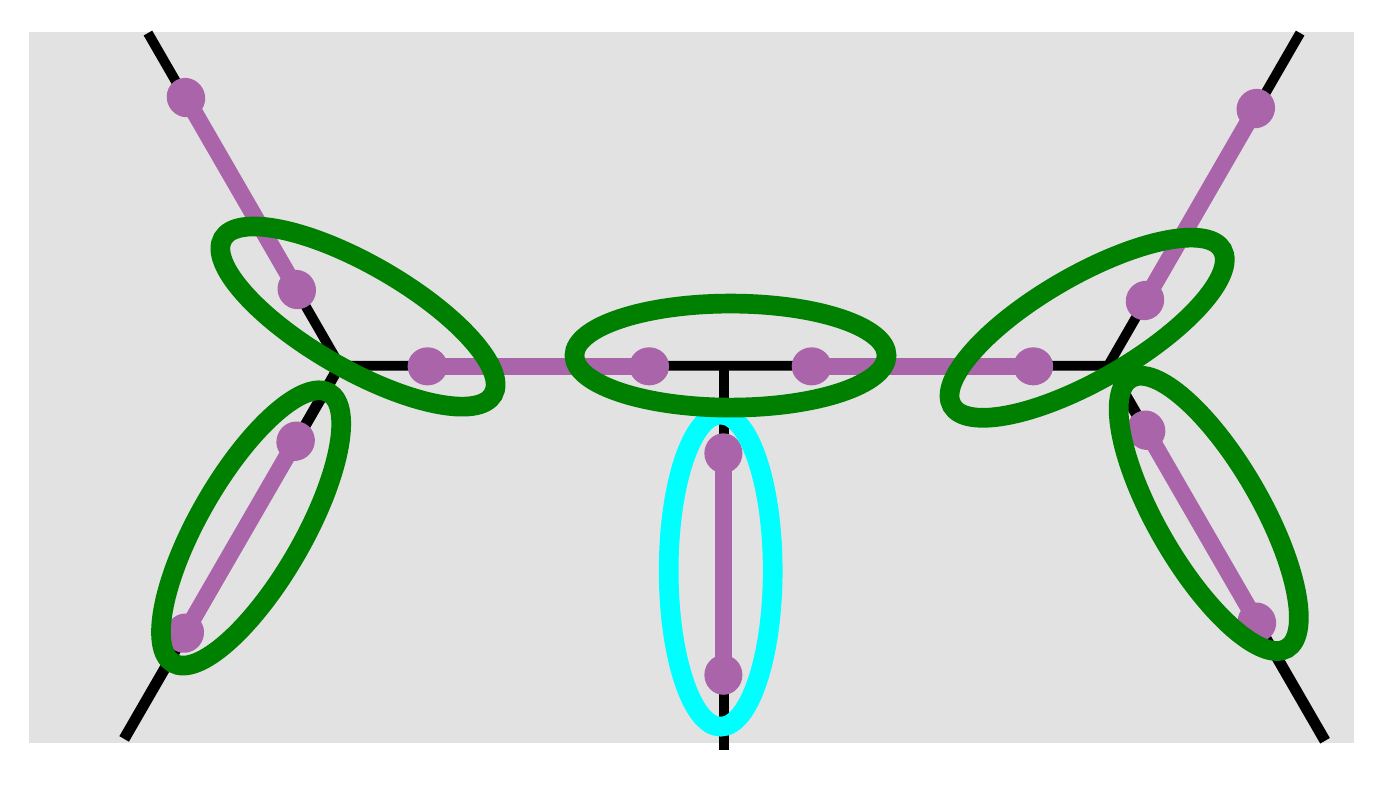}} & $\mathcal{F}_{a_{7}a_{8}}^{\dagger}$ & $\omega$ \\ \hline
\raisebox{-.5\height}{\includegraphics[width = 1.4\linewidth]{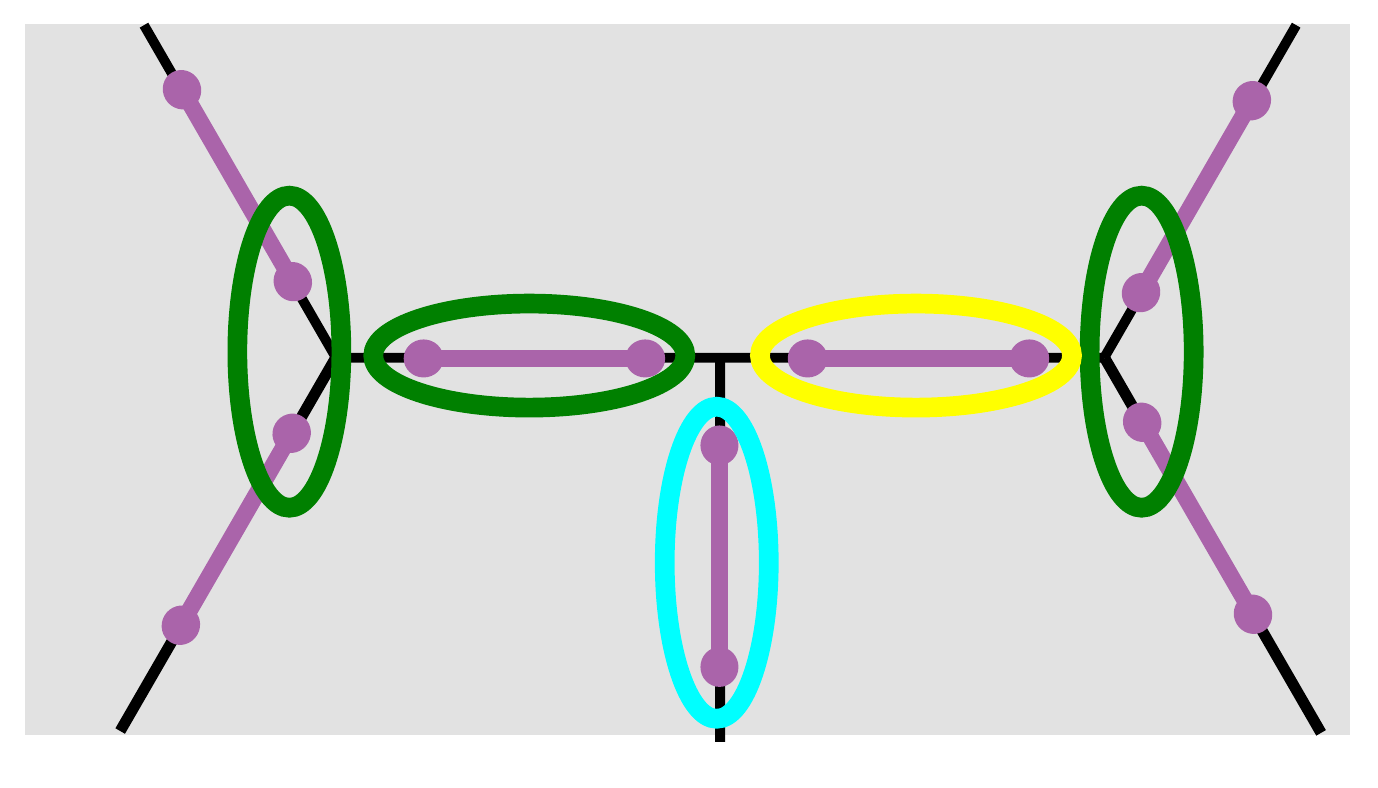}} & \raisebox{-.5\height}{\includegraphics[width = 1.4\linewidth]{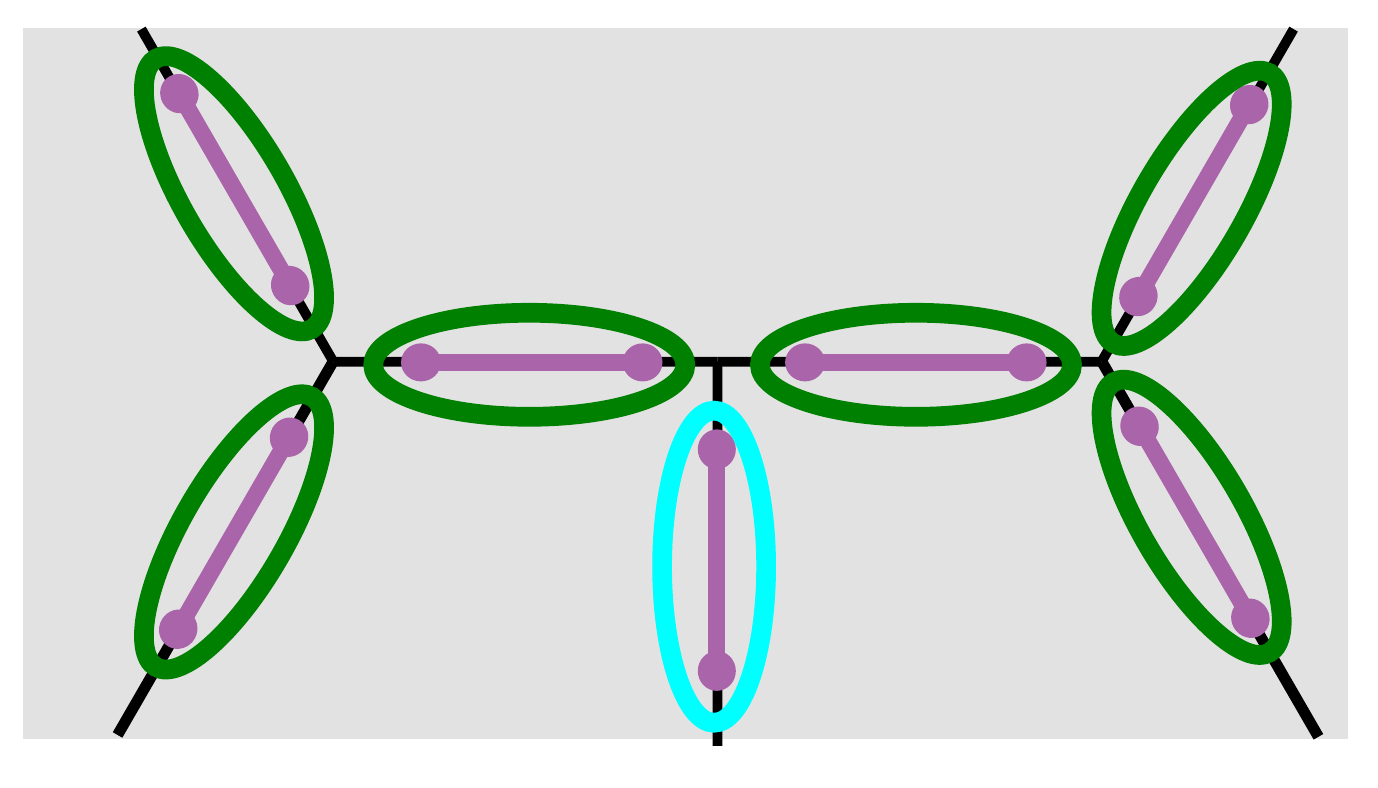}} & $\mathcal{F}_{a_{10}a_{8}}$ & $\bar{\omega}$ & \raisebox{-.5\height}{\includegraphics[width = 1.4\linewidth]{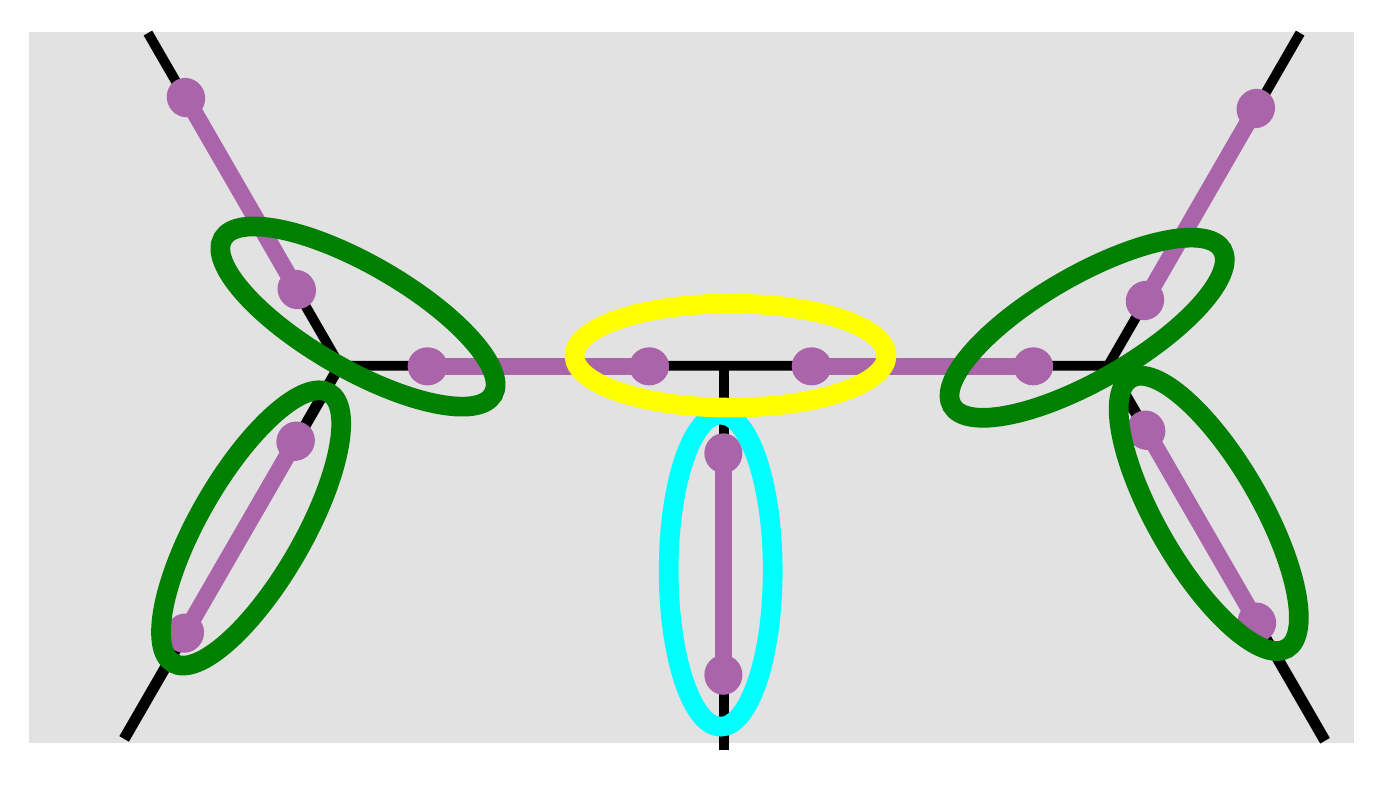}} & \raisebox{-.5\height}{\includegraphics[width = 1.4\linewidth]{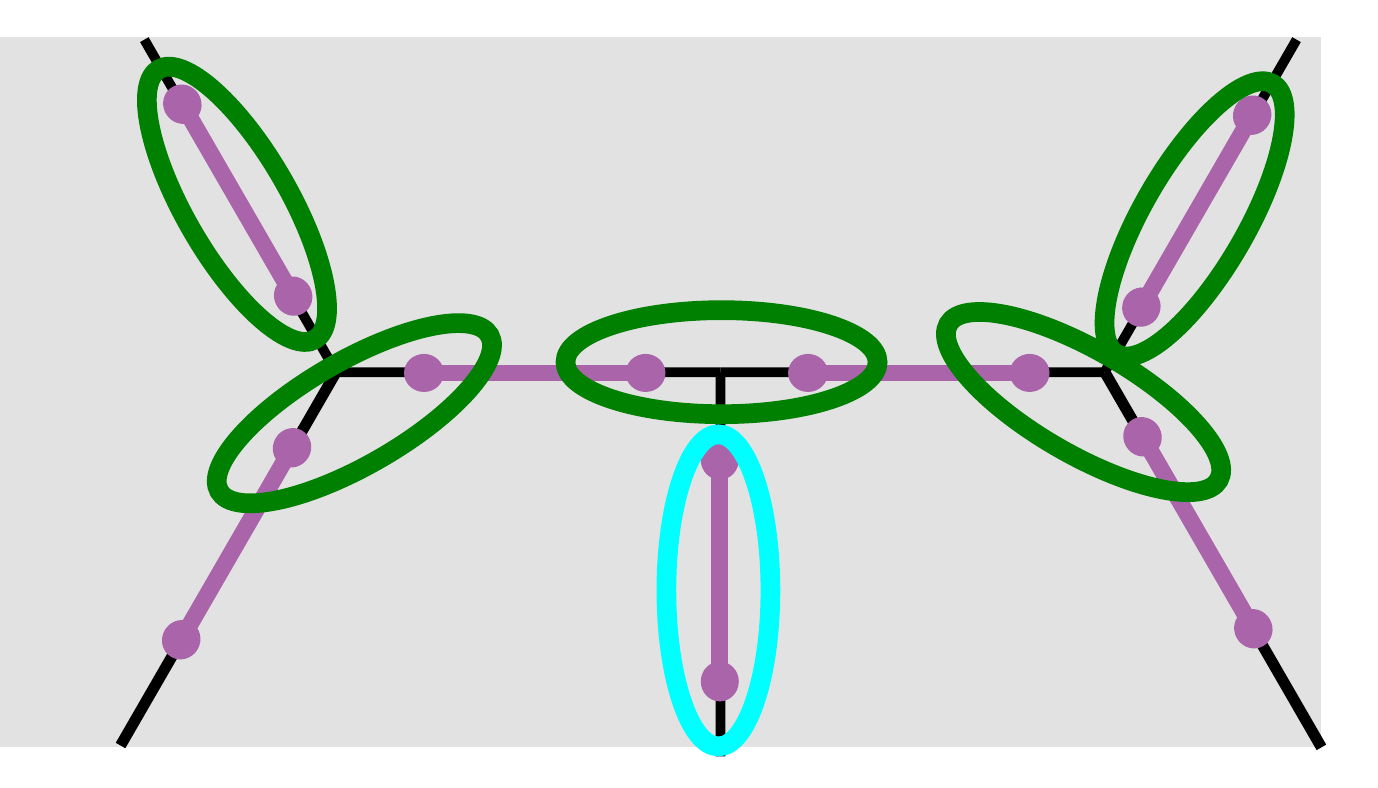}} & $\mathcal{F}_{a_{7}a_{8}}^{\dagger}$ & 1 \\ \hline
\raisebox{-.5\height}{\includegraphics[width = 1.4\linewidth]{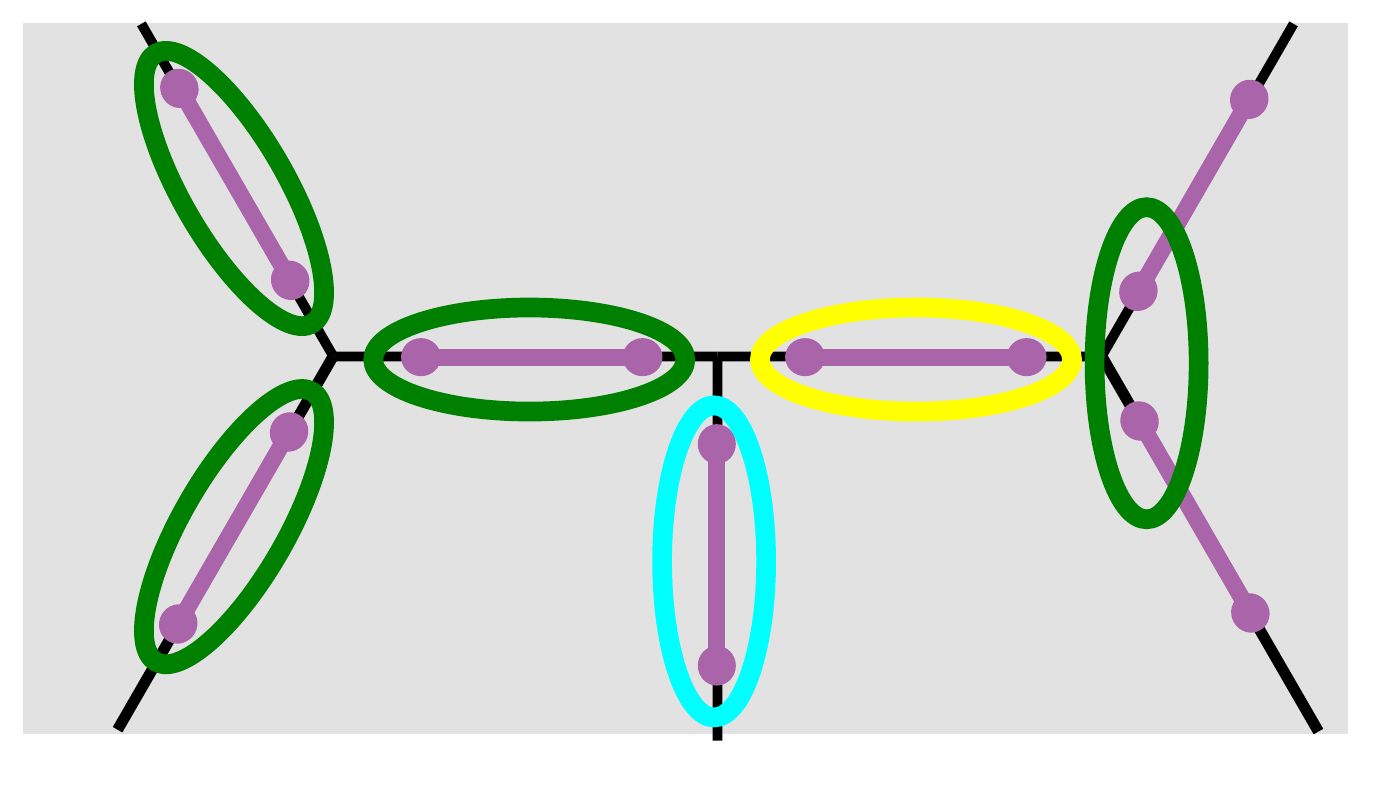}} & \raisebox{-.5\height}{\includegraphics[width = 1.4\linewidth]{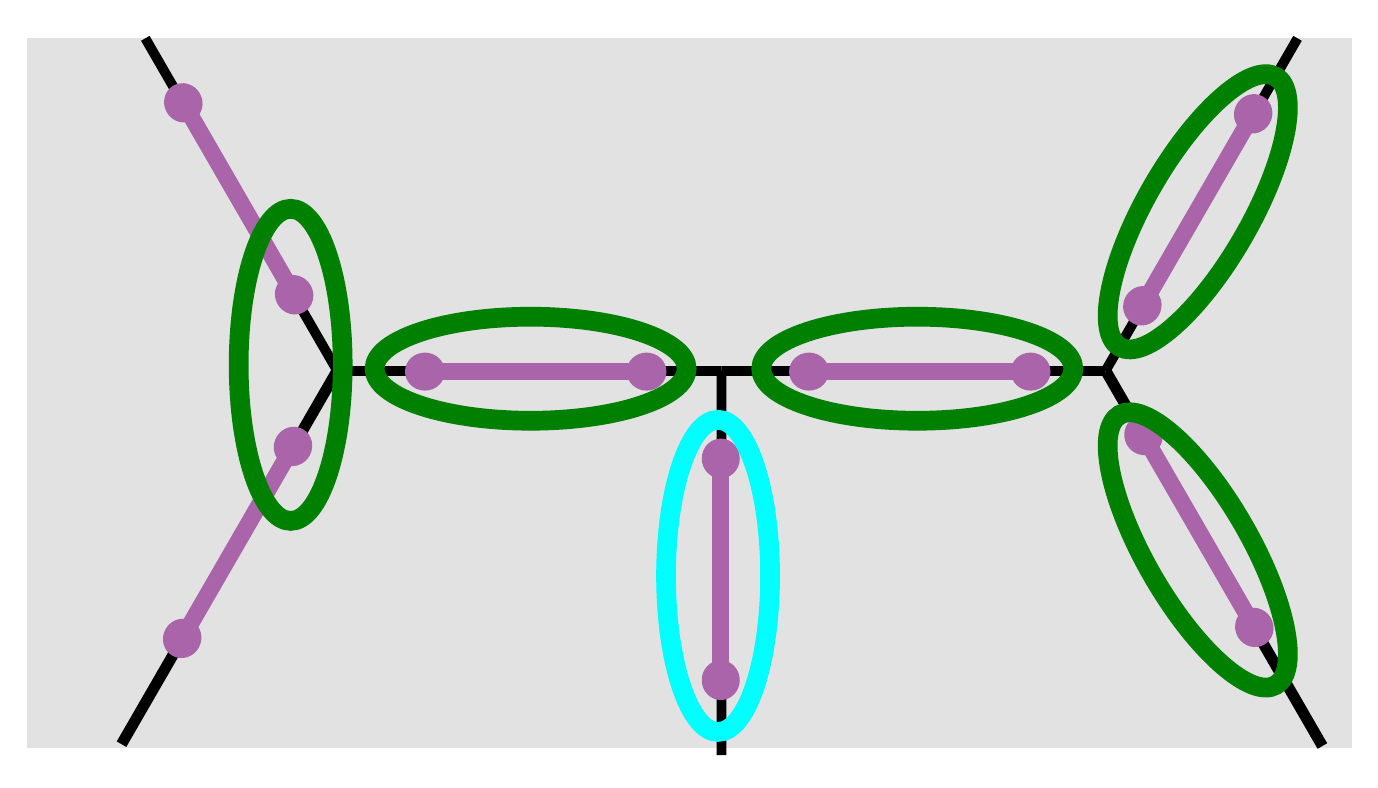}} & $\mathcal{F}_{a_{10}a_{8}}$ & $\bar{\omega}$ & \raisebox{-.5\height}{\includegraphics[width = 1.4\linewidth]{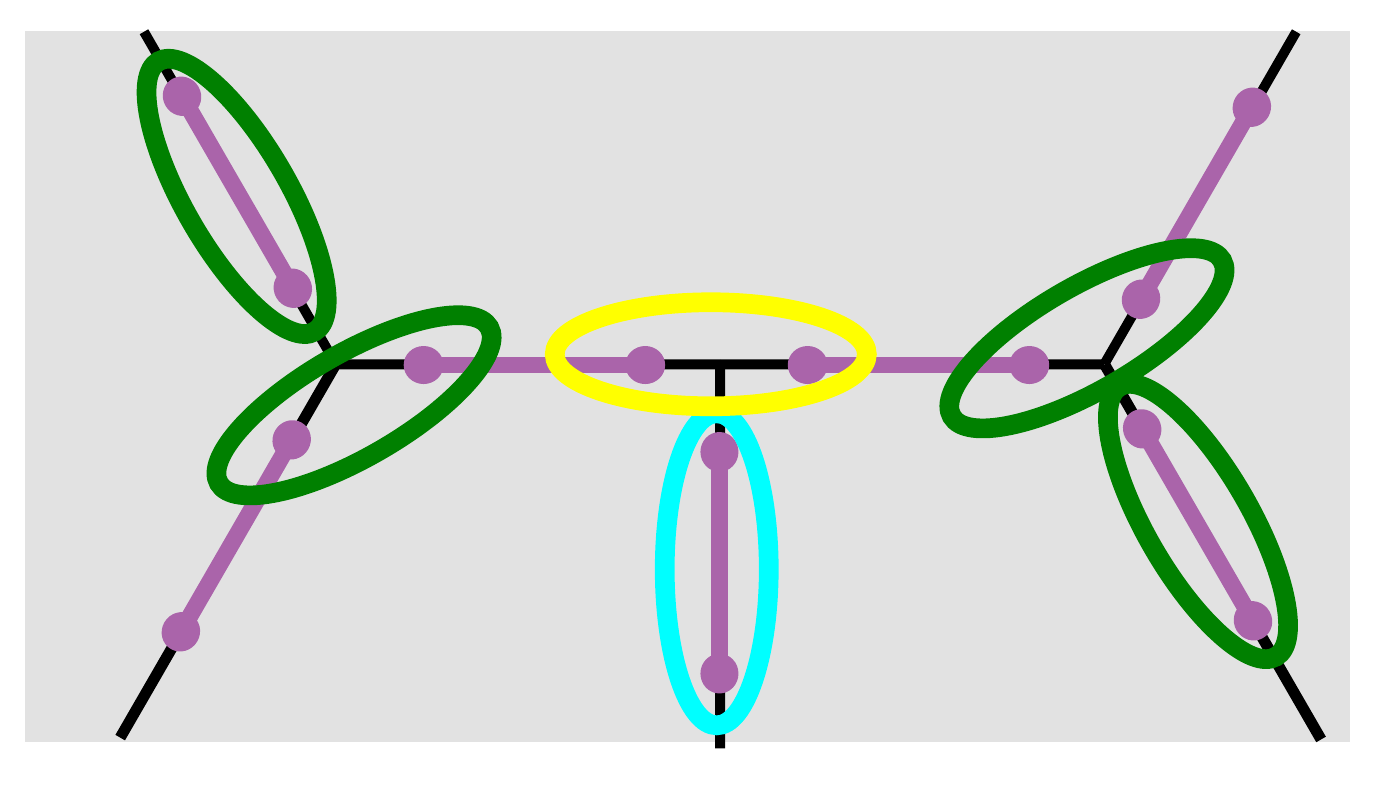}} & \raisebox{-.5\height}{\includegraphics[width = 1.4\linewidth]{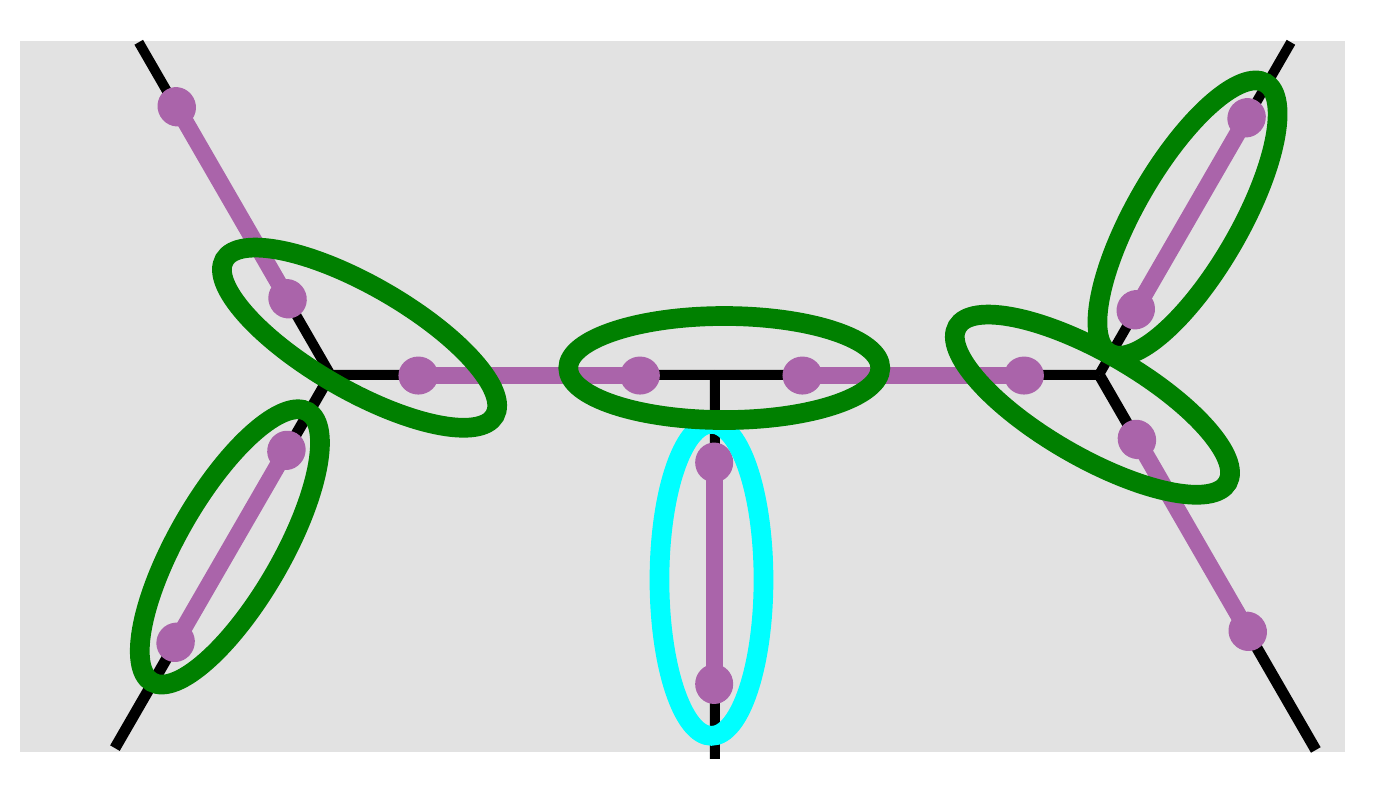}} & $\mathcal{F}_{a_{7}a_{8}}^{\dagger}$ & $i$ \\ \hline
\raisebox{-.5\height}{\includegraphics[width = 1.4\linewidth]{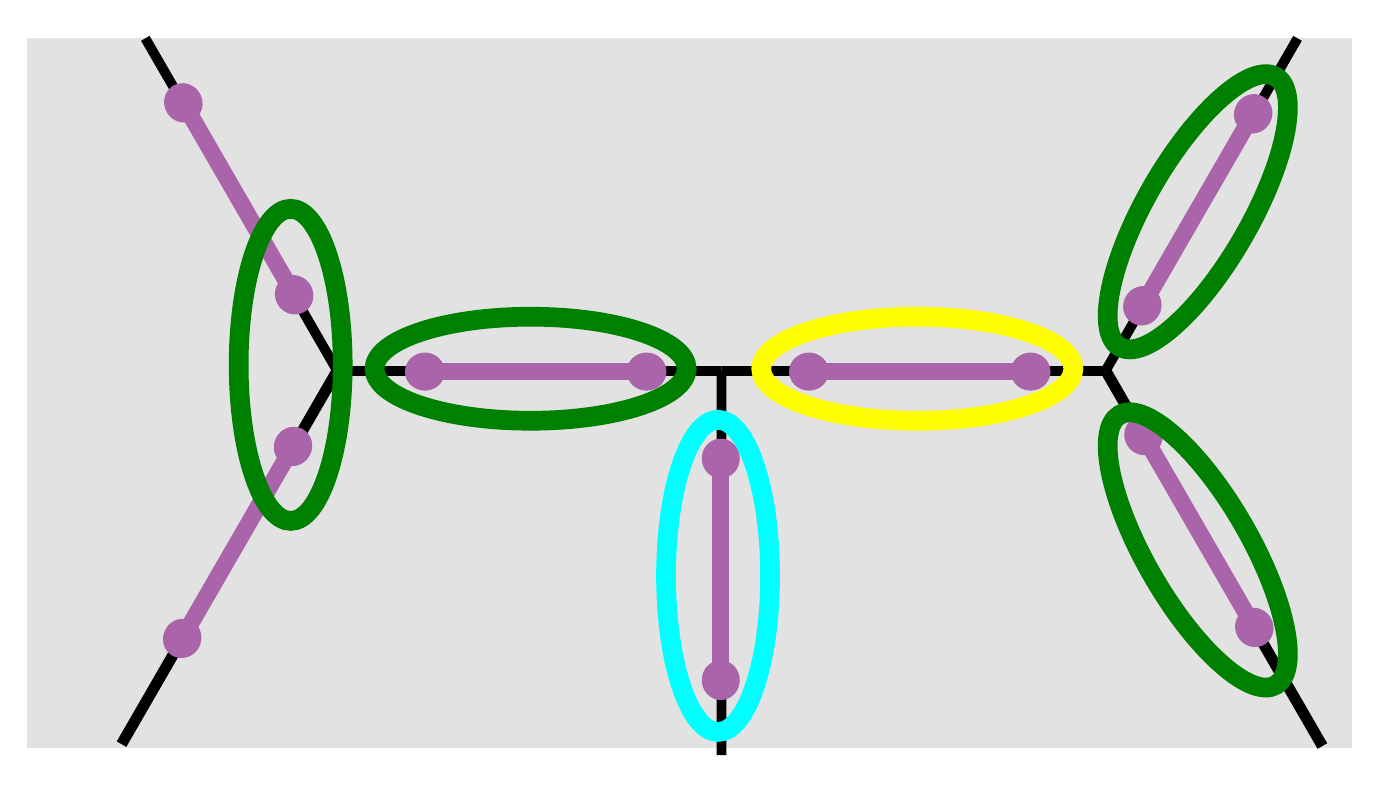}} & \raisebox{-.5\height}{\includegraphics[width = 1.4\linewidth]{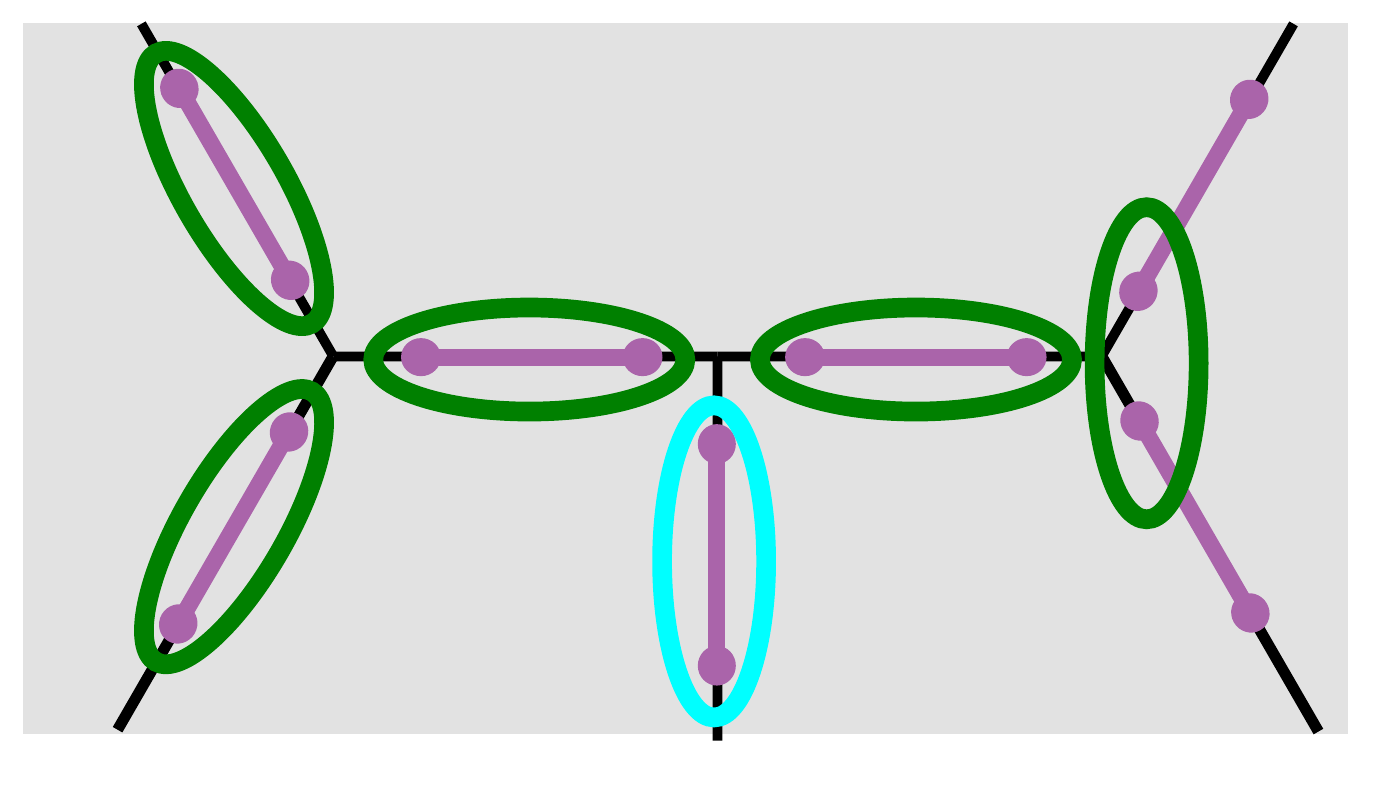}} & $\mathcal{F}_{a_{10}a_{8}}$ & $\omega$ & \raisebox{-.5\height}{\includegraphics[width = 1.4\linewidth]{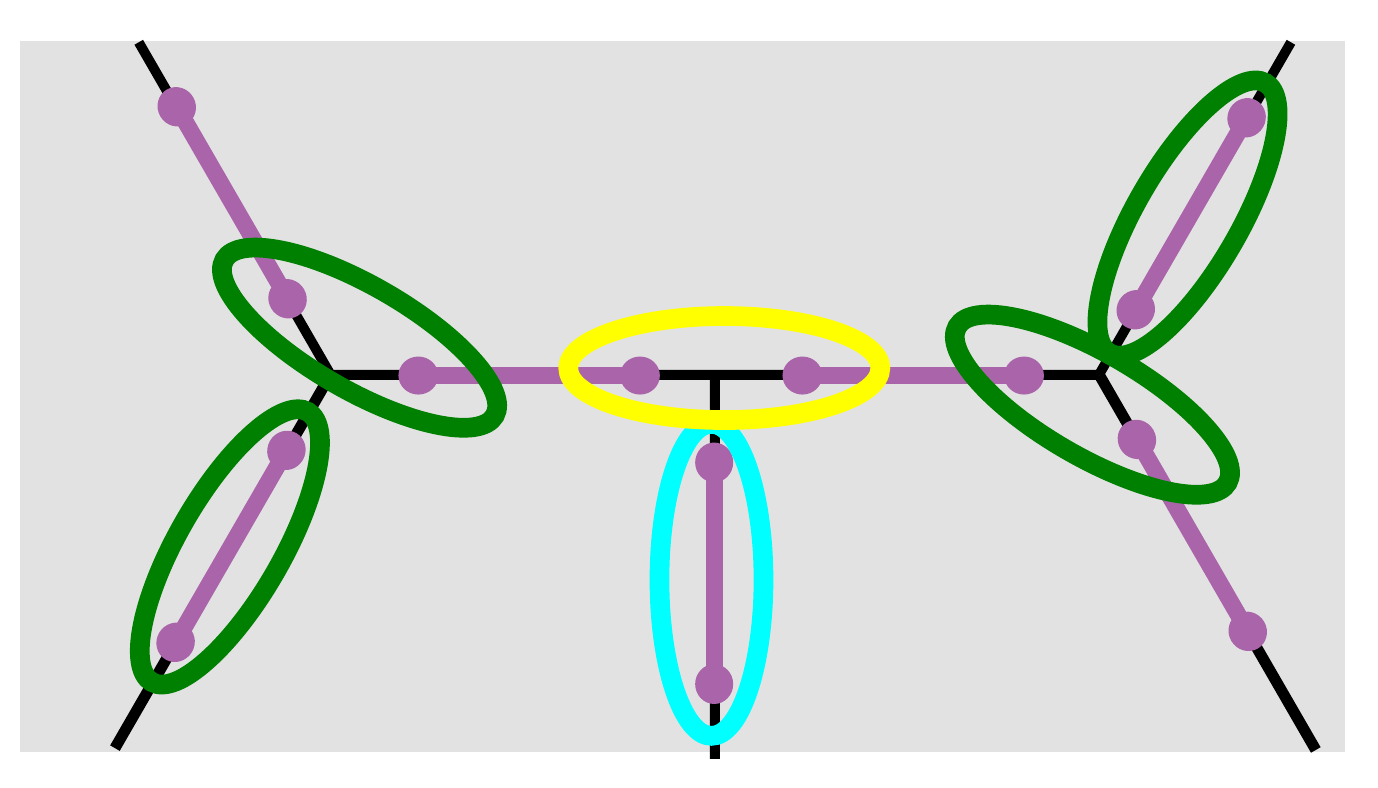}} & \raisebox{-.5\height}{\includegraphics[width = 1.4\linewidth]{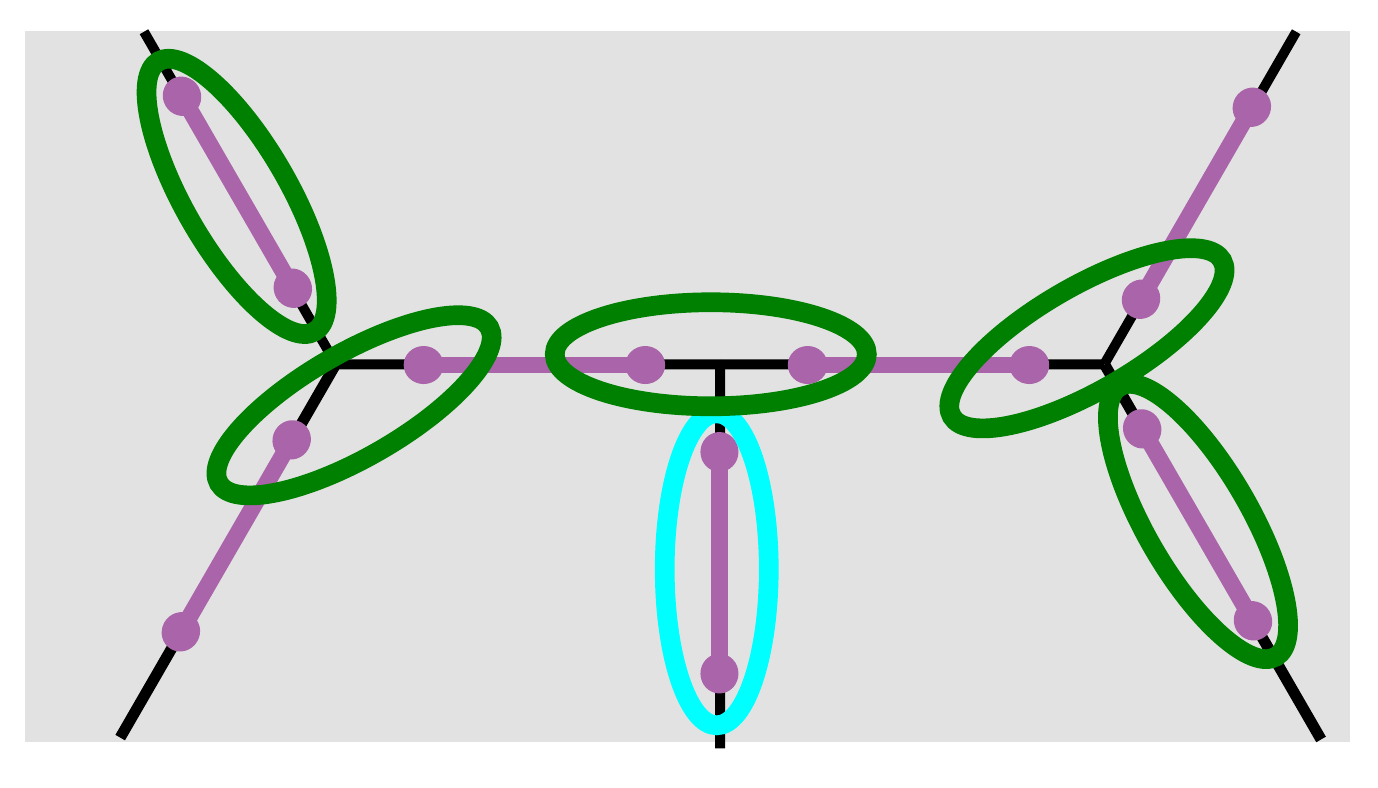}} & $\mathcal{F}_{a_{7}a_{8}}^{\dagger}$ & $-i\omega$ \\ \hline
\end{tabularx}
\end{center}

\end{table*}


\subsection{Examples}

As emphasized before, the projector algebra needed to show Eq.~\eqref{eq:comL1refined} is philosophically similar to that deployed in Appendix~\ref{app:pcom} for the proof of Eq.~\eqref{eq:comactualform}. There are, however, some subtle difference in details of the projector algebra.  In this subsection, we explicitly work out the projector manipulations for two species of starting configurations $s_{b}$.  Extending this analysis to the other six classes of $s_{b}$ proceeds straightforwardly and is left as an exercise for involved readers.  The solution is available upon request.

\subsubsection{Case 1}

Consider a starting spin configuration consistent with pairing patterns illustrated in the first row, first column of Table~\ref{t:summary2}.  In this case, $\bar{\omega}^{k_{s_{b}}} \mathcal{B}_{p_{\rho}}^{s_{i_{1}}} \mathcal{D}_{\mu_{1},\omega}^{(s_{b})}$ and $\bar{\omega}^{k_{s_{i_{2}}}} \mathcal{D}_{\mu_{1},\omega}^{(s_{i_{2}})}\mathcal{B}_{p_{\rho},\omega}^{s_{b}}$ can be written as
\begin{equation}
\label{eq:L1ex11}
\begin{split}
&\bar{\omega}^{k_{s_{b}}} \mathcal{B}_{p_{\rho}}^{s_{i_{1}}} \mathcal{D}_{\mu_{1},\omega}^{(s_{b})} = \bar{\omega} N_{1} (P_{a_{2}a_{3}}P_{a_{5}a_{6}}P_{a_{7}a_{8}}P_{a_{10}a_{11}}R_{a_{1}a_{9}}^{p_{\rho}})\\
&(P_{a_{1}a_{2}}P_{a_{3}a_{5}}P_{a_{6}a_{7}}P_{a_{8}a_{11}}P_{a_{9}a_{10}}R^{p_{\rho}})\\
& (P_{a_{3}a_{5}}P_{a_{6}a_{7}}P_{a_{8}a_{11}}R^{p_{12}}_{a_{4}a_{12}})\\
& (P_{\mu_{1}}^{\bar{\omega}}P_{a_{3}a_{4}}P_{a_{5}a_{6}}P_{a_{7}a_{8}}^{\omega}P_{a_{11}a_{12}}R^{p_{12}})
\end{split}
\end{equation}
\begin{equation}
\label{eq:L1ex12}
\begin{split}
&\bar{\omega}^{k_{s_{i_{2}}}} \mathcal{D}_{\mu_{1},\omega}^{(s_{i_{2}})}\mathcal{B}_{p_{\rho},\omega}^{s_{b}} = \omega N_{2} (P_{a_{2}a_{3}}P_{a_{5}a_{6}}P_{a_{7}a_{8}} P_{a_{10}a_{11}}R^{p_{12}}_{a_{4}a_{12}})\\
&(P_{a_{3}a_{4}}P_{a_{2}a_{5}}P_{a_{6}a_{7}}^{\omega}P_{a_{8}a_{10}}P_{a_{11}a_{12}}R^{p_{12}})\\
&(P_{a_{2}a_{5}} P_{a_{6}a_{7}}^{\omega} P_{a_{8}a_{10}} R_{a_{1}a_{9}}^{p_{\rho}})(P_{\mu_{1}}^{\bar{\omega}}P_{a_{1}a_{2}}P_{a_{5}a_{6}}P_{a_{7}a_{8}}^{\omega}P_{a_{9}a_{10}} R^{p_{\rho}}).
\end{split}
\end{equation}
Here $R^{p_{\rho}}$ and $R^{p_{12}}$ respectively denote a product of projectors involving parafermions around $p_{\rho}$ and the double plaquette consisting of $p_{1}$ and $p_{2}$, but lying outside of the geometry in Fig.~\ref{fig:appcomL1}; subscripts denote parafermion sites within the geometry of Fig.~\ref{fig:appcomL1} that are involved in those projectors.  Similar to the observation made in Appendix~\ref{app:pcom}, these parafermion projectors are largely unimportant in the proof. Also, while $\mathcal{B}_{p_{\rho},\omega}^{s_{b}}$, $\mathcal{D}_{\mu_{1},\omega}^{(s_{b})}$ and $\mathcal{D}_{\mu_{1},\omega}^{(s_{i_{1}})}$ contain projectors of the form $P^{\omega}_{ab}$, one has $P^{\omega}_{ab} \neq P_{ab}$ only when $(a,b)$ corresponds to the intra-edge pairing on bond $\rho$ or inter-edge pairing between bonds $\rho$ and $\lambda$ [recall Eq.~\eqref{Ptab}]. For the sake of making the algebra clearer, we drop $\omega$ superscripts on projectors whenever possible.  

 We will show that Eqs.~\eqref{eq:L1ex11} and \eqref{eq:L1ex12} are identical to the following effective triple-plaquette exchange term:
\begin{equation}
\label{eq:L1ex13}
\begin{split}
& e^{i\phi_{s_{b}}} \mathcal{T}^{s_{b}} \\
& = \omega N_{3}  (P_{a_{2}a_{3}}P_{a_{5}a_{6}}P_{a_{7}a_{8}} P_{a_{10}a_{11}} R^{p_{\rho}}_{a_{1}a_{9}}R^{p_{12}}_{a_{4}a_{12}}) \mathcal{F}_{a_{10}a_{8}} \\
& (P_{\mu_{1}}^{\bar{\omega}} P_{a_{1}a_{2}}P_{a_{3}a_{4}}P_{a_{5}a_{6}}P_{a_{7}a_{8}}^{\omega} P_{a_{9}a_{10}}P_{a_{11}a_{12}} R^{p_{\rho}} R^{p_{12}}).
\end{split}
\end{equation}
Let us transform Eq.~\eqref{eq:L1ex11} first. Note that $P_{a_{3}a_{5}}$, $P_{a_{6}a_{7}}$, and $P_{a_{8}a_{11}}$ in the second parenthesis can be absorbed into the third parenthesis using $P^2 = P$. Also, we have freedom to remove one projector from the third parenthesis; we choose to remove $P_{a_{6}a_{7}}$ here.  At this point we can write
\begin{equation}
\begin{split}
& \bar{\omega}^{k_{s_{b}}} \mathcal{B}_{p_{\rho}}^{s_{i_{1}}} \mathcal{D}_{\mu_{1},\omega}^{(s_{b})} \\
&= \bar{\omega} N_{1} (P_{a_{2}a_{3}}P_{a_{5}a_{6}}P_{a_{7}a_{8}} P_{a_{10}a_{11}}  R^{p_{\rho}}_{a_{1}a_{9}}R^{p_{12}}_{a_{4}a_{12}}) P_{a_{3}a_{5}}P_{a_{8}a_{11}} \\
& (P_{\mu_{1}}^{\bar{\omega}} P_{a_{1}a_{2}}P_{a_{3}a_{4}}P_{a_{5}a_{6}}P_{a_{7}a_{8}}^{\omega} P_{a_{9}a_{10}}P_{a_{11}a_{12}} R^{p_{\rho}}R^{p_{12}}).
\end{split}
\end{equation}
Using Eq.~\eqref{eq:projid3}, one obtains 
\begin{equation}
P_{a_{5}a_{6}} P_{a_{3}a_{5}} P_{a_{5}a_{6}} = \frac{1}{3} P_{a_{5}a_{6}},
\end{equation}
which allows us to remove $P_{a_{3}a_{5}}$. The second line of Eq.~\eqref{eq:projid5} allows one to additionally replace $P_{a_{8}a_{11}}$ with $\mathcal{F}_{a_{11}a_{8}}$ at the cost of changing the normalization factor.  Doing so yields
\begin{equation}
\begin{split}
& \bar{\omega}^{k_{s_{b}}} \mathcal{B}_{p_{\rho}}^{s_{i_{1}}} \mathcal{D}_{\mu_{1},\omega}^{(s_{b})} \\
&= \bar{\omega} \frac{N_{1}}{9} (P_{a_{2}a_{3}}P_{a_{5}a_{6}}P_{a_{7}a_{8}} P_{a_{10}a_{11}} R^{p_{\rho}}_{a_{1}a_{9}}R^{p_{12}}_{a_{4}a_{12}}) \mathcal{F}_{a_{11}a_{8}}\\
& (P_{\mu_{1}}^{\bar{\omega}} P_{a_{1}a_{2}}P_{a_{3}a_{4}}P_{a_{5}a_{6}}P_{a_{7}a_{8}}^{\omega} P_{a_{9}a_{10}}P_{a_{11}a_{12}} R^{p_{\rho}}R^{p_{12}}).
\end{split}
\end{equation}
Now, we observe that
\begin{equation}
\begin{split}
P_{a_{10}a_{11}} \mathcal{F}_{a_{11}a_{8}} & = P_{a_{10}a_{11}} \mathcal{F}_{a_{10}a_{11}} \mathcal{F}_{a_{11}a_{8}}\\
& = \bar{\omega} P_{a_{10}a_{11}} \mathcal{F}_{a_{10}a_{8}}
\end{split}
\end{equation}
using $P_{a_{10}a_{11}} = P_{a_{10}a_{11}} \mathcal{F}_{a_{11}a_{10}}$ and Eq.~\eqref{eq:gobrel}.  With this result $\bar{\omega}^{k_{s_{b}}} \mathcal{B}_{p_{\rho}}^{s_{i_{1}}} \mathcal{D}_{\mu_{1},\omega}^{(s_{b})}$ can be made identical to the expression in Eq.~\eqref{eq:L1ex13}.

Demonstrating equality between Eq.~\eqref{eq:L1ex12} and Eq.~\eqref{eq:L1ex13} proceeds similarly, and to some extent more easily.  Using $P^2 = P$ we can eliminate $P_{a_{2}a_{5}}$, $P_{a_{6}a_{7}}^{\omega}$, and $P_{a_{8}a_{10}}$ from the second parenthesis of Eq.~\eqref{eq:L1ex12}, and we additionally use our usual freedom to knock out $P_{a_{6}a_{7}}^{\omega}$ from the third parenthesis. After these operations and some rearrangement, one can write
\begin{equation}
\label{eq:L1ex12mid}
\begin{split}
& \bar{\omega}^{k_{s_{i_{2}}}} \mathcal{D}_{\mu_{1},\omega}^{(s_{i_{2}})}\mathcal{B}_{p_{\rho},\omega}^{s_{b}} \\
&= \omega N_{2} (P_{a_{2}a_{3}}P_{a_{5}a_{6}}P_{a_{7}a_{8}} P_{a_{10}a_{11}}  R^{p_{\rho}}_{a_{1}a_{9}}R^{p_{12}}_{a_{4}a_{12}}) P_{a_{2}a_{5}}P_{a_{8}a_{10}} \\
& (P_{\mu_{1}}^{\bar{\omega}} P_{a_{1}a_{2}}P_{a_{3}a_{4}}P_{a_{5}a_{6}}P_{a_{7}a_{8}}^{\omega} P_{a_{9}a_{10}}P_{a_{11}a_{12}} R^{p_{\rho}}R^{p_{12}}).
\end{split}
\end{equation}
Next, Eq.~\eqref{eq:projid3} implies 
\begin{equation}
P_{a_{5}a_{6}} P_{a_{2}a_{5}} P_{a_{5}a_{6}} = \frac{1}{3} P_{a_{5}a_{6}}.
\end{equation}
This identity along with the first line of Eq.~\eqref{eq:projid5} allows us to replace $P_{a_{2}a_{5}}P_{a_{8}a_{10}}$ in Eq.~\eqref{eq:L1ex12mid} with $\frac{1}{9}\mathcal{F}_{a_{10}a_{8}}$. One can then see that $\bar{\omega}^{k_{s_{i_{2}}}} \mathcal{D}_{\mu_{1},\omega}^{(s_{i_{2}})}\mathcal{B}_{p_{\rho},\omega}^{s_{b}}$ is indeed equivalent to Eq.~\eqref{eq:L1ex13}.

\subsubsection{Case 2}

As a second example we show how the projector algebra works for $s_{b}$ consistent with the parafermion pairing in the fourth column, first row of Table~\ref{t:summary2}.
In this case we have
\begin{equation}
\label{eq:L1ex21}
\begin{split}
&\bar{\omega}^{k_{s_{b}}} \mathcal{B}_{p_{\rho}}^{s_{i_{1}}} \mathcal{D}_{\mu_{1},\omega}^{(s_{b})} = \omega N_{1} (P_{a_{2}a_{5}}P_{a_{6}a_{7}}P_{a_{8}a_{10}}R_{a_{1}a_{9}}^{p_{\rho}})\\
&(P_{a_{1}a_{2}}P_{a_{5}a_{6}}P_{a_{7}a_{8}}P_{a_{9}a_{10}}R^{p_{\rho}})(P_{a_{3}a_{4}}P_{a_{5}a_{6}}P_{a_{7}a_{8}}P_{a_{11}a_{12}}R^{p_{12}})\\
&(P_{\mu_{1}}^{\bar{\omega}}P_{a_{3}a_{5}}P_{a_{6}a_{7}}^{\omega}P_{a_{8}a_{11}}R_{a_{4}a_{12}}^{p_{12}})
\end{split}
\end{equation}
\begin{equation}
\label{eq:L1ex22}
\begin{split}
&\bar{\omega}^{k_{s_{i_{2}}}} \mathcal{D}_{\mu_{1},\omega}^{(s_{i_{2}})}\mathcal{B}_{p_{\rho},\omega}^{s_{b}} = \\
&\bar{\omega} N_{2} (P_{\mu_{1}}^{\bar{\omega}}P_{a_{3}a_{4}}P_{a_{2}a_{5}}P_{a_{6}a_{7}} P_{a_{8}a_{10}}P_{a_{11}a_{12}}R^{p_{12}})\\
&(P_{a_{2}a_{3}}P_{a_{5}a_{6}}P_{a_{7}a_{8}}^{\omega}P_{a_{10}a_{11}}R_{a_{4}a_{12}}^{p_{12}})\\
&(P_{a_{2}a_{3}}P_{a_{5}a_{6}}P_{a_{7}a_{8}}^{\omega}P_{a_{10}a_{11}}R_{a_{1}a_{9}}^{p_{\rho}})\\
&(P_{a_{1}a_{2}}P_{a_{3}a_{5}}P_{a_{6}a_{7}}^{\omega}P_{a_{8}a_{11}}P_{a_{9}a_{10}}R^{p_{\rho}})
\end{split}
\end{equation}
\begin{equation}
\label{eq:L1ex23}
\begin{split}
& e^{i\phi_{s_{b}}} \mathcal{T}^{s_{b}} \\
& = \omega N_{3}  (P_{a_{3}a_{4}}P_{a_{2}a_{5}}P_{a_{6}a_{7}} P_{a_{8}a_{10}} R^{p_{\rho}}_{a_{1}a_{9}}R^{p_{12}}) \mathcal{F}_{a_{7}a_{8}}^{\dagger} \\
& (P_{\mu_{1}}^{\bar{\omega}} P_{a_{1}a_{2}}P_{a_{3}a_{5}}P_{a_{6}a_{7}}^{\omega} P_{a_{8}a_{11}} P_{a_{9}a_{10}} R^{p_{\rho}} R^{p_{12}}_{a_{4}a_{12}})
\end{split}
\end{equation}
Showing equality between Eqs.~\eqref{eq:L1ex21} and \eqref{eq:L1ex23} is relatively simple.  Using our usual tricks we can remove 
$P_{a_{5}a_{6}}$ and $P_{a_{7}a_{8}}$ from the second parenthesis (by virtue of $P^2 = P$) and additionally remove $P_{a_{5}a_{6}}$ from the third parenthesis.  Some rearrangement gives
\begin{equation}
\begin{split}
&\bar{\omega}^{k_{s_{b}}} \mathcal{B}_{p_{\rho}}^{s_{i_{1}}} \mathcal{D}_{\mu_{1},\omega}^{(s_{b})} = \omega N_{1} \\
&(P_{a_{2}a_{5}}P_{a_{3}a_{4}}P_{a_{6}a_{7}}P_{a_{8}a_{10}}P_{a_{11}a_{12}} R_{a_{1}a_{9}}^{p_{\rho}}R^{p}) (P_{a_{7}a_{8}})\\
&(P_{a_{1}a_{2}}P_{a_{9}a_{10}}P_{a_{3}a_{5}} P_{a_{6}a_{7}}^{\omega}P_{a_{8}a_{11}}R_{a_{4}a_{12}}^{p}R^{p_{\rho}}).
\end{split}
\end{equation}
Employing the third line of Eq.~\eqref{eq:projid5} allows one to further replace $P_{a_{7}a_{8}}$ with $\mathcal{F}_{a_{7}a_{8}}^{\dagger}$. This substitution establishes equality with Eq.~\eqref{eq:L1ex23}. 

Let us similarly work out the projector manipulations for Eq.~\eqref{eq:L1ex22}.  Here we can delete the repeated projectors $P_{a_{2}a_{3}}$, $P_{a_{5}a_{6}}$, $P_{a_{7}a_{8}}^{\omega}$, and $P_{a_{10}a_{11}}$ from the second parenthesis and knock out $P_{a_{5}a_{6}}$ from the third. Rearrangement yields
\begin{equation}
\begin{split}
&\bar{\omega}^{k_{s_{i_{2}}}} \mathcal{D}_{\mu_{1},\omega}^{(s_{i_{2}})}\mathcal{B}_{p_{\rho},\omega}^{s_{b}} = \bar{\omega} N_{2} \\
&(P_{a_{3}a_{4}}P_{a_{2}a_{5}}P_{a_{6}a_{7}} P_{a_{8}a_{10}}P_{a_{11}a_{12}}R^{p_{12}}R_{a_{1}a_{9}}^{p_{\rho}})(P_{a_{2}a_{3}}P_{a_{7}a_{8}}^{\omega}\\
&P_{a_{10}a_{11}}) (P_{\mu_{1}}^{\bar{\omega}} P_{a_{1}a_{2}}P_{a_{3}a_{5}}P_{a_{6}a_{7}}^{\omega}P_{a_{8}a_{11}}P_{a_{9}a_{10}}R^{p_{\rho}}R_{a_{4}a_{12}}^{p_{12}}).
\end{split}
\end{equation}
We can further remove $P_{a_{2}a_{3}}$ at the cost of introducing some phase factors by noting that
\begin{equation}
P_{a_{2}a_{5}} P_{a_{2}a_{3}} P_{a_{3}a_{5}} = \frac{i}{\sqrt{3}} P_{a_{2}a_{5}} P_{a_{3}a_{5}},
\end{equation}
which follows from Eq.~\eqref{eq:projid2}.  Using the last line of Eq.~\eqref{eq:projid4}, one can replace $P_{a_{7}a_{8}}^{\omega}$ with $\frac{\omega}{3} \mathcal{F}_{a_{7}a_{8}}^{\dagger}$, giving
\begin{equation}
\begin{split}
&\bar{\omega}^{k_{s_{i_{2}}}} \mathcal{D}_{\mu_{1},\omega}^{(s_{i_{2}})}\mathcal{B}_{p_{\rho},\omega}^{s_{b}} = \frac{i}{3\sqrt{3}} N_{2} \\
&(P_{a_{3}a_{4}}P_{a_{2}a_{5}}P_{a_{6}a_{7}} P_{a_{8}a_{10}}P_{a_{11}a_{12}}R^{p_{12}}R_{a_{1}a_{9}}^{p_{\rho}})(\mathcal{F}_{a_{7}a_{8}}^{\dagger}P_{a_{10}a_{11}})\\
&(P_{\mu_{1}}^{\bar{\omega}} P_{a_{1}a_{2}}P_{a_{3}a_{5}}P_{a_{6}a_{7}}^{\omega}P_{a_{8}a_{11}}P_{a_{9}a_{10}}R^{p_{\rho}}R_{a_{4}a_{12}}^{p_{12}}).
\end{split}
\end{equation}
Finally, one can apply the first line of Eq.~\eqref{eq:projid4} to remove $P_{a_{10}a_{11}}$ and introduce the additional factor $\frac{-i\omega}{\sqrt{3}}$ into the above expression.  The expression then becomes identical to Eq.~\eqref{eq:L1ex23}.

\subsection{Summary}

 One can straightforwardly extend the projector manipulations carried out above to the other six species of Ising configurations. The key difference comes from the fact that some projectors will become $P_{ab}^{\omega}$ instead of $P_{ab}$---necessitating the new projector identities introduced in this section to be employed.  Also, the triple plaquette flip term generically can be written as
\begin{equation}
\label{eq:tripleplaquetteex}
\begin{split}
\mathcal{T}^{s_{b}} & = \left( \mathcal{P}_{a_{2k}a_{1}}\prod_{i=1}^{k-1}\sqrt{3} P_{a_{2k}a_{2k+1}} \right) \\
&\mathcal{O}^{s_{b}} \left( \mathcal{P}_{\mu_{1}}^{\bar{\omega}}\prod_{i=1}^{k} P_{a_{2k-1}a_{2k}}^{\omega} \right).
\end{split}
\end{equation}
Note especially the $P^{\omega}_{ab}$'s in the second parenthesis and the operator $\mathcal{O}^{s_{b}}$ sandwiched between the two strings of projectors. The latter operator plays a role similar to $\mathcal{F}_{a_{\mu_{1}}a_{\rho}}$ in Eq.~\eqref{eq:ldef}---i.e., $\mathcal{O}^{s_{b}}$ moves `the excess charge' on the inter-edge pairing between $a_{\lambda}$ and $a_{\rho}$ or the intra-edge pairing within the bond $\rho$ to the outer parafermion loops being flipped by $\mathcal{T}^{s_{b}}$. One can verify through the projector algebra that $\mathcal{O}^{s_{b}} $ can always be chosen to be either $\mathcal{F}_{a_{7}a_{8}}^{\dagger}$ or $\mathcal{F}_{a_{10}a_{8}}$.  Table~\ref{t:summary2} lists the $\mathcal{O}^{s_{b}} $'s for each starting configuration, along with the $e^{i\phi_{s_b}}$ phase factors.

\section{Constraints on the Wilson loop operator $W_{l}$}
\label{app:wilsonconst}

 Recall that in Appendix~\ref{app:proofs}, we observed that there are two translation vectors $n_{1}e_{x} + m_{1} e_{y}$ and $n_{2}e_{x} + m_{2} e_{y}$ naturally defined from fixing the $T_{y}$ branch cut and the trench line. These two vectors generate all points $\{ a_{(n,m)} \}$ on the infinite plane $\mathbb{R}^{2}$ that represent the same point $a$ on the torus.  Let us constrain the possible values of $n_{1}$, $m_{1}$, $n_{2}$, and $m_{2}$. The set $\{ a_{(n,m)} \}$ is essentially the $\mathbb{Z} \times \mathbb{Z}$ lattice generated by $e_{x}$ and $e_{y}$. It is known that if $n_{1}e_{x} + m_{1} e_{y}$ and $n_{2}e_{x} + m_{2} e_{y}$ generate the same lattice as the one generated by $e_{x}$ and $e_{y}$, then $L = \begin{pmatrix} n_{1} & m_{1} \\ n_{2} & m_{2} \end{pmatrix}$ is $\textit{unimodular}$, i.e, $\det{L} = \pm 1$. 
 This property imposes a stringent condition on possible values of $n_{1}$ and $m_{1}$. First, assume that $n_{1}$ and $m_{1}$ are both nonzero. Unimodularity implies that 
\begin{equation}
n_{1} m_{2} - n_{2}m_{1} = \pm1.
\end{equation} 
Then, according to Bezout's lemma, $n_{1}$ and $m_{1}$ \textit{must be mutually prime}. Suppose that instead $n_{1}= 0$ so that $n_{2}m_{1} = \pm 1$.  Since both numbers are integer, we get $m_{1}  = \pm 1$. If one assumes $m_{1} = 0$, then one similarly obtains $n_{1} = \pm 1$. Hence $(n_{1},m_{1}) = (\pm 1 ,0 ), (0, \pm 1)$ or $\text{GCD}(n_{1},m_{1}) = 1$. One can proceed identically to constrain $(n_{2}, m_{2})$ as well, yielding $(n_{2},m_{2}) = (\pm 1 ,0 ), (0, \pm 1)$ or $\text{GCD}(n_{2},m_{2}) = 1$. Finally, since row operations preserve the determinant, $L' = \begin{pmatrix} n_{1}-n_{2} & m_{1}-m_{2} \\ n_{2} & m{2} \end{pmatrix}$ is unimodular as well. It turns out that one can place constraints on $(n_{1}-n_{2}, m_{1}-m_{2})$ that are identical to those placed on $(n_{1},m_{1})$ and $(n_{2},m_{2})$.

We will use the above results to show that $W_{l}$ defined in Eq.~\eqref{eq:wilsonloop} and associated with a loop $l$ with winding numbers $(1,0)$ $(0,1)$ or $(1,1)$ always acts nontrivially on the global degrees of freedom.  In other words, it is impossible to have $n_{l,x}=n_{l,y}=0$ in Eq.~\eqref{eq:wilsonloop}. For concreteness, we focus on the loop with winding number $(1,0)$; one can easily generalize the analysis to $(0,1)$ and $(1,1)$ loops leading to the same conclusions. 

Let us first observe how the loop $l$ with winding number $(1,0)$ on the torus is represented in the infinite plane introduced in Appendix~\ref{app:proofs}. Recall that the infinite plane $\mathbb{R}^{2}$ was generated by tiling up squares that represent the torus upon gluing top and bottom edges and left and right edges. Then, on the infinite plane, a line that connects two points whose displacement vector corresponds to $e_{x}$ \textit{faithfully represents} the loop $l$. Taking infinite copies of these lines on different copies of starting points give an infinite stack of lines, analogous to the green and blue lines shown in Fig.~\ref{fig:repeatedzone}(b). For the purpose of this appendix, however, taking only one copy that faithfully represents the loop suffices. 
 
In this picture, $n_{l,x}$ and $n_{l,y}$ are simply associated with the crossing numbers across the stacks of green and blue lines. This crossing number is homotopy invariant, so we have freedom to continuously deform the path to compute $n_{l,x}$ and $n_{l,y}$. To define a convenient deformation, note that there exists a unique pair of integers $(m_{l,x},m_{l,y})$ that satisfies $e_{x} = m_{l,x}(n_{1}e_{x}+m_{1}e_{y}) +  m_{l,y}(n_{2}e_{x}+m_{2}e_{y})$. We deform the original line to two segments that move a point by $m_{l,x}(n_{1}e_{x}+m_{1}e_{y})$ and then $m_{l,y}(n_{2}e_{x}+m_{2}e_{y})$. The crossing numbers $(n_{l,x},n_{l,y})$ are then naturally identified as $(m_{l,x},m_{l,y})$.  Finding $(n_{l,x}, n_{l,y})$ amounts to a basis change from $(e_{x},e_{y})$ to $
(n_{1}e_{x}+m_{1}e_{y}, n_{2}e_{x}+m_{2}e_{y})$, which  can be computed as
\begin{equation}
\begin{pmatrix}
n_{l,x} \\ n_{l,y}
\end{pmatrix} = L^{-1} \begin{pmatrix}
1 \\ 0
\end{pmatrix} = \begin{pmatrix}
m_{2} \\ -n_{2}
\end{pmatrix} .
\label{nxny}
\end{equation}

According to Eq.~\eqref{eq:wilsonloop}, the part of $W_{l}$ that acts on the global degrees of freedom enters as $T_{x}^{n_{l,x}}T_{y}^{n_{l,y}}$. If $n_{l,x}$ and $n_{l,y}$ are not both multiples of $3$, $W_{l}$ depends nontrivially on $T_{x}$ and/or $T_{y}$.  Equation~\eqref{nxny} along with constraints imposed previously on $(n_2,m_2)$ enforces such nontrivial action on these global degrees of freedom. 
 
 Through almost identical steps, one can establish that $(n_{l,x},n_{l,y})$ associated with winding-number $(0,1)$ and $(1,1)$ loops are $(m_{1},-n_{1})$ and $(m_{2}-m_{1},n_{1}-n_{2})$, respectively.  We already established that both numbers in each pair cannot be multiples of $3$, so $W_l$ depends nontrivially on $T_x$ and/or $T_y$ here too.

\bibliography{ref}

\begin{thebibliography}{56}%
\makeatletter
\providecommand \@ifxundefined [1]{%
 \@ifx{#1\undefined}
}%
\providecommand \@ifnum [1]{%
 \ifnum #1\expandafter \@firstoftwo
 \else \expandafter \@secondoftwo
 \fi
}%
\providecommand \@ifx [1]{%
 \ifx #1\expandafter \@firstoftwo
 \else \expandafter \@secondoftwo
 \fi
}%
\providecommand \natexlab [1]{#1}%
\providecommand \enquote  [1]{``#1''}%
\providecommand \bibnamefont  [1]{#1}%
\providecommand \bibfnamefont [1]{#1}%
\providecommand \citenamefont [1]{#1}%
\providecommand \href@noop [0]{\@secondoftwo}%
\providecommand \href [0]{\begingroup \@sanitize@url \@href}%
\providecommand \@href[1]{\@@startlink{#1}\@@href}%
\providecommand \@@href[1]{\endgroup#1\@@endlink}%
\providecommand \@sanitize@url [0]{\catcode `\\12\catcode `\$12\catcode
  `\&12\catcode `\#12\catcode `\^12\catcode `\_12\catcode `\%12\relax}%
\providecommand \@@startlink[1]{}%
\providecommand \@@endlink[0]{}%
\providecommand \url  [0]{\begingroup\@sanitize@url \@url }%
\providecommand \@url [1]{\endgroup\@href {#1}{\urlprefix }}%
\providecommand \urlprefix  [0]{URL }%
\providecommand \Eprint [0]{\href }%
\providecommand \doibase [0]{http://dx.doi.org/}%
\providecommand \selectlanguage [0]{\@gobble}%
\providecommand \bibinfo  [0]{\@secondoftwo}%
\providecommand \bibfield  [0]{\@secondoftwo}%
\providecommand \translation [1]{[#1]}%
\providecommand \BibitemOpen [0]{}%
\providecommand \bibitemStop [0]{}%
\providecommand \bibitemNoStop [0]{.\EOS\space}%
\providecommand \EOS [0]{\spacefactor3000\relax}%
\providecommand \BibitemShut  [1]{\csname bibitem#1\endcsname}%
\let\auto@bib@innerbib\@empty
\bibitem [{\citenamefont {Kitaev}(2003)}]{Kitaev2003}%
  \BibitemOpen
  \bibfield  {author} {\bibinfo {author} {\bibfnamefont {A.}~\bibnamefont
  {Kitaev}},\ }\href {\doibase http://dx.doi.org/10.1016/S0003-4916(02)00018-0}
  {\bibfield  {journal} {\bibinfo  {journal} {Annals of Physics}\ }\textbf
  {\bibinfo {volume} {303}},\ \bibinfo {pages} {2 } (\bibinfo {year}
  {2003})}\BibitemShut {NoStop}%
\bibitem [{\citenamefont {Kitaev}(2006)}]{Kitaev2006}%
  \BibitemOpen
  \bibfield  {author} {\bibinfo {author} {\bibfnamefont {A.}~\bibnamefont
  {Kitaev}},\ }\href {\doibase http://dx.doi.org/10.1016/j.aop.2005.10.005}
  {\bibfield  {journal} {\bibinfo  {journal} {Annals of Physics}\ }\textbf
  {\bibinfo {volume} {321}},\ \bibinfo {pages} {2 } (\bibinfo {year}
  {2006})}\BibitemShut {NoStop}%
\bibitem [{\citenamefont {Levin}\ and\ \citenamefont {Wen}(2005)}]{Levin2006}%
  \BibitemOpen
  \bibfield  {author} {\bibinfo {author} {\bibfnamefont {M.~A.}\ \bibnamefont
  {Levin}}\ and\ \bibinfo {author} {\bibfnamefont {X.-G.}\ \bibnamefont
  {Wen}},\ }\href {\doibase 10.1103/PhysRevB.71.045110} {\bibfield  {journal}
  {\bibinfo  {journal} {Phys. Rev. B}\ }\textbf {\bibinfo {volume} {71}},\
  \bibinfo {pages} {045110} (\bibinfo {year} {2005})}\BibitemShut {NoStop}%
\bibitem [{\citenamefont {Chen}\ \emph {et~al.}(2014)\citenamefont {Chen},
  \citenamefont {Lu},\ and\ \citenamefont {Vishwanath}}]{Chen2014}%
  \BibitemOpen
  \bibfield  {author} {\bibinfo {author} {\bibfnamefont {X.}~\bibnamefont
  {Chen}}, \bibinfo {author} {\bibfnamefont {Y.}~\bibnamefont {Lu}}, \ and\
  \bibinfo {author} {\bibfnamefont {A.}~\bibnamefont {Vishwanath}},\ }\href
  {\doibase 10.1038/ncomms4507} {\bibfield  {journal} {\bibinfo  {journal}
  {Nature Communications}\ }\textbf {\bibinfo {volume} {5}},\ \bibinfo {pages}
  {3507} (\bibinfo {year} {2014})}\BibitemShut {NoStop}%
\bibitem [{\citenamefont {Heinrich}\ \emph {et~al.}(2016)\citenamefont
  {Heinrich}, \citenamefont {Burnell}, \citenamefont {Fidkowski},\ and\
  \citenamefont {Levin}}]{Heinrich2016}%
  \BibitemOpen
  \bibfield  {author} {\bibinfo {author} {\bibfnamefont {C.}~\bibnamefont
  {Heinrich}}, \bibinfo {author} {\bibfnamefont {F.}~\bibnamefont {Burnell}},
  \bibinfo {author} {\bibfnamefont {L.}~\bibnamefont {Fidkowski}}, \ and\
  \bibinfo {author} {\bibfnamefont {M.}~\bibnamefont {Levin}},\ }\href
  {\doibase 10.1103/PhysRevB.94.235136} {\bibfield  {journal} {\bibinfo
  {journal} {Phys. Rev. B}\ }\textbf {\bibinfo {volume} {94}},\ \bibinfo
  {pages} {235136} (\bibinfo {year} {2016})}\BibitemShut {NoStop}%
\bibitem [{\citenamefont {Cheng}\ \emph {et~al.}(2017)\citenamefont {Cheng},
  \citenamefont {Gu}, \citenamefont {Jiang},\ and\ \citenamefont
  {Qi}}]{Cheng2016}%
  \BibitemOpen
  \bibfield  {author} {\bibinfo {author} {\bibfnamefont {M.}~\bibnamefont
  {Cheng}}, \bibinfo {author} {\bibfnamefont {Z.-C.}\ \bibnamefont {Gu}},
  \bibinfo {author} {\bibfnamefont {S.}~\bibnamefont {Jiang}}, \ and\ \bibinfo
  {author} {\bibfnamefont {Y.}~\bibnamefont {Qi}},\ }\href {\doibase
  10.1103/PhysRevB.96.115107} {\bibfield  {journal} {\bibinfo  {journal} {Phys.
  Rev. B}\ }\textbf {\bibinfo {volume} {96}},\ \bibinfo {pages} {115107}
  (\bibinfo {year} {2017})}\BibitemShut {NoStop}%
\bibitem [{\citenamefont {Gu}\ \emph {et~al.}(2014)\citenamefont {Gu},
  \citenamefont {Wang},\ and\ \citenamefont {Wen}}]{Gu2014}%
  \BibitemOpen
  \bibfield  {author} {\bibinfo {author} {\bibfnamefont {Z.-C.}\ \bibnamefont
  {Gu}}, \bibinfo {author} {\bibfnamefont {Z.}~\bibnamefont {Wang}}, \ and\
  \bibinfo {author} {\bibfnamefont {X.-G.}\ \bibnamefont {Wen}},\ }\href
  {\doibase 10.1103/PhysRevB.90.085140} {\bibfield  {journal} {\bibinfo
  {journal} {Phys. Rev. B}\ }\textbf {\bibinfo {volume} {90}},\ \bibinfo
  {pages} {085140} (\bibinfo {year} {2014})}\BibitemShut {NoStop}%
\bibitem [{\citenamefont {Gu}\ \emph {et~al.}(2015)\citenamefont {Gu},
  \citenamefont {Wang},\ and\ \citenamefont {Wen}}]{Gu2015}%
  \BibitemOpen
  \bibfield  {author} {\bibinfo {author} {\bibfnamefont {Z.-C.}\ \bibnamefont
  {Gu}}, \bibinfo {author} {\bibfnamefont {Z.}~\bibnamefont {Wang}}, \ and\
  \bibinfo {author} {\bibfnamefont {X.-G.}\ \bibnamefont {Wen}},\ }\href
  {\doibase 10.1103/PhysRevB.91.125149} {\bibfield  {journal} {\bibinfo
  {journal} {Phys. Rev. B}\ }\textbf {\bibinfo {volume} {91}},\ \bibinfo
  {pages} {125149} (\bibinfo {year} {2015})}\BibitemShut {NoStop}%
\bibitem [{\citenamefont {Tarantino}\ and\ \citenamefont
  {Fidkowski}(2016)}]{Tarantino2016}%
  \BibitemOpen
  \bibfield  {author} {\bibinfo {author} {\bibfnamefont {N.}~\bibnamefont
  {Tarantino}}\ and\ \bibinfo {author} {\bibfnamefont {L.}~\bibnamefont
  {Fidkowski}},\ }\href {\doibase 10.1103/PhysRevB.94.115115} {\bibfield
  {journal} {\bibinfo  {journal} {Phys. Rev. B}\ }\textbf {\bibinfo {volume}
  {94}},\ \bibinfo {pages} {115115} (\bibinfo {year} {2016})}\BibitemShut
  {NoStop}%
\bibitem [{\citenamefont {Ware}\ \emph {et~al.}(2016)\citenamefont {Ware},
  \citenamefont {Son}, \citenamefont {Cheng}, \citenamefont {Mishmash},
  \citenamefont {Alicea},\ and\ \citenamefont {Bauer}}]{Ware2016}%
  \BibitemOpen
  \bibfield  {author} {\bibinfo {author} {\bibfnamefont {B.}~\bibnamefont
  {Ware}}, \bibinfo {author} {\bibfnamefont {J.~H.}\ \bibnamefont {Son}},
  \bibinfo {author} {\bibfnamefont {M.}~\bibnamefont {Cheng}}, \bibinfo
  {author} {\bibfnamefont {R.~V.}\ \bibnamefont {Mishmash}}, \bibinfo {author}
  {\bibfnamefont {J.}~\bibnamefont {Alicea}}, \ and\ \bibinfo {author}
  {\bibfnamefont {B.}~\bibnamefont {Bauer}},\ }\href {\doibase
  10.1103/PhysRevB.94.115127} {\bibfield  {journal} {\bibinfo  {journal} {Phys.
  Rev. B}\ }\textbf {\bibinfo {volume} {94}},\ \bibinfo {pages} {115127}
  (\bibinfo {year} {2016})}\BibitemShut {NoStop}%
\bibitem [{\citenamefont {Bhardwaj}\ \emph {et~al.}(2017)\citenamefont
  {Bhardwaj}, \citenamefont {Gaiotto},\ and\ \citenamefont
  {Kapustin}}]{Bhardwaj2017}%
  \BibitemOpen
  \bibfield  {author} {\bibinfo {author} {\bibfnamefont {L.}~\bibnamefont
  {Bhardwaj}}, \bibinfo {author} {\bibfnamefont {D.}~\bibnamefont {Gaiotto}}, \
  and\ \bibinfo {author} {\bibfnamefont {A.}~\bibnamefont {Kapustin}},\ }\href
  {\doibase 10.1007/JHEP04(2017)096} {\bibfield  {journal} {\bibinfo  {journal}
  {Journal of High Energy Physics}\ }\textbf {\bibinfo {volume} {2017}},\
  \bibinfo {pages} {96} (\bibinfo {year} {2017})}\BibitemShut {NoStop}%
\bibitem [{\citenamefont {Wang}\ \emph {et~al.}(2017)\citenamefont {Wang},
  \citenamefont {Ning},\ and\ \citenamefont {Chen}}]{Wang2017}%
  \BibitemOpen
  \bibfield  {author} {\bibinfo {author} {\bibfnamefont {Z.}~\bibnamefont
  {Wang}}, \bibinfo {author} {\bibfnamefont {S.-Q.}\ \bibnamefont {Ning}}, \
  and\ \bibinfo {author} {\bibfnamefont {X.}~\bibnamefont {Chen}},\ }\href
  {https://arxiv.org/abs/1708.01684} {\enquote {\bibinfo {title} {Exactly
  solvable model for two dimensional topological superconductor},}\ } (\bibinfo
  {year} {2017}),\ \bibinfo {note} {{arXiv:1708.01684}}\BibitemShut {NoStop}%
\bibitem [{\citenamefont {Aasen}\ \emph {et~al.}(2017)\citenamefont {Aasen},
  \citenamefont {Lake},\ and\ \citenamefont {Walker}}]{Aasen2017}%
  \BibitemOpen
  \bibfield  {author} {\bibinfo {author} {\bibfnamefont {D.}~\bibnamefont
  {Aasen}}, \bibinfo {author} {\bibfnamefont {E.}~\bibnamefont {Lake}}, \ and\
  \bibinfo {author} {\bibfnamefont {K.}~\bibnamefont {Walker}},\ }\href
  {https://arxiv.org/abs/1709.01941} {\enquote {\bibinfo {title} {Fermion
  condensation and super pivotal categories},}\ } (\bibinfo {year} {2017}),\
  \bibinfo {note} {{arXiv:1709.01941}}\BibitemShut {NoStop}%
\bibitem [{\citenamefont {Bais}\ and\ \citenamefont
  {Slingerland}(2009)}]{Bais2009}%
  \BibitemOpen
  \bibfield  {author} {\bibinfo {author} {\bibfnamefont {F.~A.}\ \bibnamefont
  {Bais}}\ and\ \bibinfo {author} {\bibfnamefont {J.~K.}\ \bibnamefont
  {Slingerland}},\ }\href {\doibase 10.1103/PhysRevB.79.045316} {\bibfield
  {journal} {\bibinfo  {journal} {Phys. Rev. B}\ }\textbf {\bibinfo {volume}
  {79}},\ \bibinfo {pages} {045316} (\bibinfo {year} {2009})}\BibitemShut
  {NoStop}%
\bibitem [{\citenamefont {Ludwig}\ \emph {et~al.}(2011)\citenamefont {Ludwig},
  \citenamefont {Poilblanc}, \citenamefont {Trebst},\ and\ \citenamefont
  {Troyer}}]{Ludwig2011}%
  \BibitemOpen
  \bibfield  {author} {\bibinfo {author} {\bibfnamefont {A.~W.~W.}\
  \bibnamefont {Ludwig}}, \bibinfo {author} {\bibfnamefont {D.}~\bibnamefont
  {Poilblanc}}, \bibinfo {author} {\bibfnamefont {S.}~\bibnamefont {Trebst}}, \
  and\ \bibinfo {author} {\bibfnamefont {M.}~\bibnamefont {Troyer}},\ }\href
  {http://stacks.iop.org/1367-2630/13/i=4/a=045014} {\bibfield  {journal}
  {\bibinfo  {journal} {New Journal of Physics}\ }\textbf {\bibinfo {volume}
  {13}},\ \bibinfo {pages} {045014} (\bibinfo {year} {2011})}\BibitemShut
  {NoStop}%
\bibitem [{\citenamefont {Poilblanc}\ \emph {et~al.}(2011)\citenamefont
  {Poilblanc}, \citenamefont {Ludwig}, \citenamefont {Trebst},\ and\
  \citenamefont {Troyer}}]{Poilblanc2011}%
  \BibitemOpen
  \bibfield  {author} {\bibinfo {author} {\bibfnamefont {D.}~\bibnamefont
  {Poilblanc}}, \bibinfo {author} {\bibfnamefont {A.~W.~W.}\ \bibnamefont
  {Ludwig}}, \bibinfo {author} {\bibfnamefont {S.}~\bibnamefont {Trebst}}, \
  and\ \bibinfo {author} {\bibfnamefont {M.}~\bibnamefont {Troyer}},\ }\href
  {\doibase 10.1103/PhysRevB.83.134439} {\bibfield  {journal} {\bibinfo
  {journal} {Phys. Rev. B}\ }\textbf {\bibinfo {volume} {83}},\ \bibinfo
  {pages} {134439} (\bibinfo {year} {2011})}\BibitemShut {NoStop}%
\bibitem [{\citenamefont {Burrello}\ \emph {et~al.}(2013)\citenamefont
  {Burrello}, \citenamefont {van Heck},\ and\ \citenamefont
  {Cobanera}}]{Burrello2013}%
  \BibitemOpen
  \bibfield  {author} {\bibinfo {author} {\bibfnamefont {M.}~\bibnamefont
  {Burrello}}, \bibinfo {author} {\bibfnamefont {B.}~\bibnamefont {van Heck}},
  \ and\ \bibinfo {author} {\bibfnamefont {E.}~\bibnamefont {Cobanera}},\
  }\href {\doibase 10.1103/PhysRevB.87.195422} {\bibfield  {journal} {\bibinfo
  {journal} {Phys. Rev. B}\ }\textbf {\bibinfo {volume} {87}},\ \bibinfo
  {pages} {195422} (\bibinfo {year} {2013})}\BibitemShut {NoStop}%
\bibitem [{\citenamefont {Mong}\ \emph {et~al.}(2014)\citenamefont {Mong},
  \citenamefont {Clarke}, \citenamefont {Alicea}, \citenamefont {Lindner},
  \citenamefont {Fendley}, \citenamefont {Nayak}, \citenamefont {Oreg},
  \citenamefont {Stern}, \citenamefont {Berg}, \citenamefont {Shtengel},\ and\
  \citenamefont {Fisher}}]{Mong2014}%
  \BibitemOpen
  \bibfield  {author} {\bibinfo {author} {\bibfnamefont {R.~S.~K.}\
  \bibnamefont {Mong}}, \bibinfo {author} {\bibfnamefont {D.~J.}\ \bibnamefont
  {Clarke}}, \bibinfo {author} {\bibfnamefont {J.}~\bibnamefont {Alicea}},
  \bibinfo {author} {\bibfnamefont {N.~H.}\ \bibnamefont {Lindner}}, \bibinfo
  {author} {\bibfnamefont {P.}~\bibnamefont {Fendley}}, \bibinfo {author}
  {\bibfnamefont {C.}~\bibnamefont {Nayak}}, \bibinfo {author} {\bibfnamefont
  {Y.}~\bibnamefont {Oreg}}, \bibinfo {author} {\bibfnamefont {A.}~\bibnamefont
  {Stern}}, \bibinfo {author} {\bibfnamefont {E.}~\bibnamefont {Berg}},
  \bibinfo {author} {\bibfnamefont {K.}~\bibnamefont {Shtengel}}, \ and\
  \bibinfo {author} {\bibfnamefont {M.~P.~A.}\ \bibnamefont {Fisher}},\ }\href
  {\doibase 10.1103/PhysRevX.4.011036} {\bibfield  {journal} {\bibinfo
  {journal} {Phys. Rev. X}\ }\textbf {\bibinfo {volume} {4}},\ \bibinfo {pages}
  {011036} (\bibinfo {year} {2014})}\BibitemShut {NoStop}%
\bibitem [{\citenamefont {Stoudenmire}\ \emph {et~al.}(2015)\citenamefont
  {Stoudenmire}, \citenamefont {Clarke}, \citenamefont {Mong},\ and\
  \citenamefont {Alicea}}]{Stoudenmire2015}%
  \BibitemOpen
  \bibfield  {author} {\bibinfo {author} {\bibfnamefont {E.~M.}\ \bibnamefont
  {Stoudenmire}}, \bibinfo {author} {\bibfnamefont {D.~J.}\ \bibnamefont
  {Clarke}}, \bibinfo {author} {\bibfnamefont {R.~S.~K.}\ \bibnamefont {Mong}},
  \ and\ \bibinfo {author} {\bibfnamefont {J.}~\bibnamefont {Alicea}},\ }\href
  {\doibase 10.1103/PhysRevB.91.235112} {\bibfield  {journal} {\bibinfo
  {journal} {Phys. Rev. B}\ }\textbf {\bibinfo {volume} {91}},\ \bibinfo
  {pages} {235112} (\bibinfo {year} {2015})}\BibitemShut {NoStop}%
\bibitem [{\citenamefont {Barkeshli}\ \emph {et~al.}(2015)\citenamefont
  {Barkeshli}, \citenamefont {Jiang}, \citenamefont {Thomale},\ and\
  \citenamefont {Qi}}]{Barkeshli2015}%
  \BibitemOpen
  \bibfield  {author} {\bibinfo {author} {\bibfnamefont {M.}~\bibnamefont
  {Barkeshli}}, \bibinfo {author} {\bibfnamefont {H.-C.}\ \bibnamefont
  {Jiang}}, \bibinfo {author} {\bibfnamefont {R.}~\bibnamefont {Thomale}}, \
  and\ \bibinfo {author} {\bibfnamefont {X.-L.}\ \bibnamefont {Qi}},\ }\href
  {\doibase 10.1103/PhysRevLett.114.026401} {\bibfield  {journal} {\bibinfo
  {journal} {Phys. Rev. Lett.}\ }\textbf {\bibinfo {volume} {114}},\ \bibinfo
  {pages} {026401} (\bibinfo {year} {2015})}\BibitemShut {NoStop}%
\bibitem [{\citenamefont {Kitaev}(2001)}]{Kitaev2001}%
  \BibitemOpen
  \bibfield  {author} {\bibinfo {author} {\bibfnamefont {A.~Y.}\ \bibnamefont
  {Kitaev}},\ }\href {\doibase 10.1070/1063-7869/44/10S/S29} {\bibfield
  {journal} {\bibinfo  {journal} {Physics-Uspekhi}\ }\textbf {\bibinfo {volume}
  {44}},\ \bibinfo {pages} {131} (\bibinfo {year} {2001})}\BibitemShut
  {NoStop}%
\bibitem [{\citenamefont {Gu}\ and\ \citenamefont {Levin}(2014)}]{Gu2014a}%
  \BibitemOpen
  \bibfield  {author} {\bibinfo {author} {\bibfnamefont {Z.-C.}\ \bibnamefont
  {Gu}}\ and\ \bibinfo {author} {\bibfnamefont {M.}~\bibnamefont {Levin}},\
  }\href {\doibase 10.1103/PhysRevB.89.201113} {\bibfield  {journal} {\bibinfo
  {journal} {Phys. Rev. B}\ }\textbf {\bibinfo {volume} {89}},\ \bibinfo
  {pages} {201113} (\bibinfo {year} {2014})}\BibitemShut {NoStop}%
\bibitem [{\citenamefont {Walker}(2015)}]{Walker_unpub}%
  \BibitemOpen
  \bibfield  {author} {\bibinfo {author} {\bibfnamefont {K.}~\bibnamefont
  {Walker}},\ }\href
  {http://canyon23.net/math/talks/IPAM%20201501b%20compressed.pdf} {\enquote
  {\bibinfo {title} {Codimension 1 defects, categorified group actions, and
  condensing fermions},}\ } (\bibinfo {year} {2015}),\ \bibinfo {note} {{IPAM}
  workshop on ``Symmetry and Topology in Quantum Matter''}\BibitemShut
  {NoStop}%
\bibitem [{\citenamefont {Read}\ and\ \citenamefont {Green}(2000)}]{Read2000}%
  \BibitemOpen
  \bibfield  {author} {\bibinfo {author} {\bibfnamefont {N.}~\bibnamefont
  {Read}}\ and\ \bibinfo {author} {\bibfnamefont {D.}~\bibnamefont {Green}},\
  }\href {\doibase 10.1103/PhysRevB.61.10267} {\bibfield  {journal} {\bibinfo
  {journal} {Phys. Rev. B}\ }\textbf {\bibinfo {volume} {61}},\ \bibinfo
  {pages} {10267} (\bibinfo {year} {2000})}\BibitemShut {NoStop}%
\bibitem [{\citenamefont {Fradkin}\ and\ \citenamefont
  {Kadanoff}(1980)}]{FradkinKadanoff}%
  \BibitemOpen
  \bibfield  {author} {\bibinfo {author} {\bibfnamefont {E.}~\bibnamefont
  {Fradkin}}\ and\ \bibinfo {author} {\bibfnamefont {L.~P.}\ \bibnamefont
  {Kadanoff}},\ }\href {\doibase 10.1016/0550-3213(80)90472-1} {\bibfield
  {journal} {\bibinfo  {journal} {Nucl. Phys. B}\ }\textbf {\bibinfo {volume}
  {170}},\ \bibinfo {pages} {1} (\bibinfo {year} {1980})}\BibitemShut {NoStop}%
\bibitem [{\citenamefont {Fendley}(2012)}]{Fendley2012}%
  \BibitemOpen
  \bibfield  {author} {\bibinfo {author} {\bibfnamefont {P.}~\bibnamefont
  {Fendley}},\ }\href {\doibase 10.1088/1742-5468/2012/11/P11020} {\bibfield
  {journal} {\bibinfo  {journal} {J. Stat. Mech.}\ }\textbf {\bibinfo {volume}
  {2012}},\ \bibinfo {pages} {11020} (\bibinfo {year} {2012})}\BibitemShut
  {NoStop}%
\bibitem [{\citenamefont {Alicea}\ and\ \citenamefont
  {Fendley}(2016)}]{AliceaFendleyReview}%
  \BibitemOpen
  \bibfield  {author} {\bibinfo {author} {\bibfnamefont {J.}~\bibnamefont
  {Alicea}}\ and\ \bibinfo {author} {\bibfnamefont {P.}~\bibnamefont
  {Fendley}},\ }\href {\doibase 10.1146/annurev-conmatphys-031115-011336}
  {\bibfield  {journal} {\bibinfo  {journal} {Annual Review of Condensed Matter
  Physics}\ }\textbf {\bibinfo {volume} {7}},\ \bibinfo {pages} {119} (\bibinfo
  {year} {2016})}\BibitemShut {NoStop}%
\bibitem [{\citenamefont {Barkeshli}\ and\ \citenamefont
  {Qi}(2012)}]{Barkeshli2012}%
  \BibitemOpen
  \bibfield  {author} {\bibinfo {author} {\bibfnamefont {M.}~\bibnamefont
  {Barkeshli}}\ and\ \bibinfo {author} {\bibfnamefont {X.-L.}\ \bibnamefont
  {Qi}},\ }\href {\doibase 10.1103/PhysRevX.2.031013} {\bibfield  {journal}
  {\bibinfo  {journal} {Phys. Rev. X}\ }\textbf {\bibinfo {volume} {2}},\
  \bibinfo {pages} {031013} (\bibinfo {year} {2012})}\BibitemShut {NoStop}%
\bibitem [{\citenamefont {Barkeshli}\ \emph
  {et~al.}(2013{\natexlab{a}})\citenamefont {Barkeshli}, \citenamefont {Jian},\
  and\ \citenamefont {Qi}}]{Barkeshli2013}%
  \BibitemOpen
  \bibfield  {author} {\bibinfo {author} {\bibfnamefont {M.}~\bibnamefont
  {Barkeshli}}, \bibinfo {author} {\bibfnamefont {C.-M.}\ \bibnamefont {Jian}},
  \ and\ \bibinfo {author} {\bibfnamefont {X.-L.}\ \bibnamefont {Qi}},\ }\href
  {\doibase 10.1103/PhysRevB.88.235103} {\bibfield  {journal} {\bibinfo
  {journal} {Phys. Rev. B}\ }\textbf {\bibinfo {volume} {88}},\ \bibinfo
  {pages} {235103} (\bibinfo {year} {2013}{\natexlab{a}})}\BibitemShut
  {NoStop}%
\bibitem [{\citenamefont {Barkeshli}\ \emph
  {et~al.}(2013{\natexlab{b}})\citenamefont {Barkeshli}, \citenamefont {Jian},\
  and\ \citenamefont {Qi}}]{Barkeshli2013a}%
  \BibitemOpen
  \bibfield  {author} {\bibinfo {author} {\bibfnamefont {M.}~\bibnamefont
  {Barkeshli}}, \bibinfo {author} {\bibfnamefont {C.-M.}\ \bibnamefont {Jian}},
  \ and\ \bibinfo {author} {\bibfnamefont {X.-L.}\ \bibnamefont {Qi}},\ }\href
  {\doibase 10.1103/PhysRevB.88.241103} {\bibfield  {journal} {\bibinfo
  {journal} {Phys. Rev. B}\ }\textbf {\bibinfo {volume} {88}},\ \bibinfo
  {pages} {241103(R)} (\bibinfo {year} {2013}{\natexlab{b}})}\BibitemShut
  {NoStop}%
\bibitem [{\citenamefont {Barkeshli}\ and\ \citenamefont
  {Qi}(2014)}]{Barkeshli:2014b}%
  \BibitemOpen
  \bibfield  {author} {\bibinfo {author} {\bibfnamefont {M.}~\bibnamefont
  {Barkeshli}}\ and\ \bibinfo {author} {\bibfnamefont {X.-L.}\ \bibnamefont
  {Qi}},\ }\href {\doibase 10.1103/PhysRevX.4.041035} {\bibfield  {journal}
  {\bibinfo  {journal} {Phys. Rev. X}\ }\textbf {\bibinfo {volume} {4}},\
  \bibinfo {pages} {041035} (\bibinfo {year} {2014})}\BibitemShut {NoStop}%
\bibitem [{\citenamefont {Lindner}\ \emph {et~al.}(2012)\citenamefont
  {Lindner}, \citenamefont {Berg}, \citenamefont {Refael},\ and\ \citenamefont
  {Stern}}]{Lindner2012}%
  \BibitemOpen
  \bibfield  {author} {\bibinfo {author} {\bibfnamefont {N.~H.}\ \bibnamefont
  {Lindner}}, \bibinfo {author} {\bibfnamefont {E.}~\bibnamefont {Berg}},
  \bibinfo {author} {\bibfnamefont {G.}~\bibnamefont {Refael}}, \ and\ \bibinfo
  {author} {\bibfnamefont {A.}~\bibnamefont {Stern}},\ }\href {\doibase
  10.1103/PhysRevX.2.041002} {\bibfield  {journal} {\bibinfo  {journal} {Phys.
  Rev. X}\ }\textbf {\bibinfo {volume} {2}},\ \bibinfo {pages} {041002}
  (\bibinfo {year} {2012})}\BibitemShut {NoStop}%
\bibitem [{\citenamefont {Clarke}\ \emph {et~al.}(2013)\citenamefont {Clarke},
  \citenamefont {Alicea},\ and\ \citenamefont {Shtengel}}]{Clarke2013}%
  \BibitemOpen
  \bibfield  {author} {\bibinfo {author} {\bibfnamefont {D.~J.}\ \bibnamefont
  {Clarke}}, \bibinfo {author} {\bibfnamefont {J.}~\bibnamefont {Alicea}}, \
  and\ \bibinfo {author} {\bibfnamefont {K.}~\bibnamefont {Shtengel}},\ }\href
  {\doibase 10.1038/ncomms2340} {\bibfield  {journal} {\bibinfo  {journal}
  {Nature Comm.}\ }\textbf {\bibinfo {volume} {4}},\ \bibinfo {pages} {1348}
  (\bibinfo {year} {2013})}\BibitemShut {NoStop}%
\bibitem [{\citenamefont {Cheng}(2012)}]{Cheng2012}%
  \BibitemOpen
  \bibfield  {author} {\bibinfo {author} {\bibfnamefont {M.}~\bibnamefont
  {Cheng}},\ }\href {\doibase 10.1103/PhysRevB.86.195126} {\bibfield  {journal}
  {\bibinfo  {journal} {Phys. Rev. B}\ }\textbf {\bibinfo {volume} {86}},\
  \bibinfo {pages} {195126} (\bibinfo {year} {2012})}\BibitemShut {NoStop}%
\bibitem [{\citenamefont {Vaezi}(2013)}]{Vaezi:2013}%
  \BibitemOpen
  \bibfield  {author} {\bibinfo {author} {\bibfnamefont {A.}~\bibnamefont
  {Vaezi}},\ }\href {\doibase 10.1103/PhysRevB.87.035132} {\bibfield  {journal}
  {\bibinfo  {journal} {Phys. Rev. B}\ }\textbf {\bibinfo {volume} {87}},\
  \bibinfo {pages} {035132} (\bibinfo {year} {2013})}\BibitemShut {NoStop}%
\bibitem [{\citenamefont {Maghrebi}\ \emph {et~al.}(2015)\citenamefont
  {Maghrebi}, \citenamefont {Ganeshan}, \citenamefont {Clarke}, \citenamefont
  {Gorshkov},\ and\ \citenamefont {Sau}}]{Maghrebi}%
  \BibitemOpen
  \bibfield  {author} {\bibinfo {author} {\bibfnamefont {M.~F.}\ \bibnamefont
  {Maghrebi}}, \bibinfo {author} {\bibfnamefont {S.}~\bibnamefont {Ganeshan}},
  \bibinfo {author} {\bibfnamefont {D.~J.}\ \bibnamefont {Clarke}}, \bibinfo
  {author} {\bibfnamefont {A.~V.}\ \bibnamefont {Gorshkov}}, \ and\ \bibinfo
  {author} {\bibfnamefont {J.~D.}\ \bibnamefont {Sau}},\ }\href {\doibase
  10.1103/PhysRevLett.115.065301} {\bibfield  {journal} {\bibinfo  {journal}
  {Phys. Rev. Lett.}\ }\textbf {\bibinfo {volume} {115}},\ \bibinfo {pages}
  {065301} (\bibinfo {year} {2015})}\BibitemShut {NoStop}%
\bibitem [{\citenamefont {Fern}\ \emph {et~al.}(2017)\citenamefont {Fern},
  \citenamefont {Kombe},\ and\ \citenamefont {Simon}}]{Fern}%
  \BibitemOpen
  \bibfield  {author} {\bibinfo {author} {\bibfnamefont {R.}~\bibnamefont
  {Fern}}, \bibinfo {author} {\bibfnamefont {J.}~\bibnamefont {Kombe}}, \ and\
  \bibinfo {author} {\bibfnamefont {S.~H.}\ \bibnamefont {Simon}},\ }\href
  {\doibase 10.21468/SciPostPhys.3.6.037} {\bibfield  {journal} {\bibinfo
  {journal} {SciPost Phys.}\ }\textbf {\bibinfo {volume} {3}},\ \bibinfo
  {pages} {037} (\bibinfo {year} {2017})}\BibitemShut {NoStop}%
\bibitem [{\citenamefont {Barkeshli}\ \emph {et~al.}(2014)\citenamefont
  {Barkeshli}, \citenamefont {Bonderson}, \citenamefont {Cheng},\ and\
  \citenamefont {Wang}}]{Barkeshli2014}%
  \BibitemOpen
  \bibfield  {author} {\bibinfo {author} {\bibfnamefont {M.}~\bibnamefont
  {Barkeshli}}, \bibinfo {author} {\bibfnamefont {P.}~\bibnamefont
  {Bonderson}}, \bibinfo {author} {\bibfnamefont {M.}~\bibnamefont {Cheng}}, \
  and\ \bibinfo {author} {\bibfnamefont {Z.}~\bibnamefont {Wang}},\ }\href
  {https://arxiv.org/abs/1410.4540} {\enquote {\bibinfo {title} {Symmetry,
  defects, and gauging of topological phases},}\ } (\bibinfo {year} {2014}),\
  \bibinfo {note} {{arXiv:1410.4540}}\BibitemShut {NoStop}%
\bibitem [{\citenamefont {Teo}\ \emph {et~al.}(2015)\citenamefont {Teo},
  \citenamefont {Hughes},\ and\ \citenamefont {Fradkin}}]{Teo2015}%
  \BibitemOpen
  \bibfield  {author} {\bibinfo {author} {\bibfnamefont {J.~C.}\ \bibnamefont
  {Teo}}, \bibinfo {author} {\bibfnamefont {T.~L.}\ \bibnamefont {Hughes}}, \
  and\ \bibinfo {author} {\bibfnamefont {E.}~\bibnamefont {Fradkin}},\ }\href
  {\doibase https://doi.org/10.1016/j.aop.2015.05.012} {\bibfield  {journal}
  {\bibinfo  {journal} {Annals of Physics}\ }\textbf {\bibinfo {volume}
  {360}},\ \bibinfo {pages} {349 } (\bibinfo {year} {2015})}\BibitemShut
  {NoStop}%
\bibitem [{\citenamefont {Tarantino}\ \emph {et~al.}(2016)\citenamefont
  {Tarantino}, \citenamefont {Lindner},\ and\ \citenamefont
  {Fidkowski}}]{Tarantino2016b}%
  \BibitemOpen
  \bibfield  {author} {\bibinfo {author} {\bibfnamefont {N.}~\bibnamefont
  {Tarantino}}, \bibinfo {author} {\bibfnamefont {N.~H.}\ \bibnamefont
  {Lindner}}, \ and\ \bibinfo {author} {\bibfnamefont {L.}~\bibnamefont
  {Fidkowski}},\ }\href {http://stacks.iop.org/1367-2630/18/i=3/a=035006}
  {\bibfield  {journal} {\bibinfo  {journal} {New Journal of Physics}\ }\textbf
  {\bibinfo {volume} {18}},\ \bibinfo {pages} {035006} (\bibinfo {year}
  {2016})}\BibitemShut {NoStop}%
\bibitem [{Note1()}]{Note1}%
  \BibitemOpen
  \bibinfo {note} {Notice that in Fig.~\ref {fig:pftorus2_rev}(b) the branch
  cut slices through a tunneling-gapped region. We choose this convention
  throughout so that the cut influences $F_{ij}$ but never $e^{i \pi Q^+_j}$
  operators.}\BibitemShut {Stop}%
\bibitem [{\citenamefont {Cimasoni}\ and\ \citenamefont
  {Reshetikhin}(2007)}]{Cimasoni2007}%
  \BibitemOpen
  \bibfield  {author} {\bibinfo {author} {\bibfnamefont {D.}~\bibnamefont
  {Cimasoni}}\ and\ \bibinfo {author} {\bibfnamefont {N.}~\bibnamefont
  {Reshetikhin}},\ }\href {\doibase 10.1007/s00220-007-0302-7} {\bibfield
  {journal} {\bibinfo  {journal} {Communications in Mathematical Physics}\
  }\textbf {\bibinfo {volume} {275}},\ \bibinfo {pages} {187} (\bibinfo {year}
  {2007})}\BibitemShut {NoStop}%
\bibitem [{\citenamefont {Levin}(2013)}]{Levin2013}%
  \BibitemOpen
  \bibfield  {author} {\bibinfo {author} {\bibfnamefont {M.}~\bibnamefont
  {Levin}},\ }\href {\doibase 10.1103/PhysRevX.3.021009} {\bibfield  {journal}
  {\bibinfo  {journal} {Phys. Rev. X}\ }\textbf {\bibinfo {volume} {3}},\
  \bibinfo {pages} {021009} (\bibinfo {year} {2013})}\BibitemShut {NoStop}%
\bibitem [{\citenamefont {Lecheminant}\ \emph {et~al.}(2002)\citenamefont
  {Lecheminant}, \citenamefont {Gogolin},\ and\ \citenamefont
  {Nersesyan}}]{LGN}%
  \BibitemOpen
  \bibfield  {author} {\bibinfo {author} {\bibfnamefont {P.}~\bibnamefont
  {Lecheminant}}, \bibinfo {author} {\bibfnamefont {A.~O.}\ \bibnamefont
  {Gogolin}}, \ and\ \bibinfo {author} {\bibfnamefont {A.~A.}\ \bibnamefont
  {Nersesyan}},\ }\href {\doibase
  https://doi.org/10.1016/S0550-3213(02)00474-1} {\bibfield  {journal}
  {\bibinfo  {journal} {Nuclear Physics B}\ }\textbf {\bibinfo {volume}
  {639}},\ \bibinfo {pages} {502 } (\bibinfo {year} {2002})}\BibitemShut
  {NoStop}%
\bibitem [{\citenamefont {Grover}\ \emph {et~al.}(2014)\citenamefont {Grover},
  \citenamefont {Sheng},\ and\ \citenamefont {Vishwanath}}]{Grover}%
  \BibitemOpen
  \bibfield  {author} {\bibinfo {author} {\bibfnamefont {T.}~\bibnamefont
  {Grover}}, \bibinfo {author} {\bibfnamefont {D.~N.}\ \bibnamefont {Sheng}}, \
  and\ \bibinfo {author} {\bibfnamefont {A.}~\bibnamefont {Vishwanath}},\
  }\href {\doibase 10.1126/science.1248253} {\bibfield  {journal} {\bibinfo
  {journal} {Science}\ }\textbf {\bibinfo {volume} {344}},\ \bibinfo {pages}
  {280} (\bibinfo {year} {2014})}\BibitemShut {NoStop}%
\bibitem [{\citenamefont {Lu}\ and\ \citenamefont {Vishwanath}(2016)}]{Lu2016}%
  \BibitemOpen
  \bibfield  {author} {\bibinfo {author} {\bibfnamefont {Y.-M.}\ \bibnamefont
  {Lu}}\ and\ \bibinfo {author} {\bibfnamefont {A.}~\bibnamefont
  {Vishwanath}},\ }\href {\doibase 10.1103/PhysRevB.93.155121} {\bibfield
  {journal} {\bibinfo  {journal} {Phys. Rev. B}\ }\textbf {\bibinfo {volume}
  {93}},\ \bibinfo {pages} {155121} (\bibinfo {year} {2016})}\BibitemShut
  {NoStop}%
\bibitem [{Note2()}]{Note2}%
  \BibitemOpen
  \bibinfo {note} {We thank Meng Cheng for pointing this out.}\BibitemShut
  {Stop}%
\bibitem [{\citenamefont {Kitaev}\ and\ \citenamefont
  {Preskill}(2006)}]{Kitaev2006b}%
  \BibitemOpen
  \bibfield  {author} {\bibinfo {author} {\bibfnamefont {A.}~\bibnamefont
  {Kitaev}}\ and\ \bibinfo {author} {\bibfnamefont {J.}~\bibnamefont
  {Preskill}},\ }\href {\doibase 10.1103/PhysRevLett.96.110404} {\bibfield
  {journal} {\bibinfo  {journal} {Phys. Rev. Lett.}\ }\textbf {\bibinfo
  {volume} {96}},\ \bibinfo {pages} {110404} (\bibinfo {year}
  {2006})}\BibitemShut {NoStop}%
\bibitem [{\citenamefont {Levin}\ and\ \citenamefont {Wen}(2006)}]{Levin2006b}%
  \BibitemOpen
  \bibfield  {author} {\bibinfo {author} {\bibfnamefont {M.}~\bibnamefont
  {Levin}}\ and\ \bibinfo {author} {\bibfnamefont {X.-G.}\ \bibnamefont
  {Wen}},\ }\href {\doibase 10.1103/PhysRevLett.96.110405} {\bibfield
  {journal} {\bibinfo  {journal} {Phys. Rev. Lett.}\ }\textbf {\bibinfo
  {volume} {96}},\ \bibinfo {pages} {110405} (\bibinfo {year}
  {2006})}\BibitemShut {NoStop}%
\bibitem [{\citenamefont {Zhang}\ \emph {et~al.}(2012)\citenamefont {Zhang},
  \citenamefont {Grover}, \citenamefont {Turner}, \citenamefont {Oshikawa},\
  and\ \citenamefont {Vishwanath}}]{Zhang2012}%
  \BibitemOpen
  \bibfield  {author} {\bibinfo {author} {\bibfnamefont {Y.}~\bibnamefont
  {Zhang}}, \bibinfo {author} {\bibfnamefont {T.}~\bibnamefont {Grover}},
  \bibinfo {author} {\bibfnamefont {A.}~\bibnamefont {Turner}}, \bibinfo
  {author} {\bibfnamefont {M.}~\bibnamefont {Oshikawa}}, \ and\ \bibinfo
  {author} {\bibfnamefont {A.}~\bibnamefont {Vishwanath}},\ }\href {\doibase
  10.1103/PhysRevB.85.235151} {\bibfield  {journal} {\bibinfo  {journal} {Phys.
  Rev. B}\ }\textbf {\bibinfo {volume} {85}},\ \bibinfo {pages} {235151}
  (\bibinfo {year} {2012})}\BibitemShut {NoStop}%
\bibitem [{\citenamefont {Cincio}\ and\ \citenamefont
  {Vidal}(2013)}]{Cincio2013}%
  \BibitemOpen
  \bibfield  {author} {\bibinfo {author} {\bibfnamefont {L.}~\bibnamefont
  {Cincio}}\ and\ \bibinfo {author} {\bibfnamefont {G.}~\bibnamefont {Vidal}},\
  }\href {\doibase 10.1103/PhysRevLett.110.067208} {\bibfield  {journal}
  {\bibinfo  {journal} {Phys. Rev. Lett.}\ }\textbf {\bibinfo {volume} {110}},\
  \bibinfo {pages} {067208} (\bibinfo {year} {2013})}\BibitemShut {NoStop}%
\bibitem [{\citenamefont {Zaletel}\ \emph {et~al.}(2013)\citenamefont
  {Zaletel}, \citenamefont {Mong},\ and\ \citenamefont
  {Pollmann}}]{Zalatel2013}%
  \BibitemOpen
  \bibfield  {author} {\bibinfo {author} {\bibfnamefont {M.~P.}\ \bibnamefont
  {Zaletel}}, \bibinfo {author} {\bibfnamefont {R.~S.~K.}\ \bibnamefont
  {Mong}}, \ and\ \bibinfo {author} {\bibfnamefont {F.}~\bibnamefont
  {Pollmann}},\ }\href {\doibase 10.1103/PhysRevLett.110.236801} {\bibfield
  {journal} {\bibinfo  {journal} {Phys. Rev. Lett.}\ }\textbf {\bibinfo
  {volume} {110}},\ \bibinfo {pages} {236801} (\bibinfo {year}
  {2013})}\BibitemShut {NoStop}%
\bibitem [{\citenamefont {Buerschaper}\ \emph {et~al.}(2009)\citenamefont
  {Buerschaper}, \citenamefont {Aguado},\ and\ \citenamefont
  {Vidal}}]{Buerschaper2009}%
  \BibitemOpen
  \bibfield  {author} {\bibinfo {author} {\bibfnamefont {O.}~\bibnamefont
  {Buerschaper}}, \bibinfo {author} {\bibfnamefont {M.}~\bibnamefont {Aguado}},
  \ and\ \bibinfo {author} {\bibfnamefont {G.}~\bibnamefont {Vidal}},\ }\href
  {\doibase 10.1103/PhysRevB.79.085119} {\bibfield  {journal} {\bibinfo
  {journal} {Phys. Rev. B}\ }\textbf {\bibinfo {volume} {79}},\ \bibinfo
  {pages} {085119} (\bibinfo {year} {2009})}\BibitemShut {NoStop}%
\bibitem [{\citenamefont {Williamson}\ \emph {et~al.}(2017)\citenamefont
  {Williamson}, \citenamefont {Bultinck},\ and\ \citenamefont
  {Verstraete}}]{Williamson2017}%
  \BibitemOpen
  \bibfield  {author} {\bibinfo {author} {\bibfnamefont {D.~J.}\ \bibnamefont
  {Williamson}}, \bibinfo {author} {\bibfnamefont {N.}~\bibnamefont
  {Bultinck}}, \ and\ \bibinfo {author} {\bibfnamefont {F.}~\bibnamefont
  {Verstraete}},\ }\href {https://arxiv.org/abs/1711.07982} {\enquote {\bibinfo
  {title} {Symmetry-enriched topological order in tensor networks: Defects,
  gauging and anyon condensation},}\ } (\bibinfo {year} {2017}),\ \bibinfo
  {note} {{arXiv:1711.07982}}\BibitemShut {NoStop}%
\bibitem [{\citenamefont {Bultinck}\ \emph {et~al.}(2018)\citenamefont
  {Bultinck}, \citenamefont {Williamson}, \citenamefont {Haegeman},\ and\
  \citenamefont {Verstraete}}]{Bultinck2018}%
  \BibitemOpen
  \bibfield  {author} {\bibinfo {author} {\bibfnamefont {N.}~\bibnamefont
  {Bultinck}}, \bibinfo {author} {\bibfnamefont {D.~J.}\ \bibnamefont
  {Williamson}}, \bibinfo {author} {\bibfnamefont {J.}~\bibnamefont
  {Haegeman}}, \ and\ \bibinfo {author} {\bibfnamefont {F.}~\bibnamefont
  {Verstraete}},\ }\href {http://stacks.iop.org/1751-8121/51/i=2/a=025202}
  {\bibfield  {journal} {\bibinfo  {journal} {Journal of Physics A:
  Mathematical and Theoretical}\ }\textbf {\bibinfo {volume} {51}},\ \bibinfo
  {pages} {025202} (\bibinfo {year} {2018})}\BibitemShut {NoStop}%
\bibitem [{\citenamefont {Xu}\ and\ \citenamefont {Zhang}(2017)}]{pfMPS}%
  \BibitemOpen
  \bibfield  {author} {\bibinfo {author} {\bibfnamefont {W.-T.}\ \bibnamefont
  {Xu}}\ and\ \bibinfo {author} {\bibfnamefont {G.-M.}\ \bibnamefont {Zhang}},\
  }\href {\doibase 10.1103/PhysRevB.95.195122} {\bibfield  {journal} {\bibinfo
  {journal} {Phys. Rev. B}\ }\textbf {\bibinfo {volume} {95}},\ \bibinfo
  {pages} {195122} (\bibinfo {year} {2017})}\BibitemShut {NoStop}%
\end{thebibliography}%

\end{document}